\documentclass[12pt]{article}

\usepackage{graphicx} 
\usepackage{amsmath}  
\usepackage{commath} 
\usepackage{amsfonts} 
\usepackage{mathtools} 
\usepackage[utf8]{inputenc} 
\usepackage{booktabs} 
\usepackage{multirow} 
\usepackage{gensymb} 
\usepackage{setspace}
\usepackage{wrapfig, lipsum, booktabs} 
\usepackage{subfig}
\usepackage{subcaption}
\usepackage{lscape} 
\usepackage{rotating}
\usepackage{epstopdf}
\usepackage{dsfont}

\usepackage[dvipsnames]{xcolor}

\usepackage[parfill]{parskip}

\usepackage[page,toc,titletoc,title]{appendix}

\usepackage[margin = 2.5cm ,vmarginratio={1:1},heightrounded]{geometry}

\usepackage{amsthm} 
\usepackage[affil-it]{authblk} 
\usepackage{amsopn} 
\usepackage{enumitem}
\usepackage{amssymb} 
\usepackage{bbm} 
\usepackage{mathrsfs} 
\usepackage{booktabs} 
\usepackage[super, numbers, round, comma, sort & compress]{natbib} 
\usepackage{multibib}

\usepackage{tikz} 
\usepackage{algorithm} 
\usepackage{algorithmic}
\usepackage{float} 
\usepackage{caption} 
\usepackage{abstract} 
\usepackage{listings} 
\usepackage{color} 
\usepackage{soul} 
\usepackage{lscape} 

\usepackage[pagebackref]{hyperref} 

\newfloat{algorithm}{t}{lop}

\numberwithin{equation}{section}

\setcitestyle{authoryear}
\bibpunct{(}{)}{;}{a}{,}{,}

\newcommand{\code}[1]{\texttt{#1}} 
\DeclareMathOperator*{\argmax}{argmax} 
\newcommand{\bas}[2][]{\boldsymbol{#2}^{#1}_{\mid i}} 
\newcommand{\hbas}[2][]{\hat{\boldsymbol{#2}}^{#1}_{\mid i}} 
\newcommand{\Thetai}[1][]{\bas[#1]{\Theta}}
\newcommand{\Thetaihat}[1][]{\hbas[#1]{\Theta}}
\newcommand{\thetai}{\bas{\theta}}

\newcommand{\Sigmai}{\Sigma_{\mid i}}
\newcommand{\Gammai}{\Gamma_{\mid i}}

\newlist{todolist}{itemize}{2}
\setlist[todolist]{label=$\square$}
\usepackage{pifont}
\theoremstyle{definition}

\theoremstyle{remark}

\newcites{main}{References}
\newcites{supp}{Supplementary References}

\title{Conditional Extremes with Graphical Models}

\usepackage{xpatch}
\xpatchcmd{\author}{\relax#1\relax}{\relax\detokenize{#1}\relax}{}{}
\author[\empty]{Aiden Farrell\textsuperscript{1,}\thanks{\href{mailto:a.farrell1@lancaster.ac.uk}{a.farrell1@lancaster.ac.uk}}} 
\author[\empty]{Emma F. Eastoe\textsuperscript{1,}\thanks{\href{mailto:e.eastoe@lancaster.ac.uk}{e.eastoe@lancaster.ac.uk}}} 
\author[\empty]{Clement Lee\textsuperscript{2,}\thanks{\href{mailto:clement.lee@newcastle.ac.uk}{clement.lee@newcastle.ac.uk}}} 
\affil[1]{School of Mathematical Sciences, Fylde College, Lancaster University, UK}
\affil[2]{School of Mathematics, Statistics and Physics, Herschel Building, Newcastle University, UK}
\date{}

\begin{document}

\maketitle

\begin{abstract}
    Multivariate extreme value analysis quantifies the probability and magnitude of joint extreme events. Classical multivariate models, such as max-stable or multivariate generalised Pareto distributions, generally have a high computational cost of fitting, which limits their application. To overcome this, models based on the asymptotically dependent multivariate Pareto distribution have recently incorporated graphical models to induce sparsity and reduce the dimension of the parameter space. While this approach is computationally efficient, the assumption of asymptotic dependence is inappropriate for many applications. The conditional multivariate extreme value model (CMEVM) is a popular model for which the asymptotic dependence assumption is not required. Unfortunately, inference for this model is semi-parametric, and consequently, it has poor predictive performance in high dimensions. An extension of the CMEVM that allows both the incorporation and selection of sparse dependence structures, and fully parametric prediction is proposed. The approach fills a current gap in statistical methodology by extending graphical models to asymptotically independent multivariate extreme value models. To support inference in high dimensions, a stepwise inference procedure that is computationally efficient and loses no information or predictive power is proposed. Simulation studies show the model is highly flexible, and an application to discharges in the upper Danube River basin provides promising results.
\end{abstract}

Keywords: Extremal dependence, graphical extremes, conditional multivariate extremes, sparsity, river networks

\section{Introduction}
    \label{Section:Introduction}
    The development of statistical models to describe and predict multivariate extreme events is a crucial part of natural hazard risk assessment, especially for data arising from spatial, temporal, and spatio-temporal processes. For example, multivariate extreme value models have been adopted to predict risk from extreme snowfall \citepmain{blanchet_2011}, sea surface temperature \citepmain{Simpson_2021}, droughts \citepmain{oesting_2018}, river flows \citepmain{Keef_2013, Asadi_2015}, forest fires \citepmain{stephenson_2015}, precipitation \citepmain{westra_2011}, wind-speed \citepmain{engelke_2015}, and ocean storms \citepmain{shooter_2019}, all of which exhibit complex behaviour, which can only be effectively captured by a flexible, multi-parameter model. 


Central to multivariate extreme value modelling is the concept of extremal dependence. Let $V = \{1, \hdots, d\}$ and consider the $d$-dimensional absolutely continuous random vector $\boldsymbol{X}= \{X_{j} : j \in V\}$ with joint and marginal distributions $F_{\boldsymbol{X}}(\boldsymbol{x})= \mathbb{P}[\boldsymbol{X} \leq \boldsymbol{x}]$ and $F_{j}(x_j) = \mathbb{P}[X_j \leq x_j]$, respectively. If the quantity $\chi_A := \lim_{u \rightarrow 1} \mathbb{P}[F_{i}(X_i) > u :i\in A]/(1-u)$ for $A \subseteq V$ and $\lvert A \rvert \geq 2$ is strictly positive, then the components in $A$ are likely to experience their extremes simultaneously and are said to show asymptotic dependence (AD) \citepmain{simpson_2020}. If $\chi_A=0$ then the variables in $A$ cannot be simultaneously extreme; specifically for $|A|=2$, if $\chi_A=0$ then the two variables show asymptotic independence (AI) \citepmain{Ledford_1996}. Full AD occurs when $\chi_A>0$ for all subsets $A \subseteq V$, and full AI occurs when $\chi_{A}=0$ for all two-dimensional subsets of $V$. While independence implies AI, the converse does not hold. 
 

The strength of association between components with AI  can be quantified using the coefficient of tail dependence $\eta$ \citepmain{Ledford_1996}. The coefficient arises from a first-order approximation for the joint survivor function of $(X_i,X_j)$ for $i, j\in A$, $i \neq j$. For large $x$ the approximation is
\begin{equation}
    \mathbb{P}[F^{-1}_{F}(F_{X_{i}}(X_{i})) > x, F^{-1}_{F}(F_{X_{j}}(X_{j})) > x] \sim \mathcal{L}(x) \mathbb{P}[F^{-1}_{F}(F_{X_{i}}(X_{i})) > x]^{-\frac{1}{\eta}},
    \label{eqn:eta}
\end{equation}
where $F^{-1}_{F}(\cdot)$ is the inverse of the standard Fr\'echet distribution function and $\mathcal{L}(\cdot)$ is a slowly varying function. For pairs that exhibit AD, $\eta = 1$ and $\mathcal{L}(x) \nrightarrow 0$ as $x \to \infty$. Otherwise $X_i$ and $X_j$ are either negatively ($0<\eta<0.5)$ or positively ($0.5<\eta<1$) associated, or exactly independent ($\eta=0.5$). Estimates of $\eta$ are usually obtained over a range of finite thresholds, and the limit behaviour of $\eta(u)$, where $u$ is the $u$-th quantile of the standard Fr\'echet distribution, is used to determine the likely extremal dependence class.

\begin{figure}[t!]
    \centering
    \includegraphics[width=.48\textwidth]{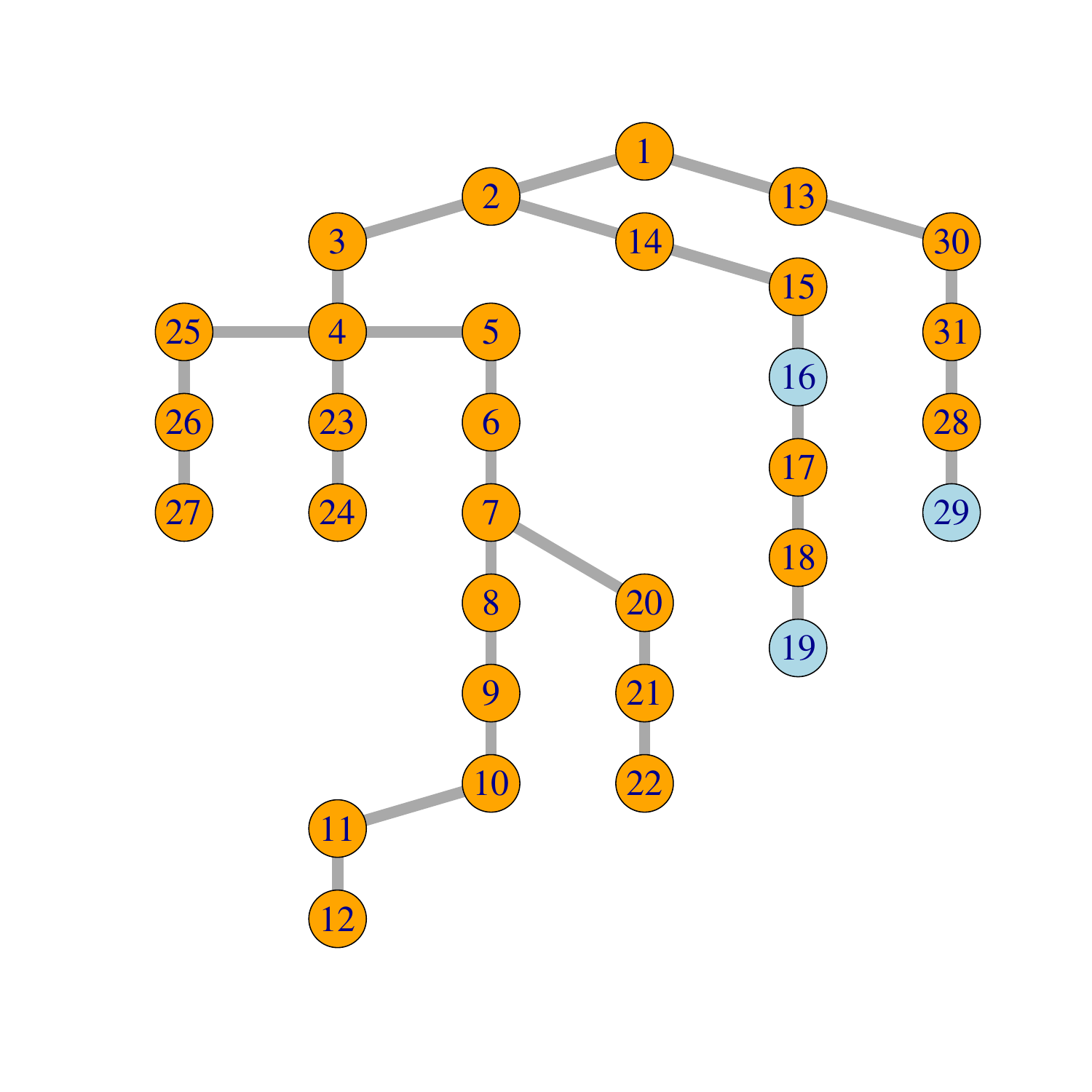} \quad
    \includegraphics[width=.48\textwidth]{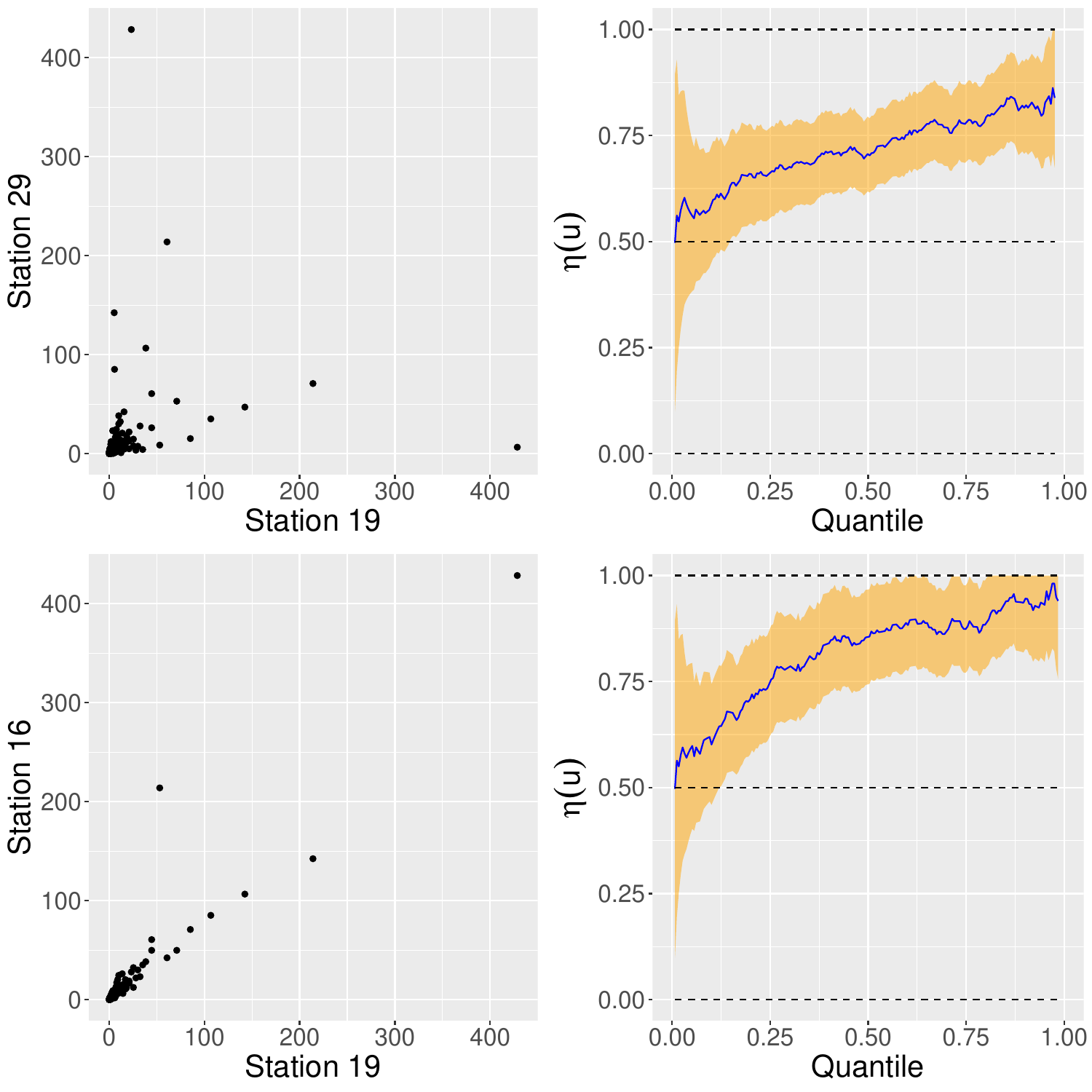}
    \caption{Undirected tree induced by the flow connections of the upper Danube River basin (left) with sites 16, 19 and 29 in blue. Scatter plots on standard Fr\'echet margins (centre) and empirical estimates of $\eta(u)$ (right) for $u \in (0,1]$ for sites 19 and 29 (top) and 16 and 29 (right). }
    \label{fig:Danube_River_EDA}
\end{figure}

We illustrate these concepts using river discharges from the upper Danube River basin. Daily discharge data at $d = 31$ gauging stations for 1960-2009 is available from the Bavarian Environmental Agency (\href{http://www.gkd.bayern.de}{http://www.gkd.bayern.de}). Figure \ref{fig:Danube_River_EDA} (left panel) shows the undirected tree implied by the flow connections of the river basin. We use the summer-only, temporally declustered dataset (see \citetmain{Asadi_2015} for details), available from the $\code{graphicalExtremes}$ package \citepmain{Engelke_2024} in $\code{R}$ \citepmain{R}. The data have previously been analysed using a max-stable Brown-Resnick process \citepmain{Asadi_2015} and a multivariate Pareto graphical model \citepmain{Engelke_2020}, both of which assume full AD. Figure \ref{fig:Danube_River_EDA} (right panel) shows scatter plots on standard Fr\'echet margins, and empirical estimates for the extremal dependence measure $\eta(u)$ \citepmain{Ledford_1996} for stations 19 and 29 (top), and 19 and 16 (bottom). Sites 19 and 29 lie on different tributaries and are flow-unconnected, while 19 and 16 both lie on the Isar tributary and are flow-connected. Stations 19 and 16 appear to exhibit AD: they experience extreme events simultaneously and $\eta(u) \rightarrow 1$ as $u \rightarrow 1$. Conversely, the scatter plots for stations 19 and 29 suggest AI, which is supported by the fact that $\eta(u) \nrightarrow 1$ as $u \rightarrow 1$. Thus, while flow-connected stations often exhibit AD, some flow-unconnected stations do not and assuming AD for such stations would lead to overestimation of their joint tail behaviour. 

Determining the class of extremal dependence is important, since not all multivariate extreme value models permit both AD and AI. Many popular models are based on max-stable distributions and processes, which are asymptotically dependent. Examples include the multivariate Pareto distribution \citepmain{rootzen_2006,rootzen_2018} and the max-stable \citepmain{smith_1990} and generalised $r$-Pareto \citepmain{ferreira_2014} processes. 
The appeal of these comes from their connection to asymptotic limit theory, and is one of the reasons why the existing graphical modelling approach for extremes \citepmain{Engelke_2020} is built on the H\"{u}sler-Reiss distribution, which belongs to the class of multivariate Pareto distributions. While the original presentation in \citetmain{Engelke_2020} allows only block graphs with cliques of size at most three, numerous subsequent extensions have been proposed \citepmain{Engelke_2024_B}. For example, \citetmain{Engelke_2025} allow for \emph{any} (sparse) graphical structure for the dependence structure, while \citetmain{Rottger_2023} introduce coloured graphs which permit symmetries into the variogram matrix. However, the underlying distributional assumption makes these models unsuitable for data which exhibit AI. 


Models for AI, such as the one proposed by \citetmain{ledford_1997}, were initially limited to the bivariate case only. While \citetmain{ramos_2011} and \citetmain{wadsworth_2013} developed extensions of this model that allowed for inference away from the diagonal, the semi-parametric conditional multivariate extreme value model (CMEVM) was the first model to provide a credible approach to data exhibiting both AD and AI \citepmain{Heffernan_2004}. The CMEVM is not based on a multivariate distribution or process; rather, conditional on one variable being large, normalising functions are defined to control the rate of growth of all other variables such that, after normalisation, the joint distribution of these ``residuals" is non-degenerate. The model has gained popularity due to the relative ease with which its parameters can be estimated and interpreted, even in high dimensions. Applications include flood risk mapping \citepmain{neal_2013,towe_2019}, and the prediction of extreme sea states \citepmain{gouldby_2014,gouldby_2017,ross_2020}, sea surface temperatures \citepmain{Simpson_2021}, heatwaves \citepmain{winter_2016}, and precipitation \citepmain{debusho_2021,Richards_2022}.

One issue with the CMEVM is that its predictive performance declines with increasing dimensionality (see next section and the Supplementary Material). The spatial CMEVM \citepmain{Wadsworth_2022, Shooter_2021, Richards_2022} overcomes this by using a fully parametric spatial kernel for the residual distribution. By construction, this model is appropriate for measurements on a continuous spatial surface but not for measurements on topographies such as road or river networks, which are represented by a graph (Figure~\ref{fig:Danube_River_EDA}, left panel). To the best of our knowledge, the only work in this area is that of \citetmain{Papastathopoulos_2017} and \citetmain{Casey_2023}, who prove numerous important theoretical results for the CMEVM when processes are observed on decomposable graphs. While the authors develop an impressive theoretical framework, developing statistical methodology was beyond the scope of their contribution.

Our contribution fills this gap. We use the multivariate asymmetric generalised Gaussian (MVAGG) to model the CMEVM residuals, increasing accuracy for full parametric prediction. We also incorporate structure into the CMEVM residuals, thereby providing a framework for (sparse) graphical structures that accommodates both extremal dependence classes. The second of these contributions builds on ideas presented by Jennifer Wadsworth in the discussion of \citetmain{Engelke_2020} and generalises the temporal Markov CMEVM of \citetmain{Winter_2017}. Finally, inference in high dimensions is achieved using stepwise optimisation that is computationally efficient without loss of information or predictive performance.

The remainder of this paper is structured as follows. Section \ref{Section:Methods} provides an overview of the CMEVM, introduces the MVAGG distribution, and describes the proposed structured CMEVM. Methods for model inference, graphical selection, and model-based predictions are provided in Section \ref{Section:Inference}. In Section \ref{Section:Simulation_Study}, we illustrate the performance of our model, the graphical selection procedure, and the utility of the stepwise inference procedures. We then apply our model to discharges in the upper Danube River basin \citepmain{Asadi_2015} in Section \ref{Section:Application} and compare it to the one proposed by \citetmain{Engelke_2020}. Finally, we outline directions for future research in Section \ref{Section:Discussion}.

\section{Methodology} 
    \label{Section:Methods}
    We review the CMEVM, introduce the MVAGG distribution, and describe a new variant of the CMEVM that incorporates sparsity into the residual distribution. Throughout, we use $V_{\mid i} := V\backslash\{i\}$ and $\boldsymbol{X}_{\mid i}:=\{X_j:j \in V_{\mid i}\}$ to refer to the set $V$ and vector $\boldsymbol{X}$ excluding their $i$th elements. Standard hat notation is used to denote parameter estimates.

\subsection{Conditional Multivariate Extreme Value Model}
\label{Sect:CMEVM}

Multivariate extreme value models are usually defined on either max-stable or heavy-tailed univariate margins. 
The use of specific margins is not restrictive since Sklar's Theorem \citepmain{Sklar_1959} allows one to transform the univariate margins of a random vector without altering the dependence structure. In what follows, $\boldsymbol{Y}$ and $\boldsymbol{Y}_{\mid i}$ denote the random vectors $\boldsymbol{X}$ and $\boldsymbol{X}_{\mid i}$ following transformation to Laplace margins. The CMEVM is based on a limiting representation of $\boldsymbol{Y}$ given a pre-selected conditioning component $Y_i$ is extreme. The key innovation is the use of normalising functions which depend on the conditioning component. Specifically for each $i\in V$, suppose that there exist functions $\{a_{j \mid i} : \mathbb{R} \rightarrow \mathbb{R}, j \in V_{\mid i}\}$ and $\{b_{j \mid i} : \mathbb{R} \rightarrow \mathbb{R}_{+}, j \in V_{\mid i}\}$ such that as $u_{Y_{i}} \rightarrow \infty$, 
\begin{equation}
    \left( \left\{ \frac{ Y_{j} - a_{j \mid i}(Y_{i}) }{ b_{j \mid i}(Y_{i}) } \right\}_{j \in V_{\mid i}}, \; Y_{i} - u_{Y_{i}} \right) \bigg\vert \; Y_{i} > u_{Y_{i}} \overset{d}{\rightarrow} \left( \{Z_{j \mid i} : j \in V_{\mid i}\}, \; E \right),
    \label{eqn:WT_2022_Limit}
\end{equation}
then the residual vector $\boldsymbol{Z}_{\mid i}=\{Z_{j \mid i} : j \in V_{\mid i}\}$ is has a non-degenerate distribution and is independent of the excesses of the conditioning component $Y_i$, which follow a standard exponential distribution $E$ in the limit \citepmain{Heffernan_2007}. 

To fit the CMEVM, limit~\eqref{eqn:WT_2022_Limit} is assumed to hold exactly above a high but finite threshold $u_{Y_{i}}$. There are currently no methods for threshold selection. Therefore, sensitivity checks should be undertaken to ensure the threshold is low enough that the resulting estimates are reliable, but not so low that limit~\eqref{eqn:WT_2022_Limit} does not hold. Once the threshold is selected, inference is undertaken separately on: (i) $Y_i - u_{Y_{i}} \mid Y_i > u_{Y_{i}}$; (ii) the normalising functions; and (iii) the residuals $\boldsymbol{Z}_{\mid i}$. The first is trivial, since $Y_i - u_{Y_{i}} \mid Y_i > u_{Y_{i}}$ is standard exponential by limit~(\ref{eqn:WT_2022_Limit}). Parts (ii) and (iii) require greater consideration. 

\citetmain{Heffernan_2004} model the normalising functions as
\begin{align*}
a_{j \mid i}(y_i)=\alpha_{j \mid i} y_i,~~~b_{j \mid i}(y_i)= y_i ^{\beta_{j \mid i}},
\end{align*}
where for Laplace margins, $\alpha_{j \mid i}\in[-1,1]$ and $\beta_{j \mid i}\in (-\infty,1]$. These flexible functions capture AD ($\alpha_{j \mid i}=1$ and $\beta_{j \mid i}=0$), complete independence ($\alpha_{j \mid i}=0$), and AI (all other parameter combinations). 
In contrast, there is no general class of distributions for modelling the residuals $\boldsymbol{Z}_{\mid i}$. \citetmain{Heffernan_2004} use the working assumption that $\boldsymbol{Z}_{\mid i}$ follows a $(d-1)$-dimensional multivariate Gaussian (MVG) distribution, denoted $ \boldsymbol{Z}_{\mid i} \sim \text{MVG}_{d-1}(\boldsymbol{\mu}_{\mid i}, \Sigma_{\mid i}^{*})$, with mean vector $\boldsymbol{\mu}_{\mid i} = \{\mu_{j \mid i} : j \in V_{\mid i} \} \in \mathbb{R}^{d-1}$ and covariance matrix $\Sigma_{\mid i}^{*}$. To further simplify, they take $\Sigma_{\mid i}^{*}$ to be diagonal so that the components of $\boldsymbol{Z}_{\mid i}$ are independent. 

Inference is performed separately for each conditioning component $Y_i$. Under the assumption that the observations $\boldsymbol{y}^{1},\ldots,\boldsymbol{y}^{n}$ are realisations of independent and identically distributed random vectors $\boldsymbol{Y}^{1},\ldots,\boldsymbol{Y}^{n}$, the parameters of the normalising functions and the residual distribution associated with the $i$th conditioning component are obtained by maximising the likelihood
\begin{equation}
    L_{\mid i} \left(\thetai \right) = \prod_{k:y_i^{k}>u_{Y_i}} \phi_{d-1}\left[ \left\{ \frac{ y_{j}^{k} - \alpha_{j \mid i}y_{i}^{k} }{ \left( y_{i}^{k} \right)^{\beta_{j \mid i}} } \right\}_{j \in V_{\mid i}} ; \boldsymbol{\mu}_{\mid i}, \Sigma_{\mid i}^{*} \right] \prod_{j \in V_{\mid i}} \left(y_{i}^{k}\right)^{-\beta_{j \mid i}},
    \label{eqn:likelihood_CMEVM}
\end{equation}
where $\phi_{d-1}(\cdot; \boldsymbol{\mu}_{\mid i}, \Sigma_{\mid i}^{*})$ is the density of the $(d-1)$-dimensional MVG distribution. The second term in the product is the Jacobian arising from the transformation of $\boldsymbol{Y}_{\mid i}$ to $\boldsymbol{Z}_{\mid i}$. While the MVG assumption is useful for parameter estimation, it is not necessarily an appropriate distributional assumption. Consequently, \citetmain{Heffernan_2004} use the empirical distribution of the fitted residuals 
\begin{equation}
    \hat{\boldsymbol{z}}_{\mid i} = \{ \hat{z}_{j \mid i} \coloneqq (y_{j \mid i} - \hat{\alpha}_{j \mid i }y_{i}) y_{i}^{-\hat{\beta}_{j \mid i}} : j \in V_{\mid i} \},
    \label{eqn:resid}
\end{equation}
to undertake model-based prediction. 

\subsection{Multivariate Asymmetric Generalised Gaussian Distribution} 
\label{sec:MVAGG}

While the CMEVM is flexible and the computational cost of parameter estimation is low, the predictive performance of the model declines in high dimensions. This is due to the well-known curse of dimensionality associated with sampling from the empirical distribution \citepmain{Nagler_2016}; see the Supplementary Material for an illustration in the specific case of the CMEVM. Overcoming this limitation requires a fully parametric model. While this approach has been taken before \citepmain{Shooter_2021, Richards_2022, Wadsworth_2022}, the model that we propose allows, for the first time, both asymmetry in the univariate marginal distributions of the residuals and a sparse dependence structure in their joint distribution.

The most commonly used fully parametric model for $\boldsymbol{Z}_{\mid i}$ is the MVG copula with generalised Gaussian margins \citepmain{Wadsworth_2022}. The generalised Gaussian distribution bridges the Gaussian and Laplace distributions, making it appropriate in cases where the pairwise dependence of the residuals is expected to vary from strong asymptotic dependence (for which the residual margins are expected to follow a Gaussian distribution) to complete independence (for which the residual margins are expected to follow a Laplace distribution). 
However, the generalised Gaussian distribution is symmetric, a property that we found is not always appropriate (see Supplementary Material). Instead, we model the marginal distribution of the residuals $\boldsymbol{Z}_{\mid i}$ using the asymmetric generalised Gaussian (AGG) distribution \citepmain{Nacereddine_2019}. This distribution has density
\begin{equation}
    f_{Z}(z) = \frac{ \delta }{ (\kappa_{1} + \kappa_{2})\Gamma(1/\delta) } 
    \begin{cases}  
        \text{exp}\left\{- \left[ \left(\nu - z \right)/\kappa_{1} \right]^{\delta} \right\} \hspace{1em} &z < \nu, \\
        \text{exp}\left\{- \left[ \left(z - \nu \right)/\kappa_{2} \right]^{\delta} \right\} \hspace{1em} &z \geq \nu, \\
    \end{cases}
    \label{eqn:AGG_density}
\end{equation}
where $z \in \mathbb{R}$, $\Gamma(\cdot)$ denotes the standard gamma function, and $\nu \in \mathbb{R}$, $\kappa_{1} > 0$, $\kappa_{2} > 0$, $\delta > 0$ are the location, left-scale, right-scale, and shape parameters, respectively. We refer to this distribution as the $\text{AGG}(\nu, \kappa_{1}, \kappa_{2}, \delta)$. When $\kappa_{1} = \kappa_{2}$, the AGG reduces to the generalised Gaussian (or delta-Laplace) distribution used by \citetmain{Wadsworth_2022}. 

Any model for the marginal distribution of the residuals $\boldsymbol{Z}_{\mid i}$ must have two properties: (i) it can adapt its shape to account for the strength of extremal dependence, and (ii) it can account for asymmetry. As discussed, the generalised Gaussian distribution satisfies (i) but not (ii). Alternatively, distributions such as the skew normal, skew-$t$, and asymmetric Laplace distributions can capture (ii) but not (i). The AGG distribution is therefore an ideal candidate as it can account for asymmetry in the margins while also containing the Gaussian (complete asymptotic dependence) and Laplace (complete independence) distributions as edge cases.

For inference, it is helpful to separate the marginal and dependence properties of the residual vector $\boldsymbol{Z}_{\mid i}$ by defining
\begin{align*}
W_{j\mid i} := \Phi^{-1}(F_{Z_{j\mid i}}(Z_{j \mid i})),~~j\in V_{\mid i},
\end{align*}
where $\Phi$ and $F_{Z_{j \mid i}}$ are the distribution functions of the standard Gaussian and AGG$(\nu,\kappa_1,\kappa_2,\delta)$ distributions, respectively. We can then assume that $\boldsymbol{W}_{\mid i} = \{W_{j \mid i} : j \in V_{\mid i}\} \sim \text{MVG}_{d-1}(\boldsymbol{0},\Sigma_{\mid i})$ where $\Sigma_{\mid i}$ is a $(d-1)$-dimensional \emph{correlation} matrix. We name this combination of marginal and joint distributions the multivariate asymmetric generalised Gaussian (MVAGG) distribution, denoted by $\boldsymbol{Z}_{\mid i}\sim\text{MVAGG}_{d-1}(\Thetai, \Thetai[\Gamma])$, where $\Thetai \coloneqq \left\{ (\nu_{j \mid i}, \kappa_{2_{j \mid i}}, \kappa_{1_{j \mid i}}, \delta_{j \mid i}) : j \in V_{\mid i} \right\}$ and $\Thetai[\Gamma]$ parameterises the (sparse) precision matrix $\Gammai \coloneqq (\Sigmai)^{-1}$. The density for this distribution is 
\begin{equation}
    f_{i}(\boldsymbol{z}_{\mid i}) = \phi_{d-1} \left[ \left\{ \Phi^{-1} \left(F_{Z_{j \mid i}} \left( z_{j \mid i} \right) \right) \right\}_{j \in V_{\mid i}} ; \; \boldsymbol{0}, \Sigmai \right] \prod_{j \in V_{\mid i}} \frac{ f_{Z_{j \mid i}} \left( z_{j \mid i} \right) }{ \phi \left[ \Phi^{-1} \left(F_{Z_{j \mid i}} \left( z_{j \mid i} \right) \right) \right] },
    \label{eqn:density_Z_given_i}
\end{equation}
where $f_{Z_{j \mid i}}$ is defined in equation \eqref{eqn:AGG_density}, and $\phi$ is the standard univariate Gaussian density.

\subsection{Structured Conditional Multivariate Extreme Value Model}
\label{Sect:graphs}

While a parametric model for the residual component of the CMEVM permits prediction in high dimensions, it has the drawback of vastly increasing the number of model parameters. Specifically, the correlation matrix $\Sigma_{\mid i}$ that parameterises the MVAGG distribution increases the number of parameters from order $d^2$ (the original CMEVM) to order $d^3$. We now explain how conditional independence structures can overcome this challenge by introducing sparsity into the precision matrix $\Gamma_{\mid i}$ associated with $\Sigma_{\mid i}$. 

We first introduce some necessary terminology. The conditional independence structure of a random vector $\boldsymbol{W} \sim \text{MVG}_{d}(\boldsymbol{0}, \Sigma)$ can be formulated by associating $\boldsymbol{W}$ with a simple undirected graph $\mathcal{G} = (V, E)$. This graph consists of vertex $V = \{1, \hdots , d\}$ and edge $E \subseteq \{ \{j, k\} \mid j, k \in V, j \neq k \}$ sets. The components $W_{j}$ and $W_{k}$ are conditionally independent given the remaining components if $\{j,k\} \notin E$. Since $\boldsymbol{W}$ additionally follows a Gaussian distribution, conditional independence of $W_{j}$ and $W_{k}$ implies that their partial correlation is zero, and hence that $\Gamma_{j,k} = \Gamma_{k,j} = 0$, where $\Gamma = (\Sigma)^{-1}$ is the precision matrix of $\boldsymbol{W}$ \citepmain{Speed_1986}. Thus, a sparse precision matrix (equivalently, a sparse graph $\mathcal{G}$) can greatly reduce the dimension of the parameter space.

To construct the structured CMEVM (SCMEVM), we begin by conditioning on a single component $Y_i$. Given that $Y_i > u_{Y_i}$, we define a conditional independence graph $\mathcal{G}_{\mid i}=\{E_{\mid i},V_{\mid i}\}$ associated with the $(d-1)$-dimensional residual vector $\boldsymbol{W}_{\mid i}$. This graph might be defined by the topology on which the process is measured, or it might be estimated empirically. Assuming that the residuals $\boldsymbol{Z}_{\mid i}$ follow a MVAGG distribution, density \eqref{eqn:density_Z_given_i} can be used to define the likelihood function such that the elements of the precision matrix $\Gamma_{\mid i}$ that correspond to edges not in the edge set $E_{\mid i}$ are set to zero. All other elements in the matrix are treated as free parameters to be estimated. 

Conditioning in turn on each component results in $d$ models, each with a graph $\mathcal{G}_{\mid i}$ that describes the dependence structure of the residual vector $\boldsymbol{W}_{\mid i}$. This presents a conundrum for model inference: how can these graphs be learnt? If we consider the limit result on which the CMEVM is constructed, all $d$ conditional dependence graphs are inherited from the graph $\mathcal{G}_{\boldsymbol{X}}$ which represents the conditional dependence structure of the generating random vector $\boldsymbol{X}$. Using this observation, it seems reasonable to learn a unified graph from which the $\mathcal{G}_{\mid i}$ are subsequently inferred. In practice, $\mathcal{G}_{\boldsymbol{X}}$ is unknown and, following the extreme value paradigm of using only data in the tails to infer tail behaviour, we prefer not to estimate it directly. This leaves two options: ignore the asymptotic self-consistency between the graphs and infer each separately, or learn a unified graph from only the tail data. The former has the advantage of providing a more flexible modelling tool at the expense of increased computational cost. The latter ensures consistency across conditioning components at the expense of model flexibility. Scalability is a major motivation for our work, so we elect for the second approach. 


Choosing a single graph presents two challenges: selection of the $d$-dimensional graph $\mathcal{G}$ and inference of the sub-graphs. The first will be addressed in Section \ref{sec:Graph_Selection}. For the second, we augment $\boldsymbol{W}_{\mid i}$ to include $W_{i\mid i}=0$, resulting in a $d$-dimensional vector. Exploiting textbook properties of the conditional MVG distribution, each $(d-1)$-dimensional precision matrix $\Gamma_{\mid i}$ can then be obtained by excluding the $i$th row and column from the matrix $\Gamma$ associated with $\mathcal{G}$. Equivalently, since $\boldsymbol{W}$ follows a MVG distribution, the graph $\mathcal{G}_{\mid i}$ associated with $\boldsymbol{W}_{\mid i}$ is found by removing the $i$th node and its incident edges from $\mathcal{G}$. In further contrast to learning the graphs individually, this approach allows \emph{explicit} conditioning on $W_{i \mid i} = 0$.

\section{Inference} 
    \label{Section:Inference}
    In this section, we describe the inference scheme for the MVAGG SCMEVM. The methods described could be easily adapted to any other residual distribution based on a Gaussian copula. We also provide algorithms for graphical selection and model-based predictions.

\subsection{Parameter estimation}
\label{sec:Inferece_MVAGG} 

Given $n$ independent and identically distributed realisations $\boldsymbol{x}^{1},\dots,\boldsymbol{x}^{n}$ of the $d$-dimensional random vector $\boldsymbol{X}$, the first step is transformation to standard Laplace margins. By double application of the probability integral transform (PIT), for each $i \in V$ and each $k \in\{1,\ldots,n\}$, 
\begin{equation*}
    y_{i}^{k}=\left\{
\begin{array}{cc}
-\log\left(2 \left[1-\tilde{F}_{i} \left(x_{i}^{k} \right) \right] \right)& \tilde{F}_{i} \left(x_{i}^{k} \right) > 0.5,\\
\log \left( 2\tilde{F}_{i} \left(x_{i}^{k} \right) \right) & \tilde{F}_{i} \left(x_{i}^{k} \right) \leq 0.5,
\end{array}
    \right.
\end{equation*}
where $\tilde{F}_{i}$ is an estimate of the marginal distribution $F_{i}$ for $X_{i}$. We use a semi-parametric estimate for $F_{i}$ consisting of the empirical distribution for $x_{i} \leq v_{X_{i}}$ and a generalised Pareto distribution for $x_{i} > v_{X_{i}}$ \citepmain{Heffernan_2004}. The threshold $v_{X_{i}}$ is selected using the automated method of \citetmain{Murphy_2024}.

Inference for the MVAGG SCMEVM is performed for each component $Y_{i}$ separately by maximising the likelihood
\begin{equation}
L_{\mid i} \left(\thetai \right)= \prod_{k : y_{i}^{k} > u_{Y_{i}} }
f_{i} \left( \left\{ \frac{ y_{j}^{k} - \alpha_{j \mid i} y_{i}^{k} }{ \left(y_{i}^{k}\right)^{ \beta_{j \mid i} } } \right\}_{j \in V_{\mid i}} ; \Thetai, \Thetai[\Gamma] \right) 
\prod_{j \in V_{\mid i}} \left( y_{i}^{k} \right)^{-\beta_{j \mid i}}, 
\label{eqn:Likelihood_function}
\end{equation}
where $f_{i}$ is given by equation~\eqref{eqn:density_Z_given_i}, $u_{Y_i}$ is the dependence threshold and $\thetai:=(\Thetai[d], \Thetai, \Thetai[\Gamma])$ combines the CMEVM dependence $\Thetai[d] \coloneqq \left\{ (\alpha_{j \mid i},\beta_{j \mid i}) : j \in V_{\mid i} \right\}$, MVAGG marginal $\Thetai$ and MVAGG correlation $\Thetai[\Gamma]$ parameters. Of the parameter vectors, only $\Thetai[\Gamma]$ is determined by the graph associated with $\boldsymbol{W}_{\mid i}$. The edge cases are the independent model, a graph with no edges, and the saturated model, a full graph. In the former, only the diagonal of the precision matrix is estimated, with all other elements set to zero. In the latter, all elements of the precision matrix are estimated. Any other combination will be referred to as a graphical model. 

The naive approach is to jointly estimate the full parameter vector $\thetai$. A ``one-step" numerical maximisation procedure for this is given in Algorithm \ref{alg:cmevm}. The algorithm iterates between maximising the profile likelihood for $\Thetai[\Gamma]$ and maximising the profile likelihood for $(\Thetai,\Thetai[d])$. Numerical optimisation of the profile likelihood for $\Thetai[\Gamma]$ is only required for non-trivial graphical structures since a closed-form expression exists for $\hat{\Sigma}_{\mid i}$, and hence also for $\hat{\Gamma}_{\mid i}$, for both the independent and saturated models (see Algorithm~\ref{alg:cmevm_Gamma} for details). For the graphical model, the graph $\mathcal{G}_{\mid i}$ is chosen \emph{a priori}; see Section \ref{sec:Graph_Selection} for details. As discussed, if $\{j, k\} \notin E_{\mid i}$ then $(\hat{\Gamma}_{\mid i})_{j,k}$ must be 0, a condition which can be enforced by using the graphical lasso \citep[Remark 2.1]{Friedman_2007} to estimate $\Gamma_{\mid i}$. 

\begin{algorithm}
\caption{One-step parameter estimation for the MVAGG SCMEVM} \label{alg:cmevm}
\begin{algorithmic}[1]
    \STATE Initialise $\Thetai[d]$, $\Thetai$, $\mathcal{G}_{\mid i} = (V_{\mid i}, E_{\mid i})$, and \texttt{tol};
    \STATE \label{step.current} The current values of $\Thetai[d]$ and $\Thetai$ are $\Thetai[d*]$ and $\Thetai[*]$, respectively;
    \STATE \label{step.Gamma} Obtain $\displaystyle\Thetaihat[\Gamma]=\argmax_{\Thetai[\Gamma]}L_{\mid i}\left(\Thetai[d*],\Thetai[*],\Thetai[\Gamma] \right)$, where $L_{\mid i}$ is likelihood~\eqref{eqn:Likelihood_function} using Algorithm~\ref{alg:cmevm_Gamma};  
    \STATE \label{step.d} Obtain $\displaystyle\left(\Thetaihat[d],\Thetaihat\right)=\argmax_{\Thetai[d],\Thetai}L_{\mid i}\left(\Thetai[d],\Thetai,\Thetaihat[\Gamma] \right)$;
    \IF{$\max( \mid \Thetaihat[d]-\Thetai[d*]\mid,~
    \mid\Thetaihat-\Thetai[*]\mid ) > \texttt{tol}$} 
    \STATE Set $\Thetai[d] = \Thetaihat[d]$ and $\Thetai=\Thetaihat$;
    \STATE Repeat steps \ref{step.current} - \ref{step.d};
    \ELSE{ \RETURN $\Thetaihat[d]$, $\Thetaihat$ and $\Thetaihat[\Gamma]$}
    \ENDIF
    \end{algorithmic}
\end{algorithm}

This procedure has limitations. Firstly, the estimates of  $\Thetai$ and $\Thetai[d]$ are not independent: the estimate of $\alpha_{j \mid i}$ ($\beta_{j \mid i}$) influences the mode (variance) of the residual distribution. Consequently, while joint estimation may result in a model with good predictive abilities (see Supplementary Material), the first-order extremal dependence structure is not entirely captured by the dependence parameters $\Thetai[d]$. Further, since AD is an edge case in the parameter space, this procedure is more likely to suggest that pairs of variables are AI when they are AD. Secondly, finding suitable initial values for the numerical optimisation becomes increasingly difficult for large $d$. Even for a sparse precision matrix $\Gammai$, the parameter space grows at least linearly in $d$. 
To address these issues, two- and three-step estimation procedures are described in Algorithm~\ref{alg:cmevm2} and Algorithm~\ref{alg:cmevm3}. The two-step approach still requires a computationally expensive numerical optimisation procedure for the MVAGG distribution, while the three-step approach does not. Hence, Algorithm~\ref{alg:cmevm3} is our preferred option. 

We make two final observations. Firstly, in contrast to the stationary spatial CMEVM \citepmain{Richards_2022,Wadsworth_2022}, in the SCMEVM, the parameter values differ with the conditioning variable. Hence, there is no information to be gained by jointly fitting the $d$ conditional models. Secondly, by construction, Algorithm~\ref{alg:cmevm3} avoids maximising the joint likelihood. Consequently, we cannot obtain uncertainty estimates using the standard asymptotic properties of the likelihood and the maximum likelihood estimators. To obtain these, we recommend using a non-parametric bootstrapping algorithm. This requires sampling with replacement from the original data to create artificial datasets, fitting the model to each dataset, and hence obtaining a bootstrap approximation to the sampling distributions of parameter estimates and model predictions.

\begin{algorithm}
\caption{Estimating $\displaystyle\Thetaihat[\Gamma]$} \label{alg:cmevm_Gamma}
\begin{algorithmic}[1]
    \STATE Initialise $\Thetaihat[d]$, $\Thetaihat$, and $\mathcal{G}_{\mid i} = (V_{\mid i}, E_{\mid i})$;
    \STATE Obtain $\hat{\boldsymbol{z}}_{\mid i}$ using equation~\eqref{eqn:resid} and $\Thetaihat[d]$;
    \STATE Obtain $\hat{\boldsymbol{w}}_{\mid i}$ such that $\hat{w}_{j\mid i} = \Phi^{-1}(F_{Z_{j \mid i}}(\hat{z}_{j \mid i} ; \hat{\boldsymbol{\Theta}}_{j \mid i}))$ for $j \in V_{\mid i}$;
    \STATE \textbf{if} $|E_{\mid i}| = 0$ \textbf{then} \hfill \textit{Independence}
    \STATE \quad $\displaystyle\Thetaihat[\Gamma] = I_{d-1}$ (the $(d-1)$-dimensional identity matrix);
    \STATE \textbf{else if} $|E_{\mid i}| = d(d-1)/2$ \textbf{then} \hfill \textit{Saturated}
    \STATE \quad $\displaystyle\Thetaihat[\Gamma] = (\text{corr}(\hat{\boldsymbol{W}}_{\mid i}))^{-1}$;
    \STATE \textbf{else} \hfill \textit{Graphical}
    \STATE \quad $\displaystyle\Thetaihat[\Gamma]$ is estimated using a graphical lasso \citepmain{glasso} on $\hat{\boldsymbol{W}}_{\mid i}$;
    \RETURN $\displaystyle\Thetaihat[\Gamma]$
    \end{algorithmic}
\end{algorithm}

\begin{algorithm}
\caption{Two-step parameter estimation for the MVAGG SCMEVM}\label{alg:cmevm2}
\begin{algorithmic}[1]
    \STATE Initialise $\Thetai[d]$, $\Thetai$, $\mathcal{G}_{\mid i} = (V_{\mid i}, E_{\mid i})$, and \texttt{tol};
    \STATE Assuming independent Gaussian residuals, obtain $\Thetaihat[d]$ by maximising likelihood~\eqref{eqn:likelihood_CMEVM};
    \STATE Using equation~\eqref{eqn:resid} and $\Thetaihat[d]$, obtain $\displaystyle\hat{\boldsymbol{z}}_{\mid i}$ and treat them as fixed;
    \STATE \label{step2.current} The current value of $\Thetai$ is $\Thetai[*]$;
    \STATE \label{step2.gamma} Obtain $\displaystyle\Thetaihat[\Gamma]=\argmax_{\Thetai[\Gamma]}L_{\mid i}\left(\Thetaihat[d],\Thetai[*],\Thetai[\Gamma] \right)$, where $L_{\mid i}$ is likelihood~\eqref{eqn:Likelihood_function} using Algorithm~\ref{alg:cmevm_Gamma};
    \STATE \label{step2.theta} Obtain $\displaystyle\Thetaihat=\argmax_{\Thetai}L_{\mid i}\left(\Thetaihat[d],\Thetai,\Thetaihat[\Gamma] \right)$;
        \IF{
            $\max(\mid \Thetaihat - \Thetai[*] \mid) > \texttt{tol}$
            } 
            \STATE set $\Thetai = \Thetaihat$;
            \STATE repeat \ref{step2.current} - \ref{step2.theta};
        \ELSE{
            \RETURN $\Thetaihat$ and $\Thetaihat[\Gamma]$.
        }
        \ENDIF
    \RETURN $\Thetaihat[d]$, $\Thetaihat$ and $\Thetaihat[\Gamma]$
\end{algorithmic}
\end{algorithm}

\begin{algorithm}
\caption{Three-step parameter estimation for the MVAGG SCMEVM}\label{alg:cmevm3}
\begin{algorithmic}[1]
    \STATE Initialise $\Thetai[d]$, $\Thetai$, $\mathcal{G}_{\mid i} = (V_{\mid i}, E_{\mid i})$, and \texttt{tol};
    \STATE Assuming independent Gaussian residuals, obtain $\Thetaihat[d]$ by maximising likelihood~\eqref{eqn:likelihood_CMEVM};
    \STATE Using equation~\eqref{eqn:resid} and $\Thetaihat[d]$, obtain $\displaystyle\hat{\boldsymbol{z}}_{\mid i}$ and treat them as fixed;
    \STATE Assuming the components of $\displaystyle\hat{\boldsymbol{Z}}_{\mid i}$ are independent, obtain $\displaystyle\Thetaihat=\argmax_{\Thetai}f_{i}(\hat{\boldsymbol{z}} _{\mid i};\Thetai,I_{d-1})$ where $f_{i}$ is given by equation~\eqref{eqn:density_Z_given_i}, and $\displaystyle I_{d-1}$ is a $(d-1)$-dimensional identity matrix; 
    \STATE \label{step3.gamma}
    Obtain $\displaystyle\Thetaihat[\Gamma]=\argmax_{\Thetai[\Gamma]}L_{\mid i}\left(\Thetaihat[d],\Thetaihat,\Thetai[\Gamma]\right)$, where $L_{\mid i}$ is likelihood~\eqref{eqn:Likelihood_function} using Algorithm~\ref{alg:cmevm_Gamma};
    \RETURN $\Thetaihat[d]$, $\Thetaihat$ and $\Thetaihat[\Gamma]$
\end{algorithmic}
\end{algorithm}


\subsection{Graph selection}
\label{sec:Graph_Selection}

We now discuss selection of the graphs $\mathcal{G}_{\mid 1},\ldots,\mathcal{G}_{\mid d}$. Recall that we assume that these graphs are all derived from a unifying graph $\mathcal{G}$. In most cases, $\mathcal{G}$ will be unknown and must be learnt. Several such learning algorithms have been proposed for multivariate generalised Pareto distributions. \citetmain{Engelke_2020} iteratively add edges to $\mathcal{G}$ to minimise the AIC, but this is costly in higher dimensions. \citetmain{Wan_2025}  present the extremal graphical lasso, an extension of the graphical lasso \citepmain{Yuan_2007, Friedman_2007}, while \citetmain{Engelke_2025} propose ``EGlearn", which combines a majority rule with either the graphical lasso or neighbourhood selection \citepmain{Meinshausen_2006}. 

For consistency with Algorithm \ref{alg:cmevm_Gamma}, our approach, detailed in Algorithm \ref{alg:cmevm_graphical_selection}, also uses the graphical lasso and a majority rule. The algorithm has two tuning parameters: the thresholds $u_{Y_{1}}, \dots, u_{Y_{d}}$ and the majority rule proportion $p$. The graphical lasso penalty parameter $\lambda$ is not a tuning parameter as it is selected objectively by comparing the composite AIC scores. While other metrics are available, the AIC gives the graph with the best predictive properties. Returning to the tuning parameters, the thresholds should not be so low that limit~\eqref{eqn:WT_2022_Limit} is a poor approximation; at the same time, if they are too high, there will be insufficient data to accurately identify the conditional dependence structure. The choice of the majority rule proportion $p$ results in a similar trade-off: too high a value of $p$ risks inferring a very sparse structure that predicts poorly; too low a value of $p$ infers a dense graph that is computationally expensive to work with. In our applications, we set $p = 0.5$. We chose this because the inferred structure is not sensitive to the choice of $p$ (see Supplementary Material) and is consistent with other majority rules used in extremal graphical selection \citepmain{Engelke_2020, Engelke_2025}.

\begin{algorithm}[t!]
\caption{Graphical selection using the MVAGG SCMEVM}
\begin{algorithmic}[1]
    \STATE Initialise $\boldsymbol{\lambda}$, $p$ and $u_{Y_{1}}, \dots, u_{Y_{d}}$.
    \FOR{$j = 1, \dots, \lvert\boldsymbol{\lambda} \rvert$}
        \FOR{$i = 1, \dots, d$}
            \STATE Assuming independent Gaussian residuals, obtain $\Thetaihat[d]$ by maximising likelihood \eqref{eqn:likelihood_CMEVM} with threshold $u_{Y_{i}}$;
            \STATE Using equation~\eqref{eqn:resid} and $\Thetaihat[d]$, obtain $\displaystyle\hat{\boldsymbol{z}}_{\mid i}$ and treat them as fixed;
            \STATE Assuming the components of $\hat{\boldsymbol{Z}}_{\mid i}$ are independent, obtain $\displaystyle\Thetaihat=\argmax_{\Thetai}f_{i}(\hat{\boldsymbol{z}}_{\mid i};\Thetai,I_{d-1})$ where $f_{i}$ is given by equation \eqref{eqn:density_Z_given_i} and $\displaystyle I_{d-1}$ is a $(d-1)$-dimensional identity matrix;
            \STATE Set $\Thetai = \Thetaihat$ and treating as fixed marginally transform $\hat{\boldsymbol{z}}_{\mid i}$ onto standard Gaussian margins $\hat{\boldsymbol{w}}_{\mid i}$;
            \STATE Apply a graphical lasso with penalisation parameter $\lambda_{j}$ to $\hat{\boldsymbol{W}}_{\mid i}$ to infer $\mathcal{G}_{\mid i}$;
        \ENDFOR
        \STATE Obtain a weighted graph $\mathcal{G}^{*}$ by combining the subgraphs $\mathcal{G}_{\mid i}$;
        \STATE Create $\mathcal{G}'$ by pruning the edges of $\mathcal{G}^{*}$ that do not occur at least $(p \times 100)$\% of the time;
        \FOR{$i = 1, \dots, d$}
            \STATE Maximise likelihood~\eqref{eqn:Likelihood_function} using Algorithm \ref{alg:cmevm3} and $\mathcal{G}_{\mid i}'$ obtained by removing the $i$th node and its incident edges from $\mathcal{G}'$;
        \ENDFOR
        \STATE \label{step:composite_AIC} Calculate and store the composite AIC
    \ENDFOR
    \RETURN $\mathcal{G}'$ that minimises the composite AIC from step \ref{step:composite_AIC}.
\end{algorithmic}
\label{alg:cmevm_graphical_selection}
\end{algorithm}

\subsection{Prediction}
\label{sec:Prediction}

By construction, the CMEVM and, by extension, the SCMEVM do not permit closed forms for either tail probabilities or quantiles. \citetmain{Heffernan_2004} use a simulation-based prediction algorithm based on the empirical distribution of the fitted residuals $\hat{\boldsymbol{Z}}_{\mid i}$. A key motivation for introducing the SCMEVM in Section~\ref{Section:Methods} was the observation that this prediction procedure fails in high dimensions. We now explain how the SCMEVM can be used to obtain fully parametric predictions using a method very similar to \citet[Section 5.2.2]{Wadsworth_2022} and \citet[Section 3.3]{Richards_2022}. 

For $u > \text{max}(u_{Y_{i}} : i \in V)$, the SCMEVM describes the distribution of $\boldsymbol{X}$ given that the largest component of $\boldsymbol{Y}$ exceeds $u$, that is 
\begin{equation}
    \left\{ \tilde{F}_{X_{i}}^{-1}(F_{L}(Y_{i})): i \in V \right\} \; \bigg\vert \left( \underset{i \in V}{\text{max}} \; Y_{i} > u \right),
    \label{eqn:joint_survival_set}
\end{equation}
where $u_{Y_{i}}$ is the SCMEVM dependence threshold for conditioning component $Y_i$, $F_{L}$ is the distribution function of the standard Laplace distribution, and $\tilde{F}_{X_{i}}$ is the estimated marginal distribution of $X_{i}$. To create realisations of $\boldsymbol{X}$, we draw samples from equation \eqref{eqn:joint_survival_set} using Algorithm \ref{alg:conditional_survival_set} with probability
\begin{equation*}
    \mathbb{P} \left( \underset{i \in V}{\text{max}} \; Y_{i} > u \right) = \frac{1}{n} \sum_{k = 1}^{n} \mathds{1}\left\{ \underset{i \in V}{\text{max}} \; y_{i}^{k} > u \right\}.
    \label{eqn:joint_survival_prob}
\end{equation*}
Otherwise, we draw realisations from the empirical distribution of $\boldsymbol{X} \; \bigg\vert \left(\underset{i \in V}{\text{max}} \; Y_{i} < u \right)$. 

\begin{algorithm}
\caption{Simulation of equation \eqref{eqn:joint_survival_set}}
\begin{algorithmic}[1]
    \STATE Initialise $u$;
    \FOR{$l = 1, \hdots N$ such that $N > n$}
        \STATE Draw a conditioning random variable $i$ from $i \in V$ with uniform probability;
        \STATE Simulate $E^l \sim \text{Exp}(1)$ and set $y_{i}^{l} = u + E^{l}$;
        \STATE \label{step.sim.z} Simulate $\boldsymbol{z}_{\mid i}^{l}$ from the distribution described in Section \ref{sec:MVAGG};
        \STATE Calculate ${y}_{j \mid i}^{l} = \hat{\alpha}_{j \mid i} y_{i}^{l} + \left(y_{i}^{k} \right)^{\hat{\beta}_{j \mid i}} z_{j \mid i}^{l}$ for $j \in V_{\mid i}$;
        \STATE Calculate an importance weight $w^{l} = \left(\frac{1}{d} \sum_{m = 1}^{d} \mathds{1}\{ y_{m}^{l} > u \}\right)^{-1}$.   
    \ENDFOR
    \STATE Sub-sample $n$ realisations from $\{\boldsymbol{y}^{1}, \hdots, \boldsymbol{y}^{N}\}$ with probabilities proportional to their importance weights;
    \STATE Transform the sub-sample $\{\boldsymbol{y}^{1}, \hdots, \boldsymbol{y}^{n}\}$ to $\{\boldsymbol{x}^{1}, \hdots, \boldsymbol{x}^{n}\}$ via a double application of the PIT;
    \RETURN $\{\boldsymbol{x}^{1}, \hdots, \boldsymbol{x}^{n}\}$
\end{algorithmic}
\label{alg:conditional_survival_set}
\end{algorithm}

\section{Simulation Study} 
    \label{Section:Simulation_Study}
    In this section, we use simulation studies to assess the performance of the SCMEVM. We compare the three stepwise inference procedures and assess the graphical selection process. Finally, we compare the SCMEVM to existing methods.

\subsection{Stepwise inference procedures}
\label{sec:Parameter_estimation}

We consider the 5-dimensional SCMEVM with dependence structure given by the graph $\mathcal{G}$ with edge set $E = \{\{1,2\}, \{1,3\}, \{2,3\}, \{3,4\}, \{3,5\}, \{4,5\}\}$. Using data simulated from this model, we compare the performance of the one-(Algorithm~\ref{alg:cmevm}), two-(Algorithm~\ref{alg:cmevm2}), and three-step (Algorithm~\ref{alg:cmevm3}) estimation procedures for each of three candidate dependence structures: independent, graphical, and saturated. 

For each $i \in V$, the conditioning variable $Y_{i} \mid Y_{i} > u_{Y_{i}}$ is simulated from a standard Laplace distribution with $u_{Y_{i}}$ the $0.80$-quantile of this distribution. To simulate the residual vector $\boldsymbol{Z}_{\mid i}$ from the MVAGG, the  PIT is used to transform the margins of $\boldsymbol{W}_{\mid i}$, which are drawn from a MVG with standard margins and correlation matrix $\Sigma_{\mid i}$. Finally, the vector $\boldsymbol{Y}_{\mid i}$ is obtained by applying the inverse normalisation of limit~\eqref{eqn:WT_2022_Limit}. We generate $200$ samples of $\boldsymbol{Y} \mid Y_{i} > u_{Y_{i}}$ and consider $n \in \{250, 500\}$. The true parameters are independently sampled from a uniform distribution on $(0.1, 0.5)$ for $\alpha_{j}$, $(0.1, 0.3)$ for $\beta_{j}$, $(-5, 5)$ for $\nu_{j}$, $(0.5, 2)$ for $\kappa_{1_{j}}$, $(1.5, 3)$ for $\kappa_{2_{j}}$, and $(0.8, 2.5)$ for $\delta_{j}$, for each $j \in V$. Correlations in $\Sigma_{\mid i}$ correspond to weak positive associations; see Supplementary Material for the cases of weak negative and strong positive associations. 

\begin{figure}[t!]
  \centering
  \subfloat{\includegraphics[width = 0.95\textwidth]{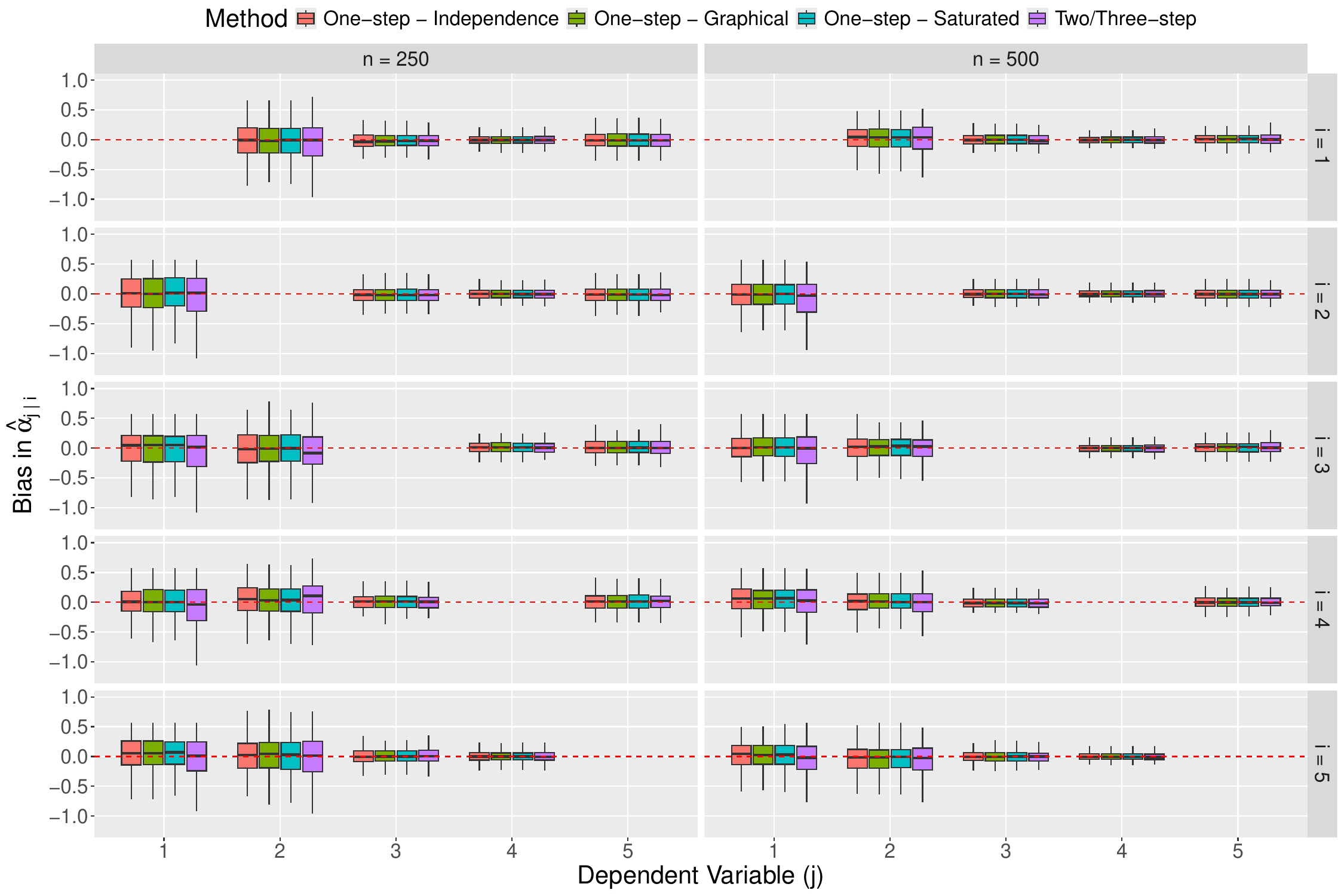}}\\
  \caption{Boxplots detailing the bias of $\hat{\alpha}_{j \mid i}$ for distinct $i, j \in V$. Each row corresponds to the conditioning variable $i$, and each column corresponds to the sample size. The fill of the boxplots denotes the different models. The red dashed line indicates the $y = 0$ line.}
  \label{fig:Sim_Study_True_Dist_Alpha}
\end{figure}

By construction, the estimates of $\Thetai[d]$ are the same under the two- and three-step procedures, regardless of the residual dependence structure. Consequently, we only compare $\Thetaihat[d]$ for the one-step (all three dependence structures) and two-step (independence only) procedures. Figure \ref{fig:Sim_Study_True_Dist_Alpha} shows the bias in $\hat{\boldsymbol{\alpha}}_{\mid i}$, with results for the other parameters found to be similar (see Supplementary Material). Reassuringly, even when using the stepwise procedure, the true parameter values are recovered. The biases from the one-step method do have a slightly narrower range, while the two-step method has fewer instances of unusually large bias. As expected, the variability in bias decreases as the sample size increases for both procedures.

Table \ref{tab:log_like_bias} presents the biases of the fitted maximum log-likelihood values. Although the SCMEVMs with graphical or saturated covariance exhibit slightly higher bias than the independent residuals model, the bias is consistent across all three stepwise procedures. Furthermore, we observe that the bias increases with sample size for the independent residual models, while for the graphical and saturated residual models, it remains similar for both $n=250$ and $n=500$. This suggests that the structured models are more robust to sample size changes. 

\begin{table}[t!]
\caption{Median (2.5\% and 97.5\% quantiles) bias in the fitted maximum log-likelihood values. Bold values denote the least biased stepwise inference procedure for each covariance structure type and conditioning variable.}
\resizebox{\textwidth}{!}{
\begin{tabular}{ccccccccccc}
\toprule
\multicolumn{2}{c}{Covariance Structure} & \multicolumn{3}{c}{Independent} & \multicolumn{3}{c}{Graphical} & \multicolumn{3}{c}{Saturated} \\
\hline
\begin{tabular}[c]{@{}c@{}}Number \\ of Excesses\end{tabular} & \begin{tabular}[c]{@{}c@{}}Conditioning\\ Variable\end{tabular} & One-step & Two-step & Three-step & One-step & Two-step & Three-step & One-step & Two-step & Three-step \\
\hline
\multirow{9}{*}{250} & 1 & \textbf{\begin{tabular}[c]{@{}c@{}}-4.8\\ (-18.3, 8.7)\end{tabular}} & \begin{tabular}[c]{@{}c@{}}-6.2\\ (-19.3, 8.2)\end{tabular} & \begin{tabular}[c]{@{}c@{}}-6.2\\ (-19.3, 8.2)\end{tabular} & \begin{tabular}[c]{@{}c@{}}14.0\\ (7.6, 22.5)\end{tabular} & \begin{tabular}[c]{@{}c@{}}12.8\\ (6.3, 21.5)\end{tabular} & \textbf{\begin{tabular}[c]{@{}c@{}}12.7\\ (6.7, 21.4)\end{tabular}} & \begin{tabular}[c]{@{}c@{}}14.8\\ (7.6, 25.1)\end{tabular} & \begin{tabular}[c]{@{}c@{}}13.6\\ (6.4, 24.1)\end{tabular} & \textbf{\begin{tabular}[c]{@{}c@{}}13.5\\ (7.0, 24.0)\end{tabular}}\\
\cline{2 - 11}
& 2 & \textbf{\begin{tabular}[c]{@{}c@{}}-4.2\\ (-20.6, 11.7)\end{tabular}} & \begin{tabular}[c]{@{}c@{}}-4.9\\ (-18.9, 10.0)\end{tabular} & \begin{tabular}[c]{@{}c@{}}-4.9\\ (-18.9, 10.0)\end{tabular} & \begin{tabular}[c]{@{}c@{}}14.2\\ (7.2, 23.6)\end{tabular} & \textbf{\begin{tabular}[c]{@{}c@{}}11.9\\ (5.4, 21.8)\end{tabular}} & \textbf{\begin{tabular}[c]{@{}c@{}}11.9\\ (5.4, 21.7)\end{tabular}} & \begin{tabular}[c]{@{}c@{}}15.3\\ (8.0, 25.1)\end{tabular} & \begin{tabular}[c]{@{}c@{}}13.2\\ (6.4, 23.0)\end{tabular} & \textbf{\begin{tabular}[c]{@{}c@{}}13.1\\ (6.2, 22.9)\end{tabular}}\\
\cline{2 - 11}
& 3 & \textbf{\begin{tabular}[c]{@{}c@{}}0.1\\ (-11.9, 13.1)\end{tabular}} & \begin{tabular}[c]{@{}c@{}}-1.6\\ (-12.3, 10.7)\end{tabular} & \begin{tabular}[c]{@{}c@{}}-1.6\\ (-12.3, 10.7)\end{tabular} & \begin{tabular}[c]{@{}c@{}}13.2\\ (7.1, 22.7)\end{tabular} & \textbf{\begin{tabular}[c]{@{}c@{}}11.1\\ (4.7, 21.3)\end{tabular}} & \textbf{\begin{tabular}[c]{@{}c@{}}11.1\\ (4.8, 21.2)\end{tabular}} & \begin{tabular}[c]{@{}c@{}}15.1\\ (7.8, 25.6)\end{tabular} & \textbf{\begin{tabular}[c]{@{}c@{}}13.2\\ (5.4, 24.3)\end{tabular}} & \begin{tabular}[c]{@{}c@{}}13.4\\ (5.6, 24.2)\end{tabular}\\
\cline{2 - 11}
& 4 & \textbf{\begin{tabular}[c]{@{}c@{}}-3.7\\ (-18.2, 11.5)\end{tabular}} & \begin{tabular}[c]{@{}c@{}}-5.2\\ (-19.4, 9.9)\end{tabular} & \begin{tabular}[c]{@{}c@{}}-5.2\\ (-19.4, 9.9)\end{tabular} & \begin{tabular}[c]{@{}c@{}}13.6\\ (8.2, 21.5)\end{tabular} & \begin{tabular}[c]{@{}c@{}}12.3\\ (5.9, 20.0)\end{tabular} & \textbf{\begin{tabular}[c]{@{}c@{}}12.2\\ (5.9, 19.9)\end{tabular}} & \begin{tabular}[c]{@{}c@{}}14.5\\ (7.7, 23.1)\end{tabular} & \textbf{\begin{tabular}[c]{@{}c@{}}13.3\\ (6.6, 21.6)\end{tabular}} & \textbf{\begin{tabular}[c]{@{}c@{}}13.3\\ (6.7, 21.6)\end{tabular}}\\
\cline{2 - 11}
& 5 & \textbf{\begin{tabular}[c]{@{}c@{}}-13.3\\ (-34.1, 3.3)\end{tabular}} & \begin{tabular}[c]{@{}c@{}}-14.8\\ (-30.0, 2.3)\end{tabular} & \begin{tabular}[c]{@{}c@{}}-14.8\\ (-30.0, 2.3)\end{tabular} & \begin{tabular}[c]{@{}c@{}}13.5\\ (7.6, 22.2)\end{tabular} & \begin{tabular}[c]{@{}c@{}}12.2\\ (4.7, 21.3)\end{tabular} & \textbf{\begin{tabular}[c]{@{}c@{}}12.0\\ (5.3, 21.0)\end{tabular}} & \begin{tabular}[c]{@{}c@{}}14.4\\ (8.7, 24.1)\end{tabular} & \textbf{\begin{tabular}[c]{@{}c@{}}12.9\\ (5.2, 22.5)\end{tabular}} & \textbf{\begin{tabular}[c]{@{}c@{}}12.9\\ (6.2, 22.4)\end{tabular}}\\
\hline
\multirow{9}{*}{500} & 1 & \textbf{\begin{tabular}[c]{@{}c@{}}-20.9\\ (-35.9, -4.9)\end{tabular}} & \begin{tabular}[c]{@{}c@{}}-22.4\\ (-39.0, -6.0)\end{tabular} & \begin{tabular}[c]{@{}c@{}}-22.4\\ (-39.0, -6.0)\end{tabular} & \begin{tabular}[c]{@{}c@{}}14.4\\ (9.0, 21.3)\end{tabular} & \textbf{\begin{tabular}[c]{@{}c@{}}12.8\\ (7.1, 20.1)\end{tabular}} & \textbf{\begin{tabular}[c]{@{}c@{}}12.8\\ (7.1, 20.0)\end{tabular}} & \begin{tabular}[c]{@{}c@{}}15.2\\ (9.1, 22.7)\end{tabular} & \textbf{\begin{tabular}[c]{@{}c@{}}13.6\\ (7.6, 21.0)\end{tabular}} & \textbf{\begin{tabular}[c]{@{}c@{}}13.6\\ (7.5, 20.9)\end{tabular}}\\
\cline{2 - 11}
& 2 & \textbf{\begin{tabular}[c]{@{}c@{}}-20.1\\ (-36.5, -0.3)\end{tabular}} & \begin{tabular}[c]{@{}c@{}}-22.0\\ (-37.0, -2.1)\end{tabular} & \begin{tabular}[c]{@{}c@{}}-22.0\\ (-37.0, -2.1)\end{tabular} & \begin{tabular}[c]{@{}c@{}}13.8\\ (7.4, 22.0)\end{tabular} & \begin{tabular}[c]{@{}c@{}}12.0\\ (4.8, 20.7)\end{tabular} & \textbf{\begin{tabular}[c]{@{}c@{}}11.9\\ (5.4, 20.6)\end{tabular}} & \begin{tabular}[c]{@{}c@{}}14.9\\ (8.4, 23.4)\end{tabular} & \begin{tabular}[c]{@{}c@{}}13.4\\ (5.9, 22.4)\end{tabular} & \textbf{\begin{tabular}[c]{@{}c@{}}13.3\\ (6.0, 22.3)\end{tabular}}\\
\cline{2 - 11}
& 3 & \textbf{\begin{tabular}[c]{@{}c@{}}-13.4\\ (-30.9, 2.6)\end{tabular}} & \begin{tabular}[c]{@{}c@{}}-16.2\\ (-31.5, 1.3)\end{tabular} & \begin{tabular}[c]{@{}c@{}}-16.2\\ (-31.5, 1.3)\end{tabular} & \begin{tabular}[c]{@{}c@{}}13.0\\ (7.1, 20.9)\end{tabular} & \textbf{\begin{tabular}[c]{@{}c@{}}10.5\\ (3.4, 19.3)\end{tabular}} & \begin{tabular}[c]{@{}c@{}}10.6\\ (3.6, 19.3)\end{tabular} & \begin{tabular}[c]{@{}c@{}}14.9\\ (8.9, 22.9)\end{tabular} & \begin{tabular}[c]{@{}c@{}}12.9\\ (5.4, 20.8)\end{tabular} & \textbf{\begin{tabular}[c]{@{}c@{}}12.8\\ (5.2, 20.7)\end{tabular}}\\
\cline{2 - 11}
& 4 & \textbf{\begin{tabular}[c]{@{}c@{}}-17.9\\ (-33.9, -0.7)\end{tabular}} & \begin{tabular}[c]{@{}c@{}}-19.9\\ (-36.2, -3.8)\end{tabular} & \begin{tabular}[c]{@{}c@{}}-19.9\\ (-36.2, -3.8)\end{tabular} & \begin{tabular}[c]{@{}c@{}}14.2\\ (8.2, 24.3)\end{tabular} & \textbf{\begin{tabular}[c]{@{}c@{}}12.1\\ (5.0, 21.3)\end{tabular}} & \textbf{\begin{tabular}[c]{@{}c@{}}12.1\\ (5.0, 21.3)\end{tabular}} & \begin{tabular}[c]{@{}c@{}}15.1\\ (8.5, 24.8)\end{tabular} & \textbf{\begin{tabular}[c]{@{}c@{}}13.2\\ (5.4, 22.1)\end{tabular}} & \textbf{\begin{tabular}[c]{@{}c@{}}13.2\\ (5.4, 22.0)\end{tabular}}\\
\cline{2 - 11}
& 5 & \textbf{\begin{tabular}[c]{@{}c@{}}-37.0\\ (-58.1, -18.8)\end{tabular}} & \begin{tabular}[c]{@{}c@{}}-39.0\\ (-60.3, -20.7)\end{tabular} & \begin{tabular}[c]{@{}c@{}}-39.0\\ (-60.3, -20.7)\end{tabular} & \begin{tabular}[c]{@{}c@{}}13.8\\ (7.5, 21.9)\end{tabular} & \textbf{\begin{tabular}[c]{@{}c@{}}12.0\\ (5.1, 20.5)\end{tabular}} & \textbf{\begin{tabular}[c]{@{}c@{}}12.0\\ (5.1, 20.4)\end{tabular}} & \begin{tabular}[c]{@{}c@{}}14.7\\ (8.6, 24.1)\end{tabular} & \textbf{\begin{tabular}[c]{@{}c@{}} 12.9\\ (6.3, 21.7)\end{tabular}} & \begin{tabular}[c]{@{}c@{}} 13.0\\ (6.4, 21.6)\end{tabular} \\
\bottomrule
\end{tabular}
}
\label{tab:log_like_bias}
\end{table}

Having assessed the accuracy of the stepwise procedures, we now evaluate computational efficiency. We consider $n \in \{250, 500, 1000, 2000, 4000\}$ excesses above the dependence threshold and dimensions $d \in \{5, 10, 15\}$. For each combination of sample size and dimension, a single sample is drawn from the SCMEVM. For comparison across dimensions, the number of edges in each graph is set to 60\% of the maximum possible number of edges. Inference is performed on a Dell Latitude 7,420 machine with 16GB of RAM and an $11$th generation Intel Core i5 processor with 8 cores. Figure \ref{fig:Time_Comp_and_Inferred_Graph} (left panel) shows the time taken to fit the one-, two-, and three-step SCMEVMs with graphical and saturated covariance structures. The one- and two-step methods are considerably slower as they require joint maximisation of likelihood functions over parameter spaces with a minimum dimension per conditioning variable of $6(d-1)$ (one-step) and $4(d-1)$ (two-step). Additionally, the higher the dimension of the parameter space, the harder it is to find initial values for the numerical optimisation. The three-step method is more efficient, with considerable time savings when $n$, $d$, or both are large.

\begin{table}[t!]
\centering
\caption{Comparison of the average time (seconds) to complete each step of the three-step model fitting procedure across different dimensions.}
\begin{tabular}{lccccc}
\toprule
& \multicolumn{5}{c}{Dimension} \\
\hline
Inference step & 100 & 200 & 300 & 400 & 500 \\
\hline
Dependence parameters & 1.34 & 2.07 & 2.84 & 3.65 & 4.42\\
AGG parameters & 0.94 & 1.71 & 2.65 & 3.64 & 4.61 \\
Graphical covariance structure & 0.10 & 0.81 & 3.61 & 11.06 & 30.25 \\
Saturated covariance structure & 0.03 & 0.10 & 0.22 & 0.38 & 0.60 \\
\bottomrule
\end{tabular}
\label{tab:Time_Comparison_Three_Step}
\end{table}

Figure \ref{fig:Time_Comp_and_Inferred_Graph} (left panel) suggests that for the three-step method, it may be faster to use the saturated covariance than its sparse graphical counterpart. We repeat the study for these two cases with $n = 4\times 10^3$ and $d \in \{100, 200, 300, 400, 500\}$, setting the proportion of edges in each graph to be 10\% of the maximum possible number of edges. Table \ref{tab:Time_Comparison_Three_Step} shows the average time taken to complete each inference step. As expected, the high cost is due to estimation of the graphical structure via the graphical lasso \citepmain{Friedman_2007}. In contrast, the saturated covariance is estimated empirically, i.e., no numerical maximisation is required, and the computational cost from inverting a $(d-1)$-dimensional correlation matrix is lower.

\subsection{Graphical selection} 
\label{sec:Graphical_selection}

\begin{figure}[t!]
  \centering
  \subfloat{\includegraphics[width = 0.48\textwidth]{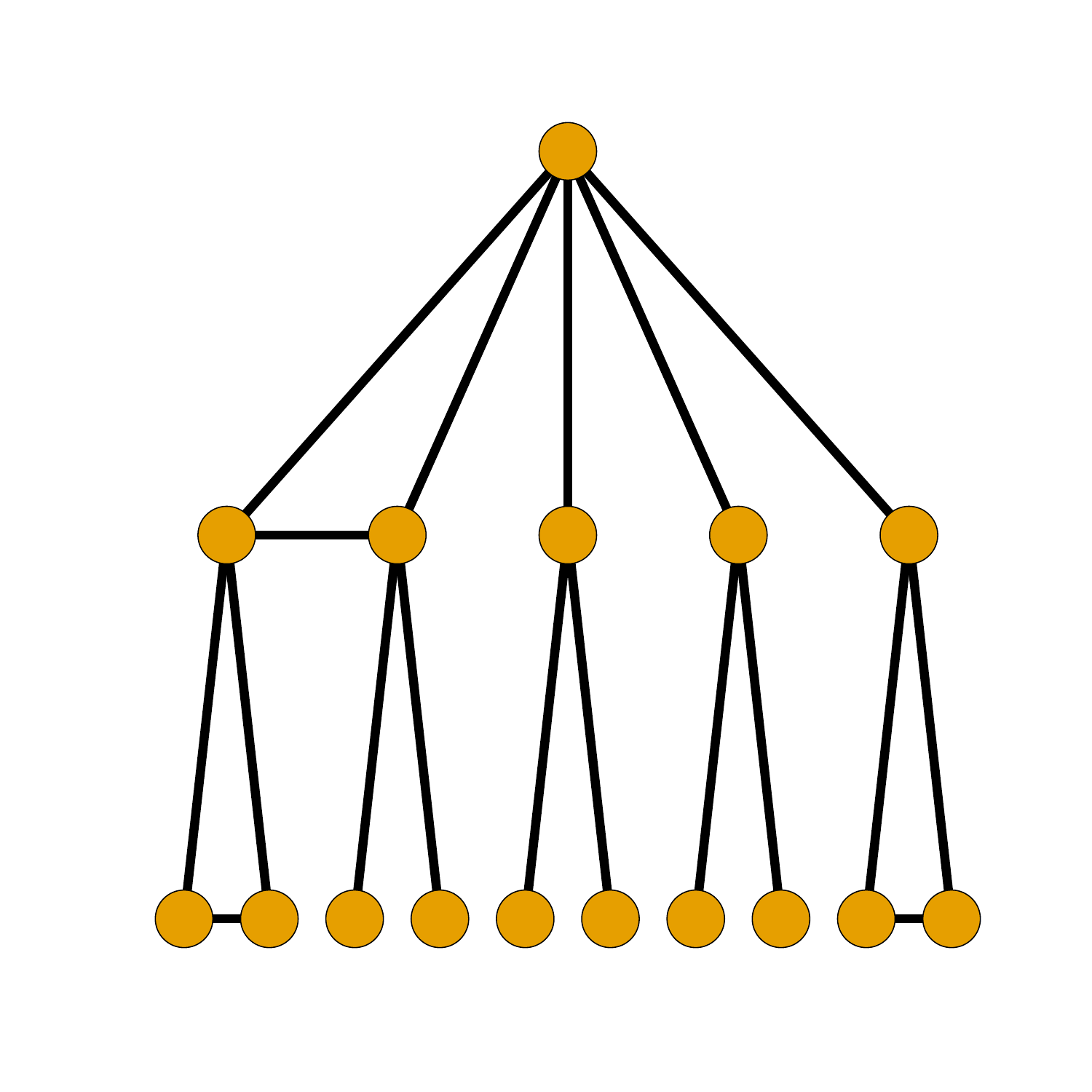}} \quad
  \subfloat{\includegraphics[width = 0.48\textwidth]{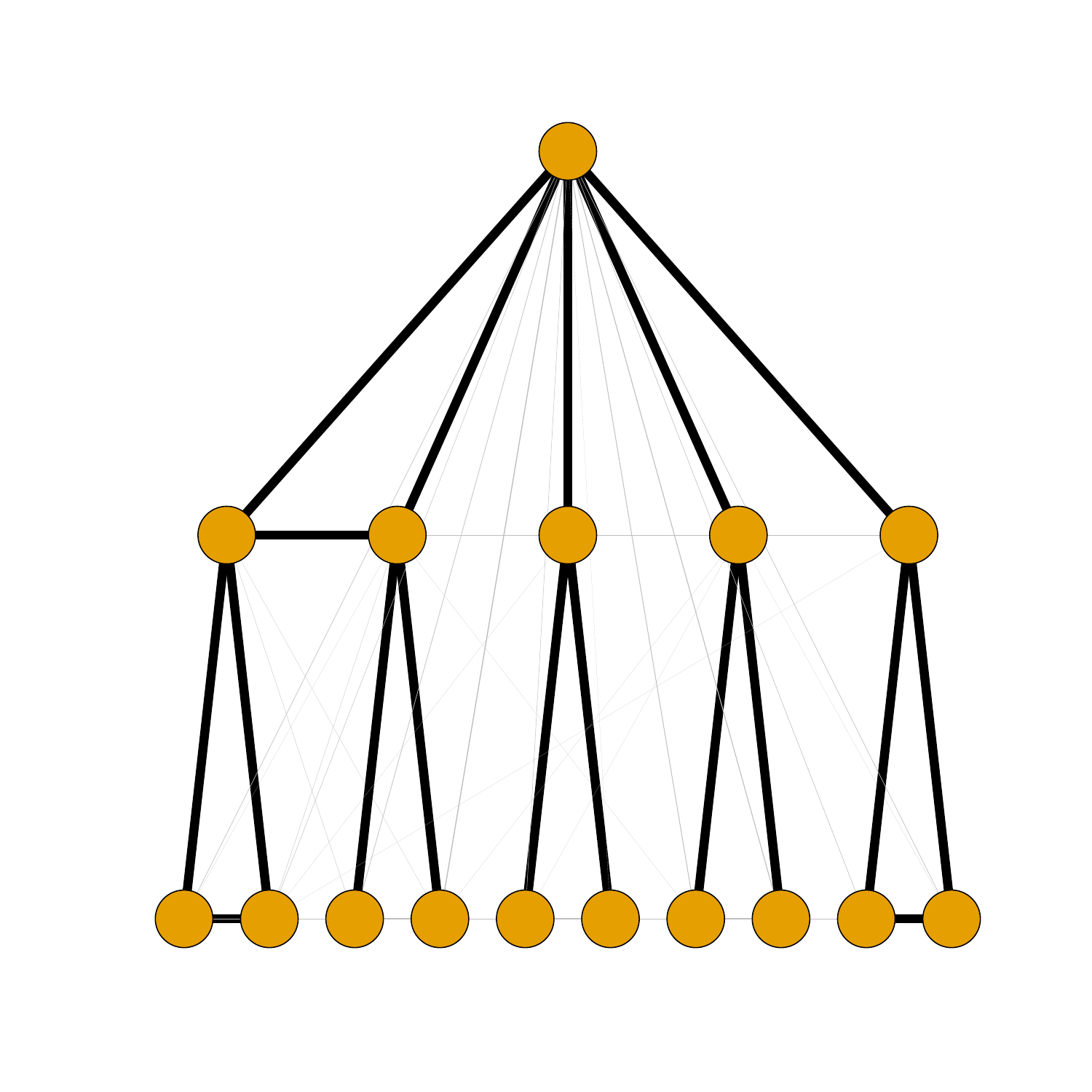}}
  \caption{True underlying graphical structure (left) and the inferred graphical structure (right), with line width and darkness indicating the number of times each edge was selected across 100 samples.}
  \label{fig:Graphical_Selection}
\end{figure}

We now replicate the simulation study of \citet[Section 5.5]{Engelke_2020} to assess how well the SCMEVM identifies a specific graphical structure. In this study, $d = 16$ and the data generating mechanism is the H\"usler-Reiss distribution with dependence structure determined by the graph $\mathcal{G}$ shown in Figure \ref{fig:Graphical_Selection} (left panel). The parameters for each of the $p = 18$ edges are sampled independently from a uniform distribution on $(0.5,1)$, subject to the constraint that the parameter matrix must be conditionally negative definite on cliques of size three. We take $100$ simulated datasets, each of size $10^3$.

Algorithm \ref{alg:cmevm_graphical_selection} is used to infer the optimal graphical structure. We take a majority rule proportion $p$ of 0.5. Setting the dependence thresholds $u_{Y_{i}}$ to the $0.90$-quantile of the standard Laplace distribution results in approximately 100 observations for inferring the graph. Figure \ref{fig:Graphical_Selection} (right panel) shows a weighted graph with line width and darkness proportional to the number of times the edge is selected across the 100 datasets. After pruning this graph using the majority rule, the true form of $\mathcal{G}$ is clearly identified.

For the graphical lasso step we set $\boldsymbol{\lambda} = \{0.4, 0.41, \dots, 0.8\}$. Across the simulated datasets, all inferred values of the penalty parameter lay between $0.61$ and $0.73$. The number of edges, equivalently the density of the dependence structure, was slightly overestimated; graphs with $\{18, 19, 20, 21, 22, 23, 24\}$ edges were selected for $\{38, 29, 14, 8, 7, 2, 2\}$ out of the 100 datasets. The results are not overly sensitive to the dependence threshold, with similar findings when using the $0.80$- or $0.95-$quantiles. Sensitivity to the majority rule proportion $p$ is also minimal (see Supplementary Material).

Despite using different underlying models and having different graphical selection processes, both our method and the \citetmain{Engelke_2020} method accurately infer the underlying graphical structure. Not unexpectedly, the \citetmain{Engelke_2020} method performs slightly better, either because their model is also the data generating mechanism \citetmain{Engelke_2020} or because the data are AD. In contrast, AD is an edge case in the SCMEVM. We could more accurately capture AD data by setting $\alpha_{j \mid i} = 1$ and $\beta_{j \mid i} = 0$, but in practice, it is usually preferable to retain the flexibility to capture both AD and AI. A second simulation study, with data generated from the asymptotically independent MVG, is provided in the Supplementary Material. The results there demonstrate that our method is more generally applicable than that of \citetmain{Engelke_2020}.

\subsection{Mixture data}
\label{sec:Simulation_mixture}
\label{sec:mixed_data}

To test the flexibility of our model, we consider data with a mixture of extremal behaviour; comparable studies for either full AI or AD can be found in the Supplementary Material. The mixture data is sampled as follows. Firstly, $(X_{1}, X_{2}, X_{3})$ are sampled from a multivariate Pareto distribution with a fully connected graph and transformed to standard Gaussian margins. Secondly, $(X_{4}, X_{5}) \mid X_{3} = x_{3}$ are sampled from a MVG distribution with a fully connected graph. Then $\boldsymbol{X}=(X_1,\ldots,X_5)$ has dependence structure consistent with $\mathcal{G}$ in Section \ref{sec:Parameter_estimation}. A total of $200$ datasets are simulated. 


Each of the one-(Algorithm~\ref{alg:cmevm}), two-(Algorithm~\ref{alg:cmevm2}), and three-step (Algorithm~\ref{alg:cmevm3}) procedures are used to fit the SCMEVM with graphical covariance. The SCMEVMs with independent and saturated covariances are only implemented for the three-step procedure. For comparison with existing methods, we fit the CMEVM as described in \citetmain{Heffernan_2004} and the graphical extremes model of \citetmain{Engelke_2020} (EHM). For the SCMEVMs with graphical covariance structure and the EHM, we use the graph $\mathcal{G}$ defined in Section \ref{sec:Parameter_estimation}.

Selection of the threshold $u_{Y_i}$ requires some consideration since subsets of the components may be fully AD, fully AI, or a mixture of both. Consequently, the rate of convergence to the limiting dependence structure varies by conditioning variable. Theoretically, it would be preferable to use a different threshold for each conditioning variable. However, since convergence rates are unknown in practice, we proceed by setting the threshold for each component to be the $0.90$-quantile of the standard Laplace distribution, giving approximately 500 excesses per conditioning variable. 

\begin{figure}[t!]
    \centering
    \subfloat{\includegraphics[width = \textwidth]{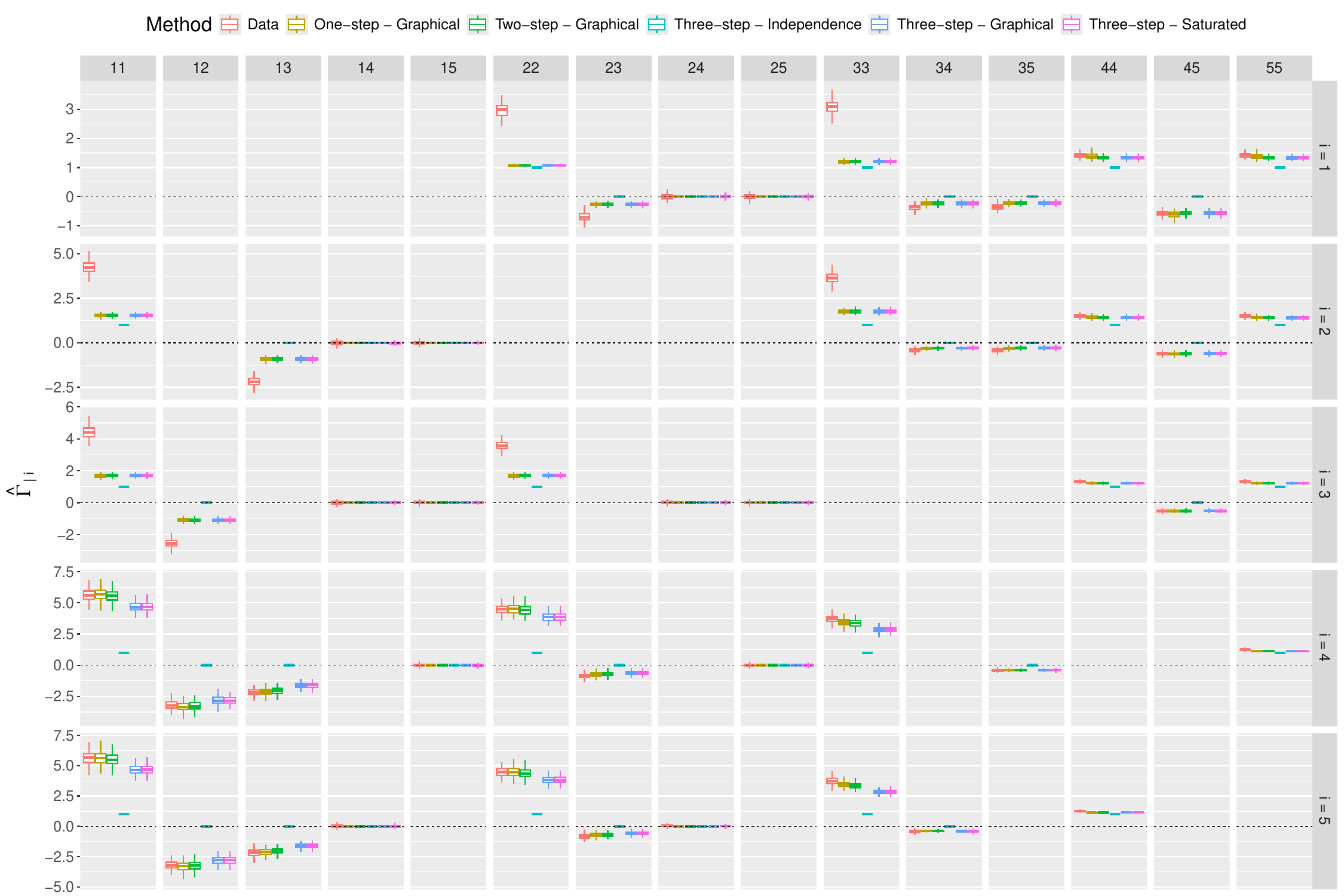}}
    \caption{Boxplots of empirical and model-based estimates of $\Gamma_{\mid i}$, for each $i \in V$, when the data is generated from a mixture distribution. Each row corresponds to the conditioning variable $i$, and each column corresponds to the correlation parameter. The colour of the boxplots distinguishes the different models. The black dashed line indicates the $y = 0$ line.}
    \label{fig:Sim_Study_Mixed_Gamma}
\end{figure}

Figure \ref{fig:Sim_Study_Mixed_Gamma} shows empirical and model-based estimates of the conditional precision matrix $\Gamma_{\mid i}$. The empirical estimates are obtained by inverting the empirical correlation matrix of $\boldsymbol{Y} \mid Y_{i} > u_{Y_{i}}$ and then excluding the $i$th row and column. We first compare the precision matrix estimates across models; the estimates are almost identical for the graphical and saturated SCMEVMs, confirming there is negligible loss in using the former. The magnitudes of the non-zero entries in the residual precision matrices cannot be directly compared with their empirical equivalents due to the non-linear transformation~\eqref{eqn:WT_2022_Limit}. We can compare the location of the zero entries and find that the graphical and saturated SCMEVM estimates are consistent with their empirical equivalents. This suggests that the graphical structure of the residual distribution is inherited from the underlying multivariate distribution, a result consistent with the theoretical findings of \citetmain{Casey_2023} and similar simulation studies for cases with full AI or full AD (see Supplementary Material). 

Figure \ref{fig:Sim_Study_Mixed_Bias_Probs} shows the bias in the estimates for the tail probabilities $p_1=\mathbb{P}[X_{1} > v_{1}, X_{2} > v_{2} \mid X_{3} > u_{3}]$ and $p_2=\mathbb{P}[X_{3} > v_{3}, X_{4} > v_{4} \mid X_{5} > u_{5}]$, where $u_{i}$ and $v_{i}$ are the $0.90$-quantile and $0.95$-quantile of $X_{i}$, respectively. Quantiles are obtained from a sample of size $10^6$ from the true distribution. Estimates from the CMEVM and the SCMEVM with graphical or saturated covariances are unbiased for $p_{1}$ and exhibit a small positive bias for $p_{2}$. The EHM estimates of $p_1$ are similarly unbiased. However, the estimates of $p_2$ are considerably more positively biased compared to (S)CMEVM models. These patterns are consistent across estimates of the remaining 73 conditional probabilities of the form $\mathbb{P}[\boldsymbol{X}_{A} > v \mid X_{i} > v]$ for $A \subseteq V_{\mid i}$ and $i \in V$. When conditioning on sites 4 and 5, estimates from the CMEVM and SCMEVM always exhibit some small positive bias. Since the bias is consistent across the two models, we believe this is due to the difference in the rate at which the different components of the generating distribution converge to the tail distribution. This bias may be reduced if lower dependence thresholds were used for these sites. 

Finally, we consider the mean absolute error (MAE) and root mean square error (RMSE) of the estimates for all 75 conditional tail probabilities. The SCMEVM outperforms both the EHM and CMEVM, minimising the MAE for 73\% of the probabilities and the RMSE for 76\% of them. In contrast, the CMEVM (EHM) minimises the MAE 15\% (12\%) and the RMSE 12\% (12\%) of the time. This suggests that even for moderate $d$, the SCMEVM has greater predictive precision and accuracy than competitor methods.

\begin{figure}[t!]
    \centering
    \includegraphics[width=0.48\textwidth]{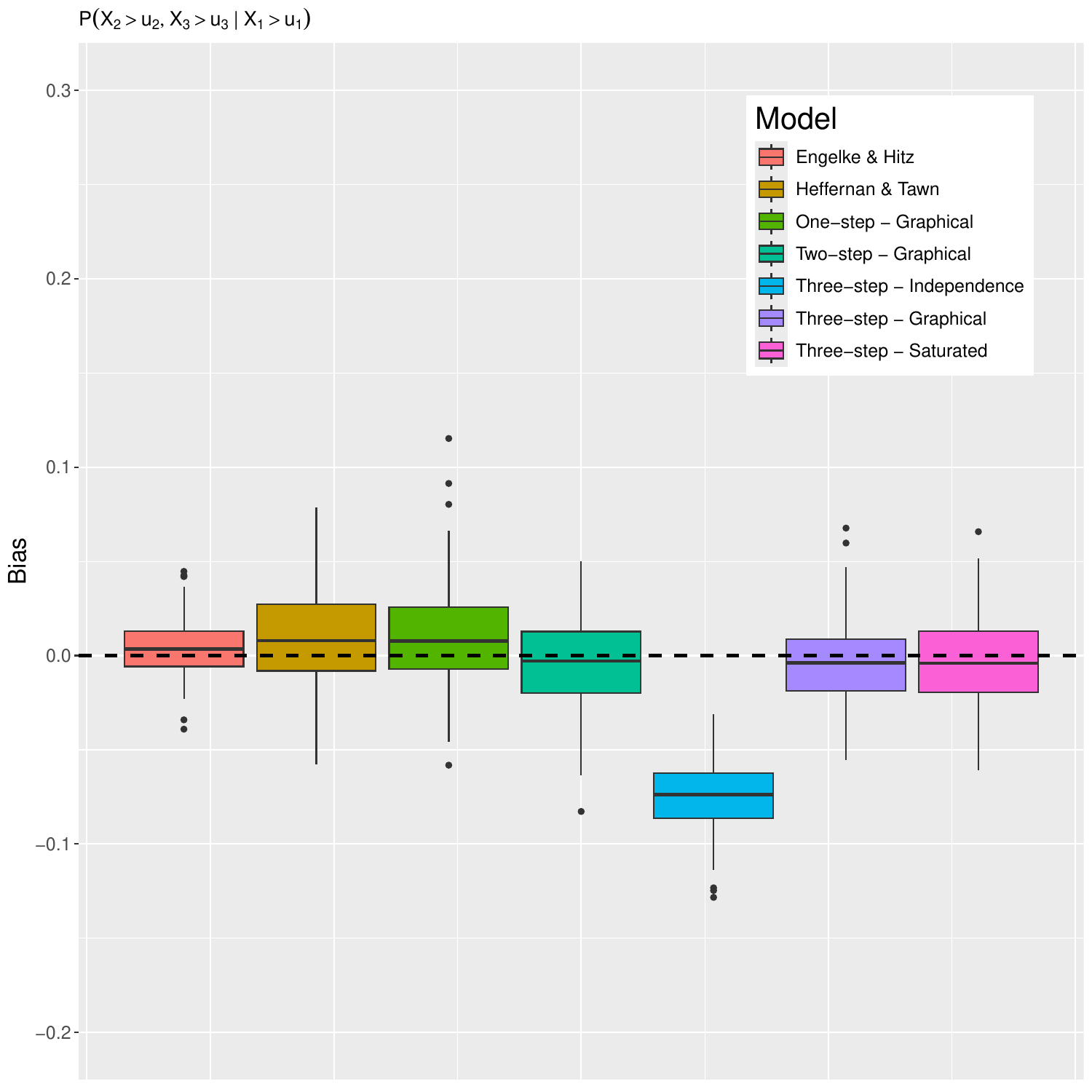} \quad
    \includegraphics[width=0.48\textwidth]{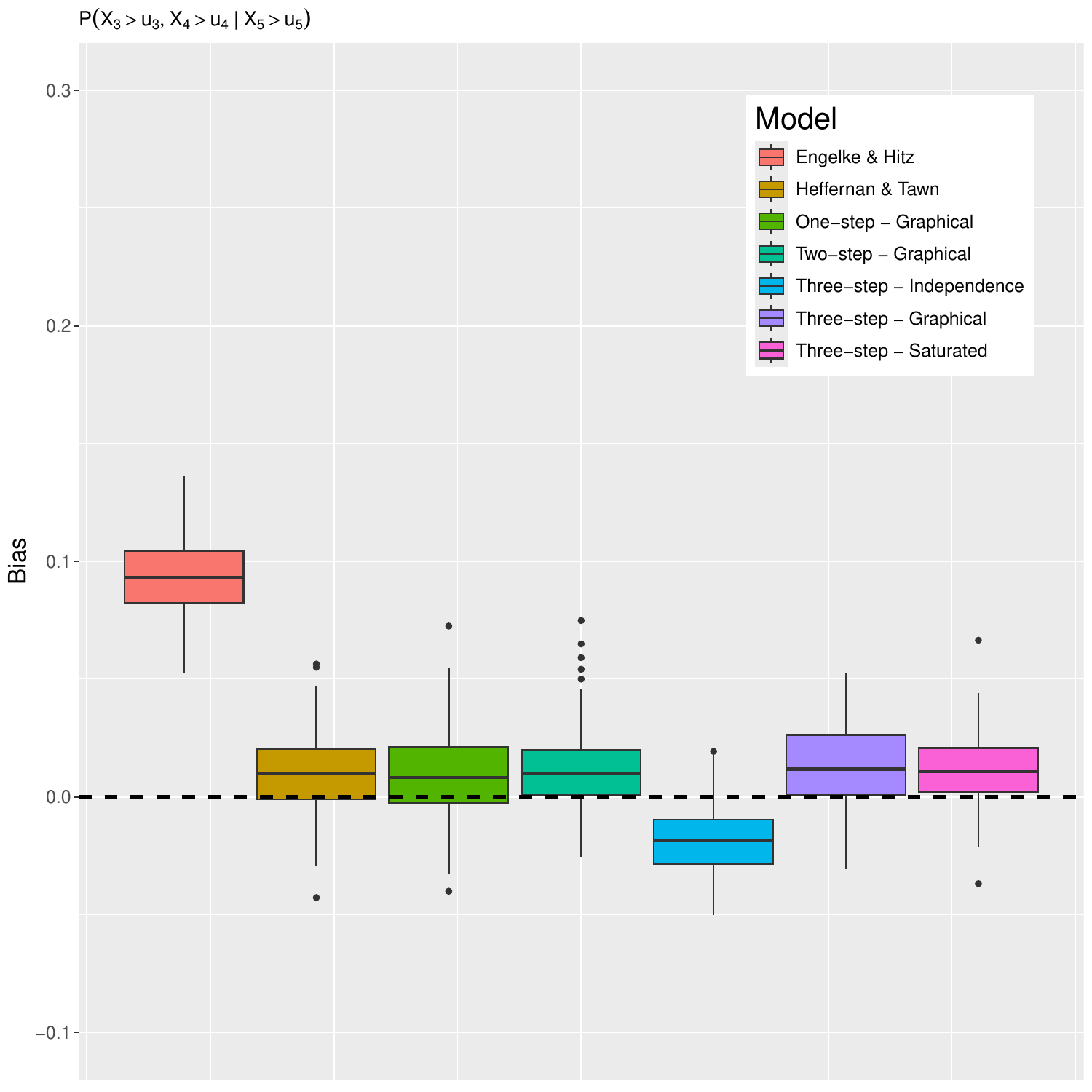}
    \caption{Boxplots of the bias in $p_1=\mathbb{P}[X_{1} > v_{1}, X_{2} > v_{2} \mid X_{3} > u_{3}]$ (left) and $p_2=\mathbb{P}[X_{3} > v_{3}, X_{4} > v_{4} \mid X_{5} > u_{5}]$ (right). The fill of the boxplots distinguishes the different models. The black dashed line indicates the $y = 0$ line.}
    \label{fig:Sim_Study_Mixed_Bias_Probs}
\end{figure}
    
\section{Application} 
    \label{Section:Application}
    We apply our model to discharge data from the upper Danube River basin described in Section \ref{Section:Introduction}. For the margins, we use the empirical-generalised Pareto distribution model from Section \ref{Section:Inference}. The dependence threshold $u_{Y_{i}}$ used is the $0.80$-quantile of the standard Laplace distribution for all $i \in V$, resulting in around 85 excesses per station.

\begin{figure}[t!]
    \centering
    \subfloat{\includegraphics[width=.48\textwidth]{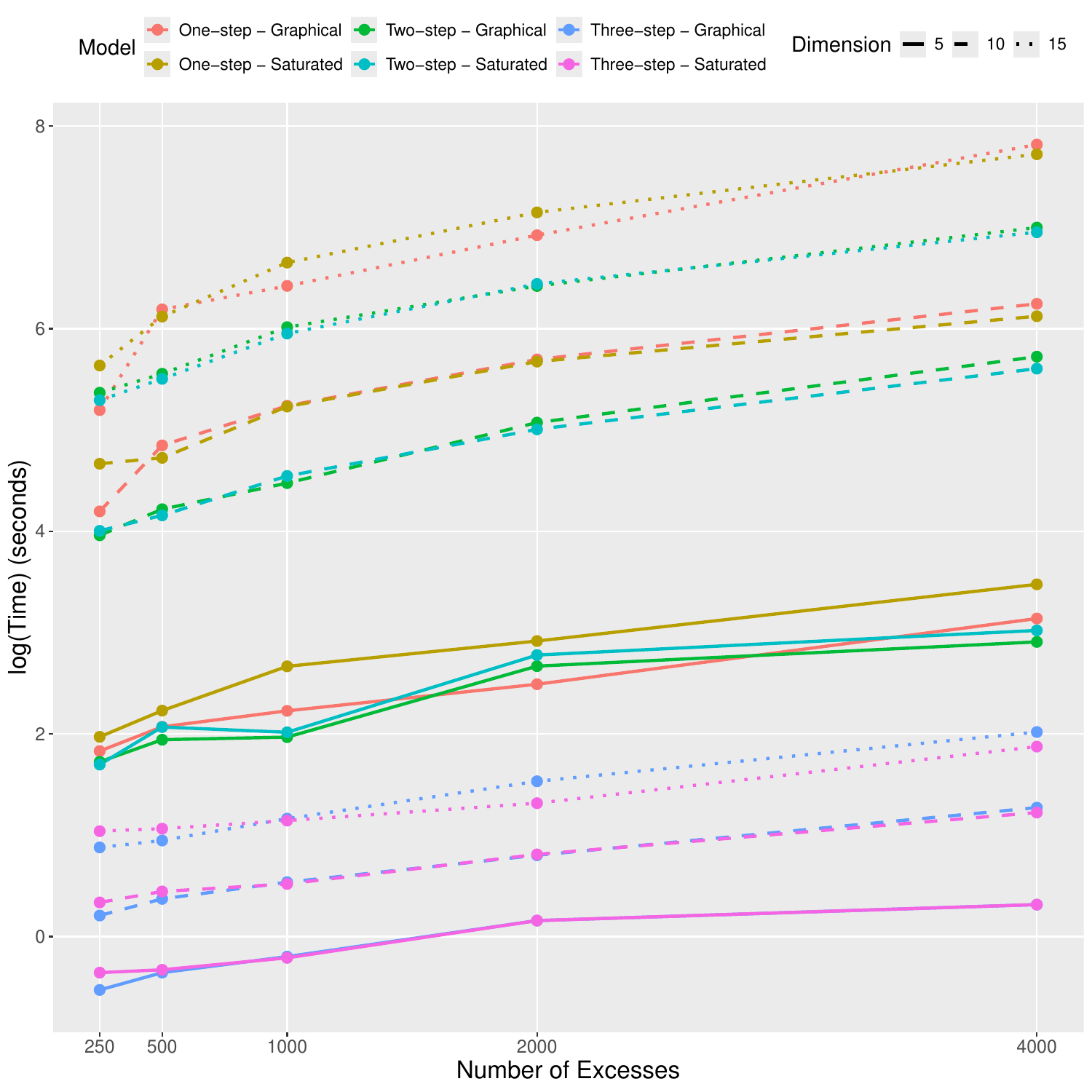}} \quad
    \subfloat{\includegraphics[width=.48\textwidth]{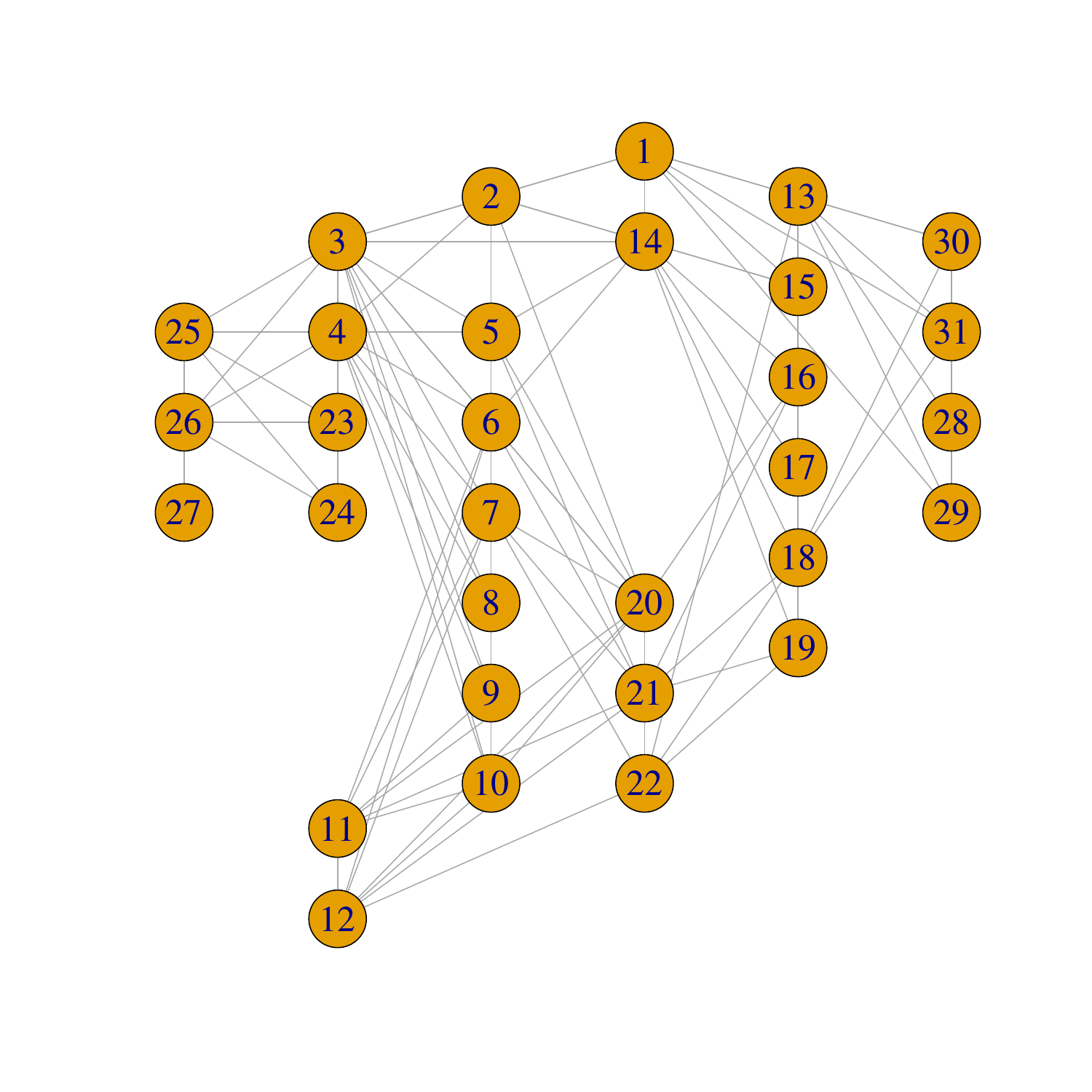}}
    \caption{Timing comparison (log scale) of the SCMEVMs (left) for various sample sizes, dimensions, denoted by the line type, and models, denoted by the line colour. Inferred graphical structure of the upper Danube River basin using Algorithm \ref{alg:cmevm_graphical_selection} (right).}
    \label{fig:Time_Comp_and_Inferred_Graph}
\end{figure}

We use the three-step procedure to fit the SCMEVM with graphical covariance structure given by the undirected tree induced by the flow connections of the river (Figure \ref{fig:Danube_River_EDA}, left panel); we also fit the EHM with the same structure. Since the flow connection tree may not be the optimal structure for describing the dependence structure of the extremes, we also apply the three-step procedure to fit both the SCMEVM with saturated covariance and the SCMEVM with an inferred graphical structure. The inferred graphical structure, obtained from Algorithm \ref{alg:cmevm_graphical_selection} is shown in Figure \ref{fig:Time_Comp_and_Inferred_Graph} (right panel). This graph has 127 edges compared to the maximum possible 465 edges. Lastly, for prediction, we simulate datasets from the fitted models using Algorithm \ref{alg:conditional_survival_set} with $N = 20n$ and $u = u_{Y_{i}}$.

\begin{figure}[t!]
    \centering
    \includegraphics[scale = 0.35]{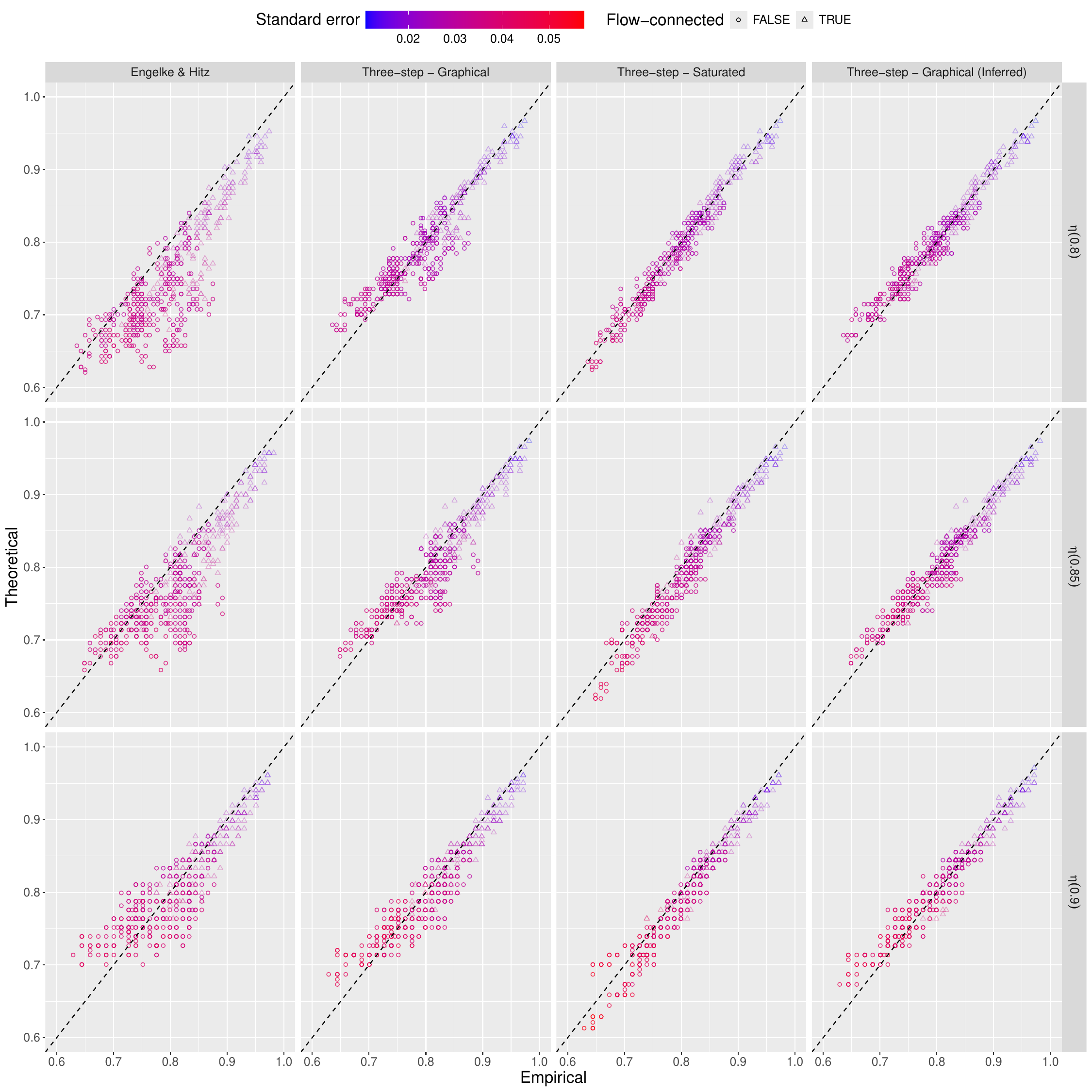}
    \caption{Empirical and model-based estimates of $\eta_{i,j}(u)$ for $u \in \{0.8, 0.85, 0.9\}$ (top to bottom), and $i, j \in V$ but $i > j$. Model-based estimates use the EHM (an AD based model) (left) and the three-step SCMEVM (an AI based model) with graphical covariance (centre left), both with structure given in Figure \ref{fig:Danube_River_EDA} (left panel), the three-step SCMEVM with saturated covariance (centre right) and graphical covariance (right) with structure given in Figure \ref{fig:Time_Comp_and_Inferred_Graph} (right panel). Black dashed lines show $y = x$. Circles (triangles) show flow-connected (flow-unconnected). The colour shows the standard error of the model-based estimates.}
    \label{fig:Danube_Eta_Comp}
\end{figure}

Bootstrapped estimates of the coefficient of tail dependence $\eta_{i,j}(u)$ for  $i, j \in V$, $i > j$, and $u \in \{0.8, 0.85, 0.9\}$, are obtained using 200 bootstrapped samples of the data. For each bootstrapped dataset, both empirical and model-based estimates of $\eta_{i,j}(u)$ are obtained. The point estimates in Figure \ref{fig:Danube_Eta_Comp} are the medians of the two sets of estimates. The SCMEVMs describe the empirical dependence better than the EHM for both flow-connected (triangles) and flow-unconnected (circles) stations, and across all values of $u$. This highlights the value of a model that captures a range of extremal dependence classes. As noted, numerous extensions to the EHM have been proposed \citepmain{Engelke_2024_B} that allow for any sparse graphical structure. Therefore, we compare both the learnt graphical structure and the predictive performance of the learnt model, for the SCMEVM and EGlearn \citepmain{Engelke_2025} in the Supplementary Material.

Figure \ref{fig:Danube_Eta_Comp} shows that all pairs of stations appear to exhibit AI with positive association, $\eta(u)\in(0.5,1)$, while analogous plots of $\chi(u)$ (see Supplementary Material) imply all pairs have AD. However, for stations with weaker (stronger) extremal dependence, the two measures decrease (stay close to 1) as $u$ increases. This supports the plausible conclusion that some pairs of stations, particularly those that are flow-unconnected, exhibit AI while others, particularly those that are flow-connected, exhibit AD.

\begin{figure}[t!]
    \centering
    \includegraphics[scale = 0.35]{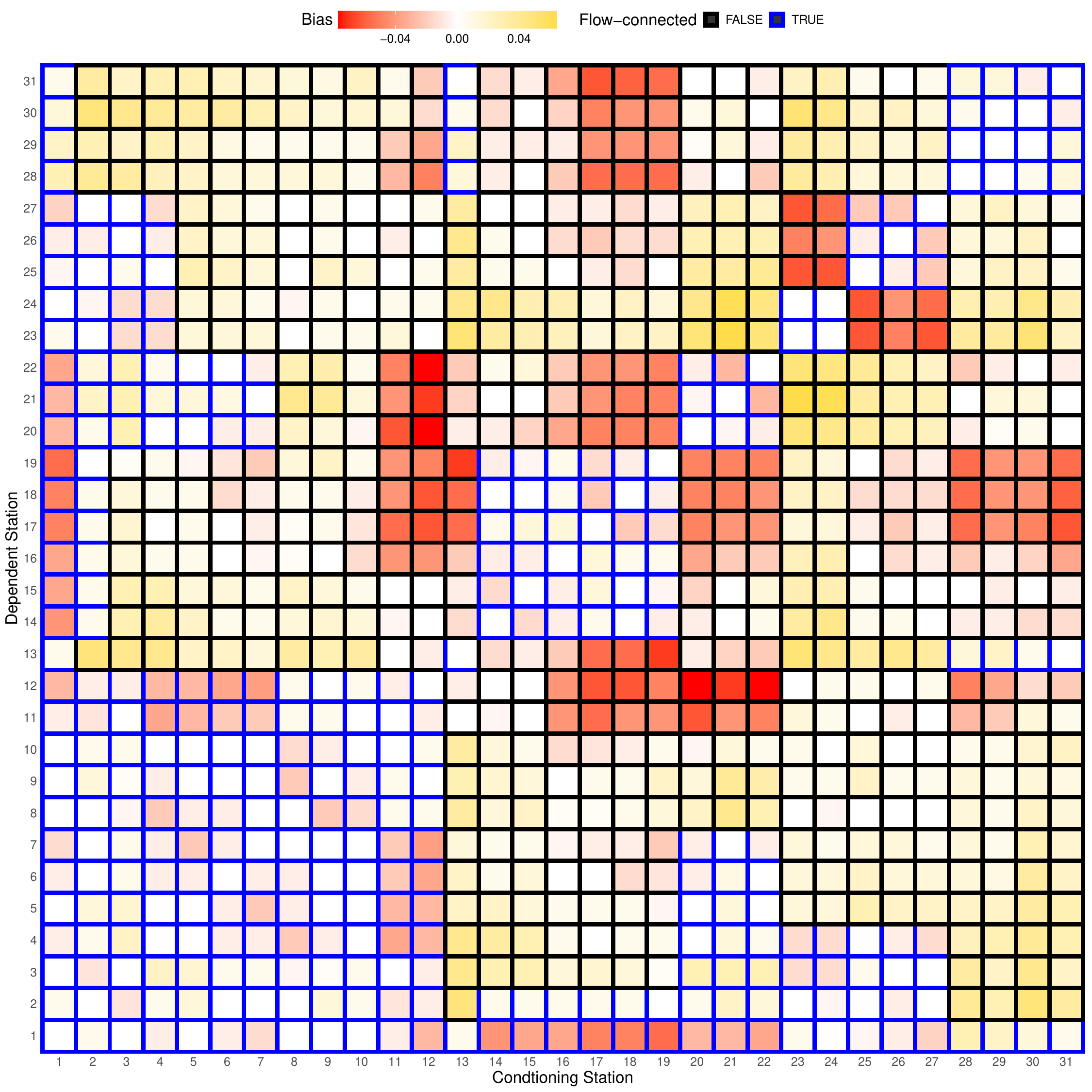}
    \caption{Difference between empirical and median of model-based estimates of $\eta_{i,j}(0.8)$ for each $i,j \in V$ for the SCMEVM with a graphical covariance, where the graphical structure is assumed to be the undirected tree induced by the flow connections in Figure \ref{fig:Danube_River_EDA} (left panel). Under- and over-estimation from the model is represented by red and gold squares, respectively. Flow-connected and flow-unconnected stations are represented by blue and black borders, respectively.}
    \label{fig:Danube_Eta_080_Bias}
\end{figure}

For $u=0.8$, the SCMEVM with saturated covariance performs noticeably better than the graphical SCMEVM with structure given by the undirected tree induced by the flow connections. This suggests that the extremal dependence is influenced by factors beyond the river structure (see also \citetmain{Asadi_2015}). To support this hypothesis, Figure \ref{fig:Danube_Eta_080_Bias} shows the difference in the empirical and model-based estimates of $\eta_{i,j}(0.8)$ for each pair $i,j \in V$, where the model-based estimates are from the SCMEVM with graphical covariance using the induced undirected tree. We observe that underestimation predominantly occurs for flow-unconnected stations. For example, the dependence between stations 11-12 and 16-22 is considerably underestimated. While these two sets of stations are flow-unconnected, the sources of their tributaries are geographically close and at similar altitudes, thus, stronger dependence than suggested by the lack of flow connection is not unexpected. Similar observations are made when comparing the Isar (stations 14 - 19) and Salzach (stations 28 - 31) tributaries, as well as stations 23 - 24 and 25 - 27. Furthermore, the inferred graphical structure in Figure \ref{fig:Time_Comp_and_Inferred_Graph} (right panel) shows many connections between sites on geographically neighbouring tributaries. Indeed, using the inferred graphical structure in the SCMEVM drastically improves the model fit (Figure \ref{fig:Danube_Eta_Comp}, right panel), and resolves the systematic underestimation caused by using a saturated covariance structure.

Returning to Figure \ref{fig:Danube_Eta_Comp}, we observe that the SCMEVMs underestimate dependence at higher thresholds, particularly for flow-unconnected stations with weaker associations. In contrast, the EHM becomes less biased as the threshold increases, although its bias at lower thresholds is larger than the bias in the SCMEVMs at higher thresholds. Moreover, the cross-site variability in the model-based estimates is higher for the EHM than the SCMEVMs, regardless of level, particularly for stations with weaker associations. In conclusion, the SCMEVMs more accurately and consistently represent the extremal dependence in the data than the EHM. 

\section{Discussion} 
    \label{Section:Discussion}
    In this paper, we have extended the conditional multivariate extreme value model \citepmain{Heffernan_2004} by replacing the non-parametric residual distribution with a flexible, fully parametric model. This overcomes the curse of dimensionality that arises when extrapolating from a non-parametric estimate of a high-dimensional distribution, resulting in more accurate and reliable predictions in high dimensions. Our proposed parametric model is the MVAGG distribution. The copula-based construction of the MVAGG, which combines asymmetric generalised Gaussian margins with a multivariate Gaussian dependence structure, facilitates efficient statistical inference, as the margins and dependence structures can be inferred separately in a stepwise manner. Further, separate estimation of the marginal and dependence parameters for the MVAGG is computationally efficient and loses no information compared to joint estimation of all parameters. 

To reduce the parameter space, we propose using a graphical structure to induce sparsity into the precision matrix of the MVG copula. In addition to reducing the number of unknown parameters to be estimated, this provides a mechanism to infer the dependence structure if it is not already known. 
Despite the sparsity induced by the graphical structure, model fitting is substantially more expensive than using a saturated covariance structure due to the required numerical optimisation. Therefore, while graphical structures can be learnt and implemented, the model in its current form may not be suitable for very large dimensions and further work is needed to address this computational hurdle. 

Our analysis of the upper Danube River basin dataset suggests the SCMEVM captures the dependence between stations more effectively than the graphical extremes model of \citetmain{Engelke_2020}, highlighting the benefit of a model that can capture both AD and AI. Furthermore, the SCMEVM, based on the undirected tree inferred from the flow connections of the river network, does not perform as well as either the graphical structure inferred from the graphical lasso or the saturated covariance matrix, highlighting the complex dependence structure in the data that is not solely captured by the data's underlying graphical structure. Thus, a possible alternative that incorporates the river network structure would be to add a second covariance matrix based on Euclidean distance \citepmain{Asadi_2015} into the MVG copula component of the SCMEVM. 

Finally, as with the graphical extremes model of \citetmain{Engelke_2020}, our model only allows predictions at measured locations. Parameterising the Gaussian copula kernel with a Mat\'ern or Whittle-Mat\'ern correlation function \citepmain{Bolin_2023}, where distance is measured along the graphical structure, would allow extrapolation to unobserved locations. The generally strong correspondence between the empirical and model-based estimates of $\eta$ for the flow-connected sites from the SCMEVM provides confidence that such a model would result in reliable extrapolations.

\section*{Declarations}

\subsection*{Funding}
Aiden Farrell was supported by the Engineering and Physical Sciences Research Council (EPSRC) grant EP/W523811/1.

\subsection*{Code}
Code that supports our findings can be found in the \href{https://github.com/AidenFarrell/Graphical_Conditional_Extremes}{AidenFarrell/Graphical\_Conditional\_Extremes} repository on GitHub.

\bibliographystylemain{apalike}
\bibliographymain{Library.bib}

\pagebreak

\section*{Supplementary Material to ``Conditional Extremes with Graphical Models"}

\setcounter{section}{0}
\setcounter{figure}{0}

\renewcommand{\thefigure}{S\arabic{figure}}
\renewcommand{\thetable}{S\arabic{table}}
\renewcommand{\theequation}{S.\arabic{equation}}
\renewcommand{\thesection}{S.\arabic{section}}

\section{Prediction from the conditional multivariate extreme value model}
\label{Sec:Pred_CMEVM}
The original conditional multivariate extreme value model (CMEVM) uses a semi-parametric algorithm for prediction to avoid over-reliance on the working distributional assumptions used for parameter estimation (see Section 2.1 of the main text). Specifically, prediction is performed by non-parametrically sampling with replacement from the empirical distribution of the fitted residuals. Such sampling suffers from the curse of the dimensionality \citepsupp{Nagler_2016_supp}, meaning the predictive performance of the CMEVM decreases as the dimension increase. To demonstrate this, we perform a simple simulation study and compare the predictive performance of the CMEVM to the structured CMEVM (SCMEVM) proposed in Section 2 of the main text.

Consider a simple undirected graph $\mathcal{G} = (V, E)$ with vertex set $V = \{1, \hdots, d\}$ and edge set $E \subseteq \{ \{j,k\} \mid j,k \in V, j \neq k \}$. We set $d = 20$ and randomly select the edges in the graph such that the number of edges is approximately $20$\% of the number of edges in the full graph. We simulate 200 datasets of size $250$ for $\boldsymbol{Y} \mid Y_{i} > u_{Y_{i}}$ as per Section 4.1 of the main text i.e., we simulate $Y_{i} | Y_{i} > u_{Y_{i}}$ from a standard exponential distribution and obtain $\boldsymbol{Y}_{\mid i} \mid Y_{i} > u_{Y_{i}}$ using equation (2.3) of the main text with $\boldsymbol{Z}_{\mid i}$ simulated from a multivariate asymmetric generalised Gaussian (MVAGG) distribution (Section 2.2 of the main text). We set the dependence threshold $u_{Y_{i}}$ to the $0.8$-quantile of the standard Laplace distribution. True dependence and asymmetric generalised Gaussian (AGG) parameters are independently sampled from uniform distributions on $(0.1, 0.3)$ for $\alpha_{j}$, $(0.1, 0.2)$ for $\beta_{j}$, $(-1, 1)$ for $\nu_{j}$, $(0.5, 1)$ for $\kappa_{1_{j}}$, $(1.5, 2)$ for $\kappa_{2_{j}}$, and $(0.8, 2.5)$ for $\delta_{j}$, for each $j \in V$. 

For computational purposes, we consider a single conditioning component $i$ selected at random from $V$; similar results can be obtained when conditioning on different components. Predictive performance is assessed on Laplace margins only since the probability integral transform (PIT) used to back-transform to the original margins does not alter the dependence structure. For each data set, we fit the (i) CMEVM, (ii) three-step SCMEVM with graphical covariance structure, and (iii) three-step SCMEVM with saturated covariance. For (ii), the graph is assumed to be known and correctly specified above. For prediction, we used datasets of size $5 \times 10^6$ for $\boldsymbol{Y} \mid Y_{i} > u_{Y_{i}}$ simulated from the fitted models using the methods described in Section 4.4 of \citetsupp{Heffernan_2004_supp} for (i) and Algorithm 1 in Section 5.1 of \citetsupp{Wadsworth_2022_supp} for (ii) and (iii). 

\begin{figure}[t!]
    \centering
    \includegraphics[width = \textwidth]{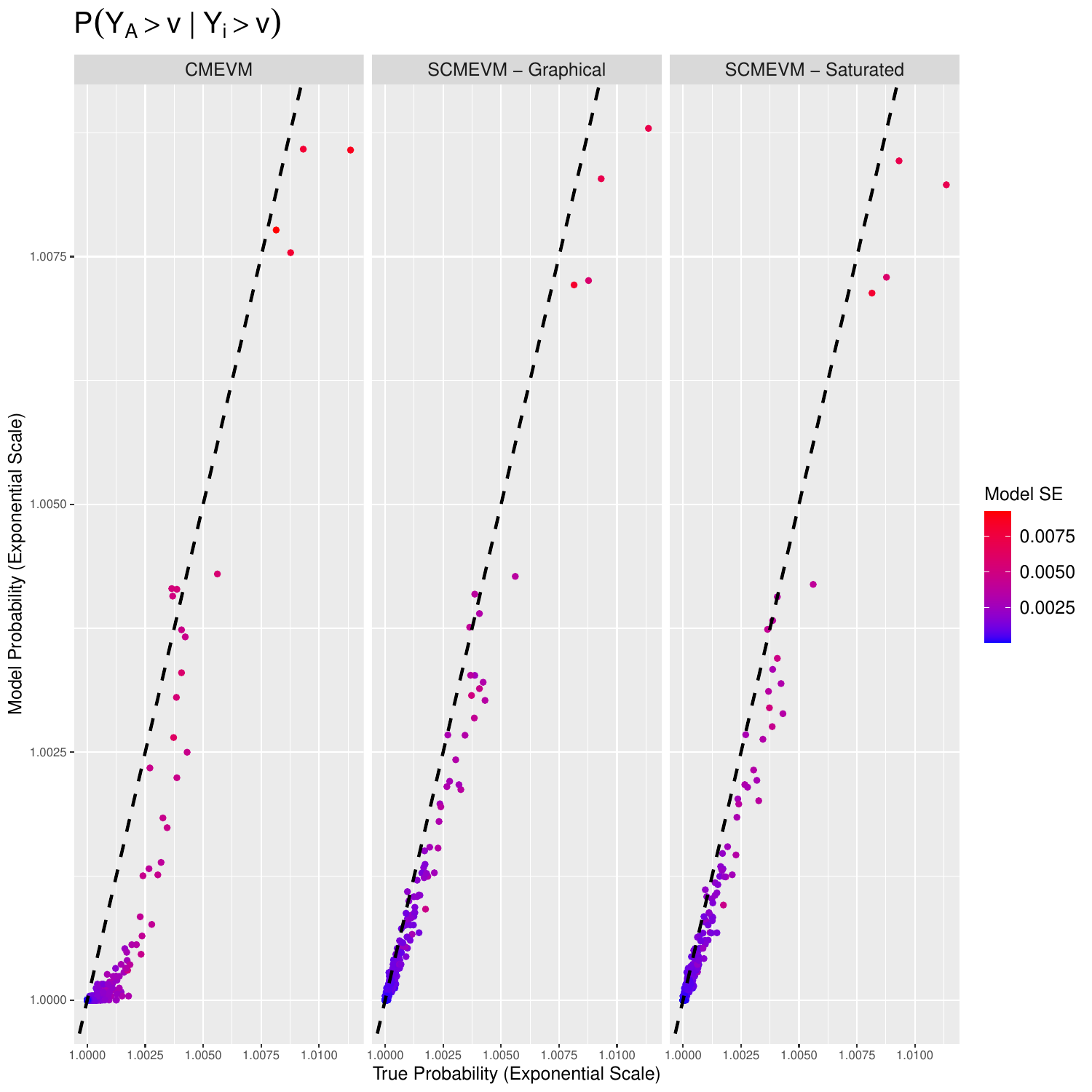}
    \caption{Empirical and model-based estimates of $\mathbb{P}[\boldsymbol{Y}_{A} > v \mid Y_{i} > v]$ for a randomly selected component $i \in V$, 500 randomly selected sets $A \subseteq V_{\mid i}$ such that $\lvert A \rvert = 3$, and $v$ is the $0.999$-quantile of the standard Laplace distribution. Model-based estimates use the CMEVM (left), the three-step SCMEVM with graphical covariance (centre) with structure described in Section \ref{Sec:Pred_CMEVM}, and the three-step SCMEVM with saturated covariance (right). The colour shows the standard error of the model-based estimates. Black dashed lines show the $y = x$.}
    \label{fig:Prob_Comp}
\end{figure}

Figure \ref{fig:Prob_Comp} shows $\mathbb{P}[\boldsymbol{Y}_{A} > v \mid Y_{i} > v]$, on the exponential scale, for 500 different sets $A \subseteq V_{\mid i} = V\backslash\{i\}$ such that $\lvert A \rvert = 3$; the sets were chosen at random. We set $v$ to be the $0.999$-quantile of the standard Laplace distribution, which would approximately correspond to a 1 in 10 year event if we had daily data. The truth is obtained empirically using a single sample of size $10^7$ from the true distribution, while the model-based estimates are the median of the model-based point estimates from each of the 200 samples. The CMEVM consistently underestimates the probabilities, whereas the SCMEVMs perform much better, particularly for those probabilities that are very close to 0. The standard error of the model-based estimates (on the original scale) appear to be lower for the SCMEVMs compared to the CMEVM. The SCMEVMs do slightly underestimate the probabilities, however, we anticipate this could be resolved by increasing the size of the prediction datasets.

\begin{figure}[t!]
    \centering
    \includegraphics[width = \textwidth]{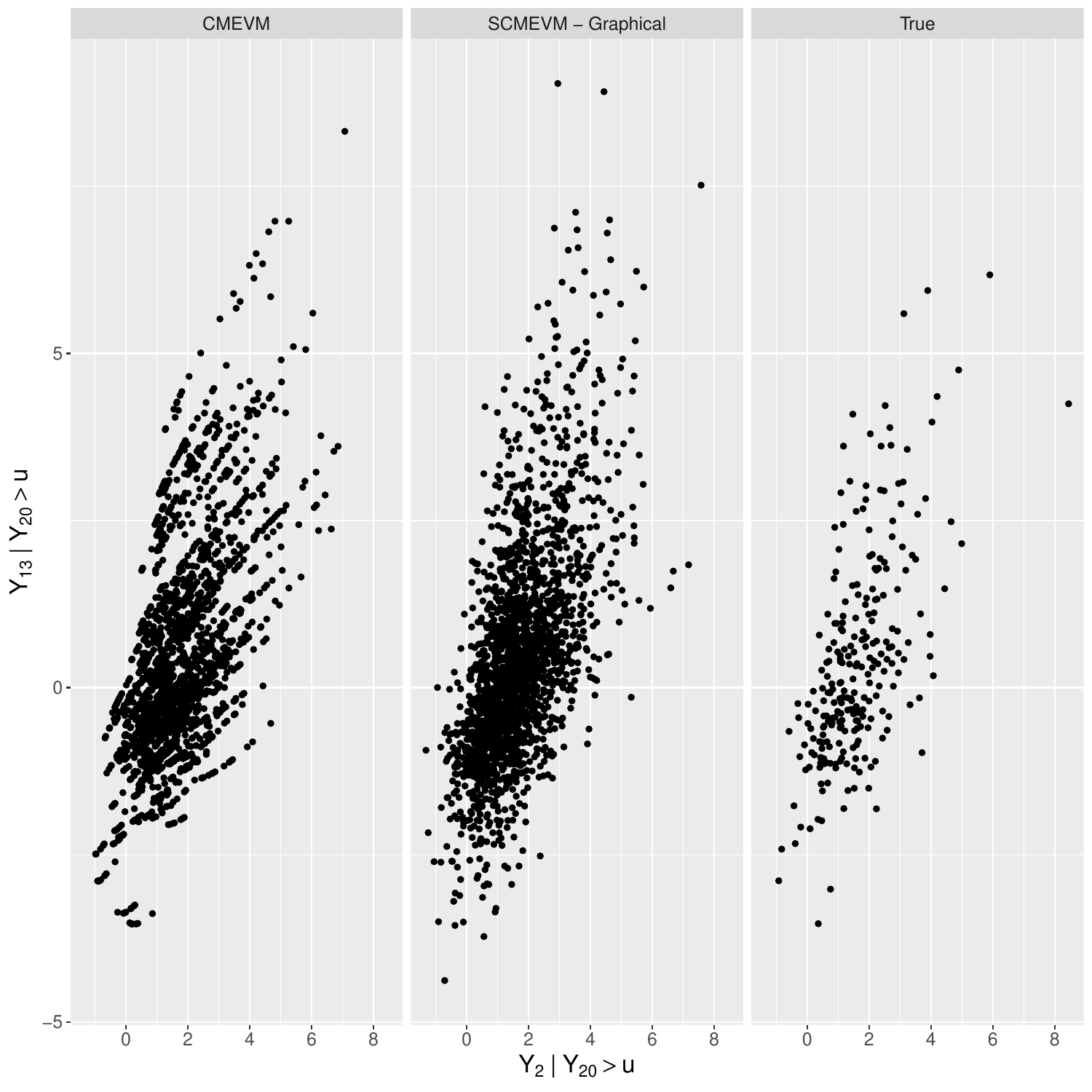}
    \caption{Scatter plots for $Y_{2}$ and $Y_{13}$ given $Y_{20} > u_{Y_{20}}$. The points correspond to $2,000$ randomly selected data points from a sample of size $5 \times 10^6$ simulated from the fitted model for the CMEVM (left), and the three-step SCMEVM with graphical covariance (centre) with structure described in Section \ref{Sec:Pred_CMEVM}. Also shown are the $250$ points used to fit the models (right).}
    \label{fig:Prediction_Sample_Cloud}
\end{figure}

The CMEVM underestimates the probabilities because it allows neither interpolation nor extrapolation of the fitted residuals, resulting in \textquotedblleft rays\textquotedblright \;in data simulated from the fitted model. This can be seen in Figure \ref{fig:Prediction_Sample_Cloud}, which shows $2,000$ randomly selected points from data simulated from the fitted models for (i) and (ii). Data used to fit the models is also shown. The CMEVM does not accurately capture dependence between components 2 and 13, but the SCMEVM does much better. This pattern will only be exacerbated as the dimension increases. Therefore, the predictive power of the CMEVM will diminish as (1) $v$ approaches the upper end-point of the distribution, (2) the size of set $A$ increases, and (3) the dimension of the problem increases. The SCMEVM overcomes such limitations by using a flexible, fully parametric distribution for the residuals.

\section{Marginal distributions for the residuals}
\label{Sec:Pred_MVAGG}
In Section 2.2 of the main text, we propose an adaptation of the generalised Gaussian model used by \citet{Wadsworth_2022_supp}. Here we show that the proposed alternative, the asymmetric generalised Gaussian distribution, improves the overall model fit. For $d = 20$ and a graph $\mathcal{G} = (V, E)$ with edges randomly selected such that the number of edges is approximately $20$\% of the number of edges in the full graph, we simulate $200$ datasets of size $5000$ from a multivariate Gaussian (MVG) distribution with mean vector $\boldsymbol{\mu}$ and correlation matrix $\Sigma$. The components $\mu_{j}$ are independently sampled from a uniform distribution on $(-5,5)$, and the correlation matrix is associated with $\mathcal{G}$. For each replicate $\boldsymbol{X}$, we transform the margins onto standard Laplace margins $\boldsymbol{Y}$, as per Section 3.1 of the main text, before fitting the SCMEVM. We set the dependence thresholds $u_{Y_{i}}$ to the 0.90-quantile of the standard Laplace distribution. 

\begin{figure}[t!]
    \centering
    \includegraphics[width = \textwidth]{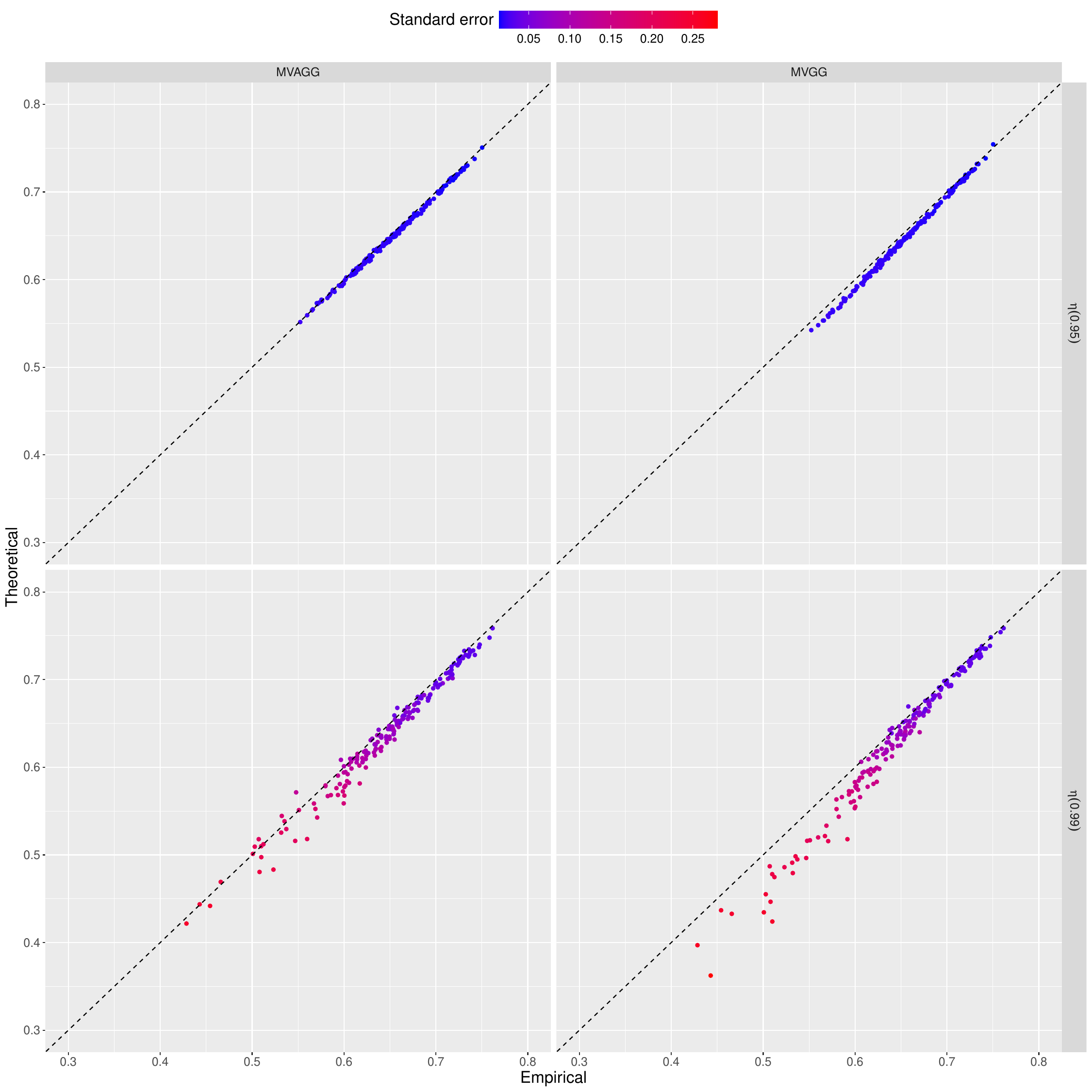}
    \caption{Empirical and model-based estimates of $\eta_{i,j}(u)$ for $u \in \{0.95, 0.99\}$ (top to bottom), and $i, j \in V$, but $i > j$. Model-based estimates use the three-step SCMEVM with residuals having a saturated covariance and either asymmetric generalised Gaussian (left) and generalised Gaussian (right) margins. Black dashed lines show $y = x$. The colour shows the standard error of the model-based estimates.}
    \label{fig:AGG_GG_ETA_Comp}
\end{figure}

To fit the SCMEVM, we use the three-step (Algorithm 3.4 of the main text) procedure and assume a saturated covariance structure. For comparison, we fit the same model but with generalised Gaussian margins for the residual distribution. 
We then simulate samples of size $N=20n$ from each of the conditional models (see Section 3.3 of the main text for the simulation algorithm). Figure \ref{fig:AGG_GG_ETA_Comp} compares the median of the empirical and model-based estimates of $\eta_{i,j}(u)$ for $i, j \in V, i > j$, and $u \in \{0.95, 0.99\}$, over the $200$ datasets. The only difference between the left and right panels is the margins used in the residual distribution, which are AGG and generalised Gaussian, respectively. The generalised Gaussian underestimates the true value of $\eta$, particularly for the pairs with weaker dependence. In contrast, the AGG tends to capture $\eta$ reasonably well with no increase in the standard error despite the additional model parameter. Therefore, the AGG is necessary to obtain accurate predictions from the model.

\section{Additional figures and simulation studies for Section 4.1}
\label{Sec:Sim_Study_True_Dist}
Section \ref{Sec:Sim_Study_True_Dist_Weak_Dependence} contains additional figures for the simulation study of Section 4.1 in the main text. Also shown are two additional simulation studies to assess the model performance in the presence of either strong positive (Section \ref{Sec:Sim_Study_True_Dist_High_Dependence}) or weak negative (Section \ref{Sec:Sim_Study_True_Dist_Negative_Dependence}) associations. Throughout this section, data are simulated from the SCMEVM with a graphical covariance structure given by $\mathcal{G} = (V, E)$, $V = \{1, \hdots, 5\}$, and $E = \{\{1,2\}, \{1,3\}, \{2,3\}, \{3,4\}, \{3,5\}, \{4,5\}\}$. 

\subsection{Weak positive dependence}
\begin{figure}[!t]
  \centering
  \includegraphics[width = \textwidth]{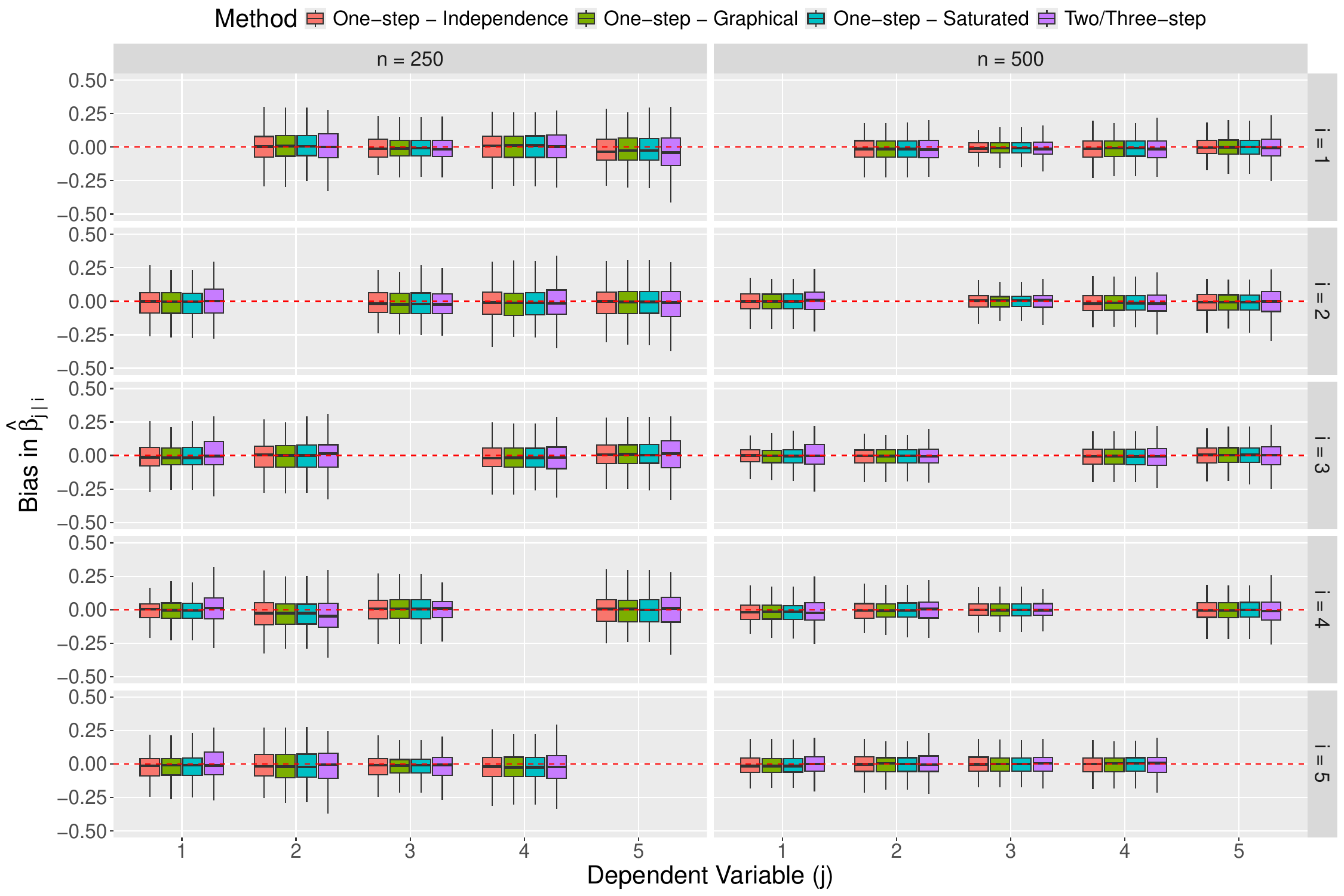}
  \caption{Boxplots detailing the bias of $\hat{\beta}_{j \mid i}$ for distinct $i, j \in V$. Each row corresponds to the conditioning variable $i$, and each column corresponds to the sample size. The different models are denoted by the fill of the boxplots. Red dashed lines show $y = 0$.}
  \label{fig:Beta_Bias}
\end{figure}

\begin{figure}[!t]
  \centering
  \subfloat{\includegraphics[width = 0.48\textwidth]{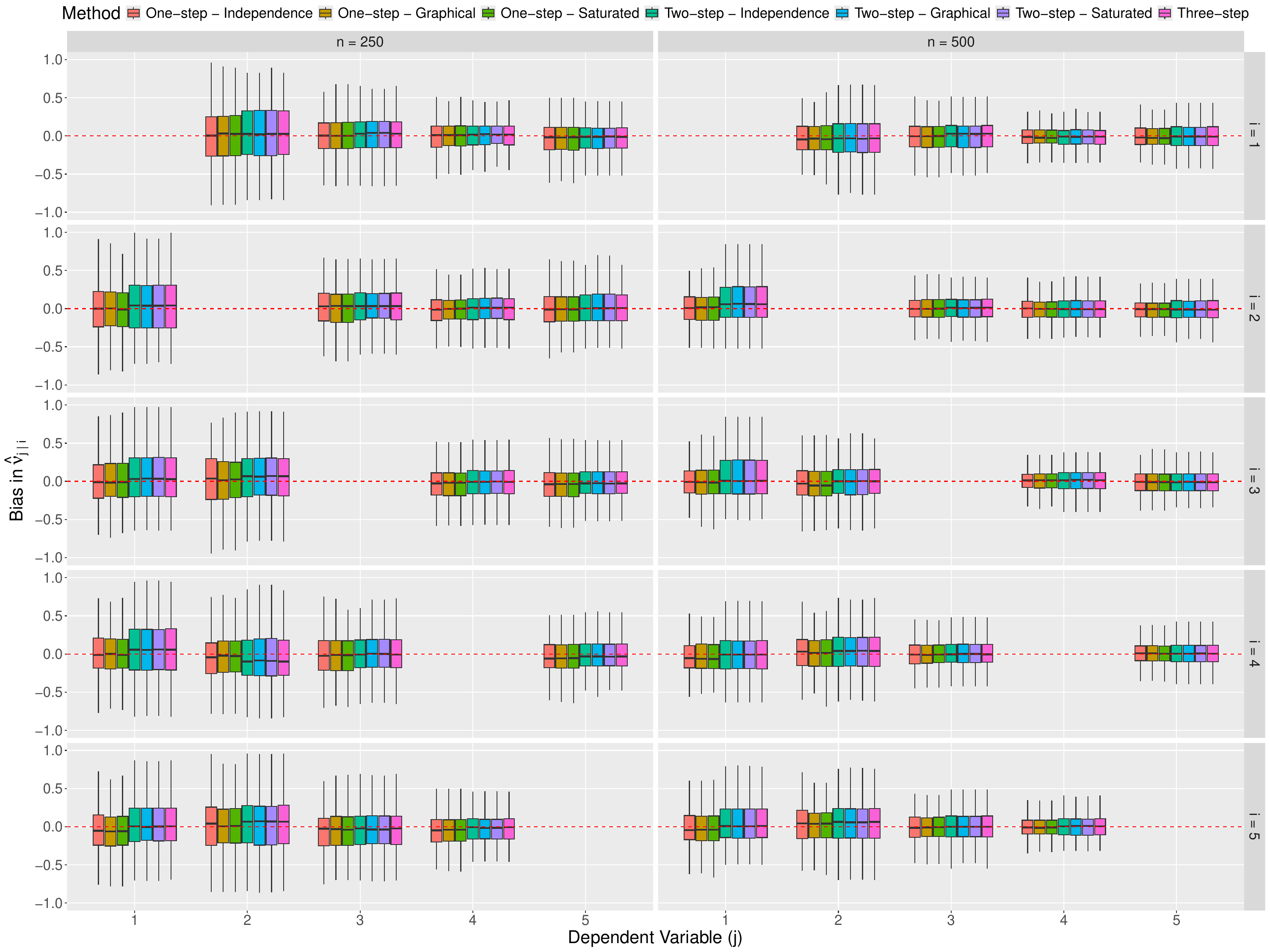}} \quad
  \subfloat{\includegraphics[width = 0.48\textwidth]{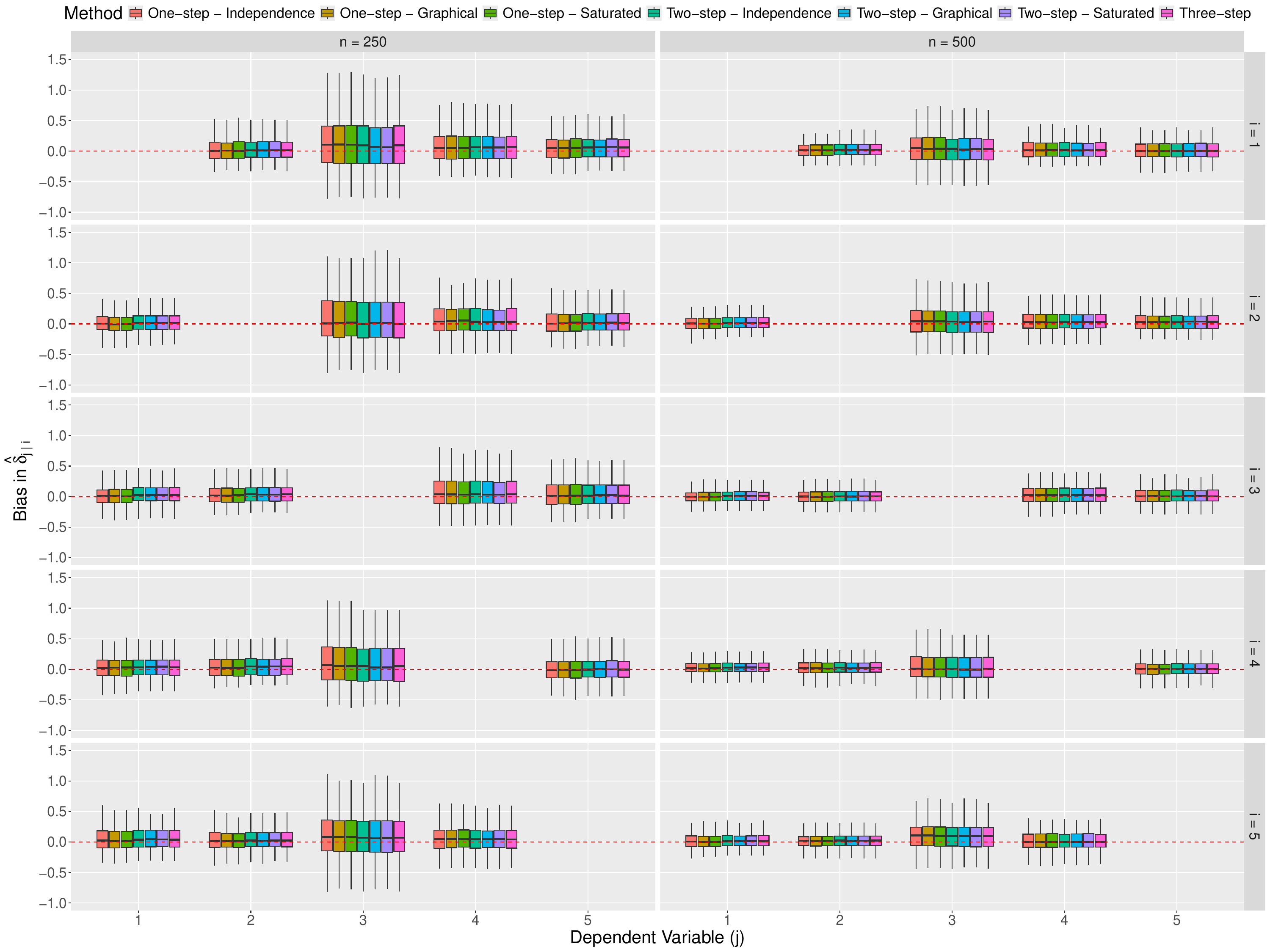}} \\
  \subfloat{\includegraphics[width = 0.48\textwidth]{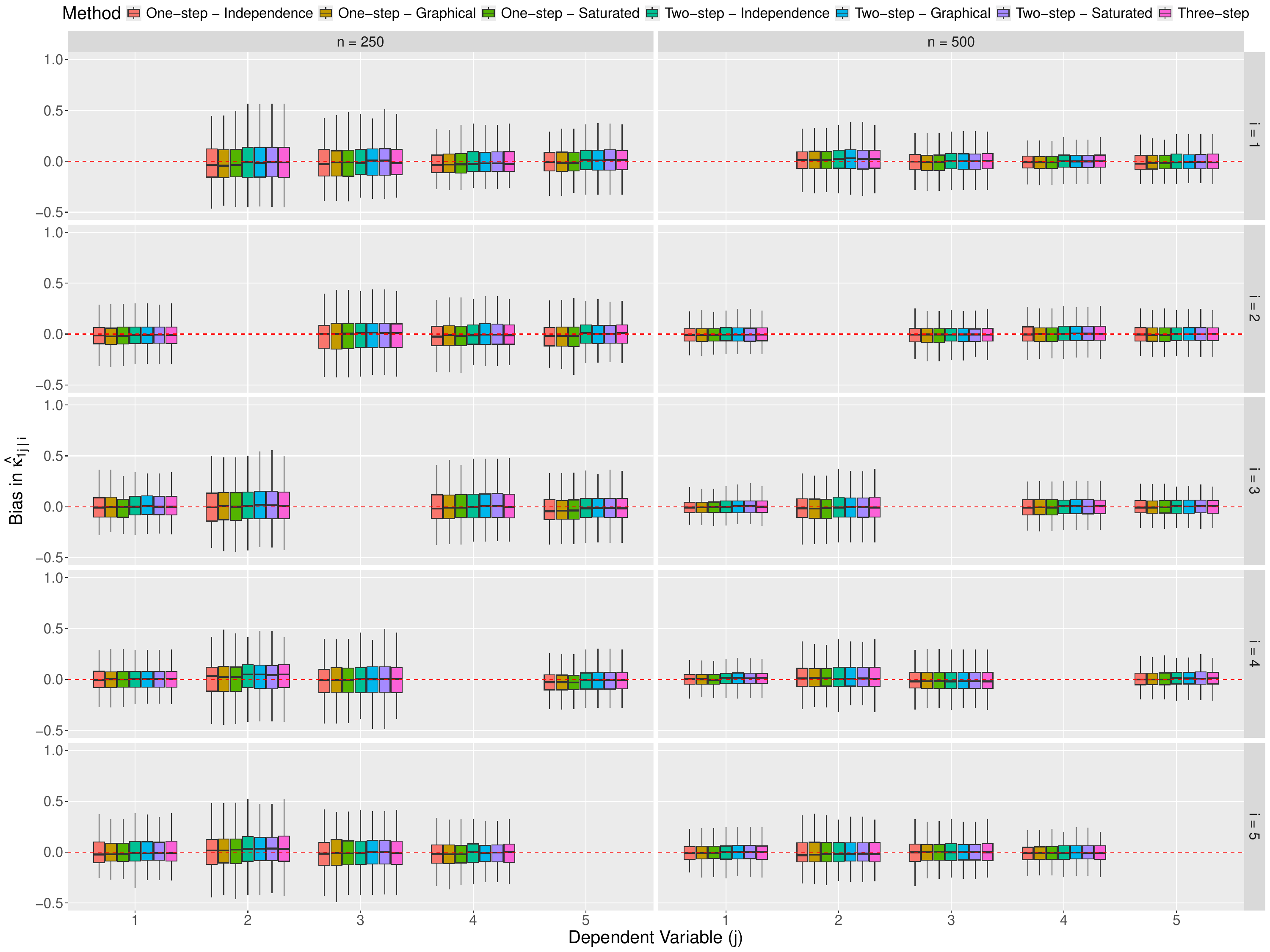}}\quad
  \subfloat{\includegraphics[width = 0.48\textwidth]{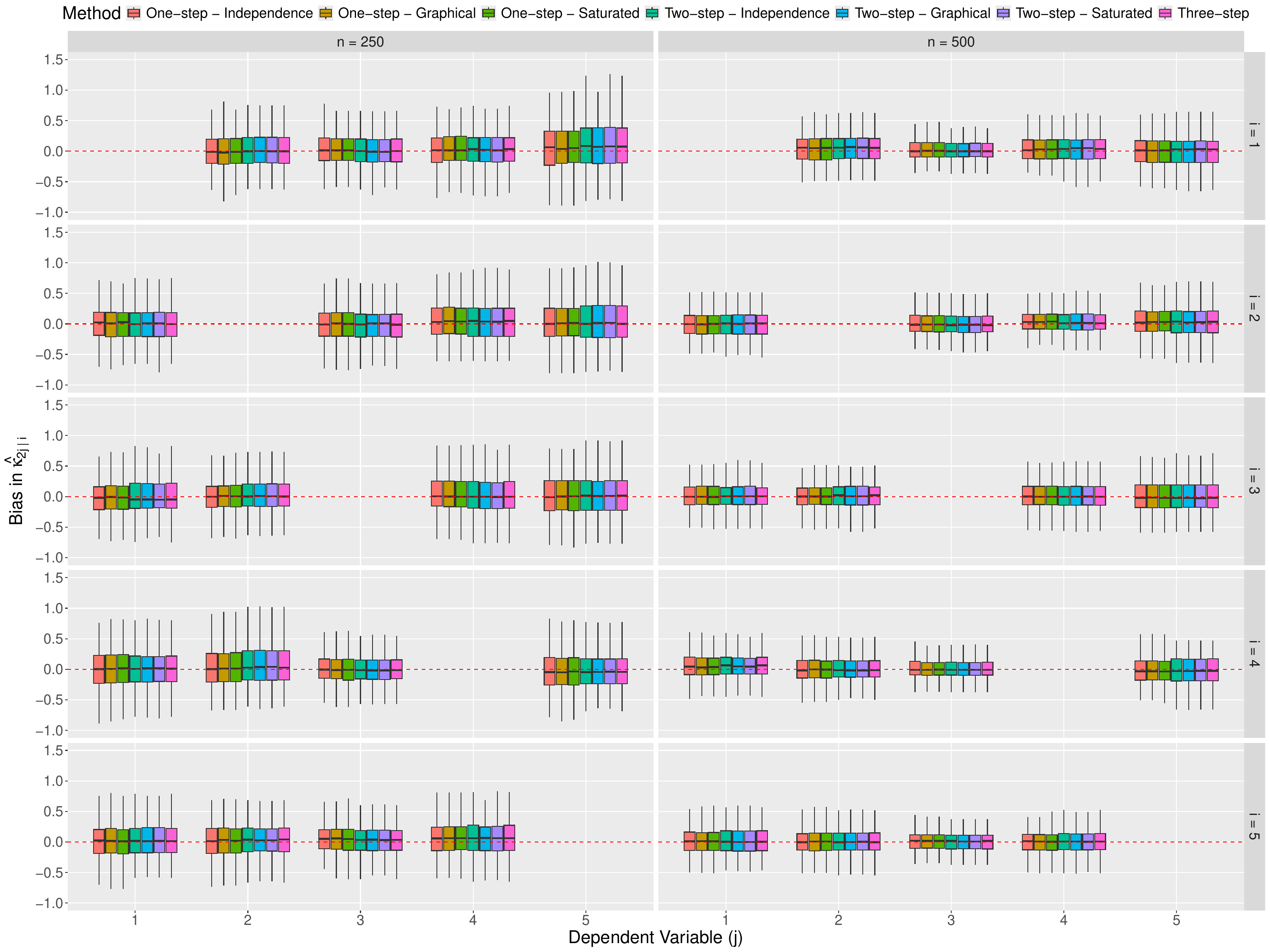}}
  \caption{Boxplots detailing the bias of $\hat{\nu}_{j \mid i}$ (top left), $\hat{\delta}_{j \mid i}$ (top right), $\hat{\kappa}_{1_{j \mid i}}$ (bottom left), and $\hat{\kappa}_{2_{j \mid i}}$ (bottom right) for distinct $i, j \in V$. Each row corresponds to the conditioning variable $i$, and each column corresponds to the sample size. The different models are denoted by the fill of the boxplots. Red dashed lines show $y = 0$.}
  \label{fig:Sim_Study_True_Dist_Low_Dependence}
\end{figure}

\begin{figure}[t!]
  \centering
  \includegraphics[width = \textwidth]{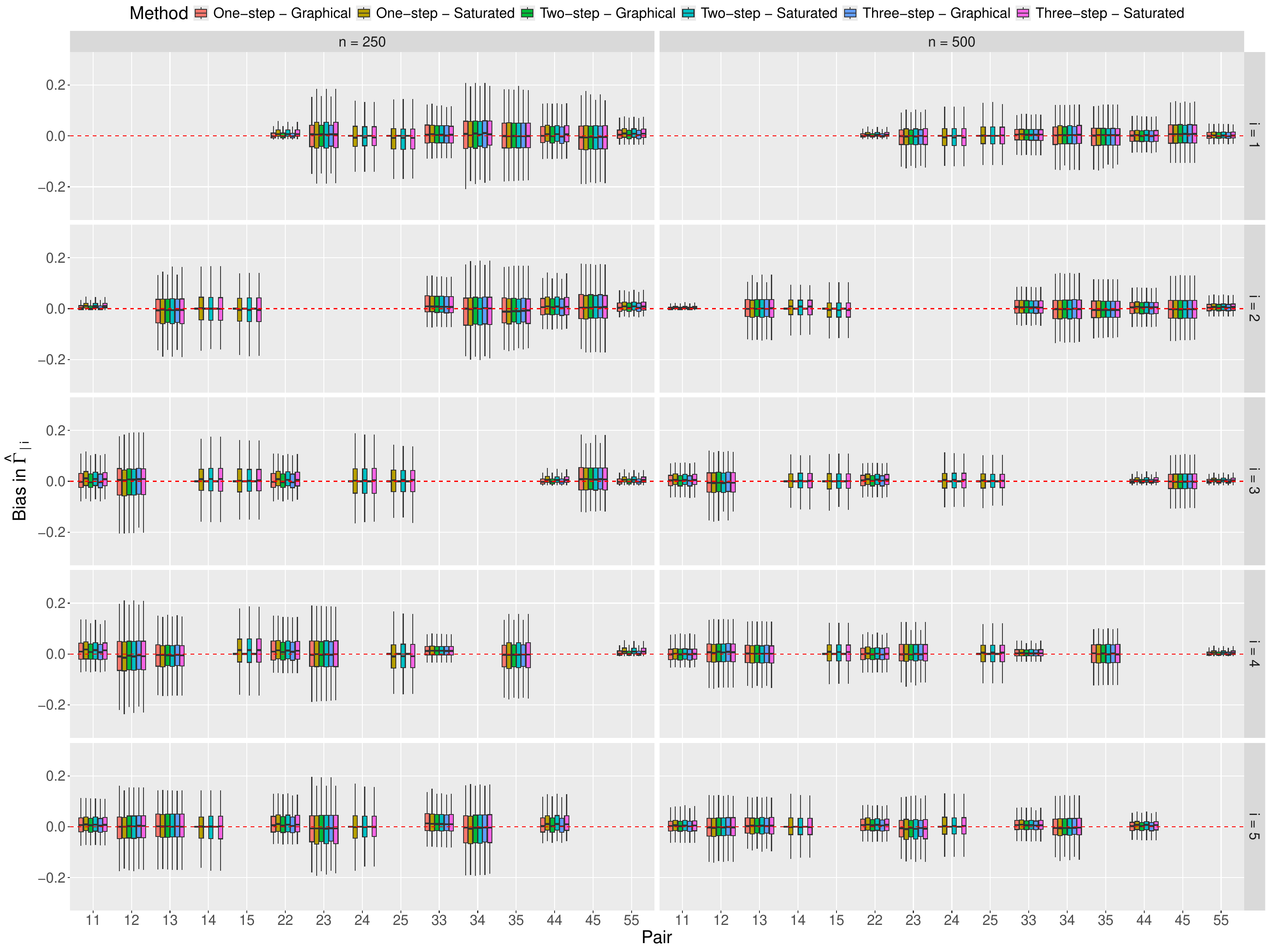}
  \caption{Boxplots for the bias of $\hat{\Gamma}_{\mid i}$ for each $i \in V$. Each row corresponds to the conditioning variable $i$, and each column corresponds to the sample size. The various models are denoted by the fill of the boxplots. Red dashed lines show $y = 0$.}
  \label{fig:Gamma_Bias}
\end{figure}

For this study, recall that the true dependence and AGG parameters were selected at random by sampling from a uniform distribution on $(0.1, 0.5)$ for $\alpha_{j}$, $(0.1, 0.3)$ for $\beta_{j}$, $(-5, 5)$ for $\nu_{j}$, $(0.5, 2)$ for $\kappa_{1_{j}}$, $(1.5, 3)$ for $\kappa_{2_{j}}$, and $(0.8, 2.5)$ for $\delta_{j}$, for each $j \in V$. Figures \ref{fig:Beta_Bias}, \ref{fig:Sim_Study_True_Dist_Low_Dependence} and \ref{fig:Gamma_Bias} show the bias in $\hat{\boldsymbol{\beta}}_{\mid i}$, the AGG parameters, and $\hat{\Gamma}_{\mid i}$ respectively. Similar to the plot for $\hat{\boldsymbol{\alpha}}_{\mid i}$ in the main text, we omit the maximum likelihood estimates (MLEs) from the stepwise methods in cases where they are, by construction, identical to estimates that are already presented. For $\hat{\Gamma}_{\mid i}$ we exclude those models that assume independent residuals since these are consistently biased. The findings are very similar to the main text: all models are unbiased across all parameters; the two- and three-step methods show slightly more cross-sample variability in their bias; variability in bias decreases as sample size increases.



\label{Sec:Sim_Study_True_Dist_Weak_Dependence}

\subsection{Strong positive dependence}
\label{Sec:Sim_Study_True_Dist_High_Dependence}
We repeat the simulation study from Section 4.1 of the main text but with strong, positive correlations $(>0.48)$. The other parameters remain unchanged from Section \ref{Sec:Sim_Study_True_Dist_Weak_Dependence}. Boxplots of the parameter estimates (not included) are almost identical to what was seen with weak positive associations. To compare the three stepwise procedures, we compare  the bias in the maximum log-likelihood values, see
Table \ref{tab:log_like_bias_high_dependence}. The models with independent residuals are biased; this is expected because the dependence structure is clearly misspecified. The bias is lower in the case when we condition on component 3 because this results in exact independence between $(W_1,W_2)$ and $(W_4,W_5)$. This result was not seen in the study from the main text due to the lower correlations used there. Similar to the results shown in the main text, models with a graphical or saturated dependence exhibit a small positive bias, but its magnitude is similar across all stepwise procedures. This supports our claim that the stepwise inference procedures result in no loss of information. Further, the three-step model is least biased when we assume the residuals have a graphical or saturated dependence structure. Finally, the bias for the models with independent residuals increases with sample size, while the bias is of similar magnitude for both sample sizes for models with graphical and saturated covariance structures. This suggests the latter models are more robust to changes in the sample size.

\begin{table}[!ht]
\caption{Median (2.5\% and 97.5\% quantiles) bias in the fitted maximum log-likelihood values for data from the SCMEVM with strong positive associations. Bold values denote the least biased stepwise inference procedure for each covariance structure type and conditioning variable.}
\resizebox{\textwidth}{!}{
\begin{tabular}{ccccccccccc}
\toprule
\multicolumn{2}{c}{Covariance Structure} & \multicolumn{3}{c}{Independent} & \multicolumn{3}{c}{Graphical} & \multicolumn{3}{c}{Saturated} \\
\hline
\begin{tabular}[c]{@{}c@{}}Number \\ of Excesses\end{tabular} & \begin{tabular}[c]{@{}c@{}}Conditioning\\ Variable\end{tabular} & One-step & Two-step & Three-step & One-step & Two-step & Three-step & One-step & Two-step & Three-step \\
\hline
\multirow{5}{*}{250} & 1 & \textbf{-135.7 (-170.2, -98.9)} & -137.3 (-171.1, -99.6) & -137.3 (-171.1, -99.6) & 14.7 (8.0, 22.5) & 12.7 (6.1, 20.8) & \textbf{11.9 (5.2, 20.6)} & 15.6 (8.4, 24.0) & 13.5 (6.6, 22.1) & \textbf{12.8 (5.9, 21.4)}\\
& 2 & \textbf{-138.5 (-176.4, -100.0)} & -139.7 (-172.4, -101.3) & -139.7 (-172.4, -101.3) & 14.5 (8.0, 23.0) & 11.8 (4.7, 21.6) & \textbf{11.3 (4.1, 21.0)} & 15.4 (9.1, 24.8) & 12.8 (5.8, 23.2) & \textbf{12.1 (4.9, 22.3)}\\
& 3 & \textbf{-39.8 (-86.1, -18.7)} & -41.4 (-59.2, -20.4) & -41.4 (-59.2, -20.4) & 13.3 (6.7, 22.8) & 11.1 (4.3, 20.6) & \textbf{11.0 (4.1, 20.3)} & 15.1 (7.8, 25.7) & 13.2 (4.8, 23.6) & \textbf{13.1 (5.2, 23.2)}\\
& 4 & \textbf{-140.0 (-174.1, -94.0)} & -140.8 (-175.2, -96.7) & -140.8 (-175.3, -96.7) & 14.1 (6.9, 22.9) & 11.9 (4.7, 19.3) & \textbf{11.3 (4.2, 18.7)} & 15.2 (8.3, 24.1) & 12.9 (5.5, 20.6) & \textbf{12.2 (4.4, 20.2)}\\
& 5 & \textbf{-137.5 (-178.1, -105.5)} & -137.8 (-174.8, -106.5) & -137.8 (-174.8, -106.5) & 14.0 (8.0, 22.6) & 11.5 (3.2, 20.8) & \textbf{10.7 (2.8, 20.5)} & 15.1 (8.9, 24.3) & 12.5 (4.8, 22.1) & \textbf{11.8 (4.4, 21.7)}\\
\hline
\multirow{5}{*}{500} & 1 & \textbf{-280.2 (-326.7, -226.3)} & -281.9 (-327.7, -228.3) & -281.9 (-327.7, -228.3) & 13.7 (7.6, 21.1) & 11.9 (5.3, 19.2) & \textbf{11.2 (4.7, 18.3)} & 14.7 (8.7, 22.2) & 13.0 (6.2, 20.7) & \textbf{12.1 (5.6, 19.9)}\\
& 2 & \textbf{-286.9 (-332.7, -240.5)} & -289.0 (-333.7, -242.0) & -289.0 (-333.7, -242.0) & 13.7 (7.7, 22.8) & 11.0 (3.3, 19.5) & \textbf{10.5 (2.6, 18.7)} & 14.7 (7.6, 24.1) & 11.7 (3.8, 21.3) & \textbf{11.3 (3.1, 20.3)}\\
& 3 & \textbf{-95.3 (-126.8, -67.7)} & -97.1 (-123.7, -71.9) & -97.1 (-123.7, -71.9) & 12.9 (7.1, 21.3) & 10.3 (3.7, 18.7) & \textbf{10.1 (3.5, 18.3)} & 14.8 (8.0, 23.4) & 12.3 (4.5, 20.6) & \textbf{12.1 (4.9, 20.1)}\\
& 4 & \textbf{-282.0 (-332.8, -229.4)} & -283.7 (-333.4, -231.2) & -283.7 (-333.4, -231.2) & 14.0 (8.5, 21.1) & 11.4 (1.9, 19.2) & \textbf{10.9 (1.0, 18.7)} & 14.9 (9.3, 23.6) & 12.5 (2.0, 20.6) & \textbf{11.8 (1.2, 19.8)}\\
& 5 & \textbf{-286.3 (-342.2, -234.9)} & -286.9 (-338.3, -236.3) & -286.9 (-338.3, -236.3) & 14.1 (7.4, 20.7) & 11.3 (2.1, 18.4) & \textbf{10.7 (1.4, 17.6)} & 14.9 (8.1, 22.3) & 12.4 (4.2, 20.4) & \textbf{11.7 (2.8, 19.2)}\\
\bottomrule
\end{tabular}
}
\label{tab:log_like_bias_high_dependence}
\end{table}

\subsection{Negative dependence}
\label{Sec:Sim_Study_True_Dist_Negative_Dependence}
Similar to Section \ref{Sec:Sim_Study_True_Dist_High_Dependence}, we repeat the simulation from Section 4.1 of the main text but with negative associations between some components. Equation~\eqref{eqn:Sim_Study_True_Dist_Matrices_Negative_Dependence} shows the true correlation matrix. All other parameters remain unchanged from Section \ref{Sec:Sim_Study_True_Dist_Weak_Dependence}.
\begin{equation}
    \Sigma = 
    \begin{bmatrix}
        1.000 & -0.308 & -0.134 & 0.034 & 0.019\\
        -0.308 & 1.000 & -0.160 & 0.041 & 0.023\\
        -0.134 & -0.160 & 1.000 & -0.254 & -0.141\\
        0.034 & 0.041 & -0.254 & 1.000 & -0.209\\
        0.019 & 0.023 & -0.141 & -0.209 & 1.000\\
    \end{bmatrix}.
    \label{eqn:Sim_Study_True_Dist_Matrices_Negative_Dependence}
\end{equation}
Parameter estimates have been omitted as they are similar to those presented for the weak positive association example in the main text. To compare the stepwise inference procedures, Table \ref{tab:log_like_bias_Negative_dependence} gives the biases of the fitted maximum log-likelihood values. As in the strong positive association study (Section \ref{Sec:Sim_Study_True_Dist_High_Dependence}), models with independent residuals have negative bias that increases with the sample size, while those with graphical or saturated dependence have small positive bias that is impervious to the sample size. Again, the magnitude of the bias is similar across all stepwise procedures, confirming no loss of information in using these.


\begin{table}[t!]
\caption{Median (2.5\% and 97.5\% quantiles) bias in the fitted maximum log-likelihood values for data from the SCMEVM with weak negative associations. Bold values denote the least biased stepwise inference procedure for each covariance structure type and conditioning variable.}
\resizebox{\textwidth}{!}{
\begin{tabular}{ccccccccccc}
\toprule
\multicolumn{2}{c}{Covariance Structure} & \multicolumn{3}{c}{Independent} & \multicolumn{3}{c}{Graphical} & \multicolumn{3}{c}{Saturated} \\
\hline
\begin{tabular}[c]{@{}c@{}}Number \\ of Excesses\end{tabular} & \begin{tabular}[c]{@{}c@{}}Conditioning\\ Variable\end{tabular} & One-step & Two-step & Three-step & One-step & Two-step & Three-step & One-step & Two-step & Three-step \\
\hline
\multirow{5}{*}{250} & 1 & \textbf{-13.3 (-26.7, 1.0)} & -14.2 (-27.6, -0.4) & -14.2 (-27.6, -0.4) & 13.9 (8.0, 22.7) & 12.5 (6.1, 21.7) & \textbf{12.4 (6.0, 21.6)} & 14.9 (8.1, 27.0) & 13.2 (6.8, 24.8) & \textbf{13.1 (7.0, 24.8)}\\
& 2 & \textbf{-11.6 (-28.4, 3.8)} & -13.3 (-27.6, 1.3) & -13.3 (-27.6, 1.3) & 14.6 (7.5, 23.4) & \textbf{12.6 (3.3, 20.6)} & 12.7 (4.8, 21.4) & 15.9 (7.6, 25.1) & \textbf{13.6 (4.6, 23.5)} & \textbf{13.6 (5.5, 23.1)}\\
& 3 & \textbf{-13.3 (-46.8, 2.3)} & -14.6 (-28.7, 2.5) & -14.6 (-28.7, 2.5) & 13.6 (6.9, 23.4) & 11.3 (4.6, 21.0) & \textbf{11.2 (4.6, 20.8)} & 15.6 (8.1, 26.3) & 13.1 (5.2, 23.9) & \textbf{13.0 (5.7, 23.7)}\\
& 4 & \textbf{-12.9 (-48.7, 1.6)} & -14.3 (-26.4, -0.2) & -14.3 (-26.4, -0.2) & 13.8 (7.3, 21.9) & 12.3 (5.4, 20.5) & \textbf{12.2 (5.9, 20.3)} & 14.6 (5.1, 23.2) & 13.0 (6.7, 21.6) & \textbf{12.9 (6.8, 21.6)}\\
& 5 & \textbf{-19.7 (-35.6, -4.7)} & -21.0 (-35.7, -5.7) & -21.0 (-35.7, -5.7) & 13.6 (7.3, 21.9) & \textbf{12.0 (4.7, 20.2)} & \textbf{12.0 (4.9, 20.0)} & 14.7 (8.1, 23.6) & 13.1 (5.7, 22.3) & \textbf{13.0 (5.6, 22.1)}\\
\hline
\multirow{5}{*}{500} & 1 & \textbf{-37.6 (-59.8, -20.8)} & -38.8 (-61.0, -22.2) & -38.8 (-61.0, -22.2) & 14.4 (8.9, 22.7) & 12.9 (6.1, 20.8) & \textbf{12.8 (6.0, 20.7)} & 15.1 (9.4, 23.2) & 13.8 (6.9, 21.5) & \textbf{13.7 (6.8, 21.4)}\\
& 2 & \textbf{-34.9 (-57.8, -12.8)} & -36.9 (-59.3, -13.1) & -36.9 (-59.3, -13.1) & 14.3 (7.7, 23.1) & 12.3 (4.4, 21.7) & \textbf{12.2 (4.3, 21.5)} & 15.1 (8.5, 24.0) & 13.2 (5.4, 22.8) & \textbf{13.1 (5.3, 22.5)}\\
& 3 & \textbf{-33.8 (-51.9, -15.4)} & -35.4 (-54.2, -17.1) & -35.4 (-54.2, -17.1) & 13.1 (6.9, 20.3) & 10.7 (2.1, 18.6) & \textbf{10.6 (2.5, 18.5)} & 14.8 (8.2, 23.0) & \textbf{12.3 (3.1, 20.3)} & \textbf{12.3 (3.0, 20.2)}\\
& 4 & \textbf{-40.5 (-57.8, -19.5)} & -41.7 (-59.7, -20.9) & -41.7 (-59.7, -20.9) & 13.9 (8.2, 22.9) & 11.9 (5.1, 20.6) & \textbf{11.7 (5.0, 20.5)} & 14.7 (9.1, 24.1) & 12.8 (5.4, 21.9) & \textbf{12.7 (5.4, 21.7)}\\
& 5 & \textbf{-51.2 (-70.9, -31.4)} & -52.8 (-70.6, -33.2) & -52.8 (-70.6, -33.2) & 13.8 (8.5, 22.6) & 12.0 (5.3, 21.3) & \textbf{11.9 (5.1, 20.9)} & 14.5 (9.3, 23.9) & 12.9 (6.4, 22.3) & \textbf{12.7 (6.2, 22.0)}\\
\bottomrule
\end{tabular}
}
\label{tab:log_like_bias_Negative_dependence}
\end{table}

\section{Majority rule proportion sensitivity analysis}
\label{Sec:Majority_Rule}
In Section 3.2 of the main text, we introduce a method for selecting the optimal graphical structure. The algorithm relies on the appropriate selection of the thresholds used to fit the CMEVMs and the majority-rule proportion used to obtain the final unified structure (step 11 of Algorithm 3.5). In Section 4.2 of the main text, we explored sensitivity to threshold choice. We now assess sensitivity to the majority-rule proportion.

\begin{figure}[t!]
    \centering
    \includegraphics[width=0.3\linewidth]{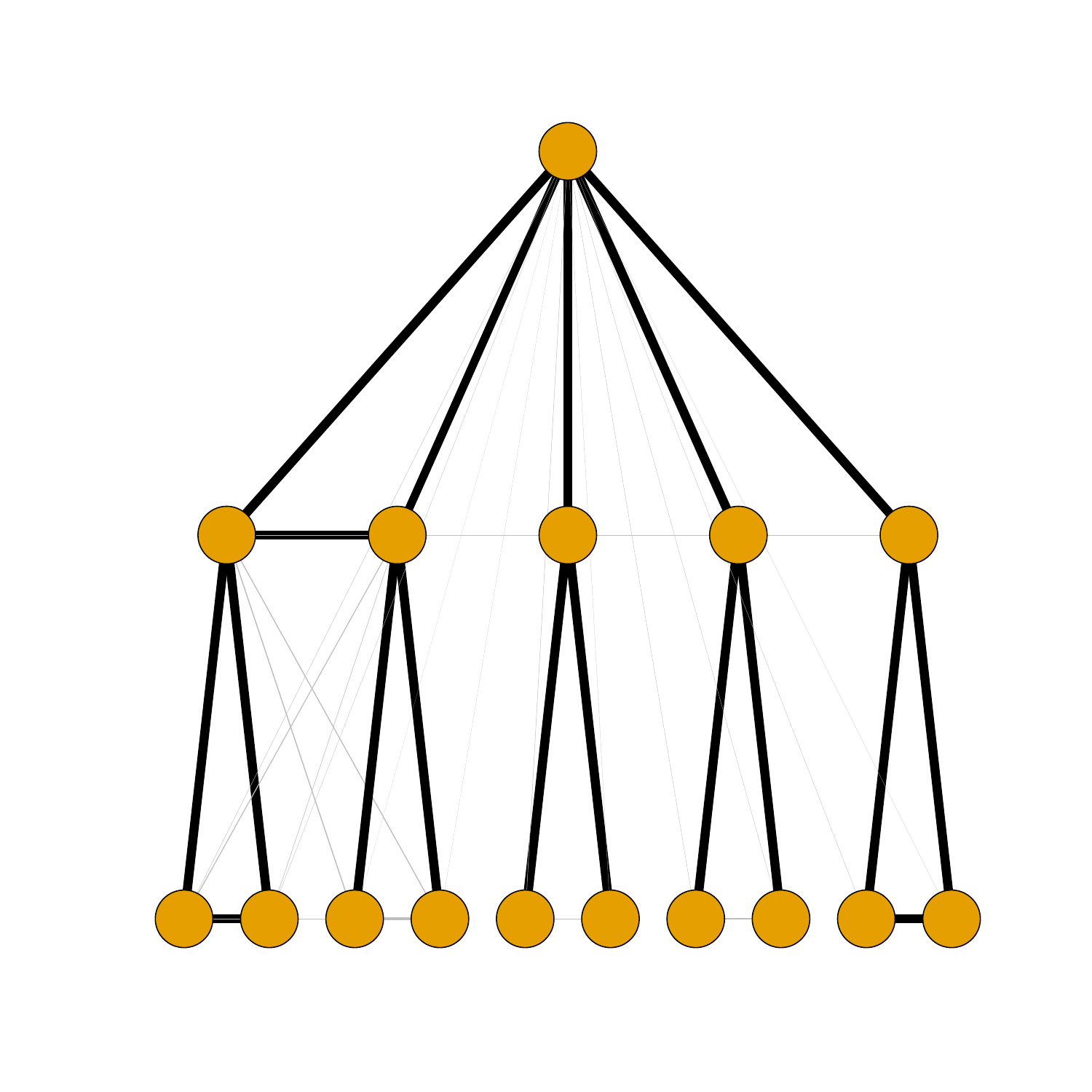} \quad
    \includegraphics[width=0.3\linewidth]{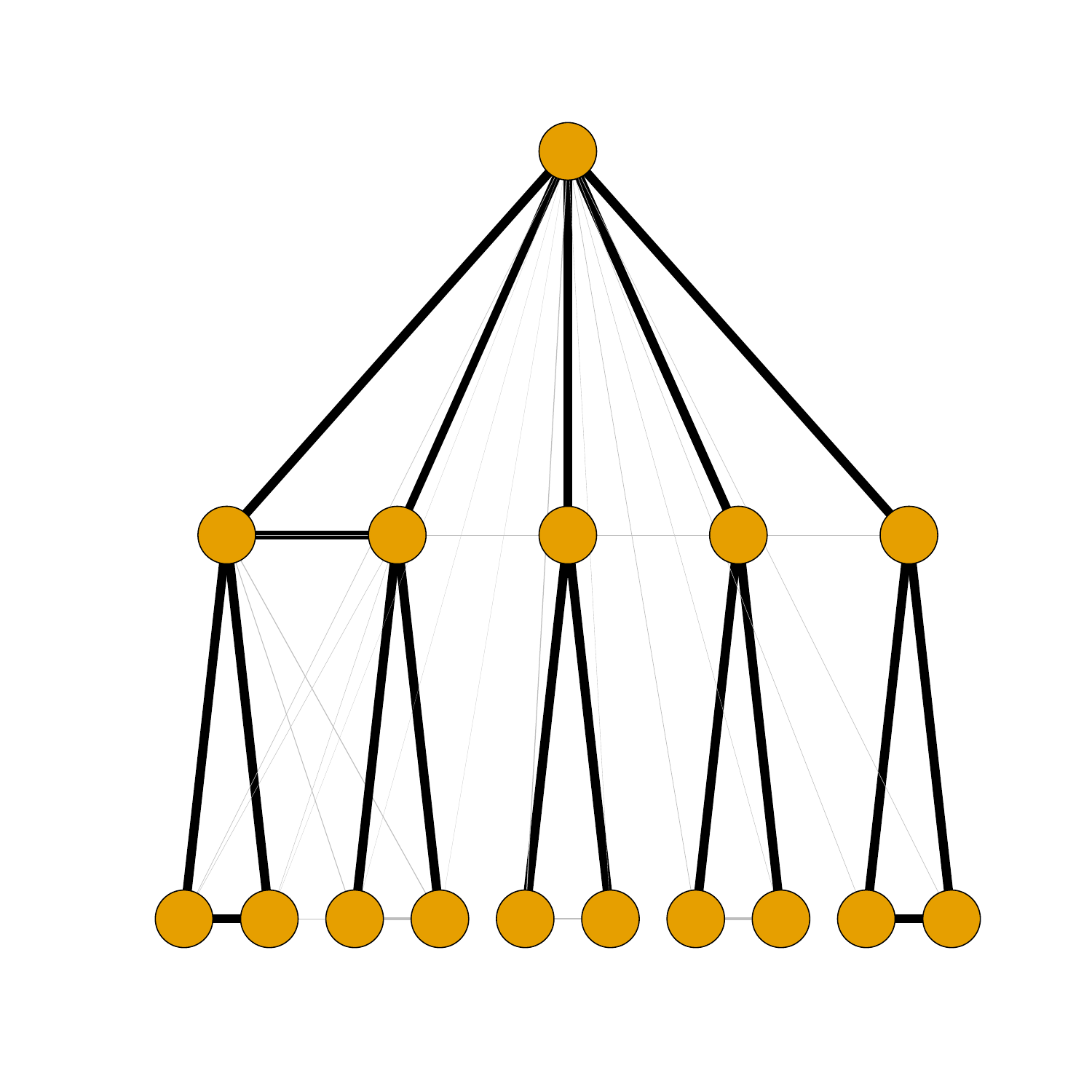} \quad
    \includegraphics[width=0.3\linewidth]{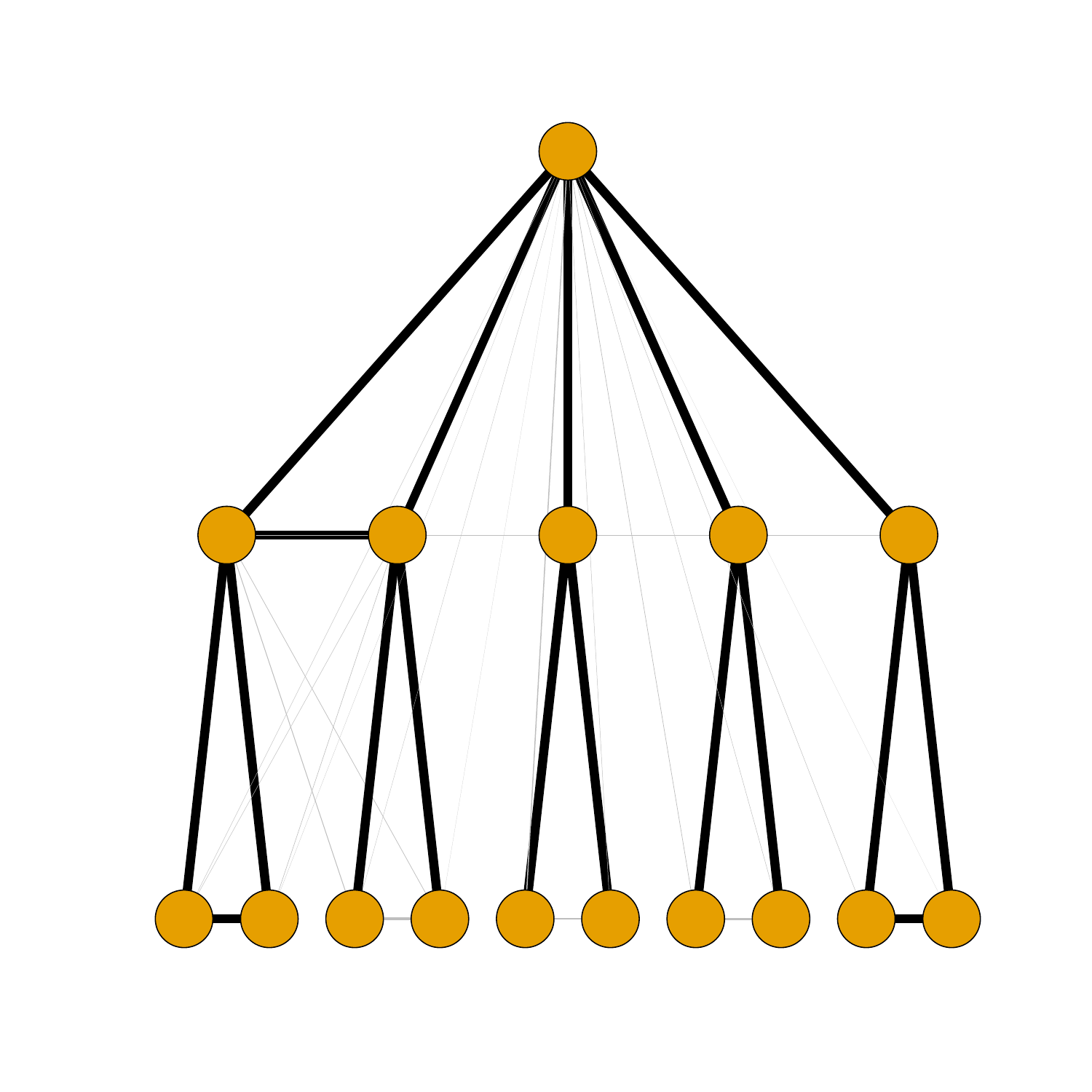}
    \caption{Inferred graphical structure for data generated from the multivariate Pareto distribution when the majority rule proportion $p$ in Algorithm 3.5 of the main text is set to $0.3$ (left), $0.5$ (centre), and $0.7$ (right). The line width and darkness in each panel correspond to the number of times each edge was selected across 100 samples. Black and grey edges correspond to true and additional edges, respectively.}
    \label{fig:MVP_MRP_Sensitivity_Analysis}
\end{figure}

Continuing the simulation study in Section 4.2 of the main text, we rerun Algorithm 3.5 with $u_{Y_{i}}$ set to the $0.8$-quantile of the standard Laplace distribution for all $i \in V$. We then compare results using three majority-rule proportions, $p \in \{0.3, 0.5, 0.7\}$. Figure \ref{fig:MVP_MRP_Sensitivity_Analysis} shows weighted graphs, with line width and darkness proportional to the number of times the edge is selected across the 100 datasets, for $p = 0.3$ (left), $p = 0.5$ (centre) and $p = 0.7$ (right). The true graph is identified in all cases with very few incorrect (grey) edges. This suggests that the graphical selection algorithm is not overly sensitive to the choice of majority-rule proportion. Nevertheless, it is important to include it to remove spurious edges that may only occur in one of the conditioned subgraphs.

To check that these results are not confined to the multivariate Pareto distribution, we repeat the simulation study, taking the underlying generating mechanism to be the multivariate Gaussian distribution. Taking $d = 16$ and $\mathcal{G}$ as in the left panel of Figure 3 in the main text, we simulate $1,000$ points from a multivariate Gaussian distribution with the components of the mean vector $\boldsymbol{\mu}$ independently sampled from a uniform distribution on $(-5, 5)$ and correlation matrix consistent with $\mathcal{G}$. We use $100$ replicates for this study. 

\begin{figure}[t!]
    \centering
    \includegraphics[width=0.3\linewidth]{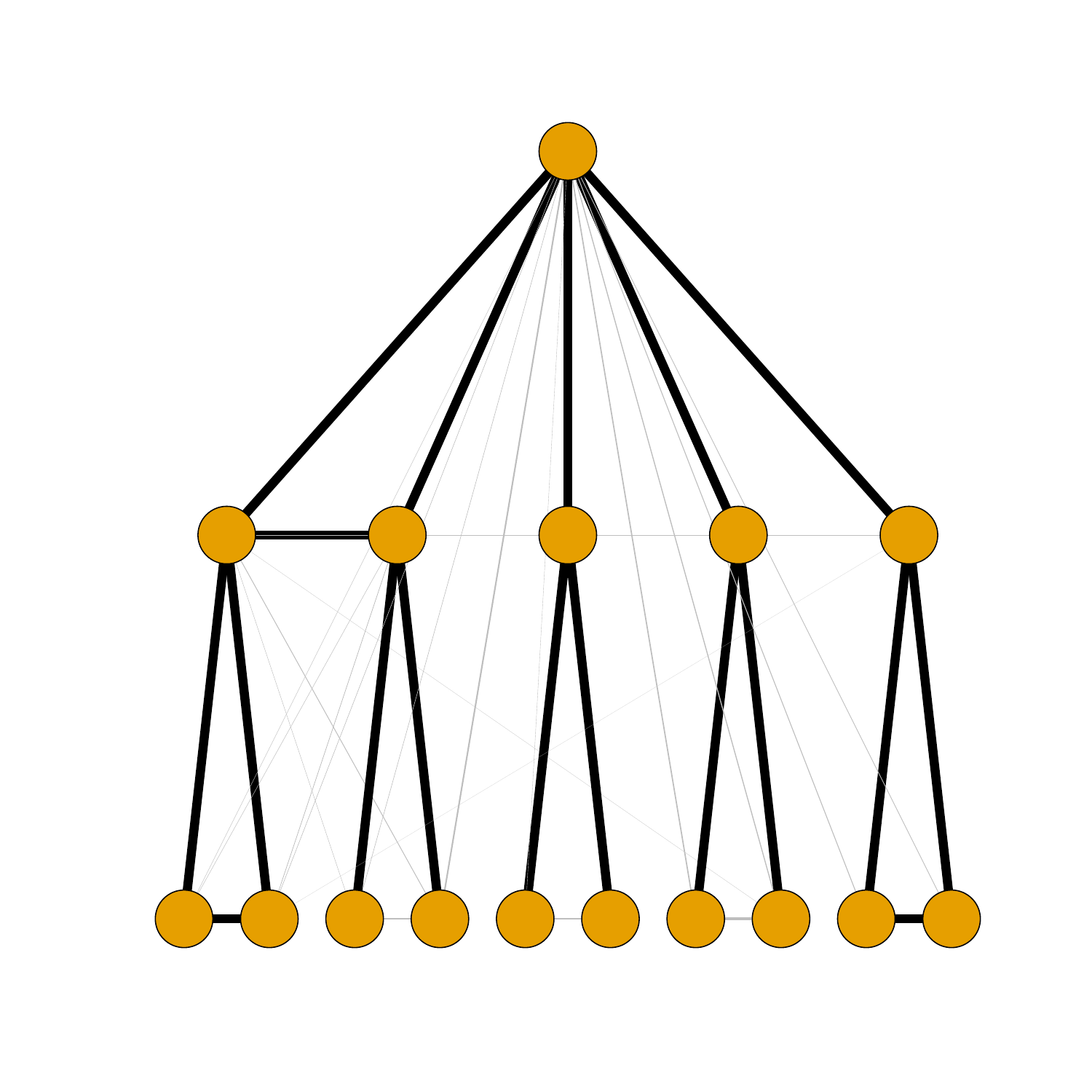} \quad
    \includegraphics[width=0.3\linewidth]{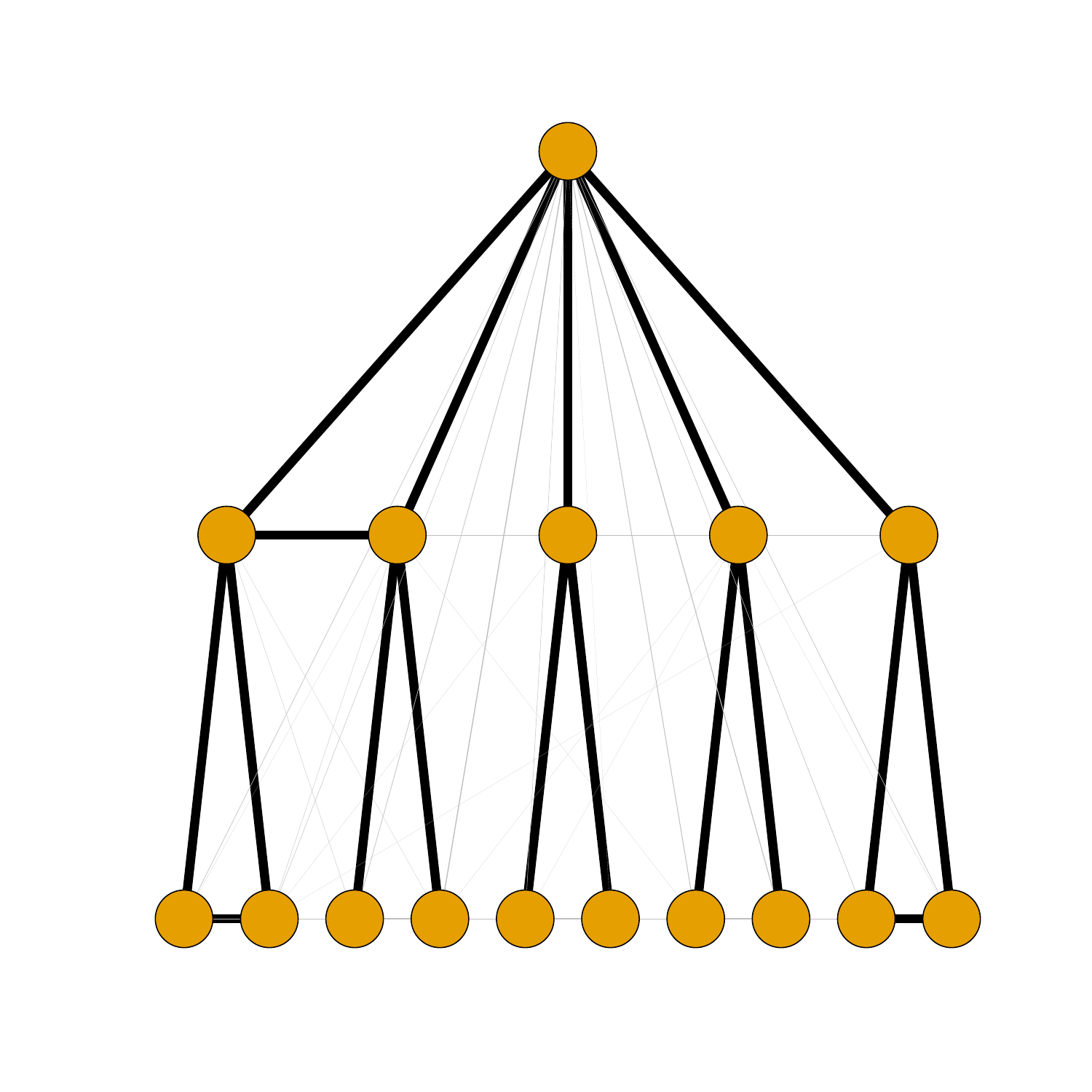} \quad
    \includegraphics[width=0.3\linewidth]{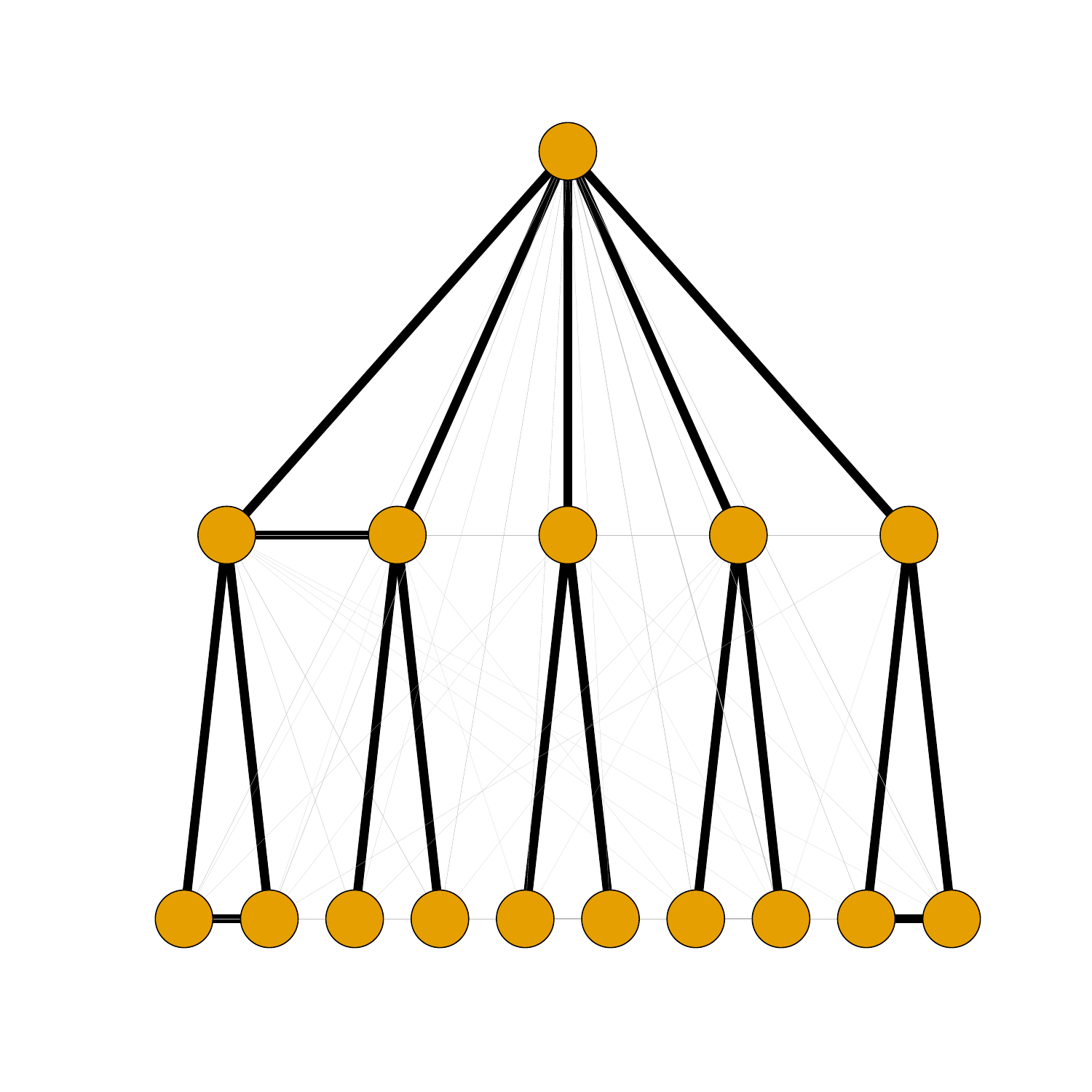}
    \caption{Inferred graphical structure for data generated from the multivariate Gaussian distribution when the majority rule proportion $p$ in Algorithm 3.5 of the main text is set to $0.3$ (left), $0.5$ (centre), and $0.7$ (right). The line width and darkness in each panel correspond to the number of times each edge was selected across 100 samples. Black and grey edges correspond to true and additional edges, respectively.}
    \label{fig:MVN_MRP_Sensitivity_Analysis}
\end{figure}

To infer the optimal graphical structure, we use Algorithm 3.5 of the main text. In the algorithm, we set the dependence thresholds $u_{Y_{i}}$ to the $0.90$-quantile of the standard Laplace distribution for all $i \in V$. Figure \ref{fig:MVN_MRP_Sensitivity_Analysis} shows weighted graphs of the $100$ inferred graphical structures, with line width and darkness proportional to the number of times the edge is selected across the 100 datasets, when the majority rule proportion is set to $0.3$ (left), $0.5$ (centre), and $0.7$ (right). Again, the true graph is well identified with very few ``incorrect'' (grey) edges.

Table \ref{tab:GraphEdgesCount} gives frequency counts of the number of edges inferred by Algorithm 3.5 of the main text, for both the multivariate Pareto and the MVG generating mechanisms. For the multivariate Pareto distribution, the distribution of edges in the selected graphs is very similar across all three proportions, suggesting the more important tuning parameter is the threshold above which the CMEVMs are fitted. For the MVG generating mechanism, increasing the majority rule proportion shifts the distribution towards the true number of edges. This is expected since edges are included in the final graph only if they are important most of the time.

\begin{table}[t!]
\centering

\caption{Number of times, out of the 100 samples, a graph with $x$ edges is inferred using Algorithm 3.5 of the main text for various majority rule proportions and underlying generating mechanisms.}

\resizebox{\textwidth}{!}{
\begin{tabular}{c c cccccccccccc}
\toprule

\multirow{2}{*}{\shortstack{Generating\\Mechanism} } 
  & \multirow{2}{*}{\shortstack{Majority Rule\\Proportion}}
  & \multicolumn{12}{c}{Number of edges in $\mathcal{G}$} \\
\cmidrule(l){3-14}
 & & 17 & 18 & 19 & 20 & 21 & 22 & 23 & 24 & 25 & 26 & 27 & 28 \\
\midrule

\multirow{3}{*}{MVP}
  & 0.3 & 6  & 33 & 24 & 16 & 15 & 4  & 2  & -  & -  & -  & -  & - \\
  & 0.5 & 7  & 21 & 27 & 24 & 9  & 9  & -  & 1  & 2  & -  & -  & - \\
  & 0.7 & 4  & 24 & 25 & 23 & 14 & 6  & 2  & 2  & -  & -  & -  & - \\
\midrule

\multirow{3}{*}{MVN}
  & 0.3 & 2  & 30 & 22 & 20 & 10 & 5  & 5  & 3  & 1  & 1  & 1  & - \\
  & 0.5 & -  & 38 & 29 & 14 & 8  & 7  & 2  & 2  & -  & -  & -  & - \\
  & 0.7 & -  & 42 & 29 & 15 & 5  & 3  & 1  & 3  & -  & 1  & -  & 1 \\
\bottomrule
\end{tabular}}
\label{tab:GraphEdgesCount}
\end{table}

\section{Additional graph selection example}
\label{Sec:Graph_Selection_Example}
In Section 4.2 of the main text, we replicate the simulation study of \citet[Section 5.5]{Engelke_2020_supp} to assess how closely the SCMEVM can identify the graphical structure of data generated from the H\"usler-Reiss distribution. Here, we repeat the study for data generated from the multivariate Gaussian distribution; see Section \ref{Sec:Majority_Rule} for details of data simulation.

\begin{figure}[!t]
  \centering
  \subfloat{\includegraphics[width = 0.3\textwidth]{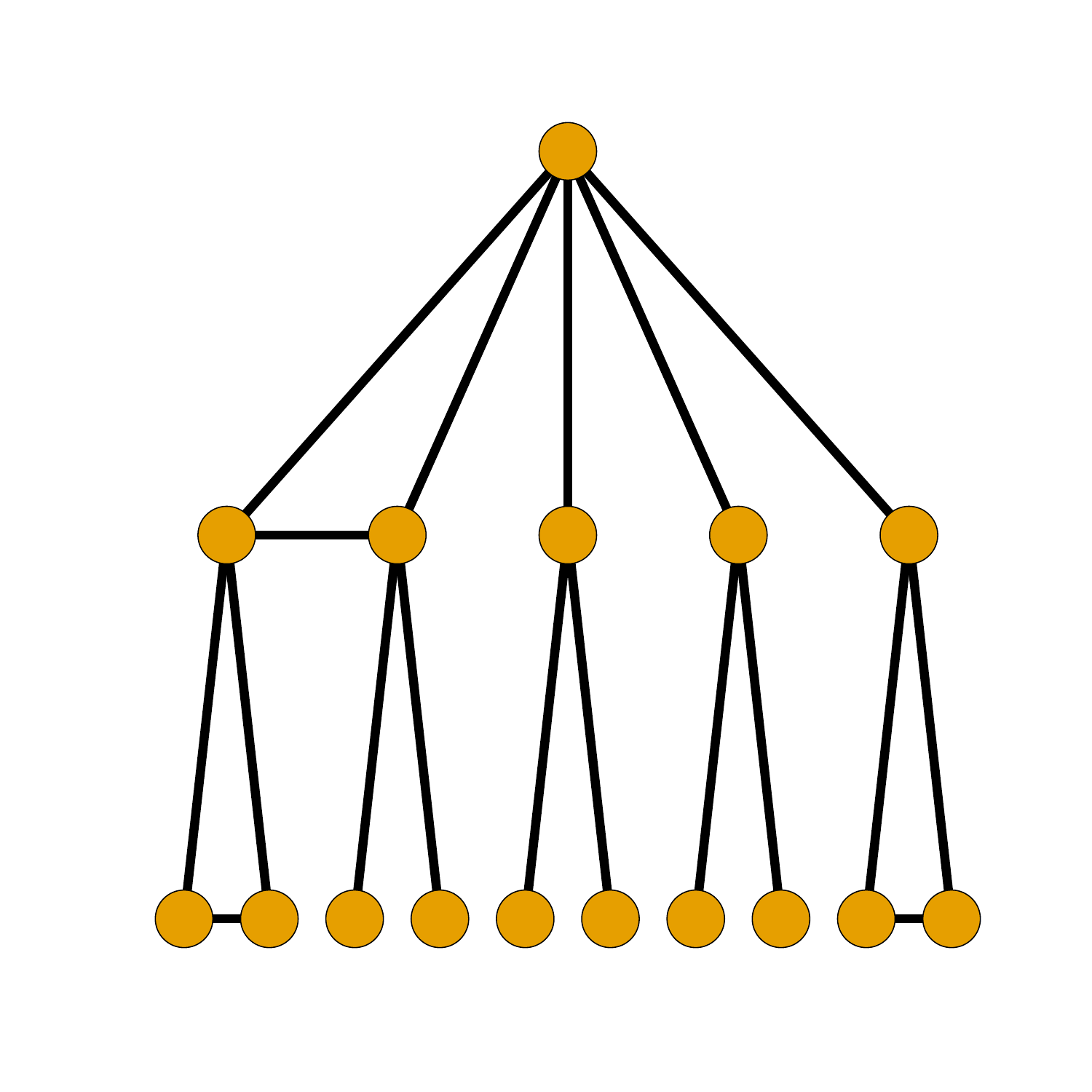}} \quad
  \subfloat{\includegraphics[width = 0.3\textwidth]{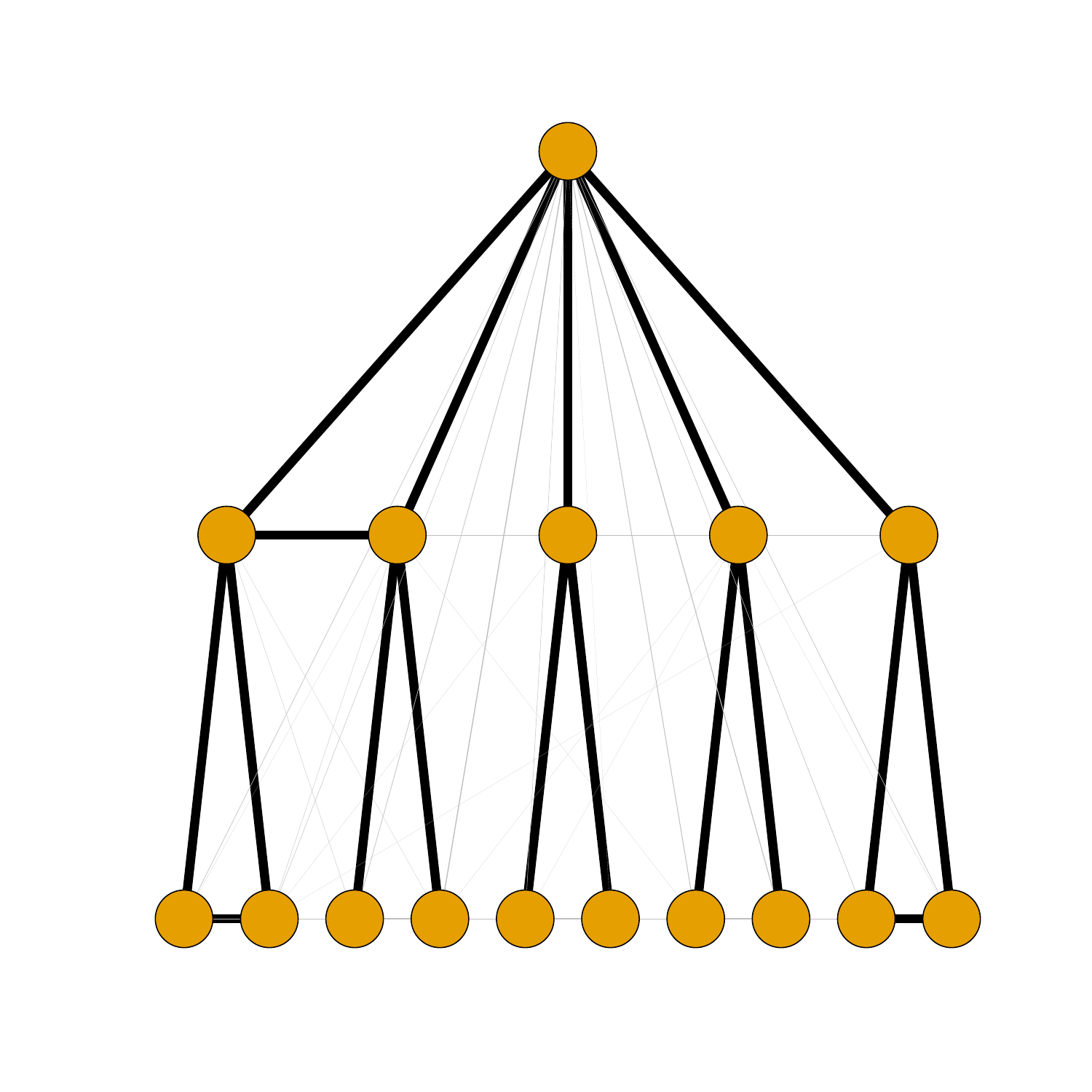}}
  \subfloat{\includegraphics[width = 0.3\textwidth]{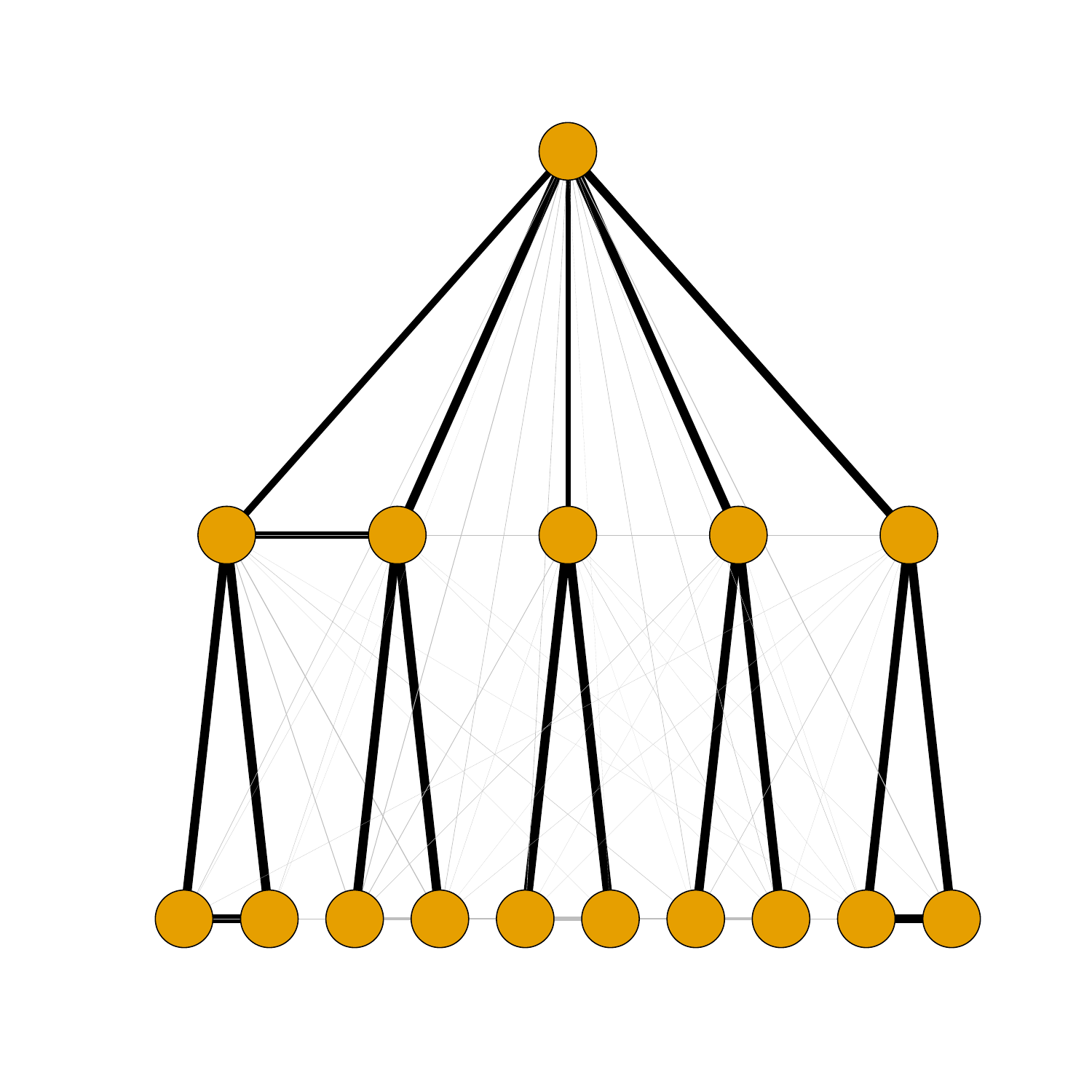}}\quad
  \caption{True underlying graphical structure (left) and inferred graphical structures using the method proposed in Section 3.2 of the main text (centre) and using ``EGlearn" (right). Line width and darkness indicate the number of times each edge was selected across 100 replicates. Black and grey edges correspond to ``true'' and ``additional'' edges, respectively.}
  \label{fig:Graphical_Selection_Supplementary}
\end{figure}

For the SCMEVM, we use Algorithm 3.5 of the main text to infer the optimal graphical structure, setting the dependence thresholds $u_{Y_{i}}$ to the $0.90$-quantile of the standard Laplace distribution for all $i \in V$ and the majority rule proportion to $0.5$. Figure \ref{fig:Graphical_Selection_Supplementary} (centre panel) shows a weighted graph of inferred graphical structures with line width and darkness proportional to the number of times the edge is selected across 100 simulated datasets. The true graphical structure is clearly recovered.  

For comparison, we also use the graphical selection method EGlearn \citepsupp{Engelke_2025_supp}. EGlearn uses a multivariate Pareto model but generalises the allowable graphical structure beyond the block structure requirement in \citetsupp{Engelke_2020_supp}. The only tuning parameter is the threshold above which the data are assumed to follow the multivariate Pareto distribution. We set this to be the $0.95$-quantile of the standard Pareto distribution. The resulting weighted graph over the 100 datasets is shown in Figure \ref{fig:Graphical_Selection_Supplementary} (right panel). The true graphical structure is largely identified, but there are more ``incorrect" edges compared to the SCMEVM method. 
This result seems counter-intuitive since EGlearn is designed to learn the graphical structure for AD data and not AI data. That being said, the results are somewhat threshold sensitive. Using a lower threshold of the $0.8$-quantile of the standard Pareto distribution results in 5 additional edges in the pruned weighted graph, while a higher threshold of the $0.99$-quantile of the standard Pareto distribution results in six missing edges in the pruned weighted graph and the algorithm inferring fewer than 18 edges in 97\% of cases. In contrast, our method is relatively stable with respect to the threshold, as it identifies the true graph when $u_{Y_{i}}$ is set to the $0.8-$ and $0.95-$quantile of the standard Laplace distribution for all $i \in V$.

These results suggest that the \citetsupp{Engelke_2025_supp} method can struggle to correctly identify the underlying structure of the data when the data-generating mechanism does not have complete AD, while the method proposed in the main text does not have this limitation. 

\section{Additional figures and simulation studies for Section 4.3}
In the main text, we considered data with a mixture of extremal dependence structures. We now repeat the simulation study in Section 4.3 of the main text for data that exhibits either full asymptotic independence (AI) or full asymptotic dependence (AD). For AI, we simulate from each of the (a) multivariate Gaussian (MVG), (b) symmetric multivariate Laplace (MVL), and (c) multivariate $t$- (MVT) distributions. In all cases, both positive and negative associations are investigated. For AD, we simulate from a multivariate Pareto (MVP) distribution.

All simulation studies follow a similar pattern. For each true distribution, 200 datasets are sampled using a dependence structure consistent with $\mathcal{G}$ in Section \ref{Sec:Sim_Study_True_Dist}. Data are transformed from their canonical margins (e.g., Gaussian for the MVG distribution) to standard Laplace margins as per Section 3.1 of the main text. For all datasets, each of the one-, two- and three-step procedures is used to fit the SCMEVM with graphical covariance, where the graph is assumed to be known and correctly specified. The three-step procedure is also used to fit the SCMEVM with independent and saturated covariances. For comparison, we also fit the original CMEVM \citepsupp{Heffernan_2004_supp} and the graphical extremes model \citepsupp{Engelke_2020_supp} (EHM). The mean absolute error (MAE) and root mean squared error (RMSE) of the model-based estimates form the basis of model comparison. Such metrics are calculated via probabilities of the form $\mathbb{P}[\boldsymbol{X}_{A} > u_{\boldsymbol{X}_{A}} \mid X_{i} > u_{X_{i}}]$ ($\mathbb{P}[\boldsymbol{X}_{A} < u_{\boldsymbol{X}_{A}} \mid X_{i} > u_{X_{i}}]$) for simulations that have positive (negative) associations and for all sets $A \subseteq V_{\mid i}$ and $i \in V$

\subsection{Multivariate Gaussian distribution}

In this section, we assume $\boldsymbol{X}$ follows a MVG distribution with mean vector $\boldsymbol{\mu}$, where each $\mu_{j}$ is independently sampled from a uniform distribution on $(-5, 5)$, and correlation matrix $\Sigma$. Various strengths of correlation are considered, however $\Gamma = \Sigma^{-1}$ is always consistent with $\mathcal{G}$ in Section \ref{Sec:Sim_Study_True_Dist}. We set dependence thresholds $u_{Y_{i}}$ to the $0.90$-quantile of the standard Laplace distribution for all $i \in V$. For prediction, we set $u_{X_{i}}$ to the $0.95$-quantile for the true distribution of $X_{i}$ for each $i \in V$.

\subsubsection{Weak positive dependence}
\label{Sec:MVN_Low_Dependence}
\begin{figure}[t!]
  \centering
  \subfloat{\includegraphics[width = 0.48\textwidth]{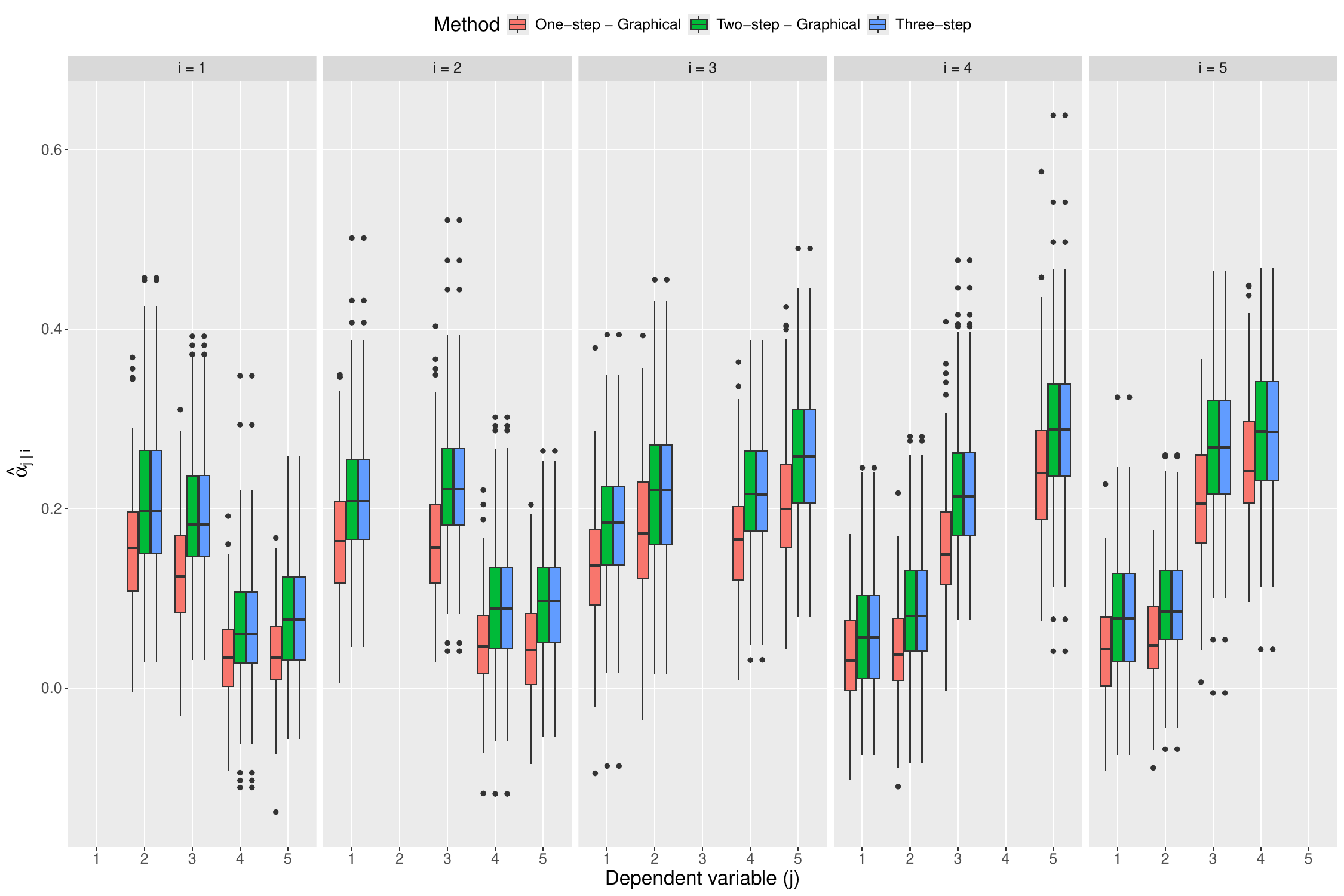}} \quad
  \subfloat{\includegraphics[width=.48\textwidth]{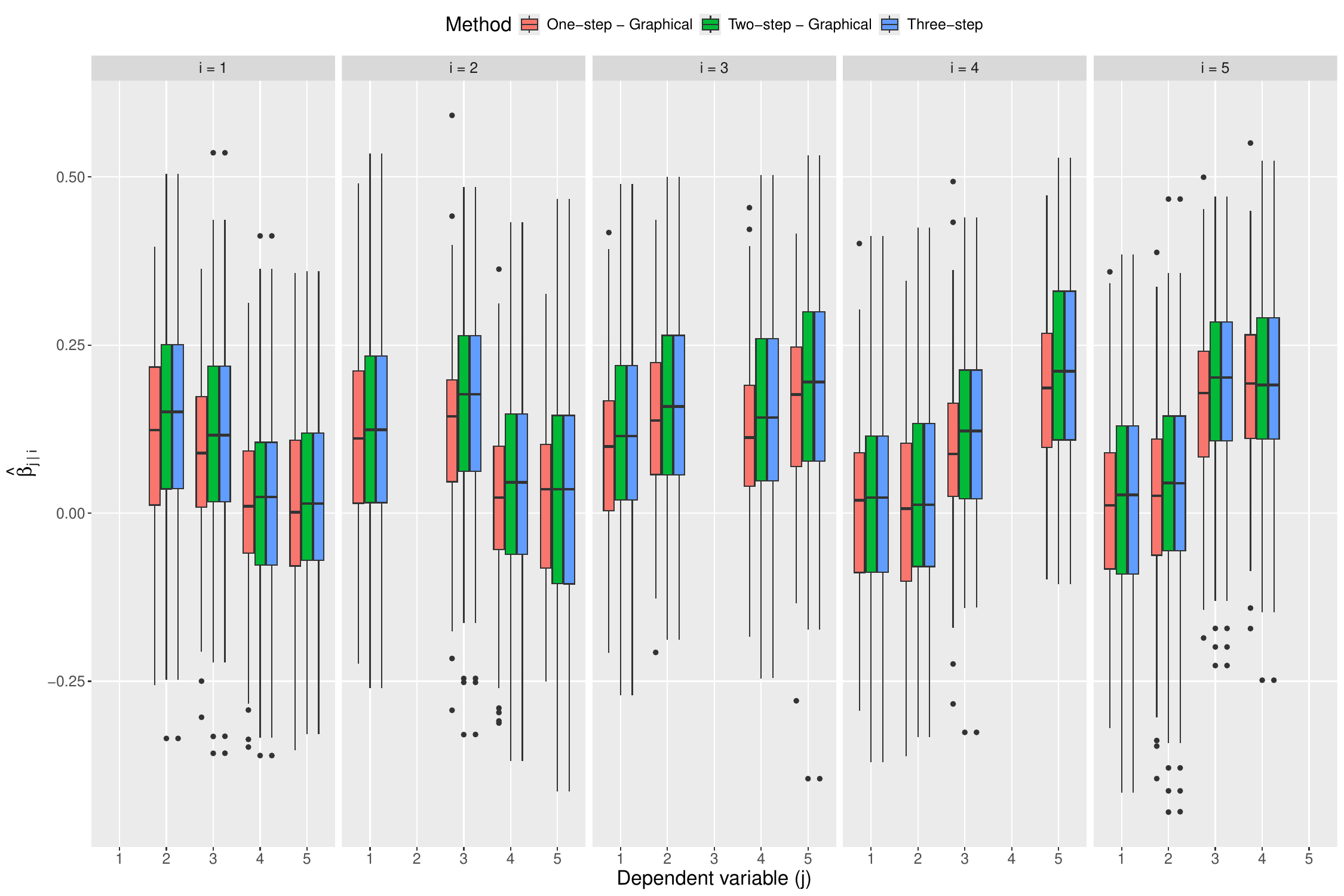}} \\
  \subfloat{\includegraphics[width = 0.48\textwidth]{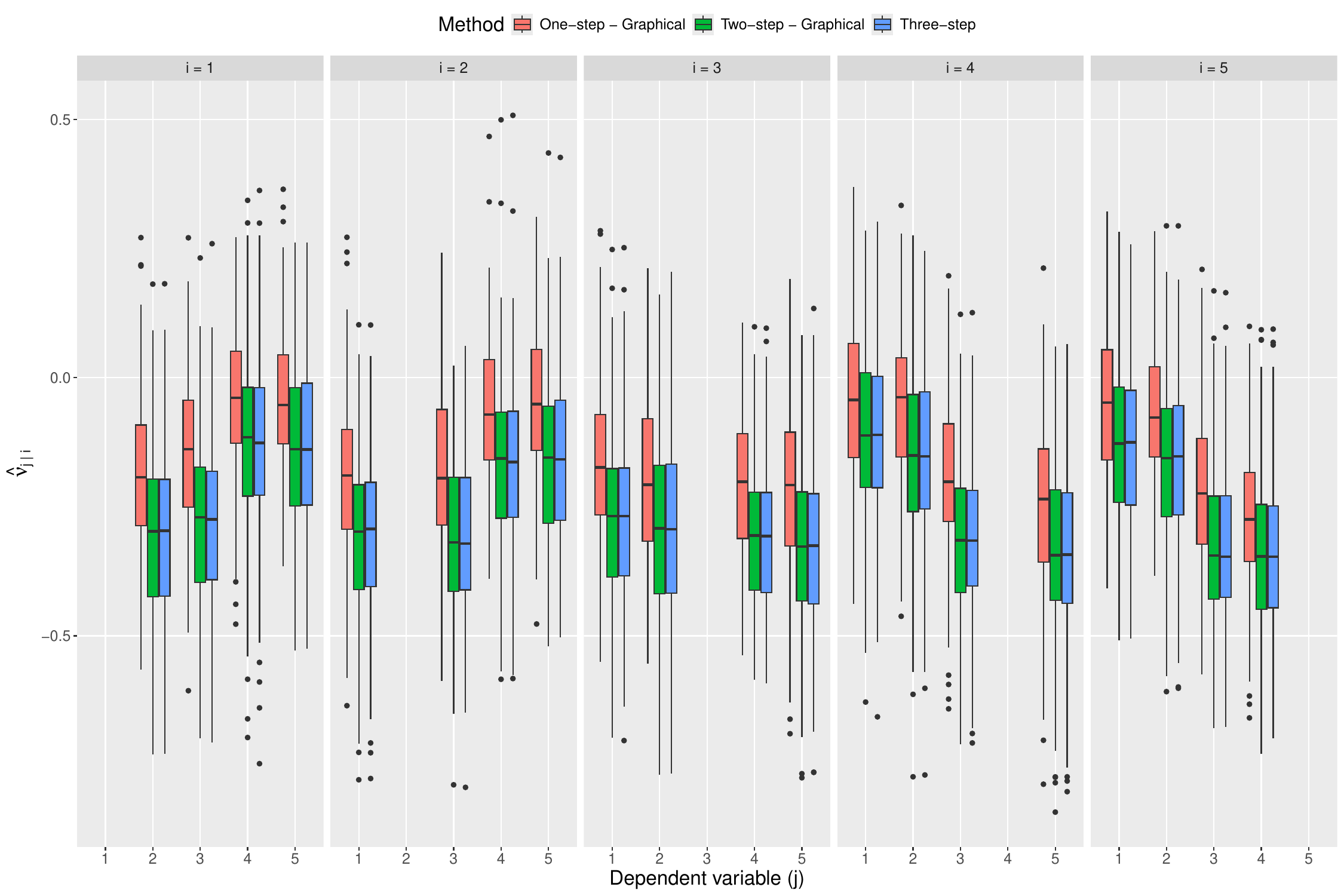}} \quad
  \subfloat{\includegraphics[width=.48\textwidth]{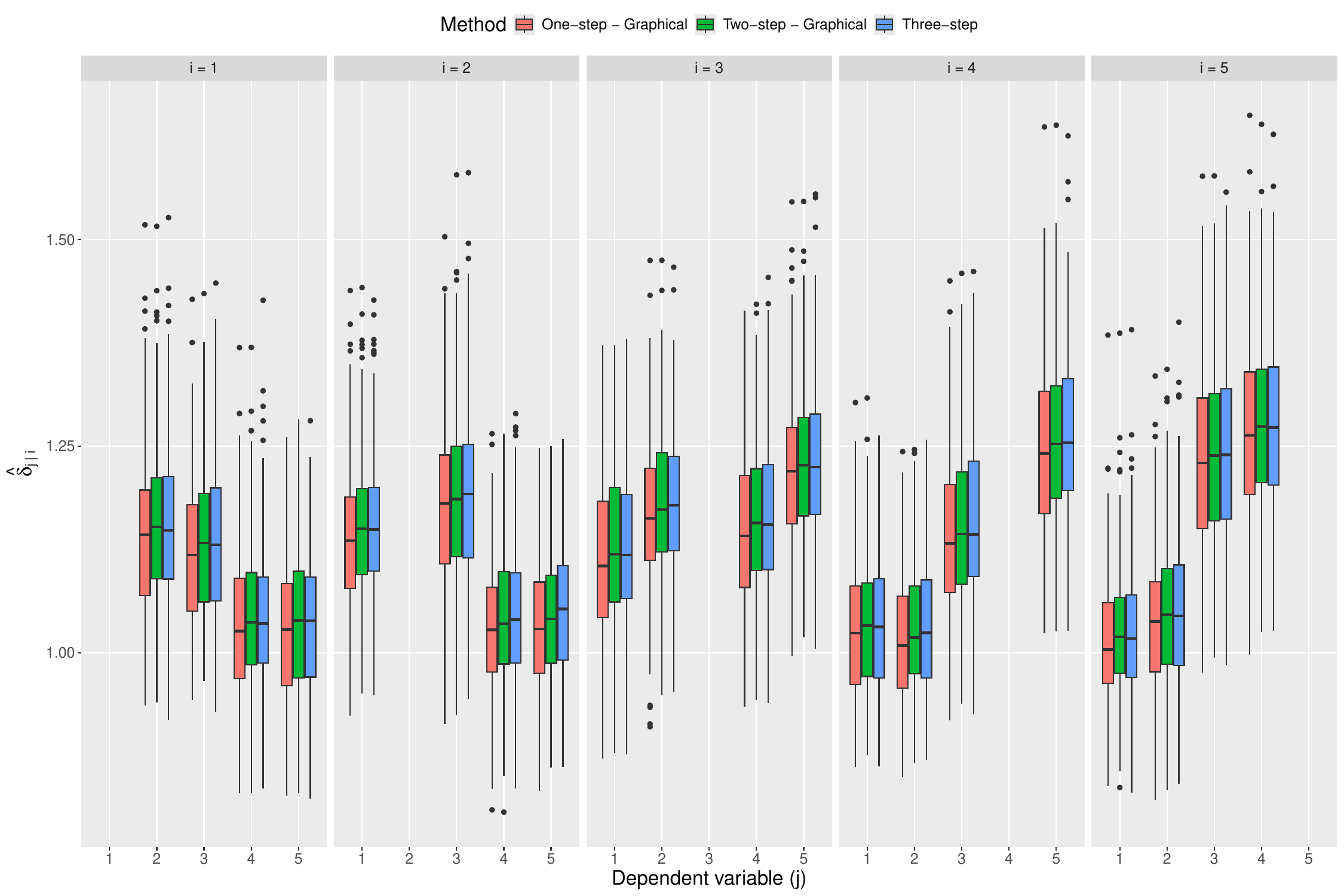}} \\
  \subfloat{\includegraphics[width=.48\textwidth]{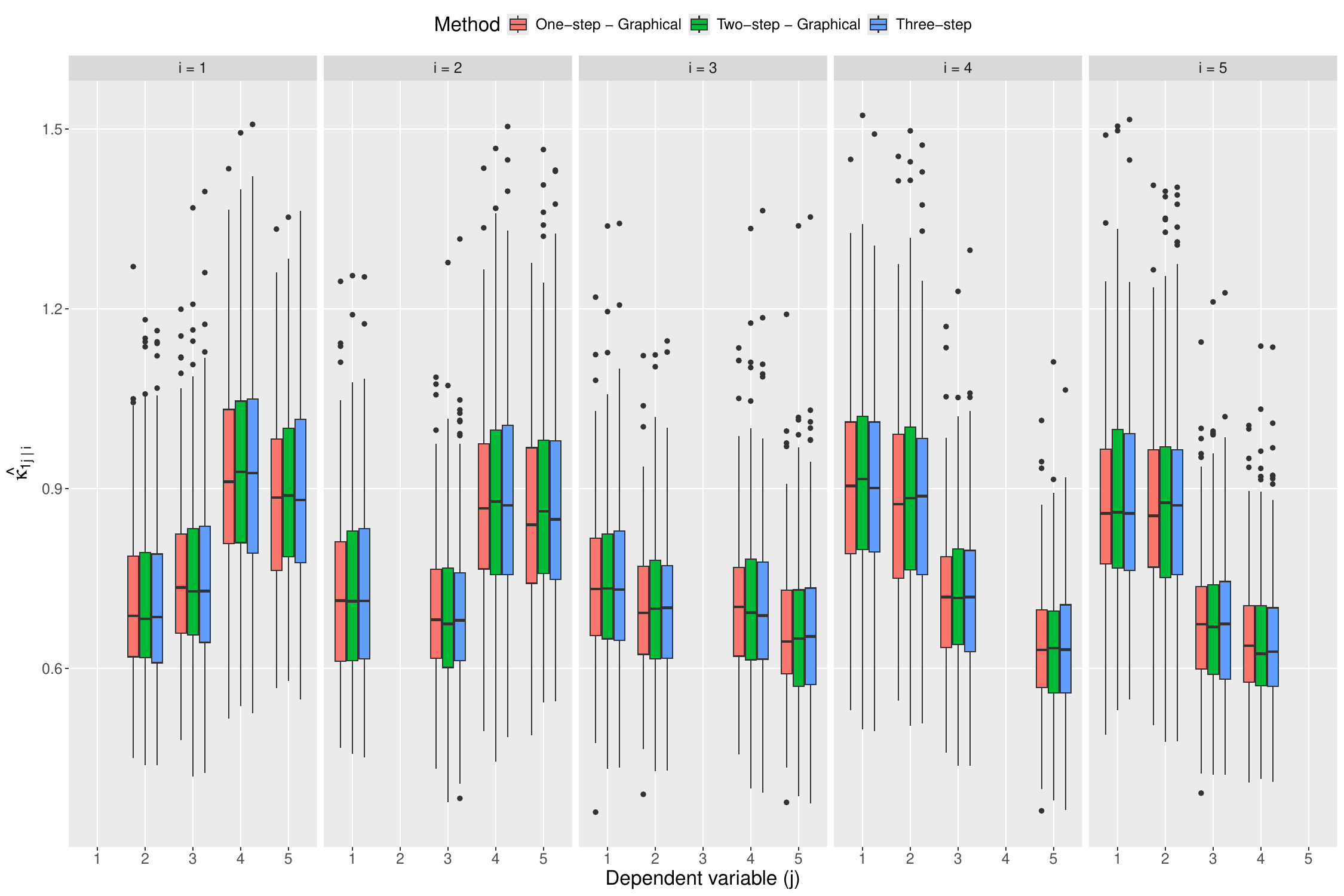}} \quad
  \subfloat{\includegraphics[width=.48\textwidth]{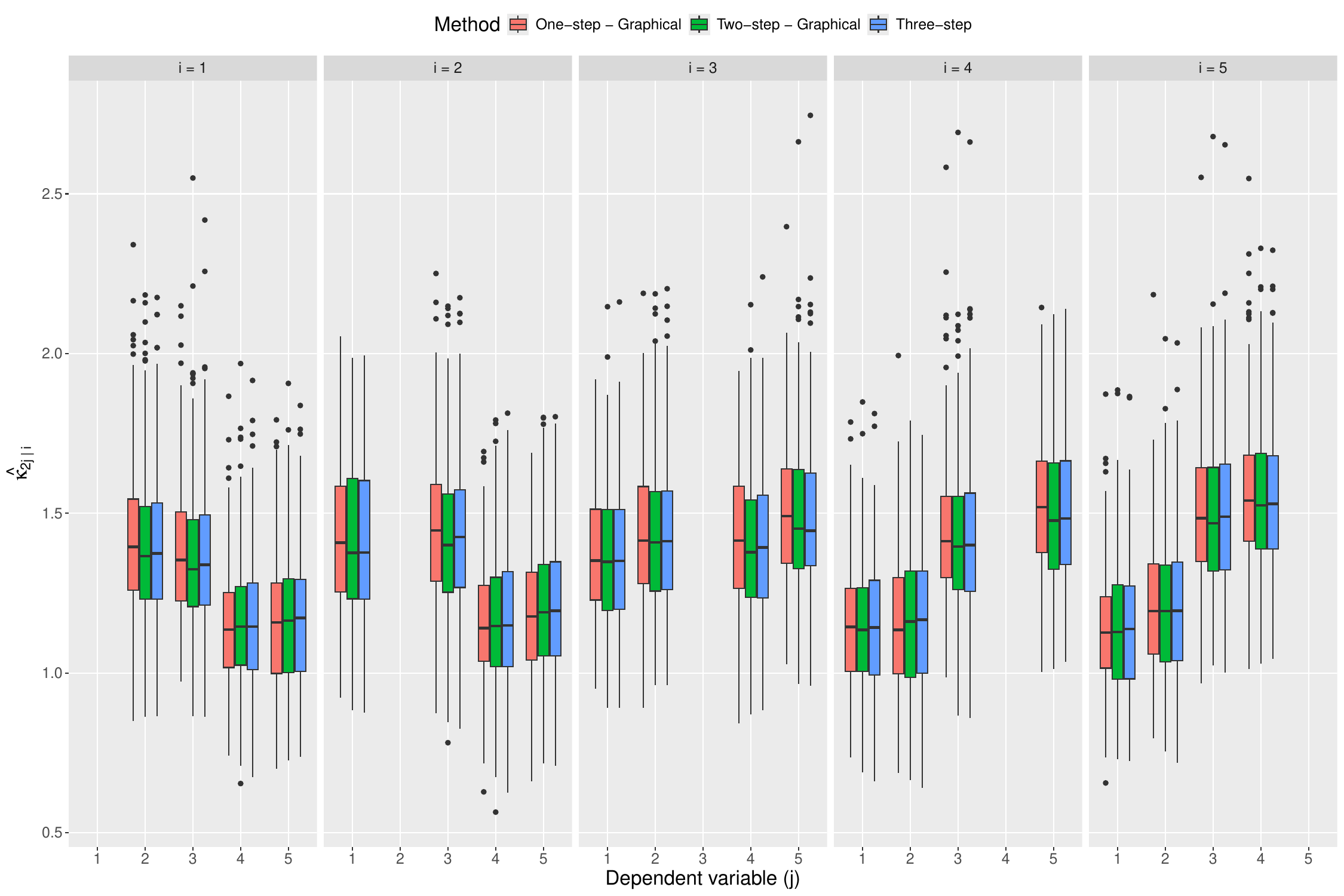}} \\
  \caption{Boxplots of MLEs for $\alpha_{j \mid i}$ (top left), $\beta_{j \mid i}$ (top right), $\nu_{j \mid i}$ (centre left), $\delta_{j \mid i}$ (centre right), $\kappa_{1_{j \mid i}}$ (bottom left), and $\kappa_{2_{j \mid i}}$ (bottom right) for distinct $i, j \in V$. Each column corresponds to the conditioning variable $i$. The different models are denoted by the fill of the boxplots.}
  \label{fig:MVN_Low_Dependence_MLEs}
\end{figure}

In the first study, correlations between all components lie in $(0,0.47)$. Figure \ref{fig:MVN_Low_Dependence_MLEs} shows MLEs of the dependence and AGG parameters. Here, and in the other studies in this section, estimates from the three-step SCMEVM with graphical and saturated covariance structures are omitted as they are identical to results for the three-step SCMEVM with independent residuals. Also note that the MLEs for CMEVM dependence parameters are the same for the two- and three-step methods. The MLEs of $\boldsymbol{\alpha}_{\mid i}$ ($\boldsymbol{\nu}_{\mid i}$) from the one-step procedure are consistently lower (higher) than the MLEs from the stepwise approaches, confirming that the one-step method does not guarantee that the first-order extremal dependence structure will be captured by the dependence parameters. At best, by attributing some extremal dependence structure to the residual distribution, the interpretability of the SCMEVMs fitted with the one-step procedure is reduced. Potentially, it also makes the models less reliable. In contrast, the MLEs of $\boldsymbol{\beta}_{\mid i}$, $\boldsymbol{\kappa_{1}}_{\mid i}$, $\boldsymbol{\kappa_{2}}_{\mid i}$, and $\boldsymbol{\delta}_{\mid i}$ are similar across all the models and fitting procedures. Note that the estimates for $\beta_{j \mid i}$ should be close to $0.5$ when the underlying generating mechanism is the MVG, and there is a positive association between components $i$ and $j$. From the top right panel of Figure \ref{fig:MVN_Low_Dependence_MLEs}, this is clearly not the case. Since the estimated parameters for the one- and two-/three-step fits are similar, the underestimation is likely due to the slow convergence of the MVG distribution to its asymptotic limit. Finally, the right-scale parameter $\kappa_{2_{j \mid i}}$ is almost always estimated to be larger than the left-scale parameter $\kappa_{1_{j \mid i}}$, supporting the choice of an asymmetric marginal distribution for $\boldsymbol{Z}_{\mid i}$. 

\begin{figure}[t!]
    \centering
    \includegraphics[width = \textwidth]{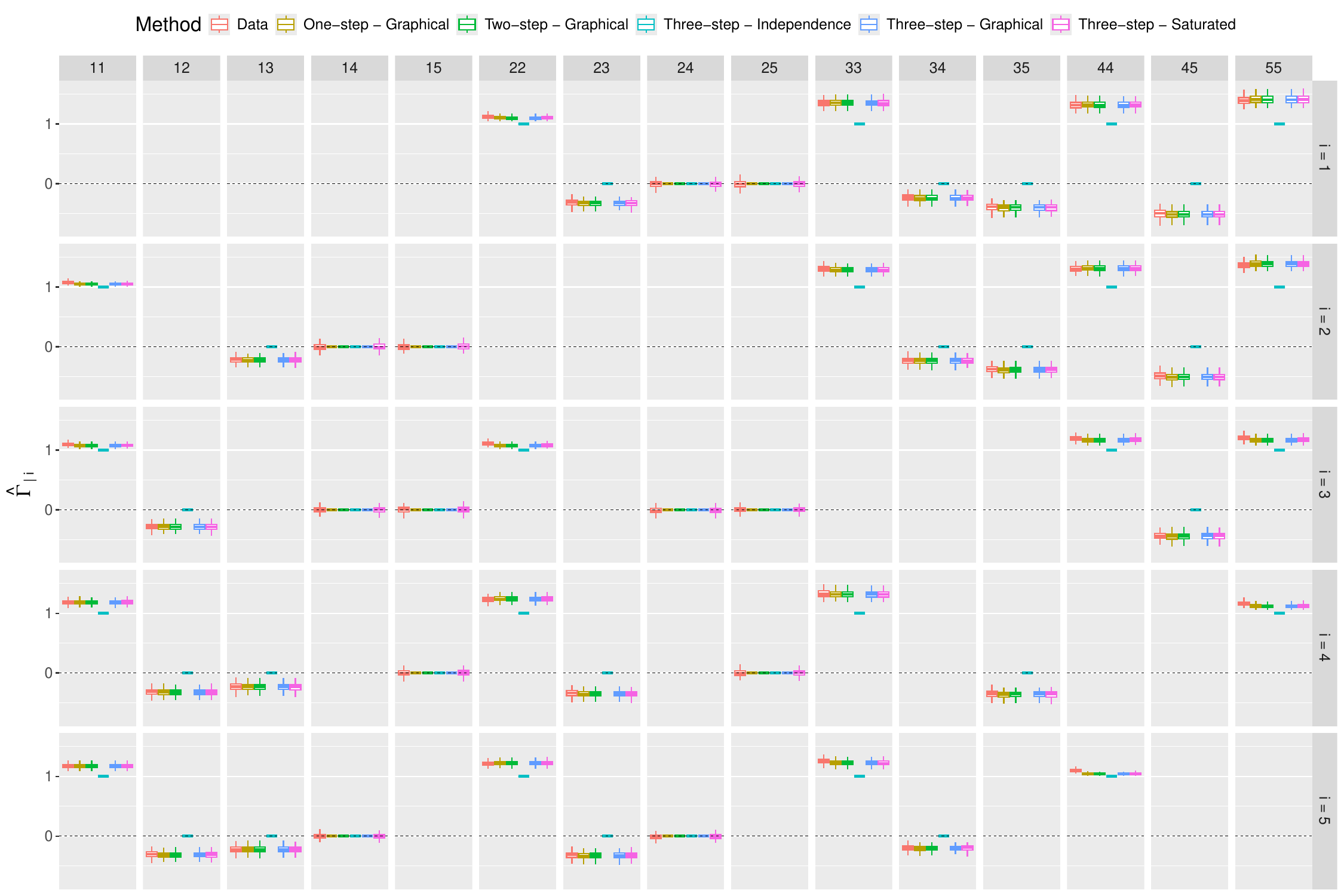}
    \caption{Boxplots of empirical and model-based estimates of $\Gamma_{\mid i}$, for each $i \in V$, when the data is generated from a MVG distribution with weak positive associations. Each row corresponds to the conditioning variable $i$, and each column corresponds to the correlation parameter. The different models are denoted by the colour of the boxplots. Black dashed lines show $y = 0$.}
    \label{fig:Sim_Study_MVN_Gamma}
\end{figure}

Figure \ref{fig:Sim_Study_MVN_Gamma} shows empirical and model-based estimates of the conditional precision matrix $\Gamma_{\mid i}$. Empirical estimates are the inverse of the conditional correlation matrix for $\boldsymbol{Y} \mid Y_{i} = y_{i}$, such that $y_{i} > u_{Y_{i}}$, equivalently the inverse correlation matrix of $\boldsymbol{Y} \mid Y_{i} > u_{Y_{i}}$ excluding the $i$th row and column. Similar to the study in the main text, the estimated matrices are the same for the graphical and saturated SCMEVMs, confirming there is negligible loss in using the former. Further, the estimated structure of the conditional precision matrices for the graphical and saturated SCMEVMs is consistent with the empirical version. While it is plausible that the results here are specific to the MVG generating mechanism, similar patterns are observed for the other multivariate distributions.

\begin{figure}[t!]
  \centering
  \subfloat{\includegraphics[width = 0.48\textwidth]{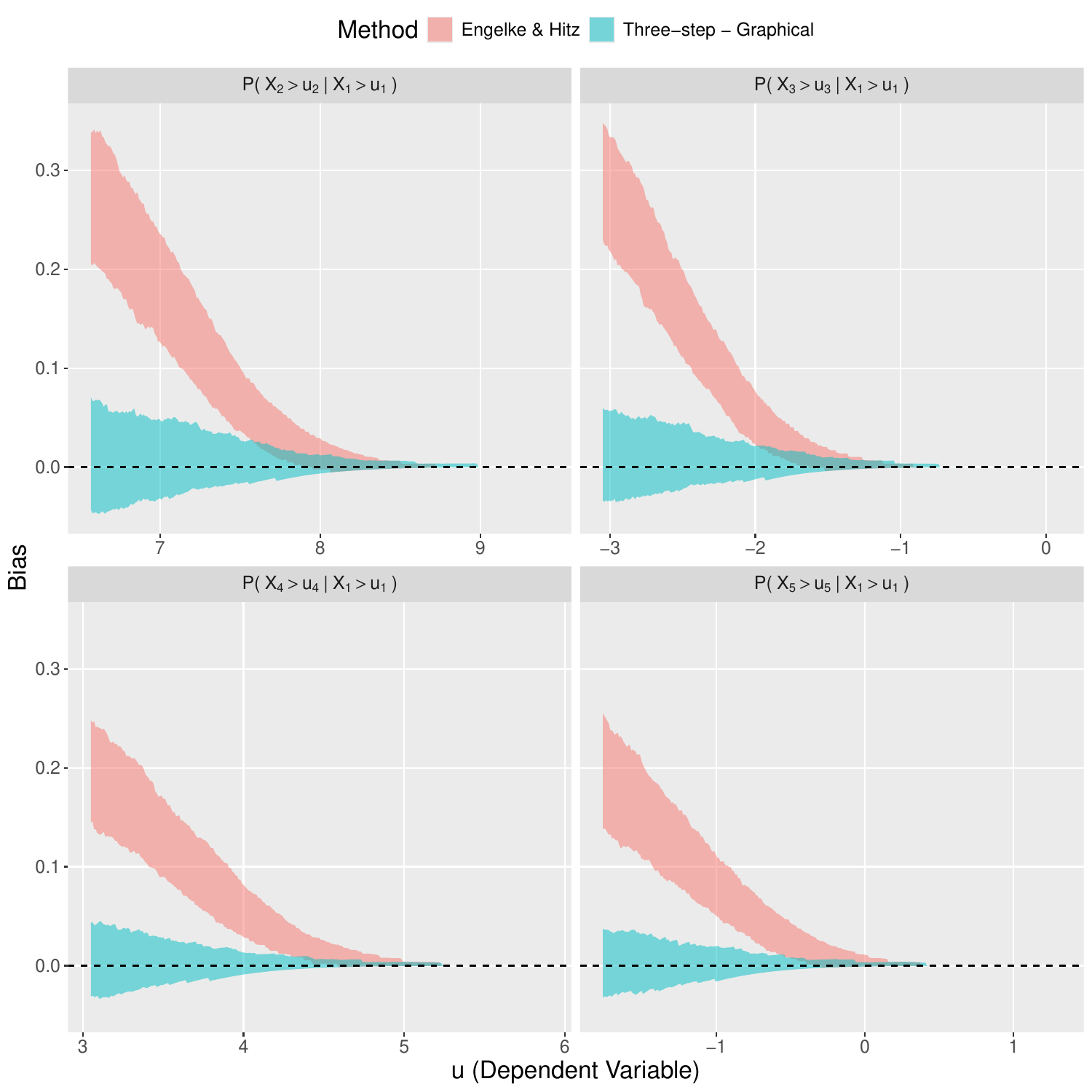}} \quad
  \subfloat{\includegraphics[width = 0.48\textwidth]{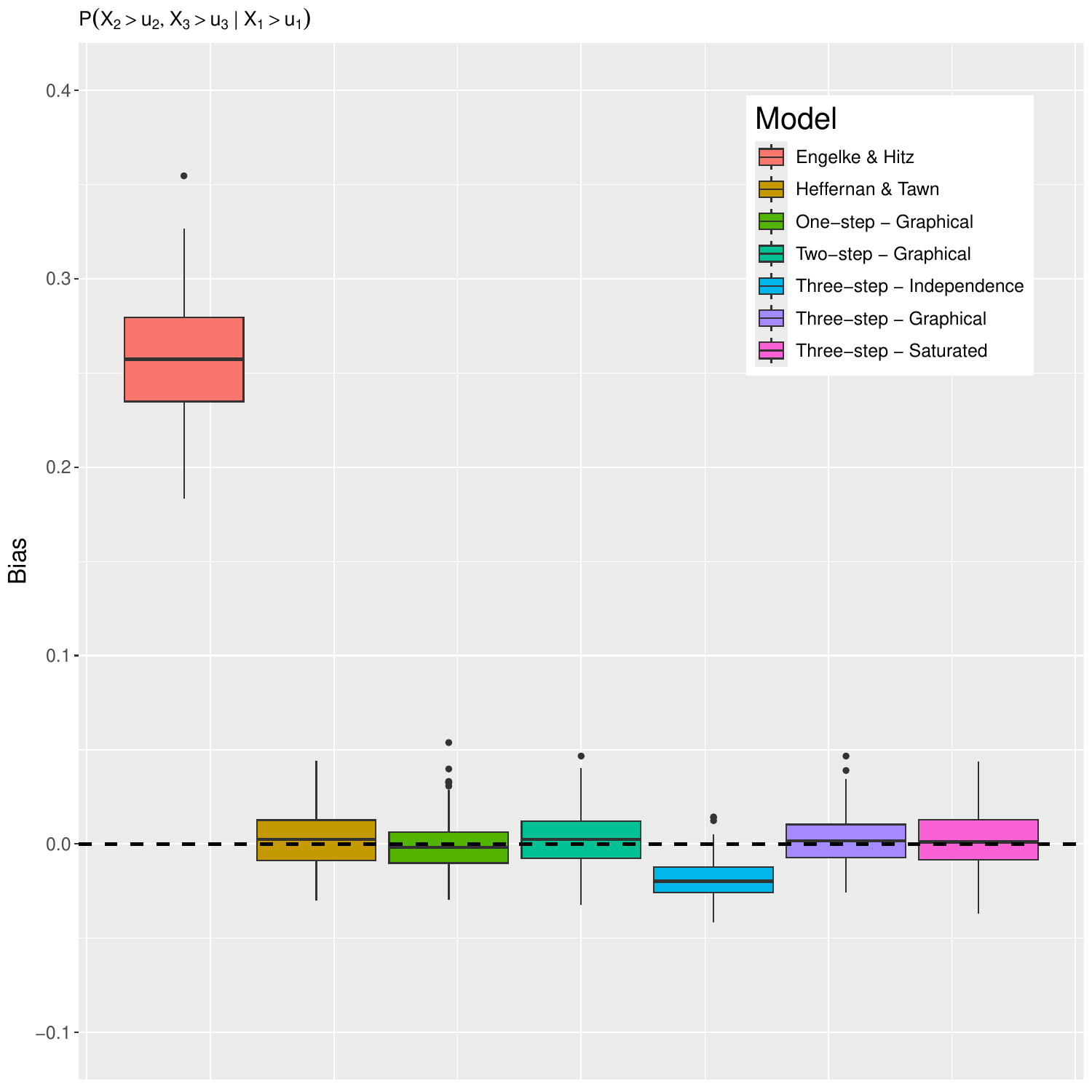}}
  \caption{Polygon plots detailing 95\% confidence intervals, over 200 samples, of the bias in $\mathbb{P}[X_{j} > u_{X_{j}} \mid X_{1} > u_{X_{1}}]$, for each $j \in V_{\mid 1}$, where $\boldsymbol{X}$ follows a MVG distribution with weak positive associations (left). The bias from the EHM and the three-step SCMEVM, assuming a graphical covariance structure for the residuals, are in pink and blue, respectively. Boxplots of the bias in $\mathbb{P}[X_{2} > u_{X_{2}}, X_{3} > u_{X_{3}} \mid X_{1} > u_{X_{1}}]$ (right). The bias from the various models is denoted by the fill of the boxplots. Black dashed lines show $y = 0$.}
  \label{fig:MVN_Prediction}
\end{figure}

We now compare predictions from the EHM and three-step SCMEVM with graphical covariance. Figure \ref{fig:MVN_Prediction} (left panel) shows the bias in the conditional survival curves of $X_{j} \mid X_{1} > u_{X_{1}}$ for each $j \in V_{\mid 1}$. The SCMVEM is unbiased for all curves, whereas the EHM has positive bias for lower values of $u_{X_{j}}$; this decreases as $u_{X_{j}}$ increases. The positive bias of the EHM persists in bivariate conditional survival probabilities. Figure \ref{fig:MVN_Prediction} (right panel) shows the bias in $\mathbb{P}[X_{2} > u_{X_{2}}, X_{3} > u_{X_{3}} \mid X_{1} > u_{X_{1}}]$. The three-step SCMEVM with independent residuals exhibits negative bias because $X_{2}$ is not conditionally independent of $X_{3}$ given $X_{1}$. In contrast, the SCMEVMs with graphical and saturated covariances are unbiased. The CMEVM predictions are also unbiased due to the low dimension $d$. Lastly, the SCMEVMs with graphical covariance exhibit the least amount of bias and variability, minimising the MAE and RMSE for 87\% and 77\% of the 75 conditional probabilities, respectively. This confirms that there is no loss in performance when using a graphical structure over the more flexible saturated one and that the fully parametric SCMEVM outperforms the semi-parametric CMEVM. The EHM performs poorly in this case because the true data have AI. 

\subsubsection{Strong positive dependence}
\label{sec:MVN_High_Dependence}
\begin{figure}[!t]
  \centering
  \subfloat{\includegraphics[width=.48\textwidth]{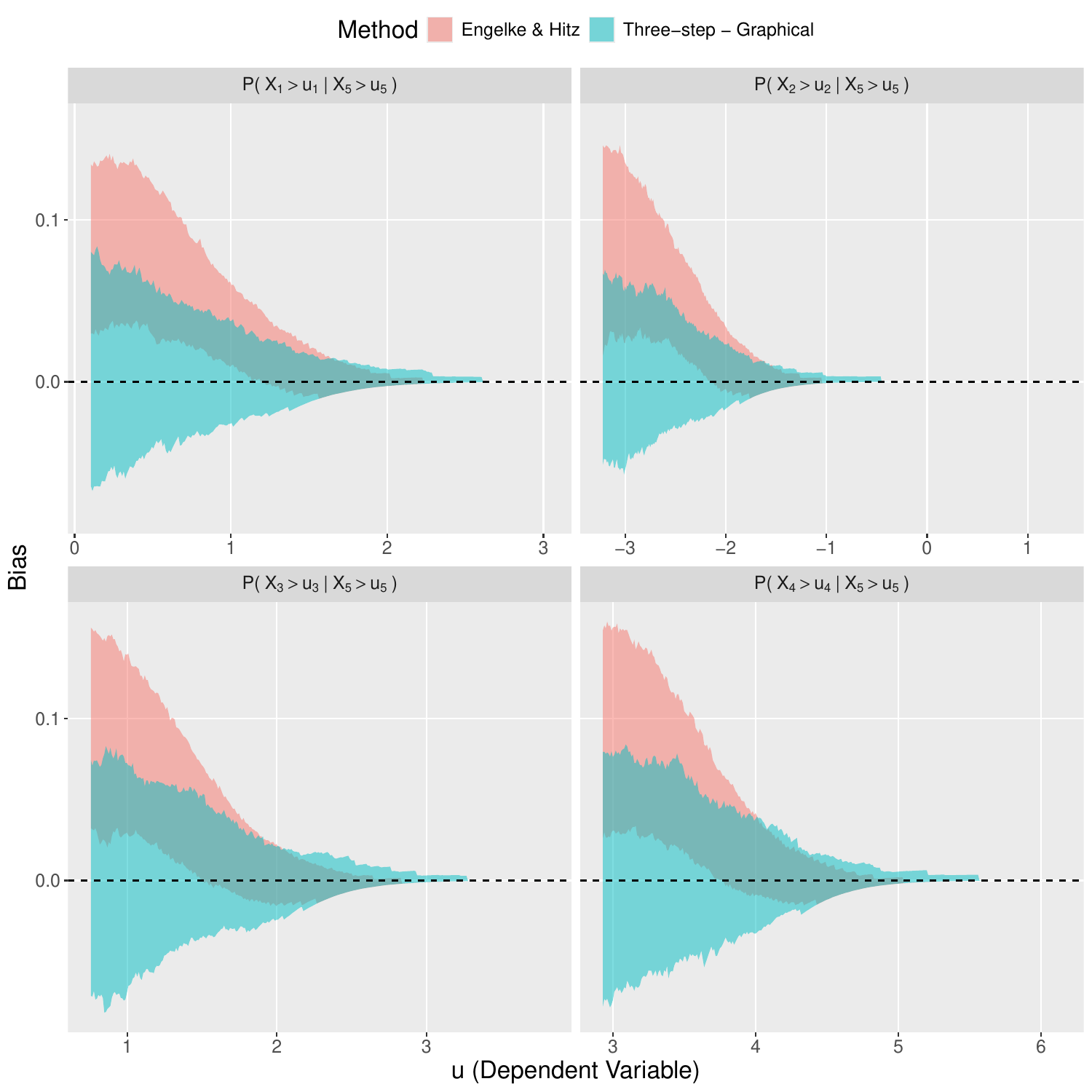}} \quad
  \subfloat{\includegraphics[width=.48\textwidth]{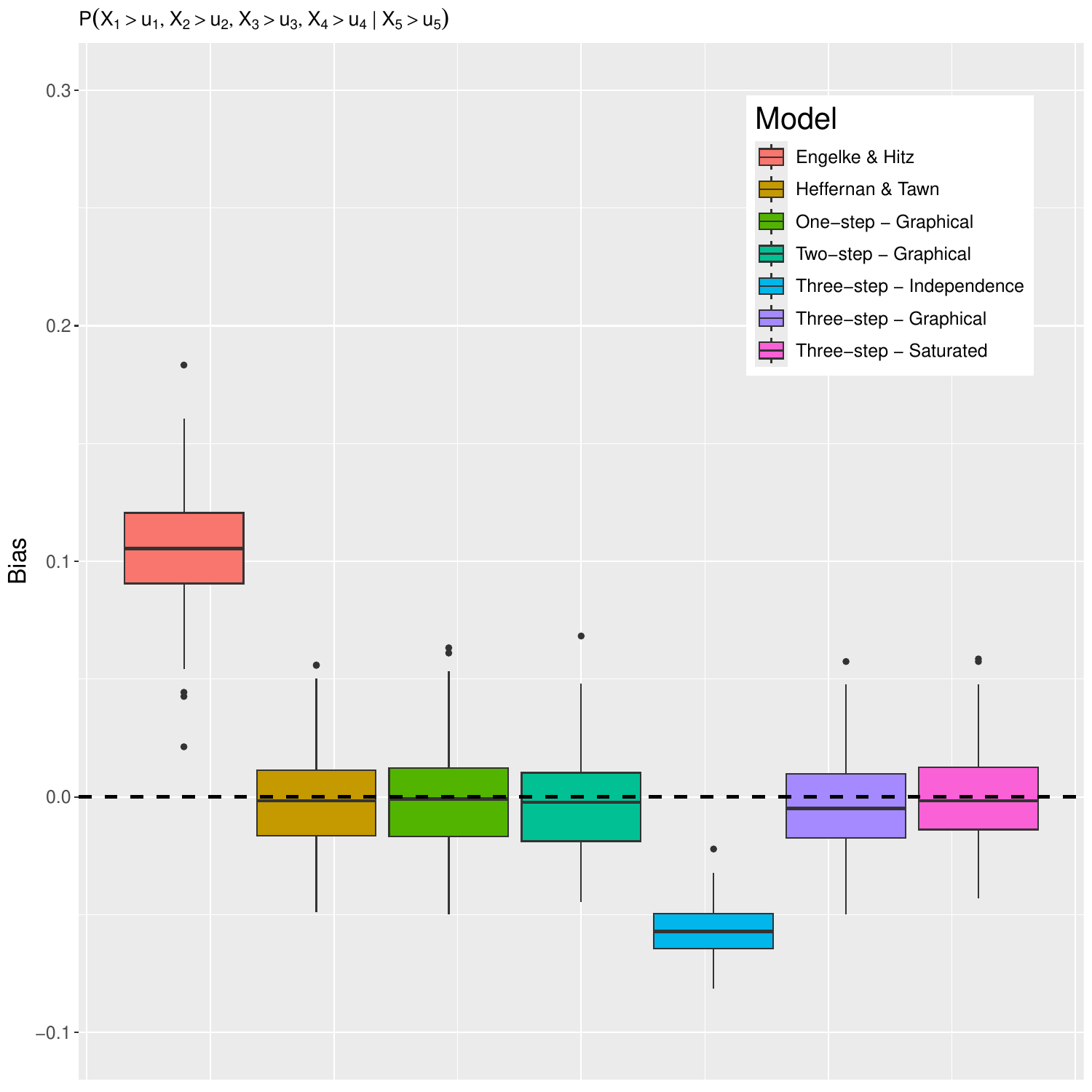}} \\
  \caption{Polygon plots detailing 95\% confidence intervals, over 200 samples, of the bias in $\mathbb{P}[X_{j} > u_{X_{j}} \mid X_{5} > u_{X_{5}}]$, for $j \in V_{\mid 5}$ where $\boldsymbol{X}$ follows a MVG distribution with strong positive associations (left). The bias from the EHM and the three-step SCMEVM with a graphical covariance structure are in pink and blue, respectively. Boxplots of the bias in $\mathbb{P}[\boldsymbol{X}_{\mid 5} > u_{\boldsymbol{X}_{\mid 5}} \mid X_{5} > u_{X_{5}}]$ (right). The bias from the various models is denoted by the fill of the boxplots. Black dashed lines show $y = 0$.}
  \label{fig:MVN_High_Dependence_Probs}
\end{figure}

We repeat the simulation study in Section \ref{Sec:MVN_Low_Dependence}, but the associations between the components of $\boldsymbol{X}$ are strong and positive $(>0.52)$. We present only the predictive performances, as the parameter estimates show similar patterns to those seen in Section \ref{Sec:MVN_Low_Dependence}. Figure \ref{fig:MVN_High_Dependence_Probs} (left panel) shows bias in the conditional survivor curves for $X_{j} \mid X_{5} > u_{X_{5}}$ such that $j \in V_{\mid 5}$ from both the EHM and the three-step SCMEVM with a graphical covariance structure. Again, the EHM is biased for low values of $u_{X_{j}}$, but this diminishes as $u_{X_{j}}$ increases; the three-step SCMEVM with graphical structure is unbiased for all $u_{X_{j}}$. 
Figure \ref{fig:MVN_High_Dependence_Probs} (right panel) shows the bias in $\mathbb{P}[\boldsymbol{X}_{\mid 5} > u_{\boldsymbol{X}_{\mid 5}} \mid X_{5} > u_{X_{5}}]$. The EHM has positive bias, whereas both the CMEVM and the SCMEVMs with graphical or saturated covariance structures are unbiased. The three-step SCMEVM with independent residuals exhibits negative bias; this is expected since the components of $\boldsymbol{X}_{\mid 5}$ are not independent given $X_{5}$ is large. Assessing overall predictive performance, the SCMEVMs with graphical covariance structure are again the least biased and variable, minimising the MAE and RMSE metrics 81\% and 84\% of the time, respectively.

\subsubsection{Negative dependence}
\label{sec:MVN_Negative_Dependence}
\begin{figure}[t!]
    \centering
    \includegraphics[width = \textwidth]{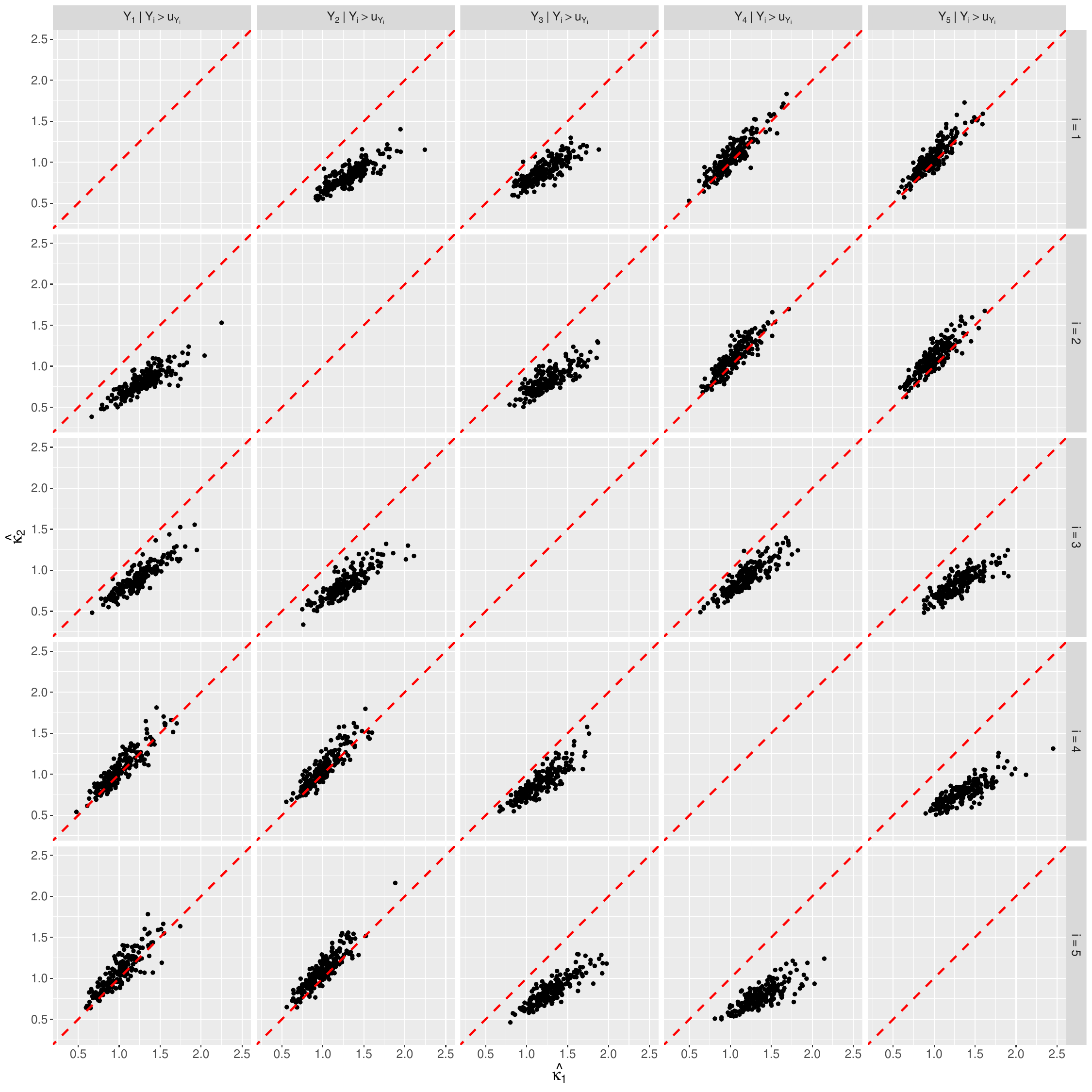}
    \caption{Scatter plots comparing $\hat{\kappa}_{1_{j \mid i}}$ and $\hat{\kappa}_{2_{j \mid i}}$ from the three-step SCMEVM with graphical covariance structure for distinct $i, j \in V$. Red dashed lines show $y = x$.}
    \label{fig:MVN_Neg_Dependence_Scale_Comp}
\end{figure}

We repeat the simulation study in Section \ref{Sec:MVN_Low_Dependence}, but the association between the components of $\boldsymbol{X}$ are now allowed to be negative. The correlation matrix $\Sigma$ is given in equation~\eqref{eqn:MVN_Negative_Dependence_Matrices}.
\begin{equation}
    \Sigma = 
    \begin{bmatrix}
        1.000 & -0.468 & -0.370 & -0.136 & 0.134 \\
        -0.468 & 1.000 & 0.390 & 0.144 & -0.141 \\
        -0.370 & 0.390 & 1.000 & 0.369 & -0.362 \\
        -0.136 & 0.144 & 0.369 & 1.000 & -0.346 \\
        0.134 & -0.141 & -0.362 & -0.346 & 1.000 \\
    \end{bmatrix}.
    \label{eqn:MVN_Negative_Dependence_Matrices}
\end{equation}
Figure \ref{fig:MVN_Neg_Dependence_Scale_Comp} compares the MLEs of $\kappa_{1_{j \mid i}}$ and $\kappa_{2_{j \mid i}}$ from the three-step SCMEVM with a graphical covariance structure, for distinct $i, j \in V$. In the other MVG examples, the right-scale parameter is generally larger than the left-scale parameter, whereas here a range of behaviour is observed. This further justifies the need for a flexible, asymmetric distribution for $\boldsymbol{Z}_{\mid i}$.

\begin{figure}[t!]
  \centering
  \subfloat{\includegraphics[width=.48\textwidth]{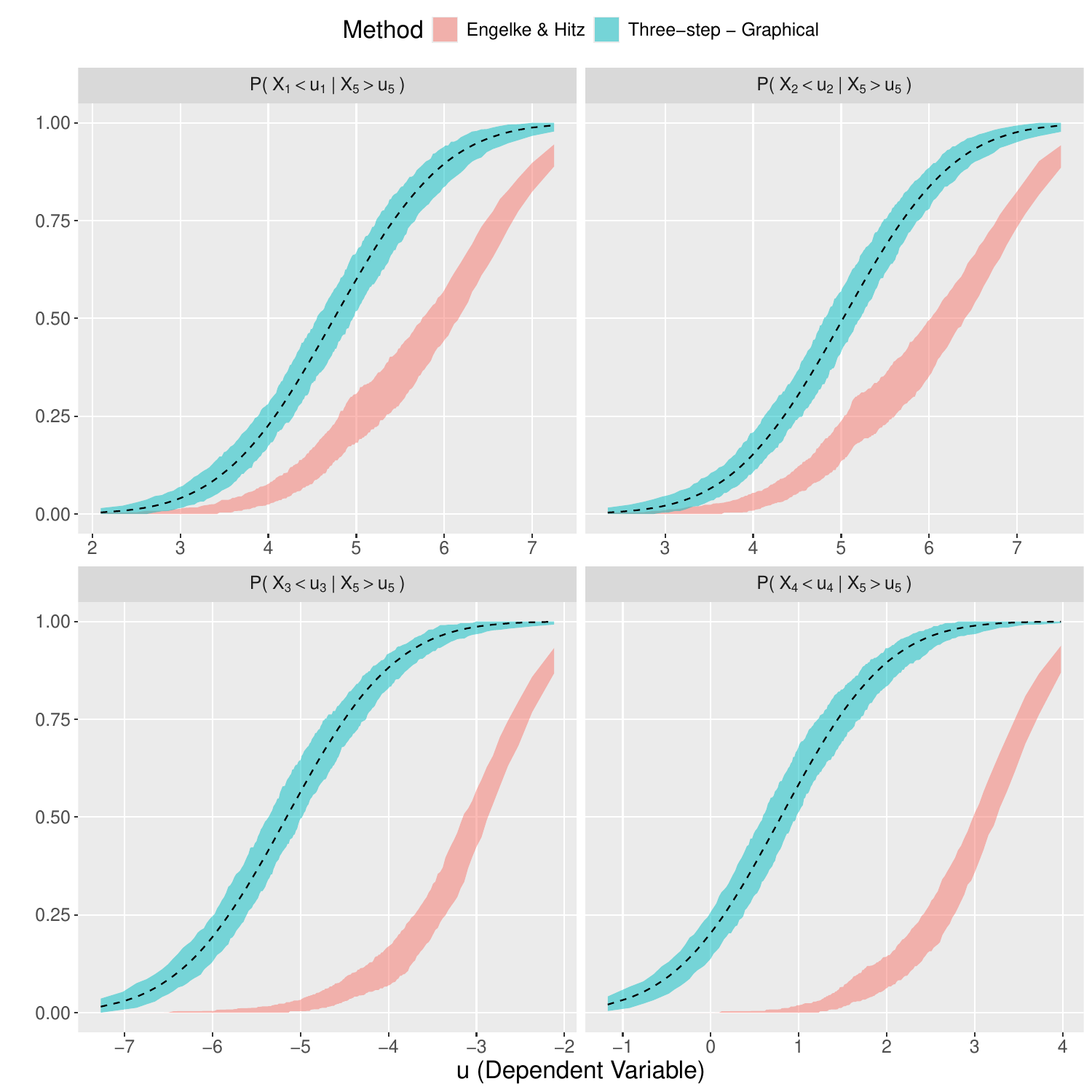}} \quad
  \subfloat{\includegraphics[width=.48\textwidth]{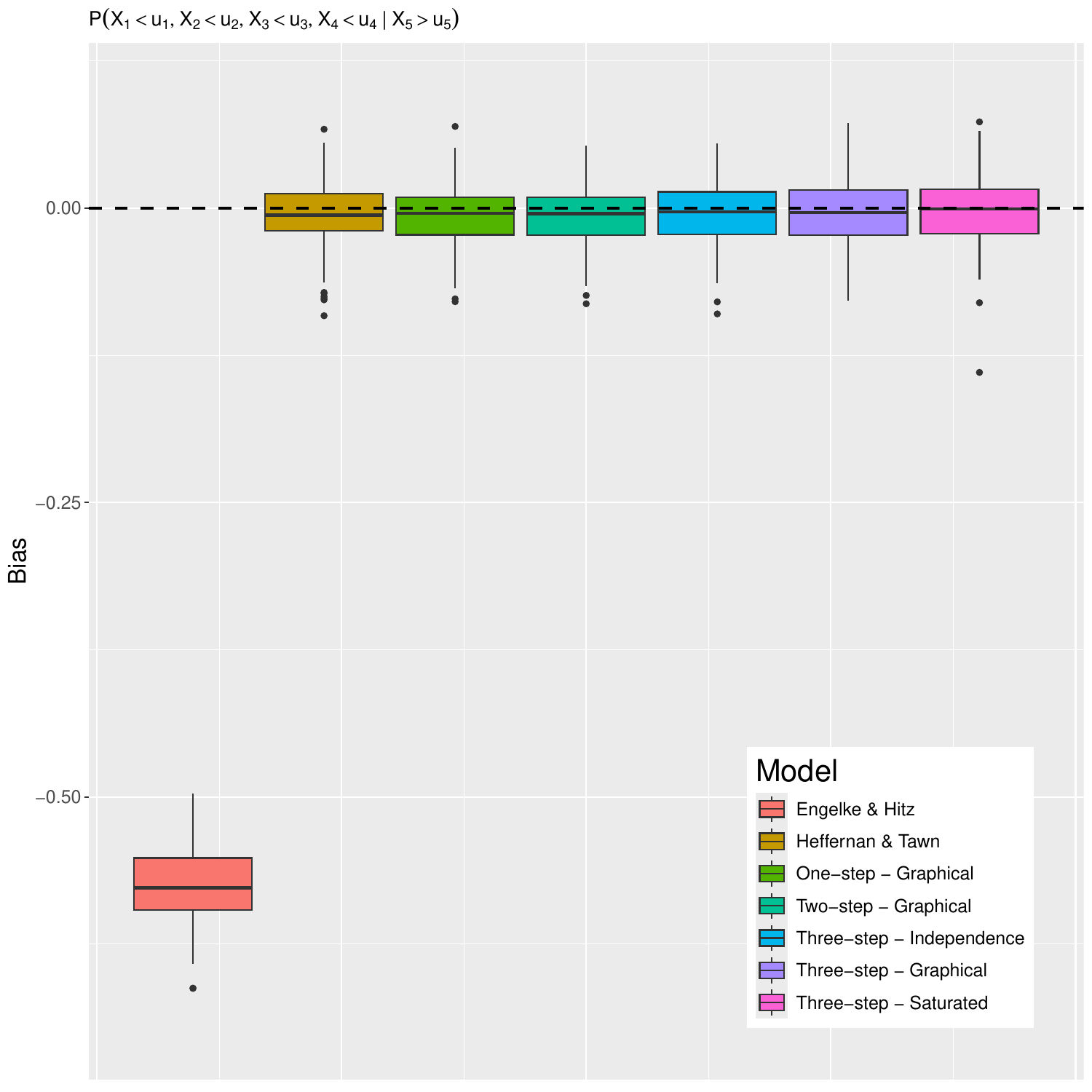}} \\
  \caption{Polygon plots detailing 95\% confidence intervals, over 200 samples, of $\mathbb{P}[X_{j} < u_{X_{j}} \mid X_{5} > u_{X_{5}}]$ for $j \in V_{\mid 5}$, when $\boldsymbol{X}$ follows a MVG distribution with negative associations (left). The estimated curves from the EHM and the three-step SCMEVM with a graphical covariance structure are in pink and blue, respectively. The true conditional cumulative distribution curves are given by the black dashed lines. Boxplots of the bias in $\mathbb{P}[\boldsymbol{X}_{\mid 5} < u_{\boldsymbol{X}_{\mid 5}} \mid X_{5} > u_{X_{5}}]$ (right). The bias from the various models is denoted by the fill of the boxplots. The $y = 0$ line is indicated by the black dashed line.}
  \label{fig:MVN_Negative_Dependence_Probs}
\end{figure}

Figure \ref{fig:MVN_Negative_Dependence_Probs} (left panel) shows the bias in the conditional cumulative distribution curves for $X_{j} \mid X_{5} > u_{X_{5}}$ for the EHM and the three-step SCMEVM with graphical covariance structure. The three-step SCMEVM exhibits no bias, but the EHM underestimates the curve over the entire range. Again, this is not surprising, as the AD assumption is not satisfied by the data. Figure \ref{fig:MVN_Negative_Dependence_Probs} (right panel) considers the bias in $\mathbb{P}[\boldsymbol{X}_{\mid 5} < u_{\boldsymbol{X}_{\mid 5}} \mid X_{5} > u_{X_{5}}]$. The EHM exhibits negative bias, while the CMEVM and SCMEVMs are unbiased. Finally, as in the previous studies, the SCMEVMs with graphical covariance structures are the least biased and variable, minimising the MAE and RMSE 76\% and 87\% of the time, respectively.

\subsection{Multivariate Laplace distribution}
In this study, $\boldsymbol{X}$ follows a MVL distribution with mean vector $\boldsymbol{\mu}$, where $\mu_{j}$ are independently sampled from a uniform distribution on $(-5, 5)$, and precision matrix consistent with $\mathcal{G}$ in Section \ref{Sec:Sim_Study_True_Dist}.

\subsubsection{Weak positive dependence}
\label{sec:MVL_Weak_Dependence}
\begin{figure}[t!]
    \centering
    \includegraphics[width = \textwidth]{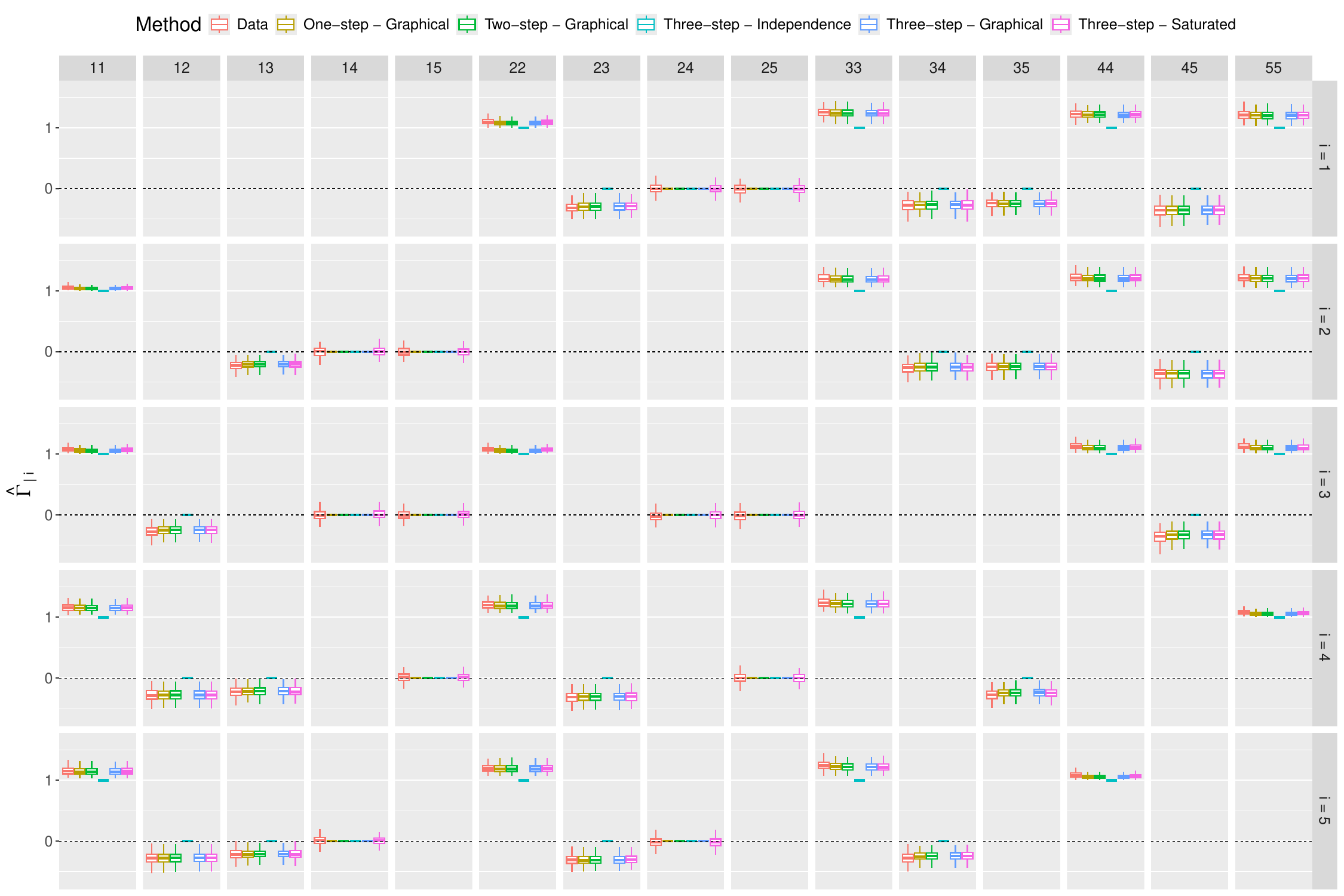}
    \caption{Boxplots of empirical and model-based estimates of $\Gamma_{\mid i}$, for each $i \in V$, when the data is generated from a MVL distribution with weak positive associations. Each row corresponds to the conditioning variable $i$, and each column corresponds to the correlation parameter. The different models are distinguished by the colour of the boxplots. Black dashed lines show $y = 0$.}
    \label{fig:Sim_Study_MVL_Gamma}
\end{figure}

In this simulation, associations between components are weakly positive i.e., the elements of the true correlation matrix are strictly positive but less than $0.37$. We set the dependence threshold $u_{Y_{i}}$ to the $0.95$-quantile of the standard Laplace distribution for all $i \in V$, resulting in approximately $250$ excesses per conditioning variable. For prediction, $u_{X_{i}}$ is set to the $0.95$-quantile from a single sample of $10^6$ from the true distribution. 
Figure \ref{fig:Sim_Study_MVL_Gamma} shows empirical and model-based estimates of the conditional precision matrix $\Gamma_{\mid i}$. The estimated structure of the conditional precision matrix from both the graphical and the saturated
SCMEVM is consistent with the empirical version. Analysis of other parameter estimates has been omitted, but they are very similar across all three stepwise procedures. The only point to note is that estimates for the left- and right-scale parameters in the MVAGG are very similar, raising the question of whether the generalised Gaussian would be a better choice of marginal residual distribution. However, other examples do have very different scale parameters (see Figures \ref{fig:MVN_Low_Dependence_MLEs} and \ref{fig:MVN_Neg_Dependence_Scale_Comp}) and the more flexible asymmetric generalised Gaussian distribution is therefore recommended.

\begin{figure}[t!]
  \centering
  \subfloat{\includegraphics[width = 0.48\textwidth]{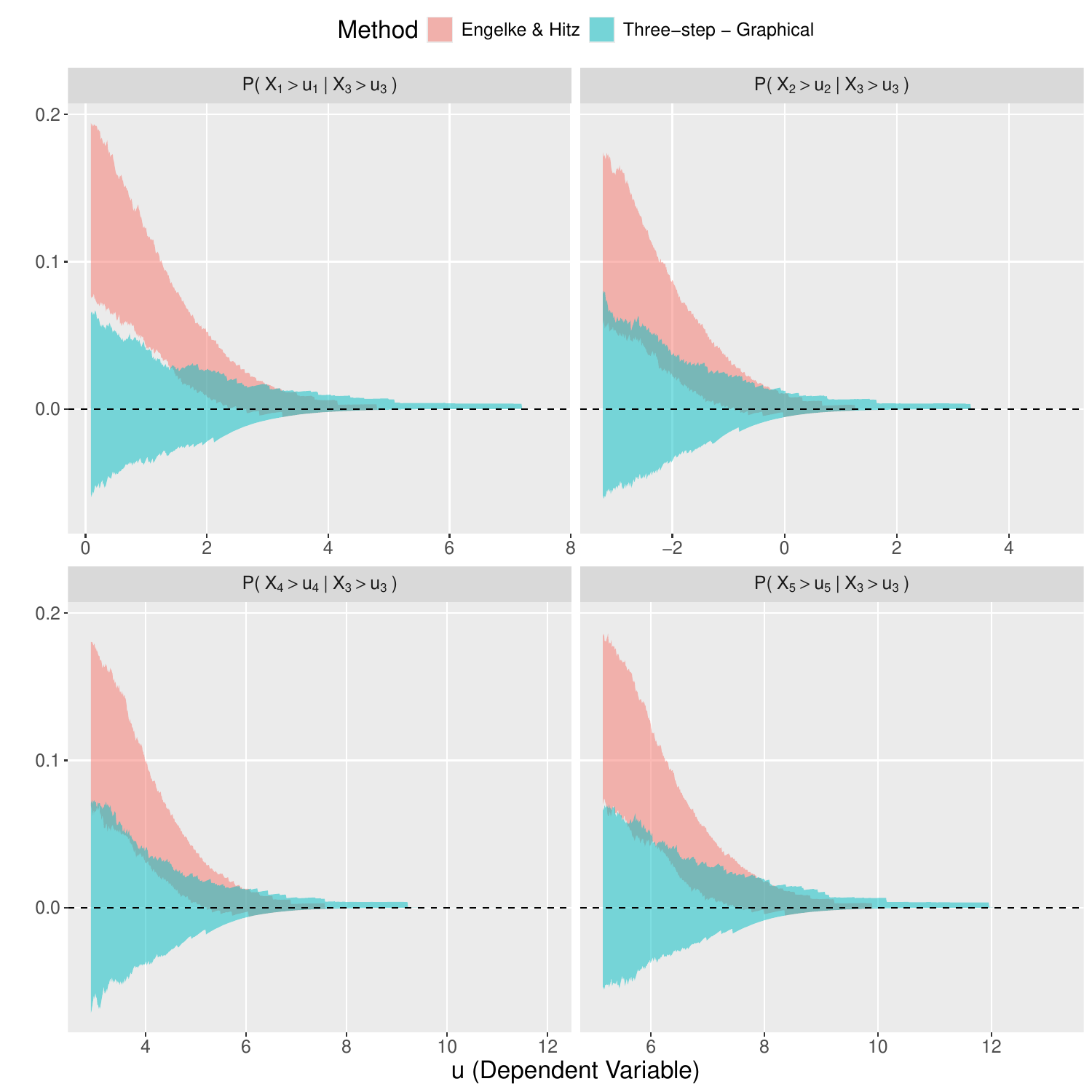}} \quad
  \subfloat{\includegraphics[width = 0.48\textwidth]{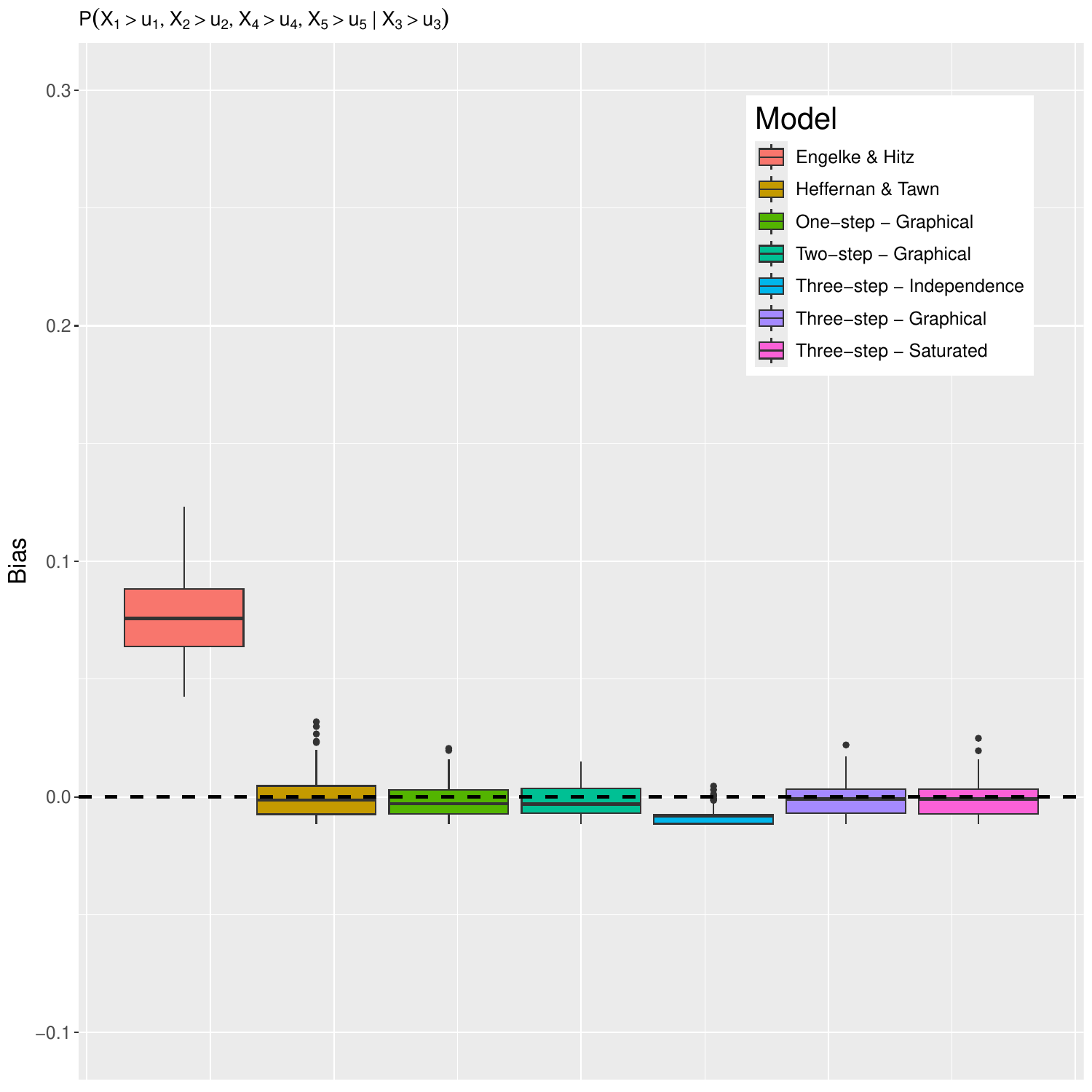}}
  \caption{Polygon plots detailing 95\% confidence intervals, over 200 samples, of the bias in $\mathbb{P}[X_{j} > u_{X_{j}} \mid X_{3} > u_{X_{3}}]$ for $j \in V_{\mid 3}$, where $\boldsymbol{X}$ follows a MVL distribution with weak positive associations (left). The bias from the EHM and the three-step SCMEVM, assuming a graphical covariance structure for the residuals are in pink and blue, respectively. Boxplots of the bias in $\mathbb{P}[\boldsymbol{X}_{\mid 3} > u_{\boldsymbol{X}_{\mid 3}} \mid X_{3} > u_{X_{3}}]$ (right). The bias from the various models is denoted by the fill of the boxplots. Black dashed lines show $y = 0$.}
  \label{fig:MVL_Prediction_Weak}
\end{figure}

To compare predictive performance, Figure \ref{fig:MVL_Prediction_Weak} (left panel) shows the bias in the conditional survival curves of $X_{j} \mid X_{3} > u_{X_{3}}$ for $j \in V_{\mid 3}$. Similar to the MVG examples, the SCMEVM with graphical covariance structure is unbiased for all curves, whereas the EHM has positive bias for low values of $u_{X_{j}}$, which decreases as $u_{X_{j}}$ increases. Figure \ref{fig:MVL_Prediction_Weak} (right panel) shows the bias in $\mathbb{P}[\boldsymbol{X}_{\mid 3} > u_{\boldsymbol{X} \mid 3} \mid X_{3} > u_{X_{3}}]$. The CMEVM and the SCMEVMs with graphical or saturated covariance structures are unbiased, whereas the EHM has positive bias. Again, the three-step SCMEVM with independent residuals has negative
bias because not all components of $\boldsymbol{X}_{\mid 3}$ are independent given $X_{3}$. Similar findings are made when assessing other conditional probabilities of the form $\mathbb{P}[\boldsymbol{X}_{A} > u_{A} \mid X_{i} > u_{i}]$ for all $A \subseteq V_{\mid i}$ and $i \in V$. Lastly, the SCMEVMs with graphical covariance have the least amount of bias and variability, minimising the MAE and RMSE for 86\% and 77\% of the 75 conditional probabilities, respectively.  

\subsubsection{Strong positive dependence}
\label{sec:MVL_High_Dependence}
We repeat the simulation study in Section \ref{sec:MVL_Weak_Dependence} with strong positive association between the components, i.e., the entries of the true correlation matrix are all greater than $0.69$. The dependence threshold $u_{Y_{i}}$ is set to the $0.9$-quantile for the standard Laplace distribution, for all $i \in V$. For prediction, we set $u_{X_{i}}$ to the $0.95$-quantile from a dataset of size $10^6$ simulated from the true distribution for each $i \in V$. We omit parameter estimates since the only point to note is that a comparison of the estimates from the one-, two-, and three-step SCMEVMs shows that the dependence parameters are slightly larger for the one-step method, while the location and scale parameters are slightly lower. 

\begin{figure}[t!]
  \centering
  \subfloat{\includegraphics[width=.48\textwidth]{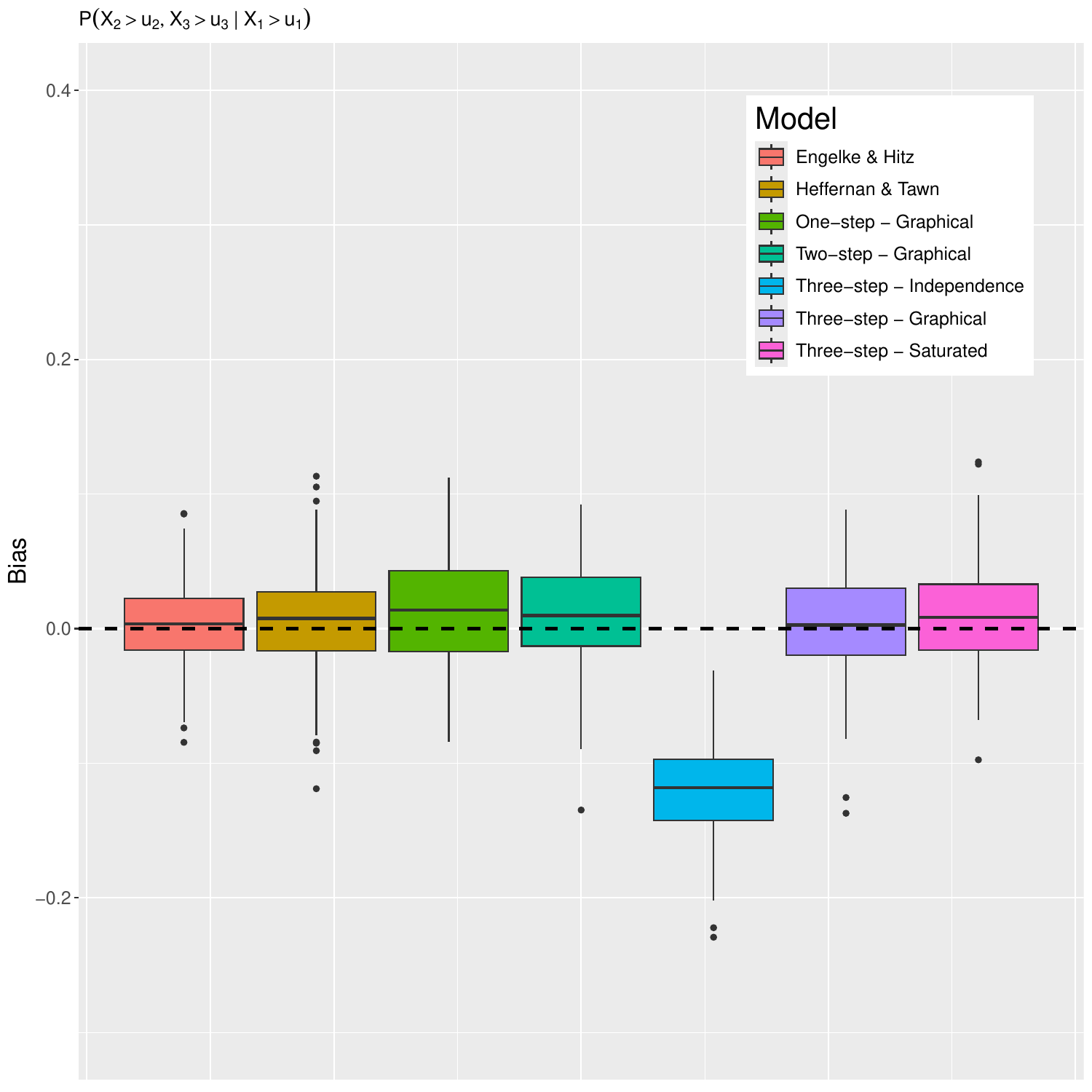}} \quad
  \subfloat{\includegraphics[width=.48\textwidth]{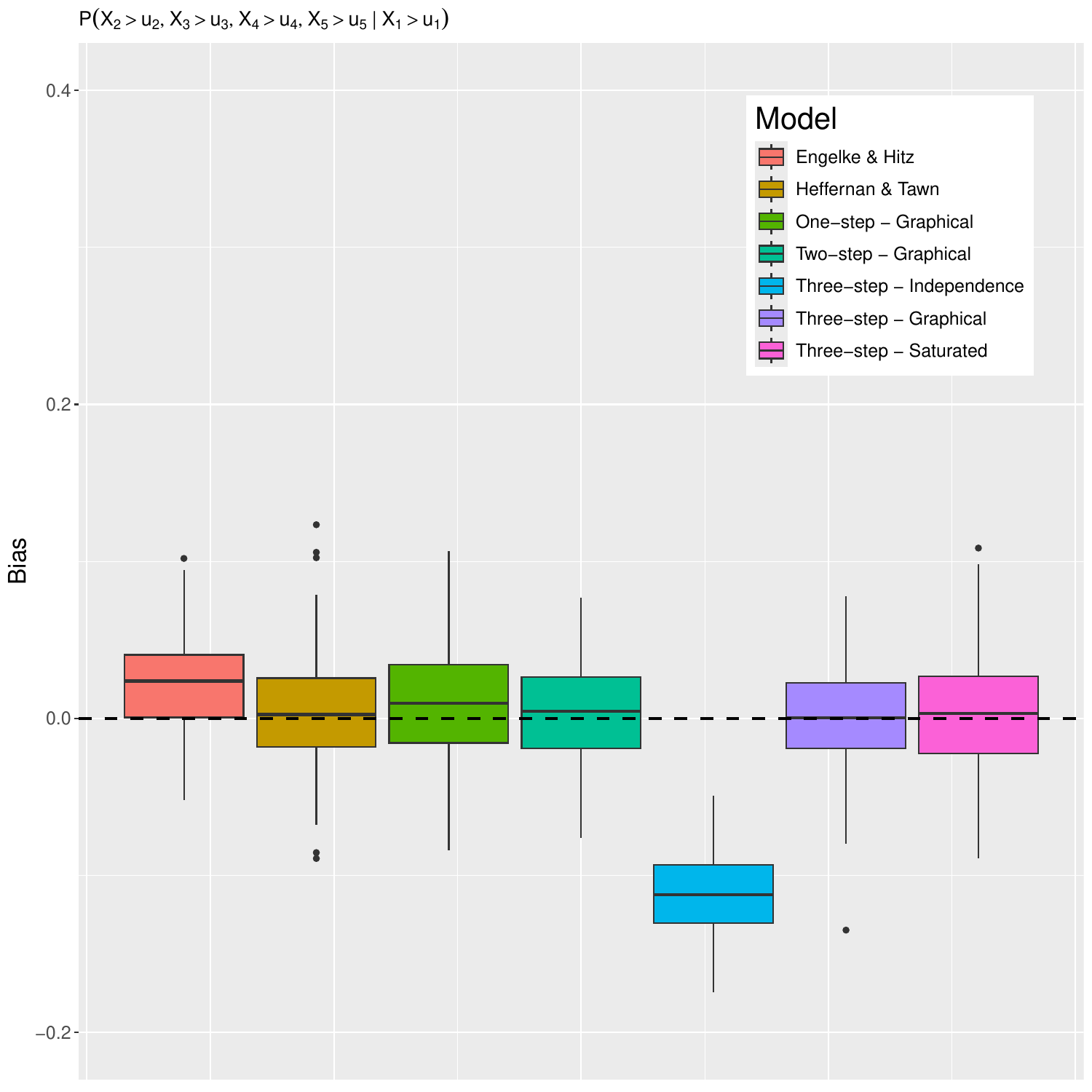}} \\
  \caption{Boxplots of the bias in $p_{1} = \mathbb{P}[X_{2} > u_{X_{2}}, X_{3} > u_{X_{3}} \mid X_{1} > u_{X_{1}}]$ (left) and $p_{2} = \mathbb{P}[\boldsymbol{X}_{\mid 1} > u_{\boldsymbol{X}_{\mid 1}} \mid X_{1} > u_{X_{1}}]$ (right) when $\boldsymbol{X}$ follows a MVL distribution with strong positive associations. The bias from the various models is denoted by the fill of the boxplots. Black dashed lines show $y = 0$.}
  \label{fig:MVL_High_Dependence_Probs}
\end{figure}

Figure \ref{fig:MVL_High_Dependence_Probs} shows the bias in two tail probabilities, $p_{1} = \mathbb{P}[X_{2} > u_{X_{2}}, X_{3} > u_{X_{3}} \mid X_{1} > u_{X_{1}}]$ and $p_{2} = \mathbb{P}[\boldsymbol{X}_{\mid 1} > u_{\boldsymbol{X}_{\mid 1}} \mid X_{1} > u_{X_{1}}]$. In this case, the EHM performs more similarly to the CMEVM and SCMEVMs. However, the model exhibits a small positive bias for $p_2$. The EHM minimises the MAE and RMSE for 37 and 42 of the 75 conditional probabilities, respectively. For comparison, the three-step SCMEVM with graphical covariance structure minimises the metrics 31 and 25 times, respectively. Despite poorer performance on the metrics, the CMEVM and SCMEVMs with graphical or saturated covariance structures are unbiased for both probabilities in Figure \ref{fig:MVL_High_Dependence_Probs}, suggesting these models scale better with dimension compared to the EHM.

\subsubsection{Negative dependence}
Finally, we consider negative associations between components. The true correlation matrix is provided in equation~\eqref{eqn:MVL_Negative_Dependence_Matrices}. The dependence threshold $u_{Y_{i}}$ is set to the $0.8$-quantile for the standard Laplace distribution for all $i \in V$. For prediction, we set $u_{X_{i}}$ to the $0.9$-quantile from a dataset of size $10^6$ simulated from the true distribution for each $i \in V$.

\begin{equation}
    \Sigma = 
    \begin{bmatrix}
        1.000 & -0.200 & -0.139 & 0.026 & 0.022\\
        -0.200 & 1.000 & -0.243 & 0.045 & 0.038\\
        -0.139 & -0.243 & 1.000 & -0.185 & -0.158\\
        0.026 & 0.045 & -0.185 & 1.000 & -0.276\\
        0.022 & 0.038 & -0.158 & -0.276 & 1.000\\
    \end{bmatrix}.
  \label{eqn:MVL_Negative_Dependence_Matrices}
\end{equation}

\begin{figure}[t!]
  \centering
  \subfloat{\includegraphics[width=.48\textwidth]{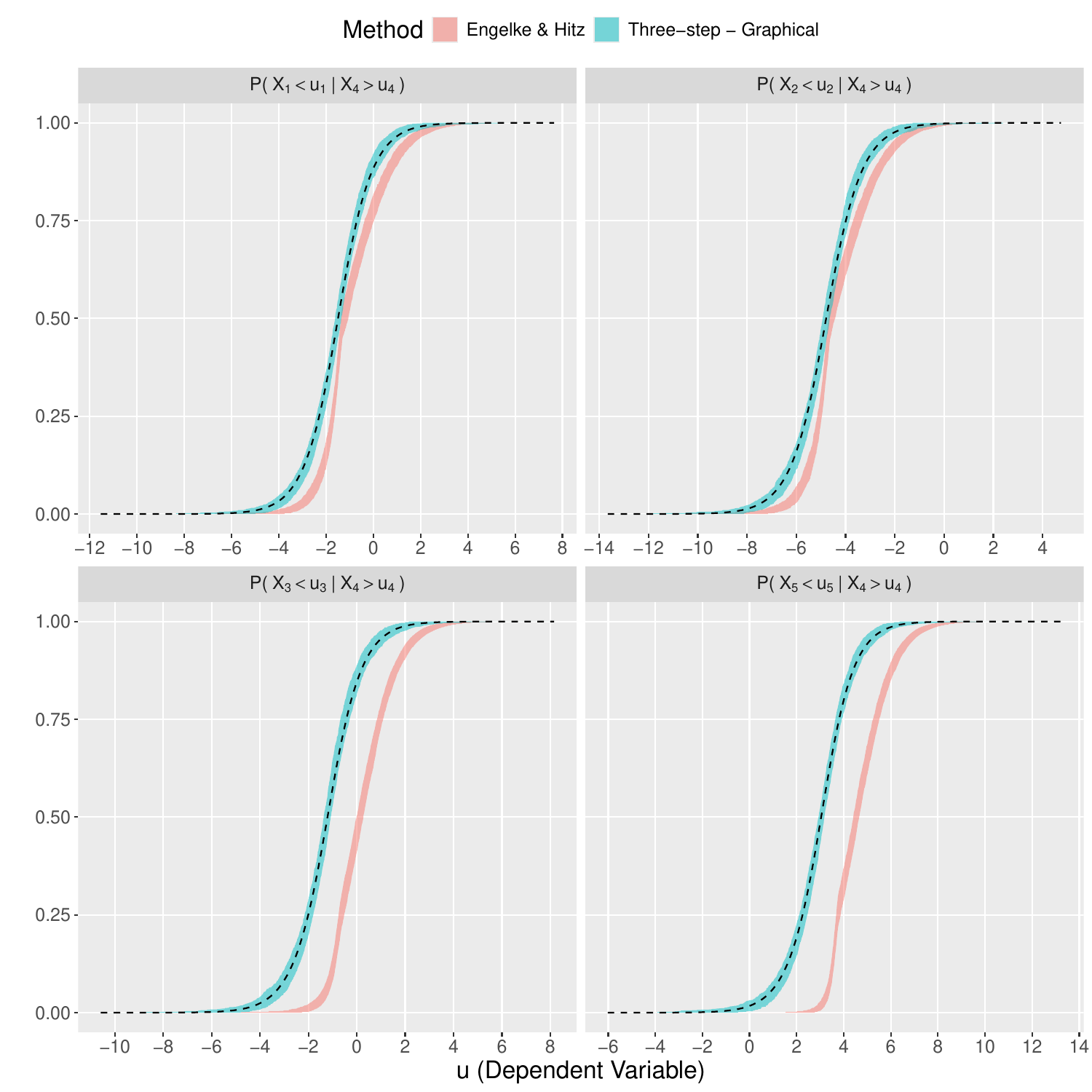}} \quad
  \subfloat{\includegraphics[width=.48\textwidth]{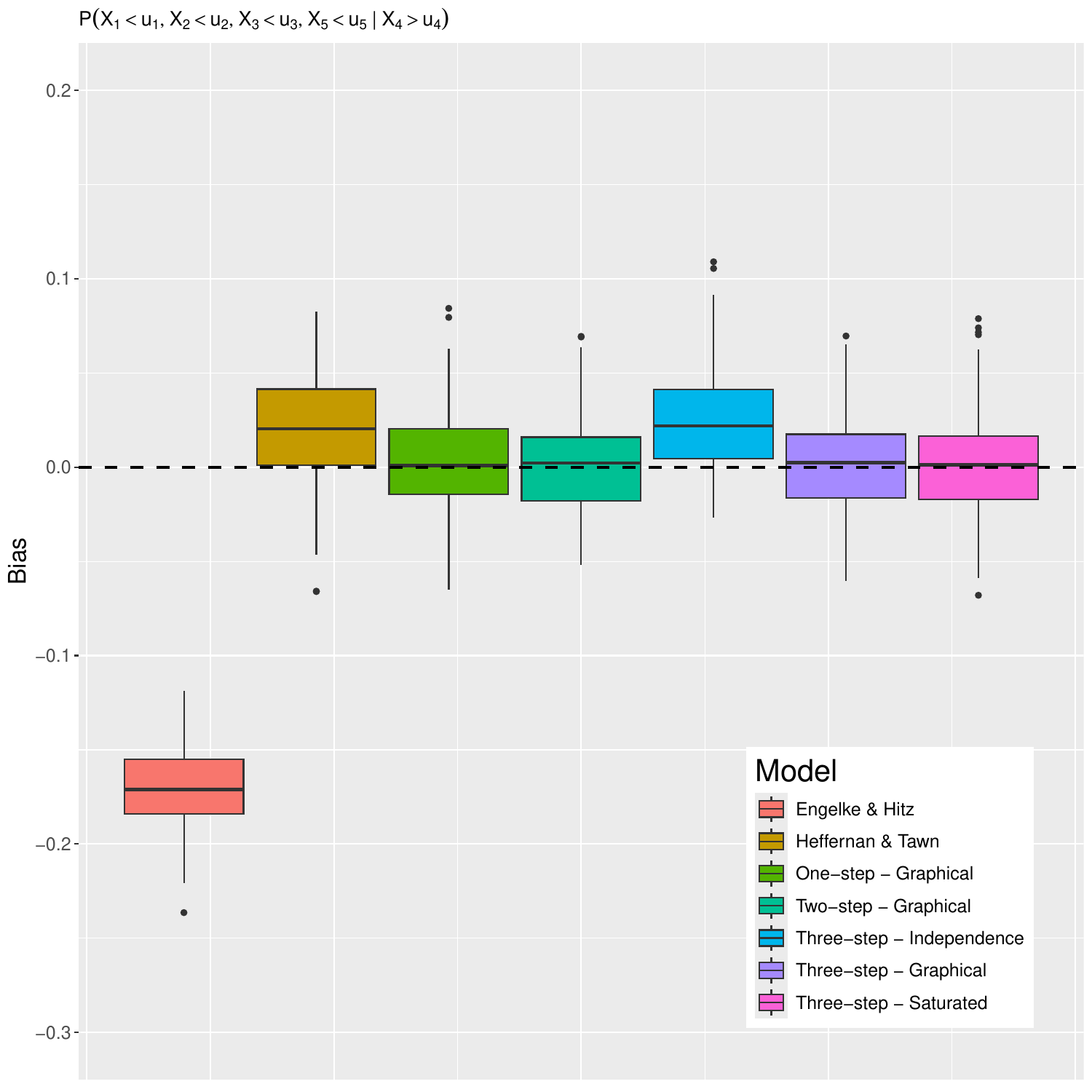}} \\
  \caption{Polygon plots detailing 95\% confidence intervals, over 200 samples, of the bias in $\mathbb{P}[X_{j} < u_{X_{j}} \mid X_{4} > u_{X_{4}}]$ for $j \in V_{\mid 4}$. where $\boldsymbol{X}$ follows a MVL distribution with negative associations (left). The bias from the EHM and the three-step SCMEVM, assuming a graphical covariance structure for the residuals are in pink and blue, respectively. The true conditional cumulative distribution curves are given by the black dashed lines. Boxplots of the bias in $\mathbb{P}[\boldsymbol{X}_{\mid 4} < u_{\boldsymbol{X}_{\mid 4}} \mid X_{4} > u_{X_{4}}]$ (right). The bias from the various models is denoted by the fill of the boxplots. The $y = 0$ line is given by the black dashed line.}
  \label{fig:MVL_Negative_Dependence_Probs}
\end{figure}

The only point to note on the parameter estimates is that $\beta_{j \mid i}$ tends to always be slightly higher for the one-step SCMEVM than for the two- and three-step SCMEVMs. Figure \ref{fig:MVL_Negative_Dependence_Probs} (left panel) shows 95\% confidence intervals for the conditional cumulative distribution curves of $X_{j} \mid X_{4} > u_{X_{4}}$ from the EHM and the three-step SCMEVM with graphical covariance structure for $j \in V_{\mid 4}$. As with the MVG distribution with negative association, the SCMEVM captures the curves perfectly, while the EHM underestimates all curves. The SCMEVMs with graphical covariance structure minimise the MAE and RMSE 63\% and 59\% of the time, respectively. The three-step SCMEVM with saturated covariance structure also performs very well, minimising the metrics 23\% and 28\% of the time, respectively. However, the numerical values of the metrics are almost indistinguishable for the two models as shown in Figure \ref{fig:MVL_Negative_Dependence_Probs} (right panel) where we plot the bias in $\mathbb{P}[\boldsymbol{X}_{\mid 4} < u_{\boldsymbol{X}_{\mid 4}} \mid X_{4} > u_{X_{4}}]$. The EHM and SCMEVM with independent residuals are both biased, while the SCMEVMs with graphical or saturated covariance structure are unbiased. The CMEVM predictions exhibit a small positive bias, although the reason for this is unclear.

\subsection{Multivariate $t$-distribution}
For this study, $\boldsymbol{X}$ follows a MVT distribution with mean $\boldsymbol{\mu} = \boldsymbol{0}$, $k = 5$ degrees of freedom, and a dispersion matrix with inverse consistent with $\mathcal{G}$ in Section \ref{Sec:Sim_Study_True_Dist_Weak_Dependence}. We consider weak positive, strong positive, and negative associations in $\boldsymbol{X}$ in Sections \ref{sec:MVT_Weak_Dependence}, \ref{sec:MVT_High_Dependence}, and \ref{sec:MVN_Negative_Dependence}, respectively. For all simulations, the dependence threshold $u_{Y_{i}}$ is set to the $0.8$-quantile of the standard Laplace distribution for all $i \in V$, resulting in approximately $1,000$ excesses per conditioning variable. As with the previous examples, parameter estimates are omitted unless of specific interest.

\subsubsection{Weak positive dependence}
\label{sec:MVT_Weak_Dependence}
In this simulation, we ensure the associations between components are weakly positive i.e., entries in the dispersion matrix are strictly positive but less than $0.17$. For prediction, we set $u_{X_{i}} = 0.75$ for each $i \in V$.

\begin{figure}[t!]
    \centering
    \includegraphics[width = \textwidth]{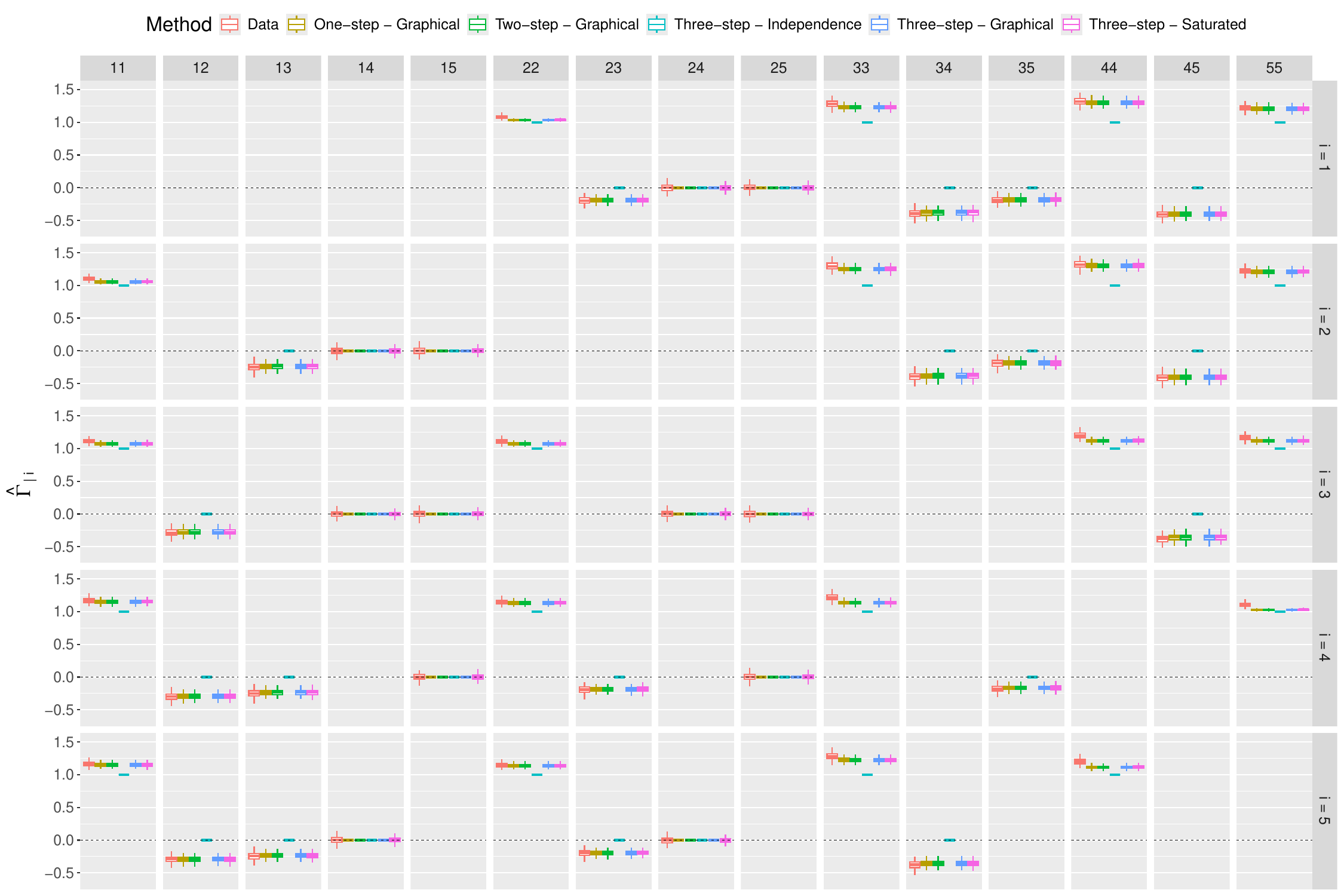}
    \caption{Boxplots of empirical and model-based estimates of $\Gamma_{\mid i}$, for each $i \in V$, when the data is generated from a MVT distribution with weak positive associations. Each row corresponds to the conditioning variable $i$ and each column corresponds to the correlation parameter. The colour of the boxplots distinguishes the different models. Black dashed lines show $y = 0$.}
    \label{fig:Sim_Study_MVT_Gamma}
\end{figure}

Figure \ref{fig:Sim_Study_MVT_Gamma} shows empirical and model-based estimates of the conditional precision matrix $\Gamma_{\mid i}$. The estimated structure of the conditional precision matrix from both the graphical and saturated SCMEVMs is again consistent with the empirical version. Analysis of other parameter estimates has been omitted, other than noting that estimates of $\beta_{j \mid i}$ from the one-step SCMVEM are generally larger than corresponding estimates from the two- and three-step SCMEVMs.

\begin{figure}[t]
  \centering
  \subfloat{\includegraphics[width = 0.48\textwidth]{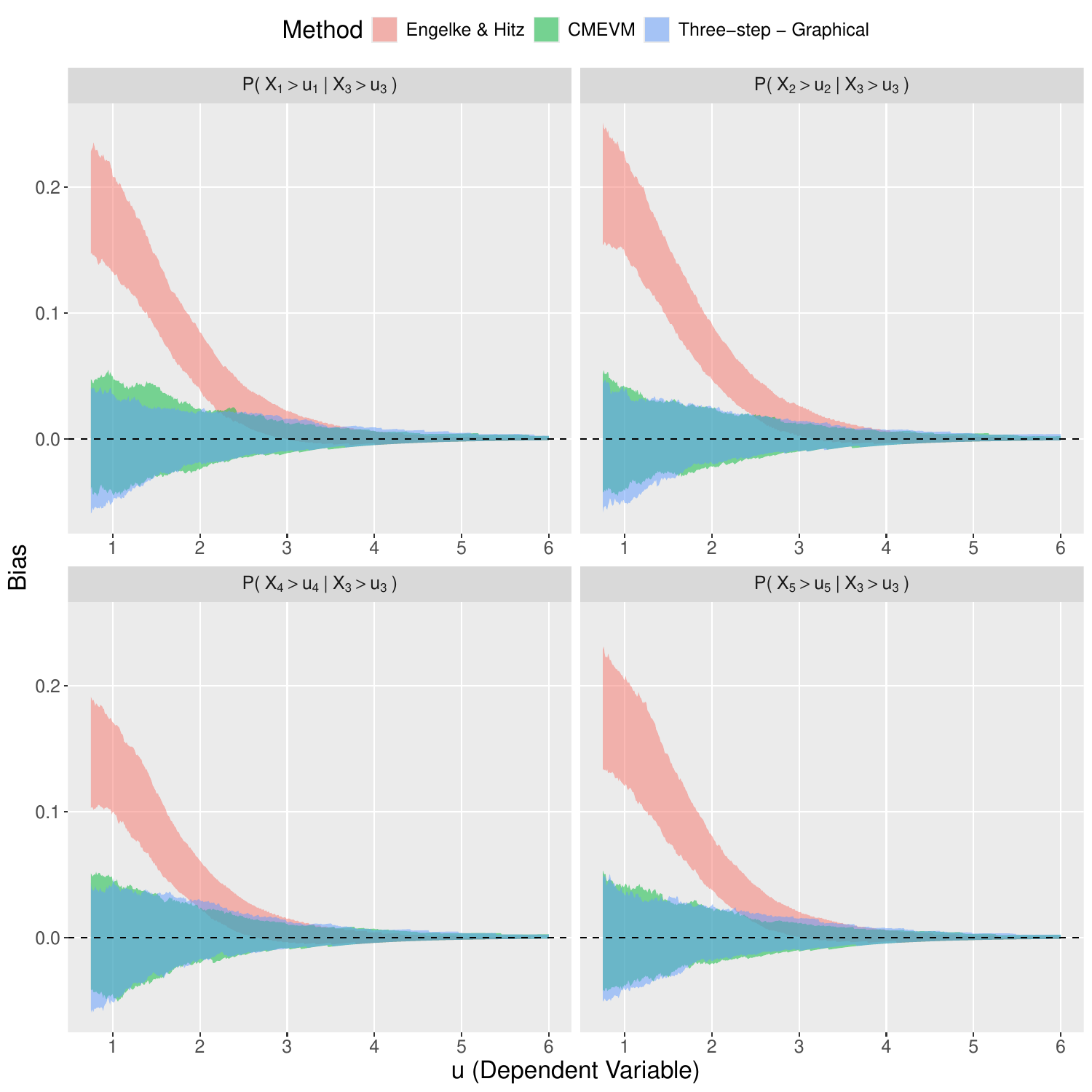}} \quad
  \subfloat{\includegraphics[width = 0.48\textwidth]{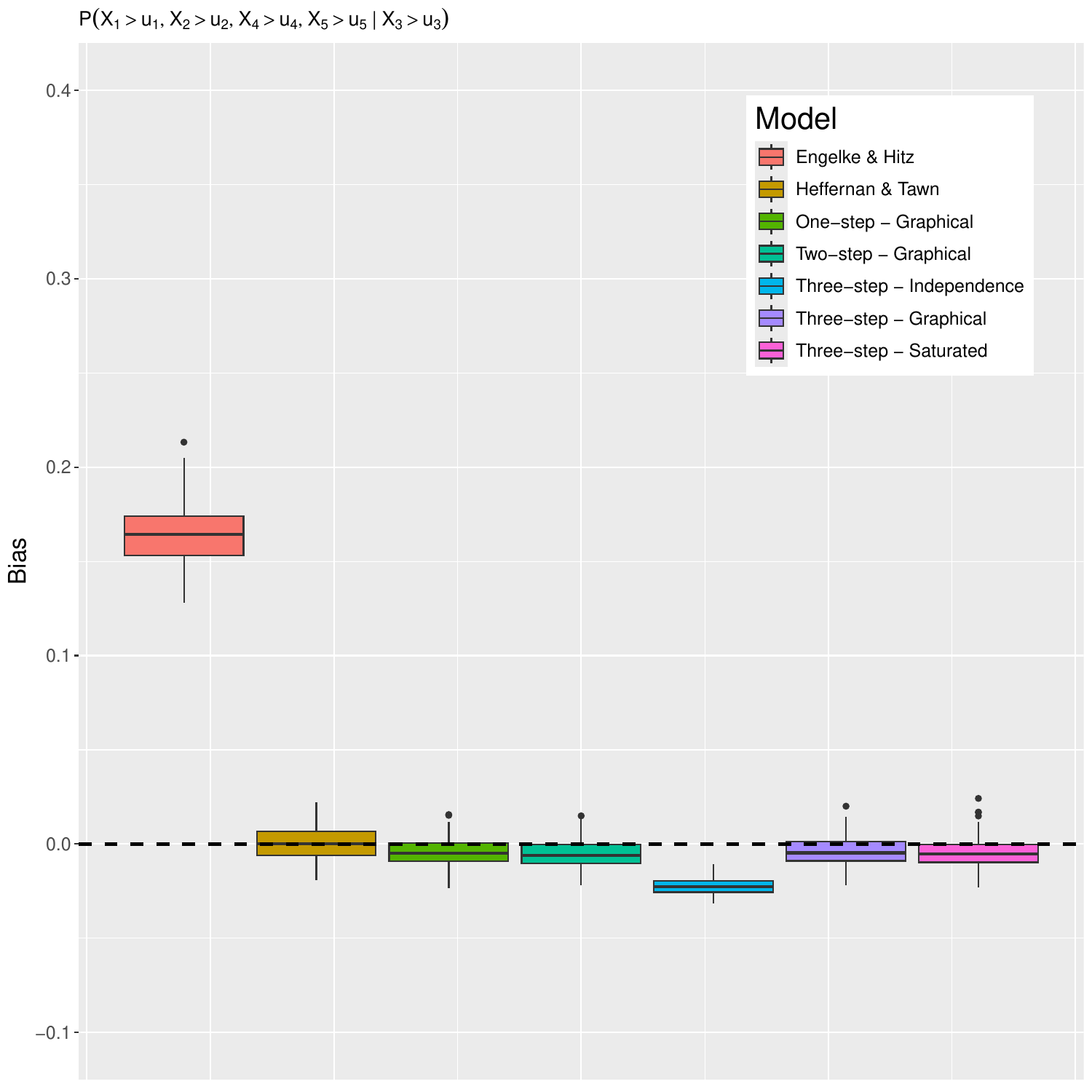}}
  \caption{Polygon plots detailing 95\% confidence intervals, over 200 samples, of the bias in $\mathbb{P}[X_{j} > u_{X_{j}} \mid X_{3} > u_{X_{3}}]$, for each $j \in V_{\mid 3}$, where $\boldsymbol{X}$ follows a MVT distribution with weak positive associations (left). The bias from the EHM, the CMEVM, and the three-step SCMEVM, assuming a graphical covariance structure for the residuals are in pink, green, and blue, respectively. Boxplots of the bias in $\mathbb{P}[\boldsymbol{X}_{\mid 3} > u_{\boldsymbol{X}_{\mid 3}} \mid X_{3} > u_{X_{3}}]$ (right). The fill of the boxplots distinguishes the different models. Black dashed lines show $y = 0$.}
  \label{fig:MVT_Prediction_Weak}
\end{figure}

Figure \ref{fig:MVT_Prediction_Weak} (left panel) shows the bias in the conditional survival curves of $X_{j} \mid X_{3} > u_{X_{3}}$ for $j \in V_{\mid 3}$. Similar to the MVG and MVL examples, the SCMVEM with graphical covariance structure is unbiased for all curves, and the EHM exhibits positive bias for low values of $u_{X_{j}}$ that decreases as $u_{X_{j}}$ increases. We have also included the estimated curve from the CMEVM to show that there is little difference between the estimates from the CMEVM and the SCMEVM with a graphical covariance structure. 

Figure \ref{fig:MVT_Prediction_Weak} (right panel) shows the bias in $\mathbb{P}[\boldsymbol{X}_{\mid 3} > u_{\boldsymbol{X}_{\mid 3}} \mid X_{3} > u_{X_{3}}]$. As expected, the CMEVM estimates are unbiased, the EHM exhibits positive bias, and the three-step SCMEVM with independent residuals exhibits negative bias because not all the components of $\boldsymbol{X}_{\mid 3}$ are independent given $X_{3}$. Interestingly, the fully parametric SCMEVMs with graphical or saturated covariance structures exhibit a very small negative bias. Despite this, the three-step SCMEVM with graphical structure is the least biased and variable model as it minimises the MAE and RMSE for 38 and 46 of the 75 conditional tail probabilities, respectively. For comparison, the CMEVM minimises the metrics 21 and 13 times, respectively. This suggests there is little difference between the CMEVM and the SCMEVM in this scenario. 

Assessing diagnostic plots from the SCMEVM raises no concerns about the model fit. Therefore, to fix the slight underestimation from the SCMEVMs in Figure \ref{fig:MVT_Prediction_Weak} (right panel), we may need to increase $N$ used in Algorithm 3.6 of the main text (we let $N = 250,000$ in this simulation). Alternatively, we may wish to simulate data from the fitted model for $\boldsymbol{X} \mid X_{i} > u_{X_{i}}$ for each $i \in V$, rather than using the method outlined in Section 3.3 in the main text which simulates data for the entire domain, with the extreme region corresponding to at least one component being extreme.

\subsubsection{Strong positive dependence}
\label{sec:MVT_High_Dependence}

\begin{figure}[!t]
  \centering
  \subfloat{\includegraphics[width=.48\textwidth]{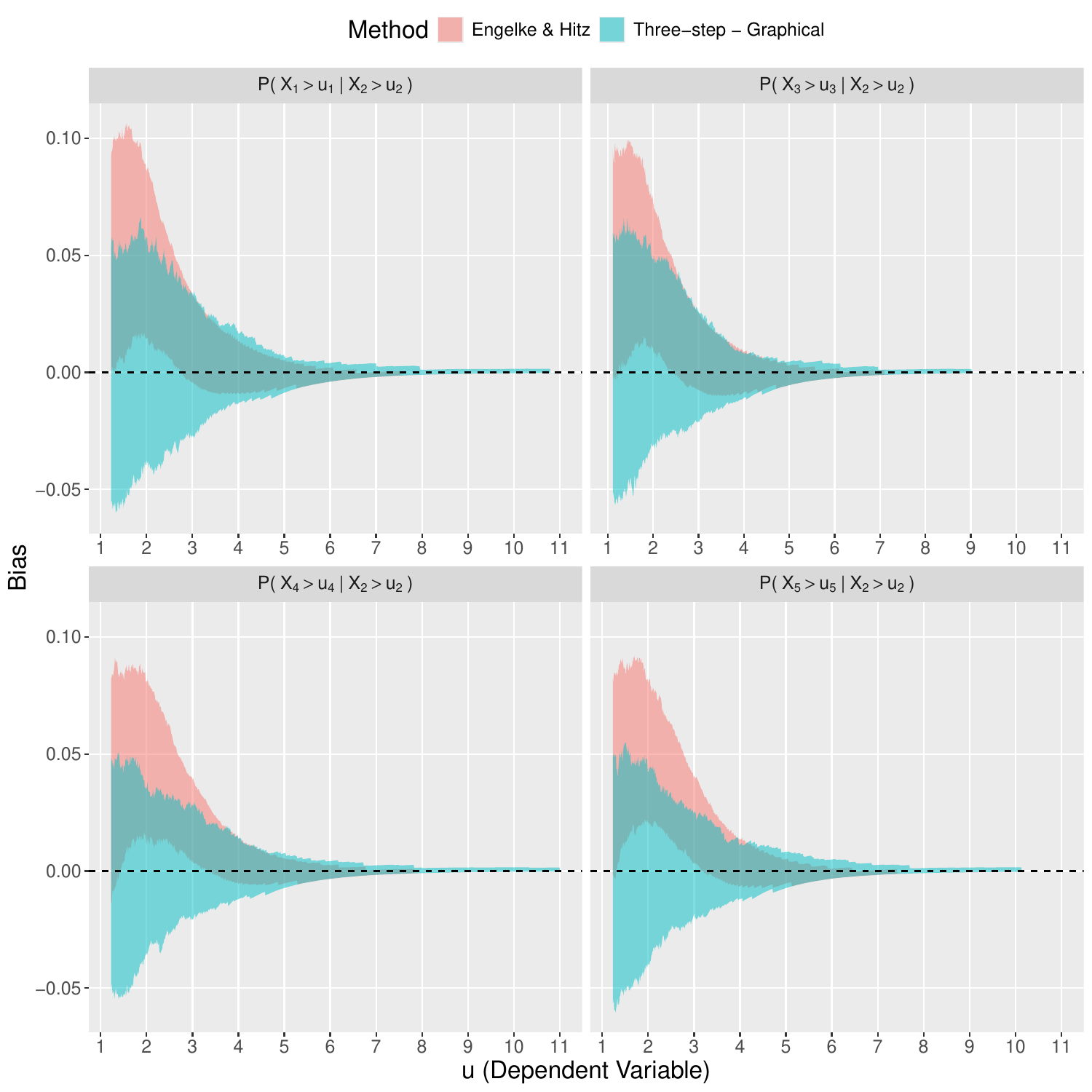}} \quad
  \subfloat{\includegraphics[width=.48\textwidth]{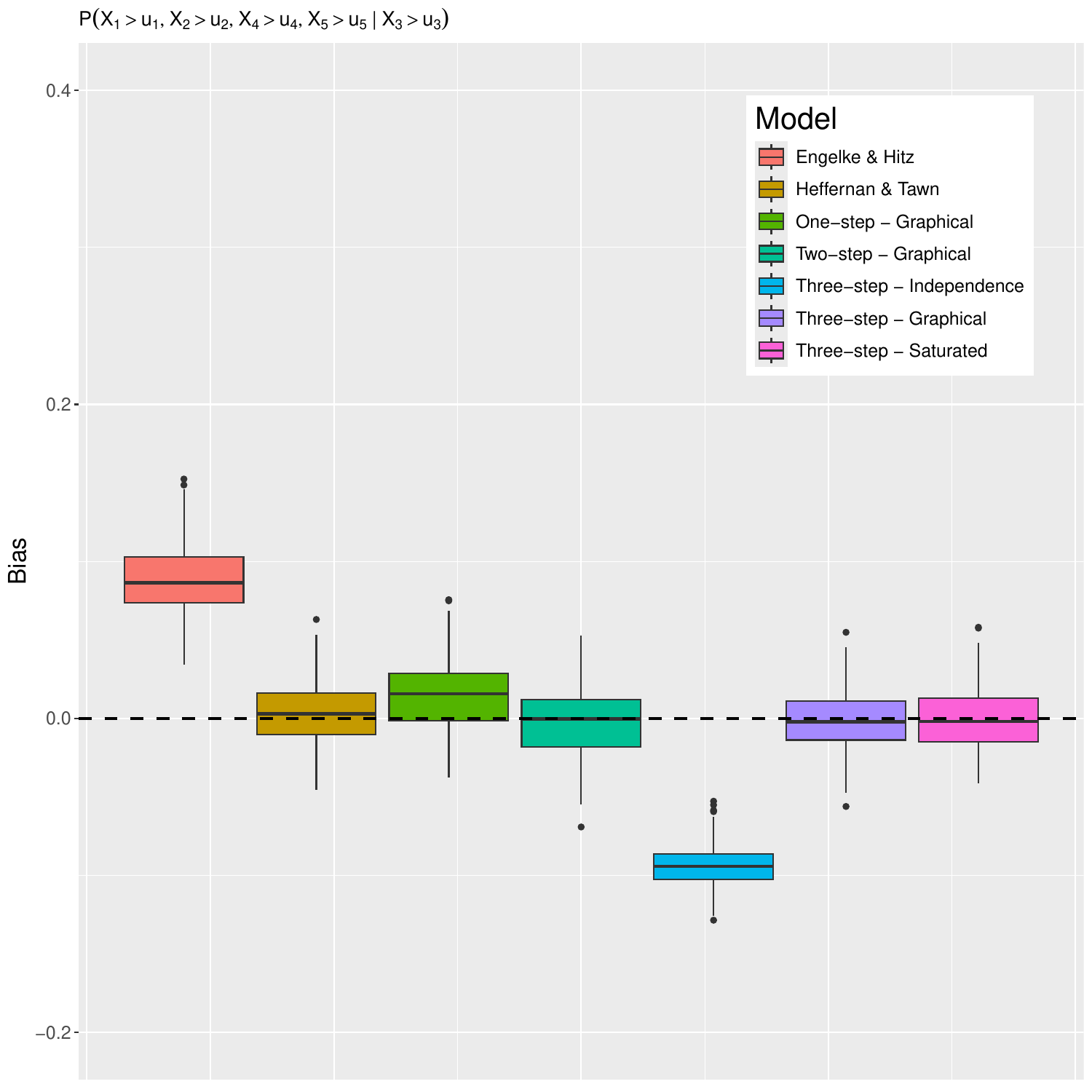}} \\
  \caption{Polygon plots detailing 95\% confidence intervals, over 200 samples, of the bias in $\mathbb{P}[X_{j} > u_{X_{j}} \mid X_{2} > u_{X_{2}}]$ for $j \in V_{\mid 2}$, where $X$ follows a MVT distribution with strong positive associations (left). The bias from the EHM and the three-step SCMEVM, assuming a graphical covariance structure for the residuals are in pink and blue, respectively. Boxplots of the bias in $\mathbb{P}[\boldsymbol{X}_{\mid 3} > u_{\boldsymbol{X}_{\mid 3}} \mid X_{3} > u_{X_{3}}]$ (right). The bias from the various models is denoted by the fill of the boxplots. Black dashed lines show $y = 0$.}
  \label{fig:MVT_High_Dependence_Probs}
\end{figure}

We now allow strong positive associations between the components. The only note on the parameter estimates is that the CMEVM dependence parameters tend to be higher for the one-step SCMEVM compared to the two- and three-stepwise SCMEVMs, and the marginal AGG parameters (excluding the shape) tend to be lower for the one-step SCMEVM.
Setting $u_{X_{i}} = 1.25$ for each $i \in V$, Figure \ref{fig:MVT_High_Dependence_Probs} (left panel) shows the bias in the conditional survivor curves of $X_{j} \mid X_{2} > u_{X_{2}}$ for $j \in V_{\mid 2}$, and for the EHM and the three-step SCMEVM with graphical covariance structure. The EHM is biased for low values of $u_{X_{j}}$, but this diminishes as $u_{X_{j}}$ increases. In contrast, the three-step SCMEVM with a graphical covariance structure is unbiased across all curves.
Figure \ref{fig:MVT_High_Dependence_Probs} (right panel) shows the bias in $\mathbb{P}[\boldsymbol{X}_{\mid 3} > u_{\boldsymbol{X}_{\mid 3}} \mid X_{3} > u_{X_{3}}]$. The EHM has positive bias, while the CMEVM and the stepwise SCMEVMs with graphical or saturated covariance structures are unbiased. Once again, the SCMEVMs with graphical covariance structure are the least biased and the least variable, minimising both the MAE and RMSE for 48 of the 75 conditional tail probabilities.

\subsubsection{Negative dependence}
\begin{figure}[!t]
  \centering
  \subfloat{\includegraphics[width=.48\textwidth]{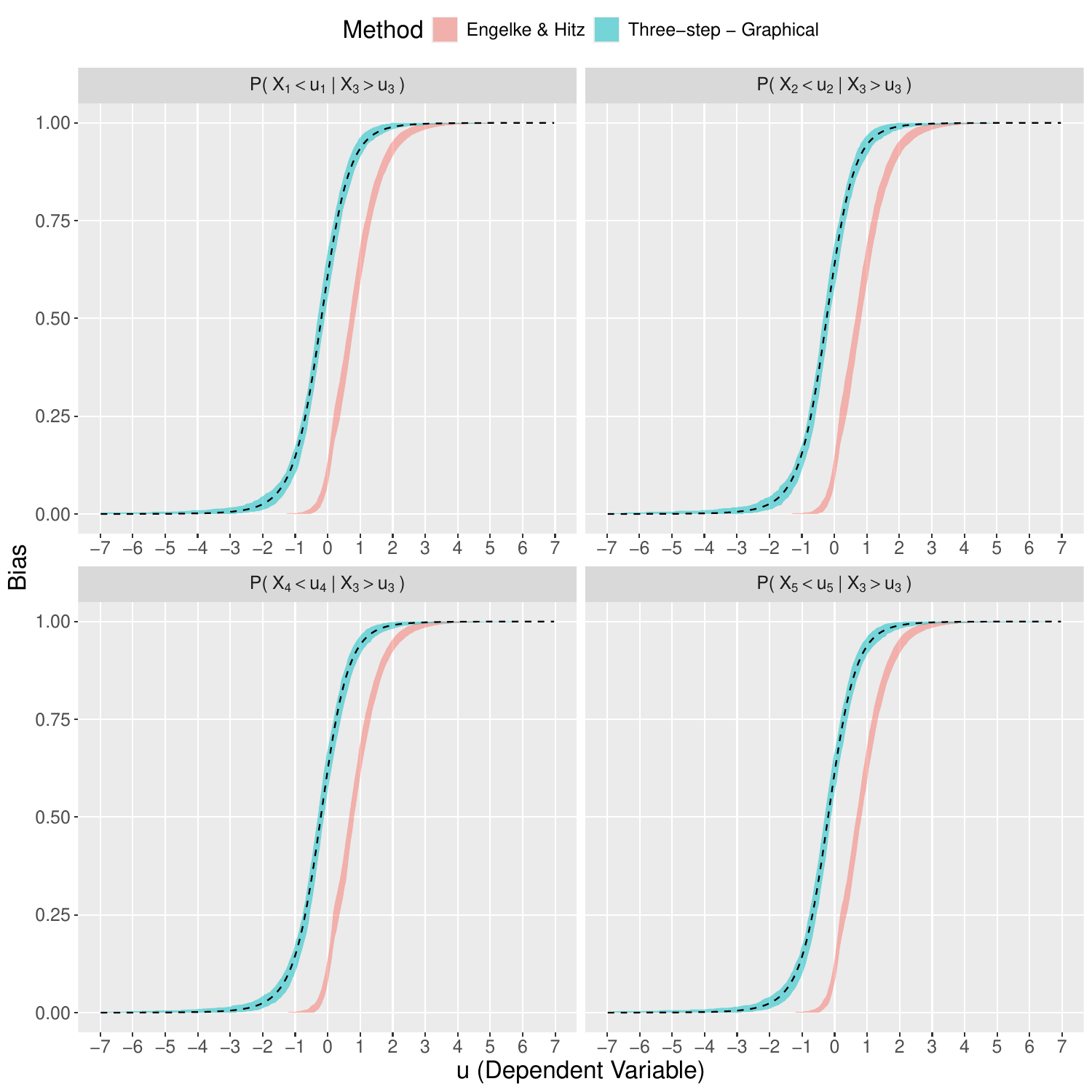}} \quad
  \subfloat{\includegraphics[width=.48\textwidth]{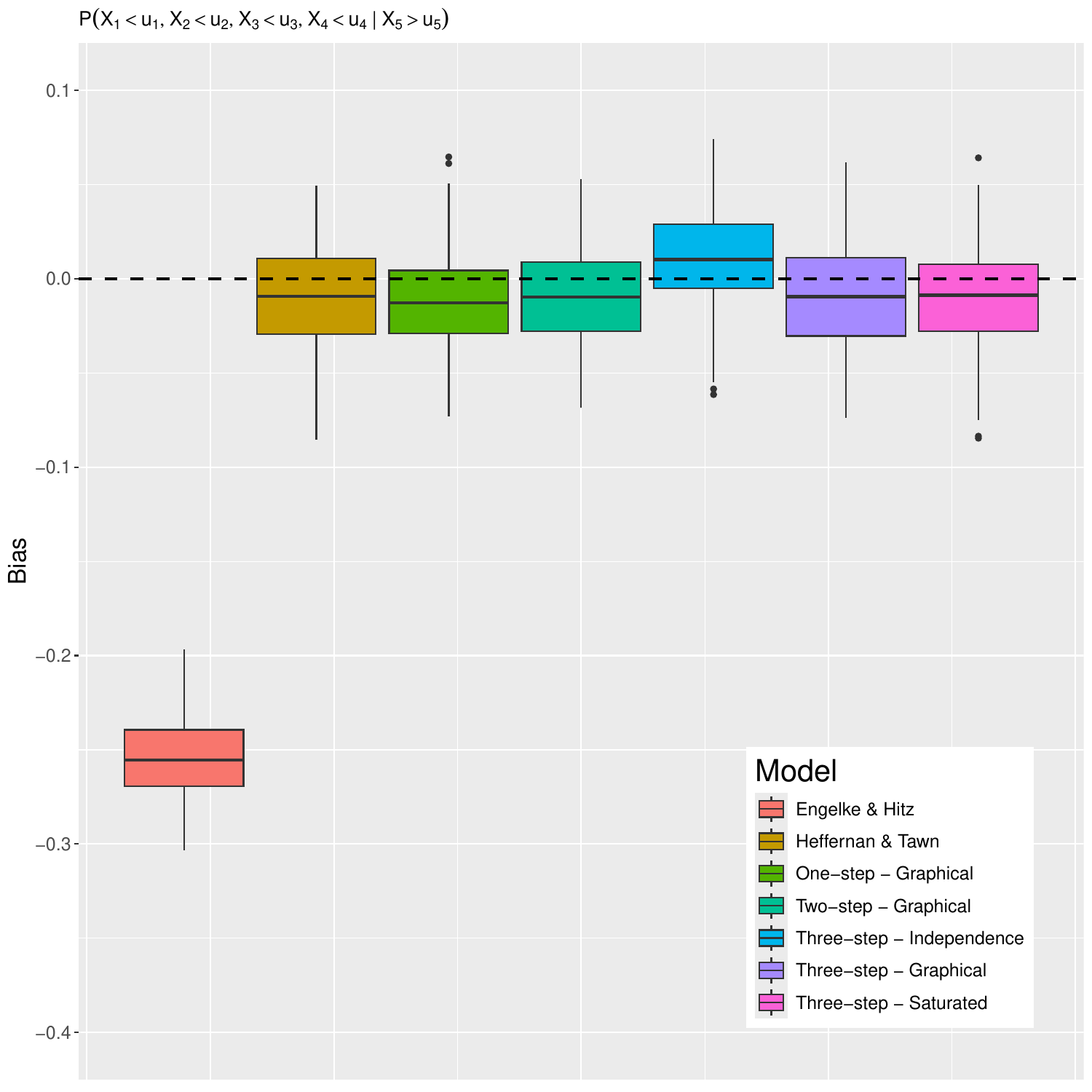}} \\
  \caption{Polygon plots detailing 95\% confidence intervals, over 200 samples, of the bias in $\mathbb{P}[X_{j} < u_{X_{j}} \mid X_{3} > u_{X_{3}}]$ for $j \in V_{\mid 3}$, where $\boldsymbol{X}$ follows a MVT distribution with negative associations (left). The bias from the EHM and the three-step SCMEVM, assuming a graphical covariance structure for the residuals are in pink and blue, respectively. The true conditional cumulative distribution curves are given by the black dashed lines. Boxplots of the bias in $\mathbb{P}[\boldsymbol{X}_{\mid 5} < u_{\boldsymbol{X}_{\mid 5}} \mid X_{5} > u_{X_{5}}]$ (right). The bias from the various models is denoted by the fill of the boxplots. The $y = 0$ line is given by the black dashed line.}
  \label{fig:MVT_Negative_Dependence_Probs}
\end{figure}

Finally, we allow for weak negative associations between the components. For prediction, we set $u_{X_{i}}$ to the $0.9$-quantile from a dataset of size $10^6$ simulated from the true distribution for each $i \in V$. Figure \ref{fig:MVT_Negative_Dependence_Probs} (left panel) shows 95\% confidence intervals for the conditional cumulative distribution curves of $X_{j} \mid X_{3} > u_{X_{3}}$ from the EHM and the three-step SCMEVM with graphical covariance structure for $j \in V_{\mid 3}$. As with the MVG and MVL distributions with negative associations, the three-step SCMEVM captures all curves perfectly, whereas the EHM always underestimates. Figure \ref{fig:MVT_Negative_Dependence_Probs} (right panel) shows the bias in $\mathbb{P}[\boldsymbol{X}_{\mid 5} < u_{\boldsymbol{X}_{\mid 5}} \mid X_{5} > u_{X_{5}}]$. The EHM performs poorly due to its inability to capture the negative dependence, while predictions from the CMEVM and the SCMEVMs with graphical or saturated covariance structures are unbiased. The SCMEVMs with a graphical covariance structure minimise the MAE and RMSE metrics for 79\% and 72\% of all the conditional tail probabilities, respectively.

\subsection{Multivariate Pareto distribution}
To test the SCMEVM under AD, we repeat the simulation study for $\boldsymbol{X}$ with a MVP distribution such that the parameter matrix is consistent with $\mathcal{G}$ in Section \ref{Sec:Sim_Study_True_Dist}. For the CMEVM and SCMEVMs, we only use data above the dependence threshold $u_{Y_{i}}$ set at the $0.90$-quantile of the standard Laplace distribution for all $i \in V$. The EHM uses all data since, by construction, the data is on standard Pareto margins. For prediction, $u_{X_{i}} = 11$ for each $i \in V$.

\begin{figure}[t!]
  \centering
  \subfloat{\includegraphics[width = 0.48\textwidth]{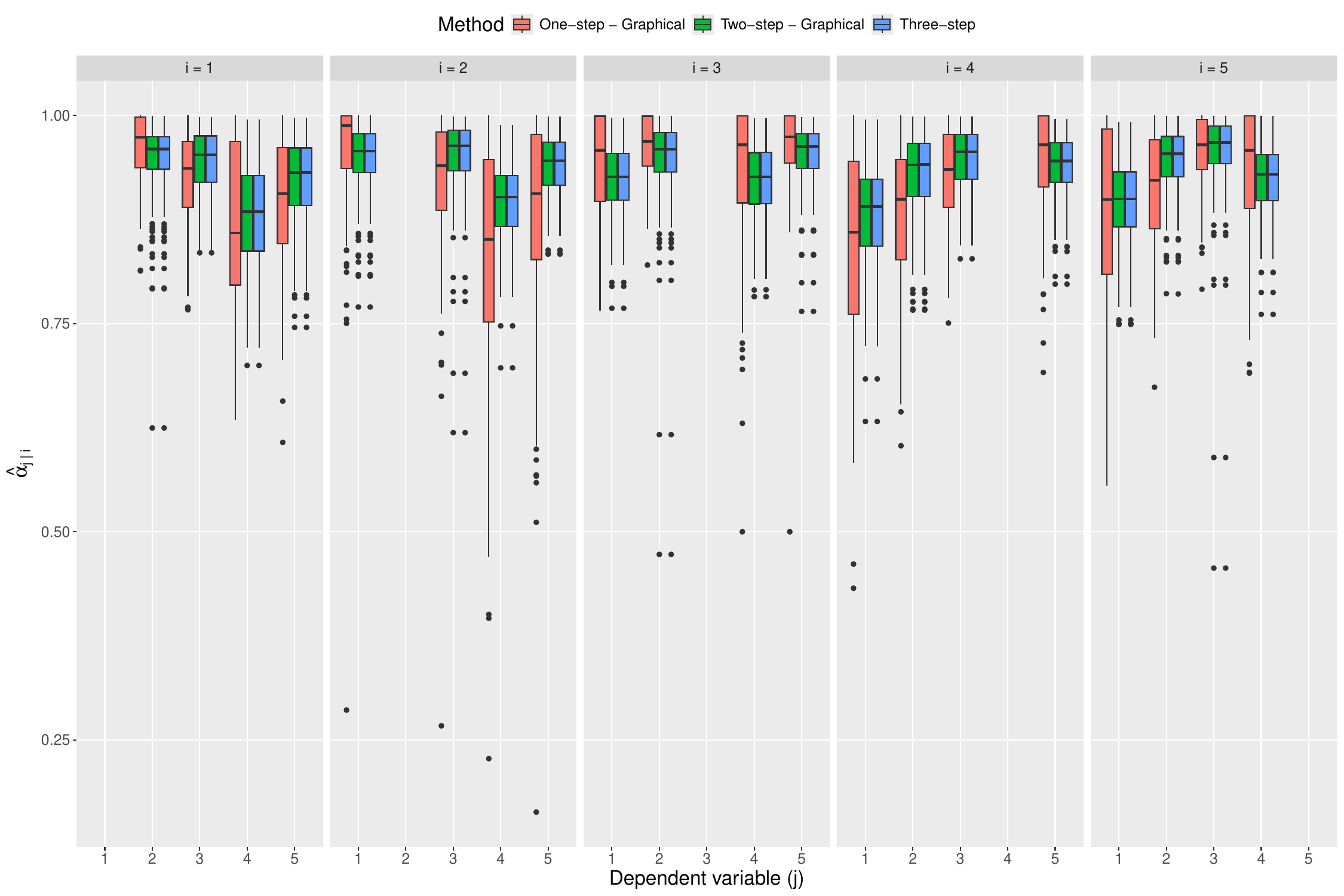}} \quad
  \subfloat{\includegraphics[width=.48\textwidth]{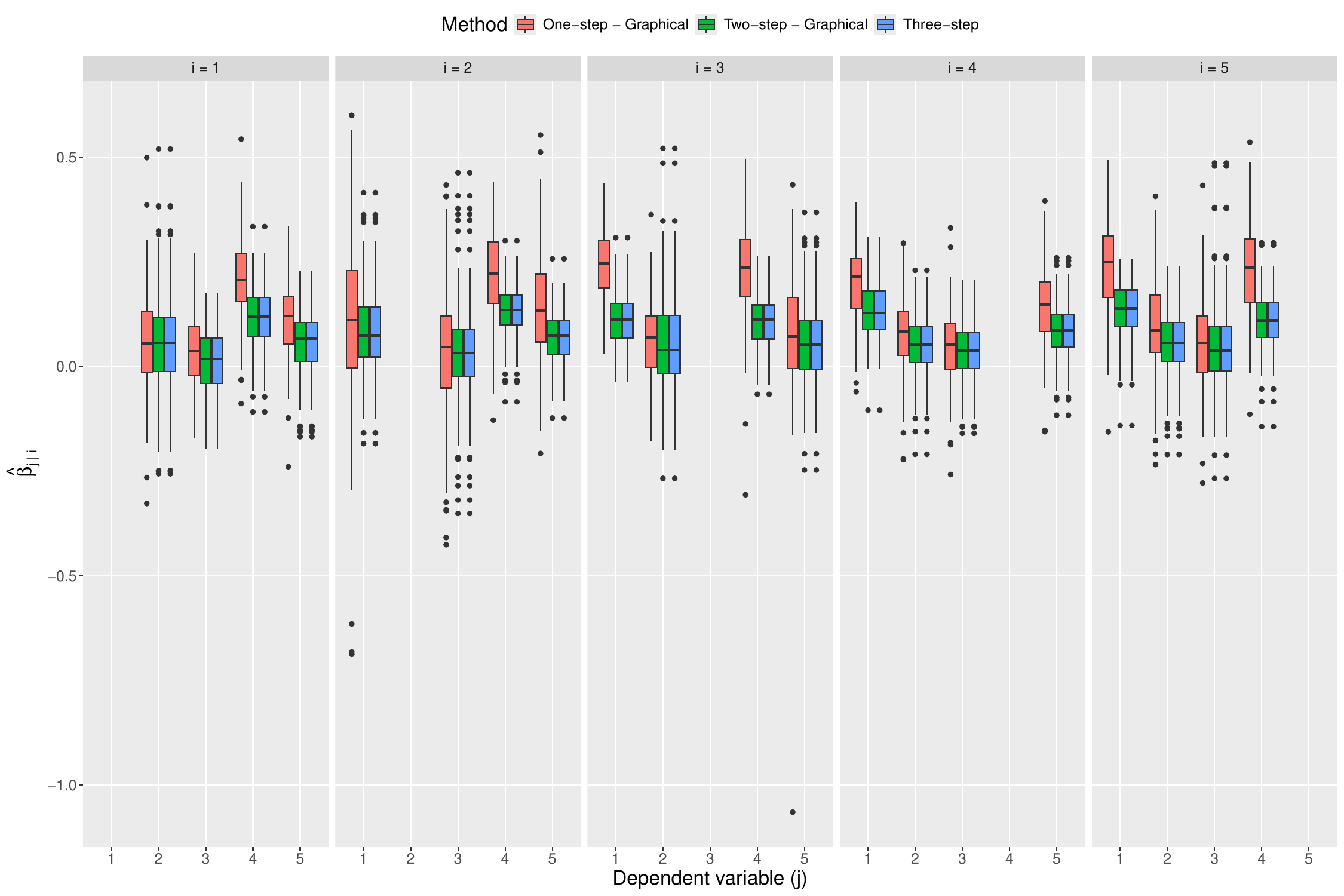}} \\
  \subfloat{\includegraphics[width = 0.48\textwidth]{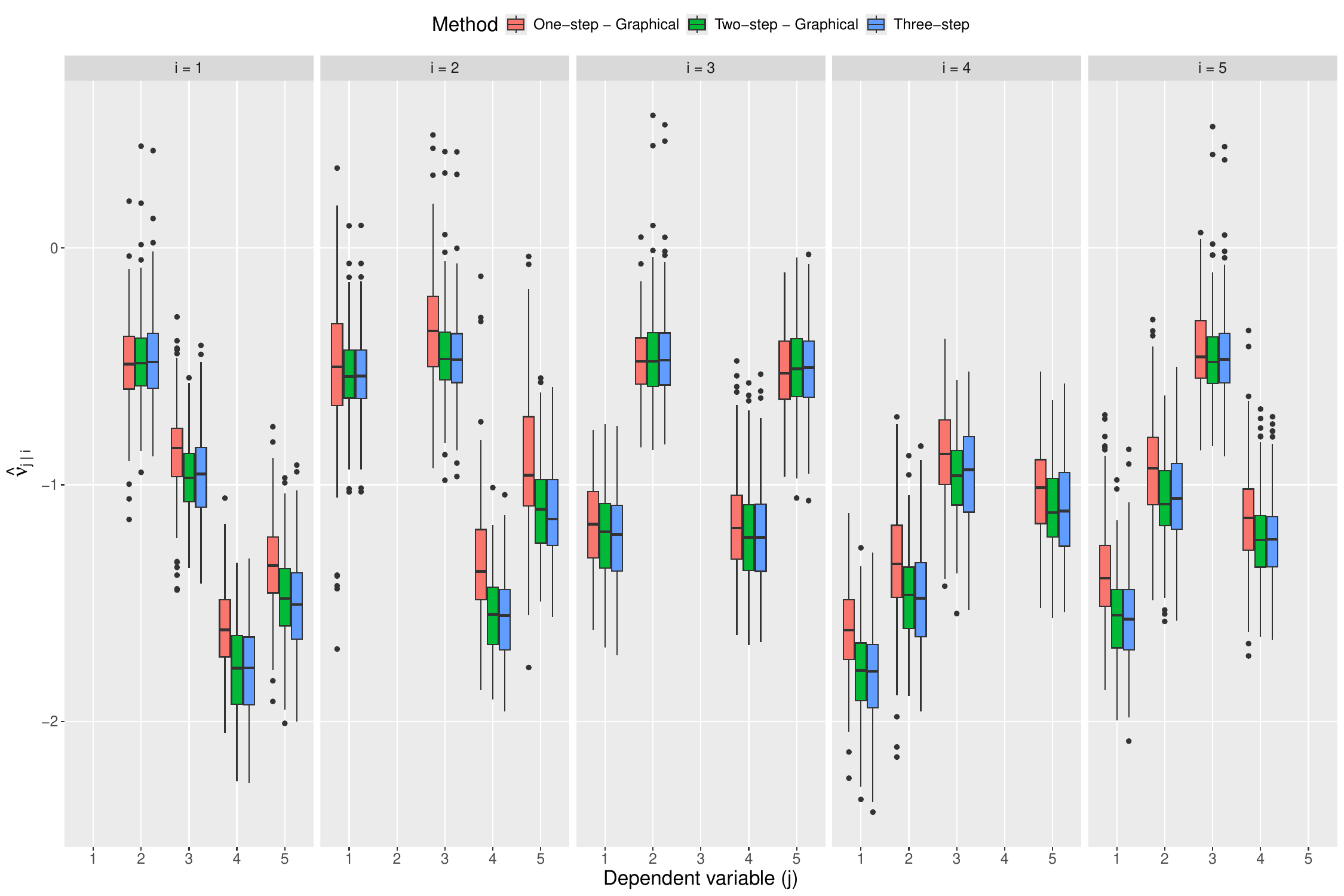}} \quad
  \subfloat{\includegraphics[width=.48\textwidth]{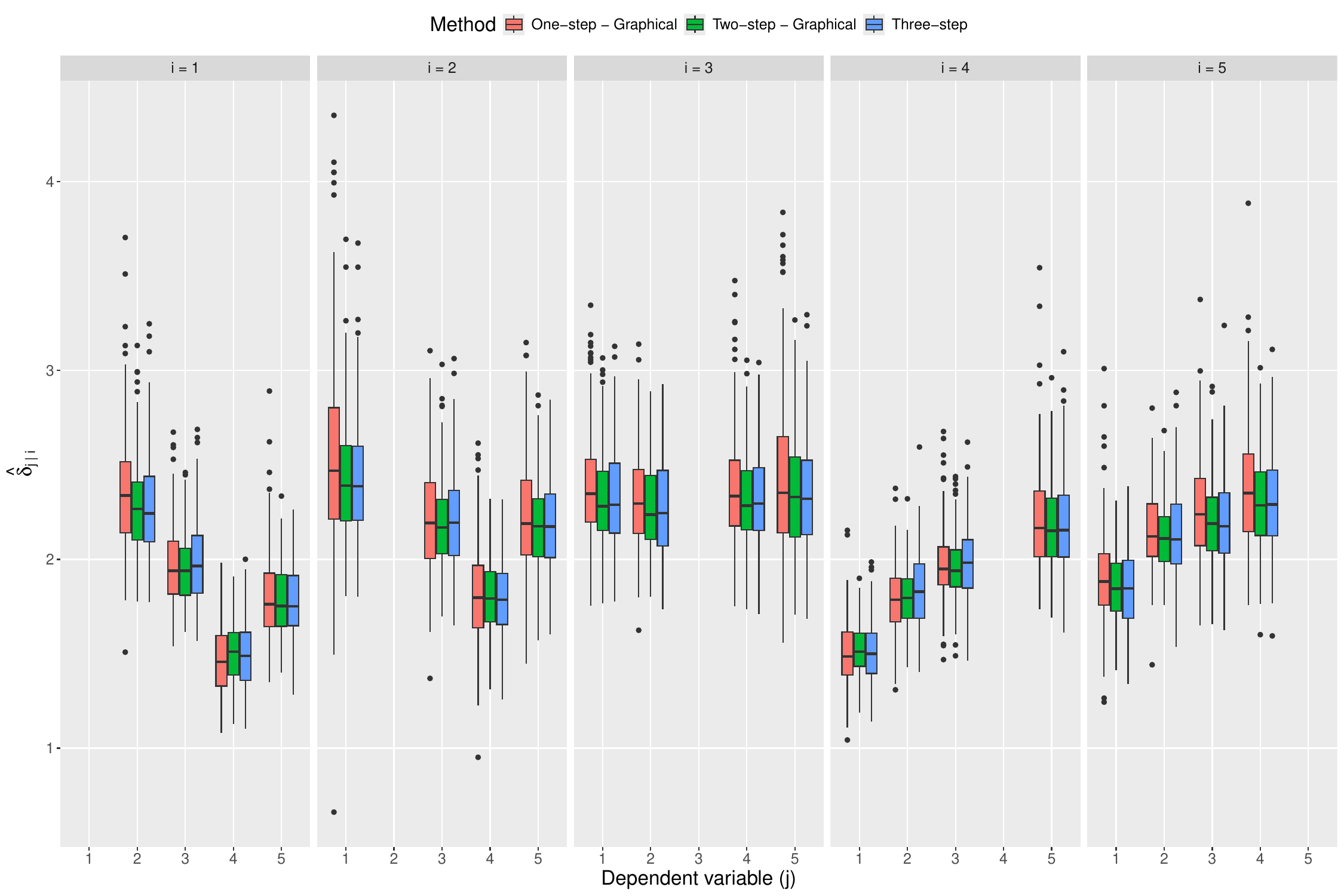}} \\
  \subfloat{\includegraphics[width=.48\textwidth]{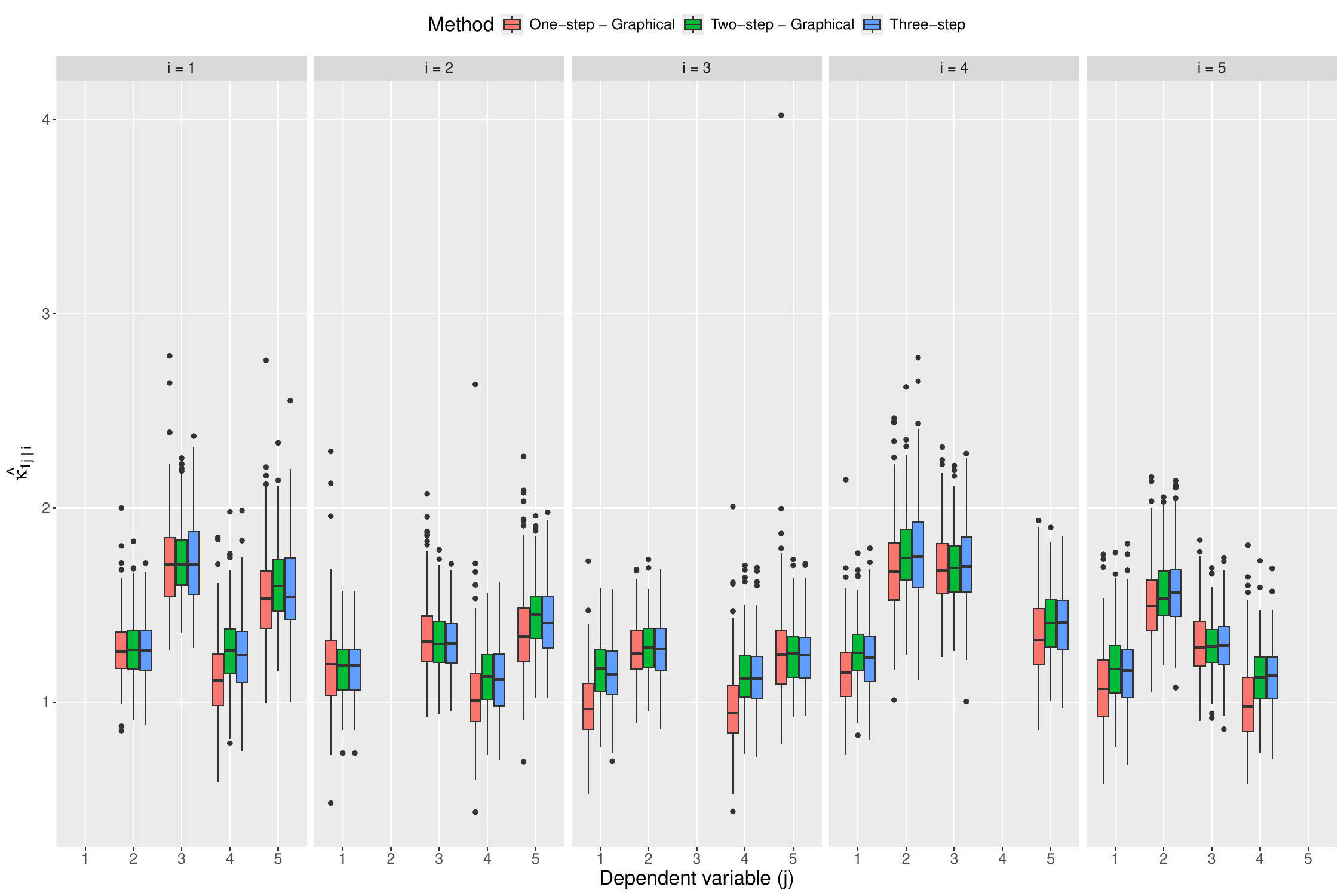}} \quad
  \subfloat{\includegraphics[width=.48\textwidth]{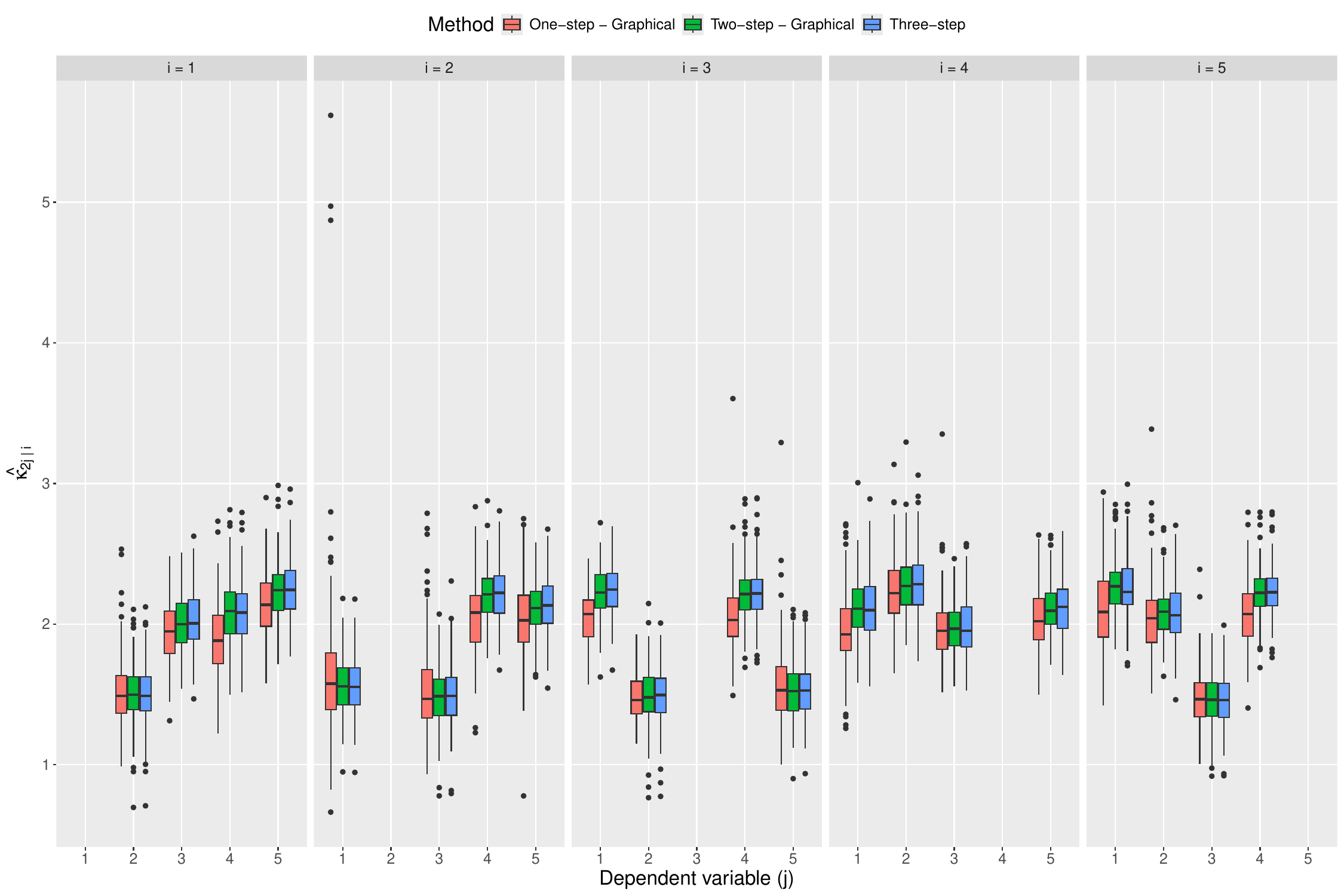}} \\
  \caption{Boxplots of MLEs for $\alpha_{j \mid i}$ (top left), $\beta_{j \mid i}$ (top right), $\nu_{j \mid i}$ (centre left), $\delta_{j \mid i}$ (centre right), $\kappa_{1_{j \mid i}}$ (bottom left), and $\kappa_{2_{j \mid i}}$ (bottom right) for distinct $i, j \in V$. Each column corresponds to the conditioning variable $i$. The different models are denoted by the fill of the boxplots.}
  \label{fig:MVP_MLEs}
\end{figure}

To illustrate the difference between the one-, two-, and three-step parameter estimation procedures, Figure \ref{fig:MVP_MLEs} shows boxplots of MLEs of the dependence and AGG parameters. Here, we would expect $\hat{\alpha}_{j \mid i} = 1$ and $\hat{\beta}_{j \mid i} = 0$ since the data are asymptotically dependent. While $\hat{\alpha}_{j \mid i} \approx 1$ for all the methods, the variability from the one-step method is much greater. This is linked to the variability in the AGG parameters, particularly the location parameter, being larger under the one-step method. Although some MLEs of $\beta_{j \mid i} > 0$ under the two- and three-step methods, these are closer to 0 than the corresponding estimates under the one-step method. This highlights that the one-step method cannot guarantee that the first-order extremal dependence is restricted to the CMEVM dependence parameters, and why we prefer the three-step estimation method.

\begin{figure}[t!]
    \centering
    \includegraphics[width = \textwidth]{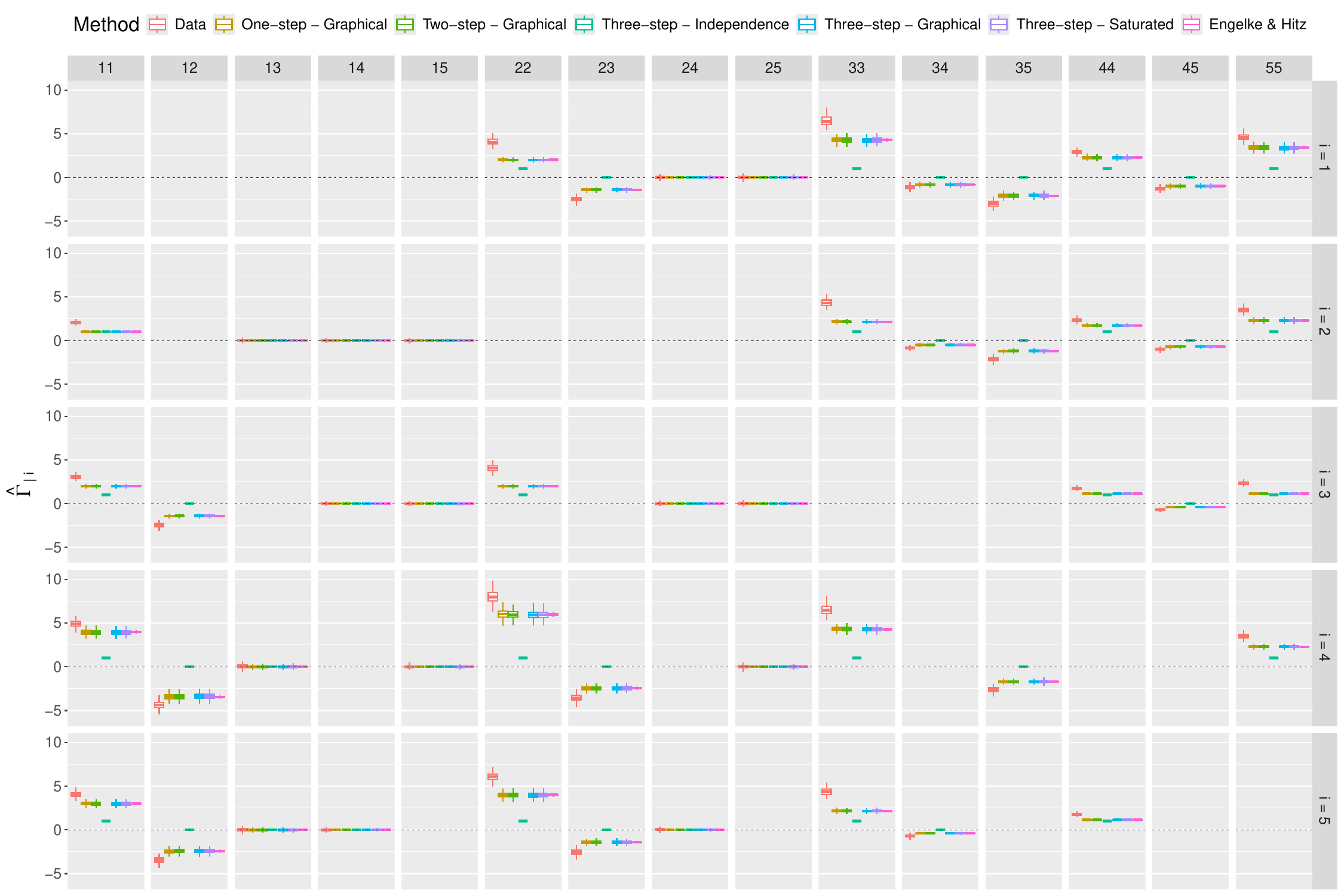}
    \caption{Boxplots of empirical and model-based estimates of $\Gamma_{\mid i}$, for each $i \in V$, when the data is generated from a MVP distribution. Each row corresponds to the conditioning variable $i$, and each column corresponds to the correlation parameter. The various models are denoted by the colour of the boxplots. Black dashed lines show $y = 0$.}
    \label{fig:Sim_Study_MVP_Gamma}
\end{figure}

\begin{figure}[t!]
    \centering
    \includegraphics[width=0.48\textwidth]{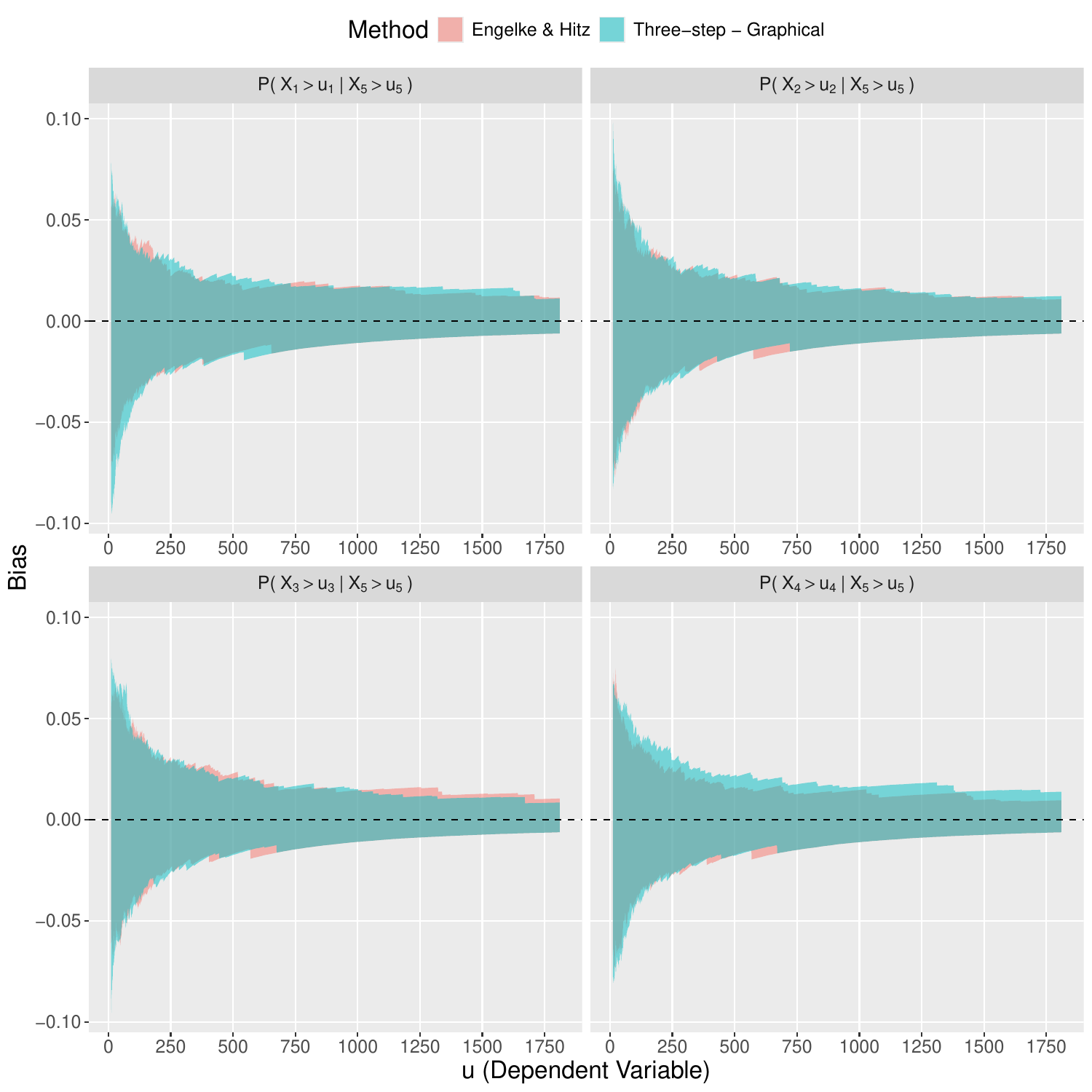} \quad
    \includegraphics[width=0.48\textwidth]{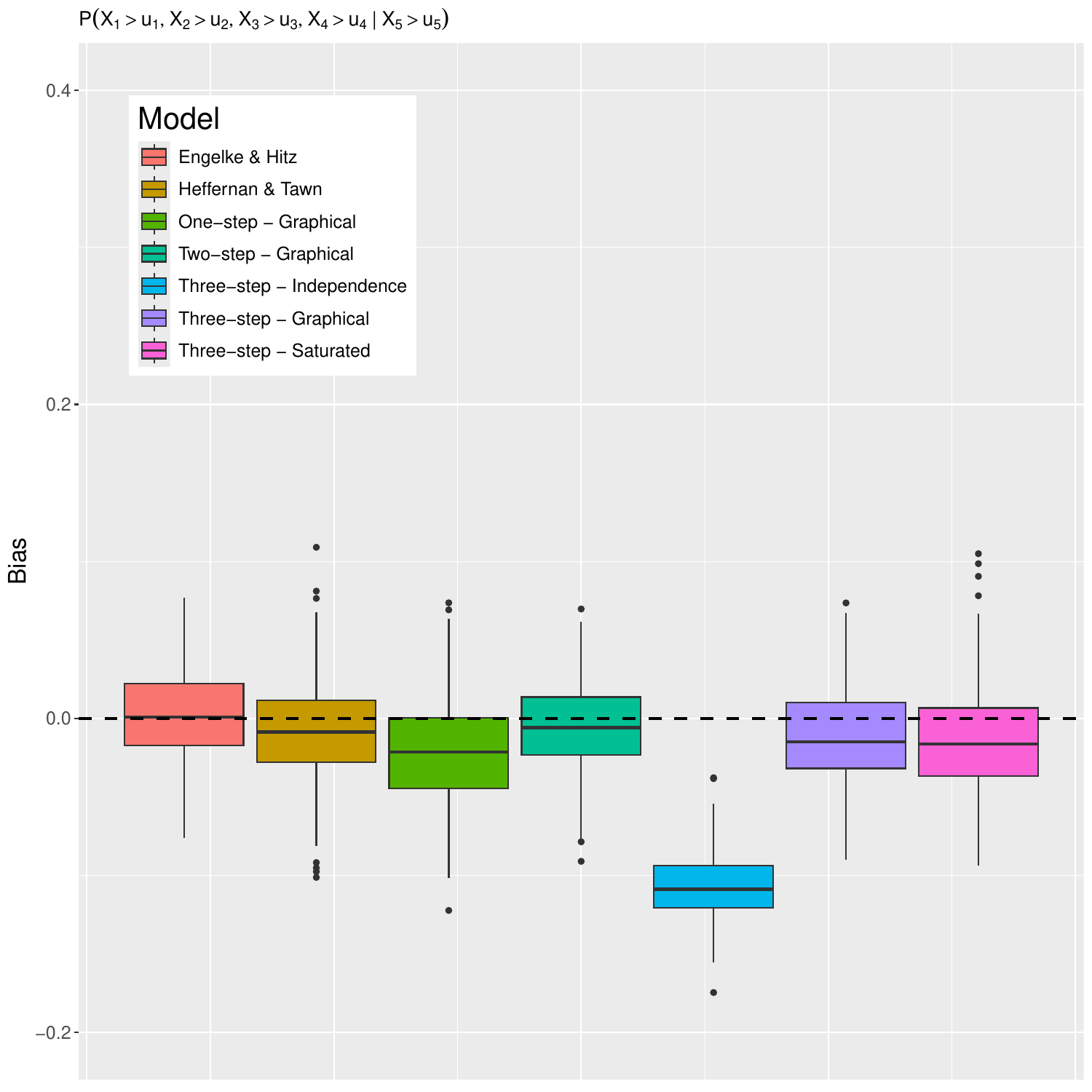}  
    \caption{Polygon plots detailing 95\% confidence intervals, over 200 samples, of the bias in $\mathbb{P}[X_{j} > u_{X_{j}} \mid X_{5} > u_{X_{5}}]$ for $j \in V_{\mid 5}$, where $\boldsymbol{X}$ follows a MVP distribution (left). The bias from the EHM and the three-step SCMEVM, assuming a graphical covariance structure for the residuals are in pink and blue, respectively. Boxplots of the bias in $\mathbb{P}[\boldsymbol{X}_{\mid 5} > u_{\boldsymbol{X}_{\mid 5}} \mid X_{5} > u_{X_{5}}]$ (right). The bias from the various models is denoted by the fill of the boxplots. Black dashed lines show $y = 0$.}
    \label{fig:Sim_Study_MVP_Bias_Probs}
\end{figure}

Figure \ref{fig:Sim_Study_MVP_Gamma} shows empirical and SCMEVM model-based estimates of $\Gamma_{\mid i}$ and transformed model-based estimates from the EHM. Again, the empirical structure is retained in all models. The EHM and SCMEVMs have a very close correspondence, although the former has less variability due to the larger sample size.
Figure \ref{fig:Sim_Study_MVP_Bias_Probs} (left panel) shows the bias in the conditional survivor curves of $X_{j} \mid X_{5} > u_{X_{5}}$ for $j \in V_{\mid 5}$. Both the EHM and the three-step SCMEVM with graphical covariance are unbiased, but estimates from the former are slightly less variable due to the larger sample size. Figure \ref{fig:Sim_Study_MVP_Bias_Probs} (right panel) shows the bias in $\mathbb{P}[\boldsymbol{X}_{\mid 5} > u_{\boldsymbol{X}_{\mid 5}} \mid X_{5} > u_{X_{5}}]$. As expected, the EHM is unbiased while the SCMEVMs exhibit a slight negative bias, perhaps due to the $\alpha$ parameter not converging to the true value, $\alpha=1$, that lies on the edge of the parameter space. However, the bias in the SCMEVMs with graphical and saturated covariances is small. Considering all 75 conditional tail probabilities, the EHM minimises both the MAE and RMSE metrics the majority of the time. Excluding the EHM, which is the only model specifically designed for AD data, the SCMEVMs with graphical covariance minimise the metrics 83\% and 89\% of the time, respectively, suggesting that the SCMEVM is an acceptable alternative to the EHM when the extremal dependence class cannot be pre-determined.

\section{Application to the upper Danube River basin}
\label{Sec:Danube_River}
We first present additional figures for the model comparison made in Section 5 of the main text. We then use EGlearn \citepsupp{Engelke_2025_supp} to learn the graphical structure for the upper Danube River basin under the assumption of asymptotic dependence. Using the learnt structure, we then assess the predictive qualities of the model.

\subsection{Additional figures for Section 5}
\label{subsec:Danube_River_Figures}

We obtain 200 non-parametric bootstrap samples of the declustered river discharge data from the upper Daube River basin. For each bootstrapped dataset, we fit: (i) the EHM and (ii) the three-step SCMEVM, both with a graphical covariance structure given by the undirected tree induced by the flow connections of the upper Danube River basin (Figure 1, left panel, of the main text); (iii) the three-step SCMEVM with saturated covariance structure; (iv) the three-step SCMEVM with a graphical covariance structure inferred from the data (Figure 6, right panel, of the main text). 

\begin{figure}[t!]
    \centering
    \includegraphics[width = \textwidth]{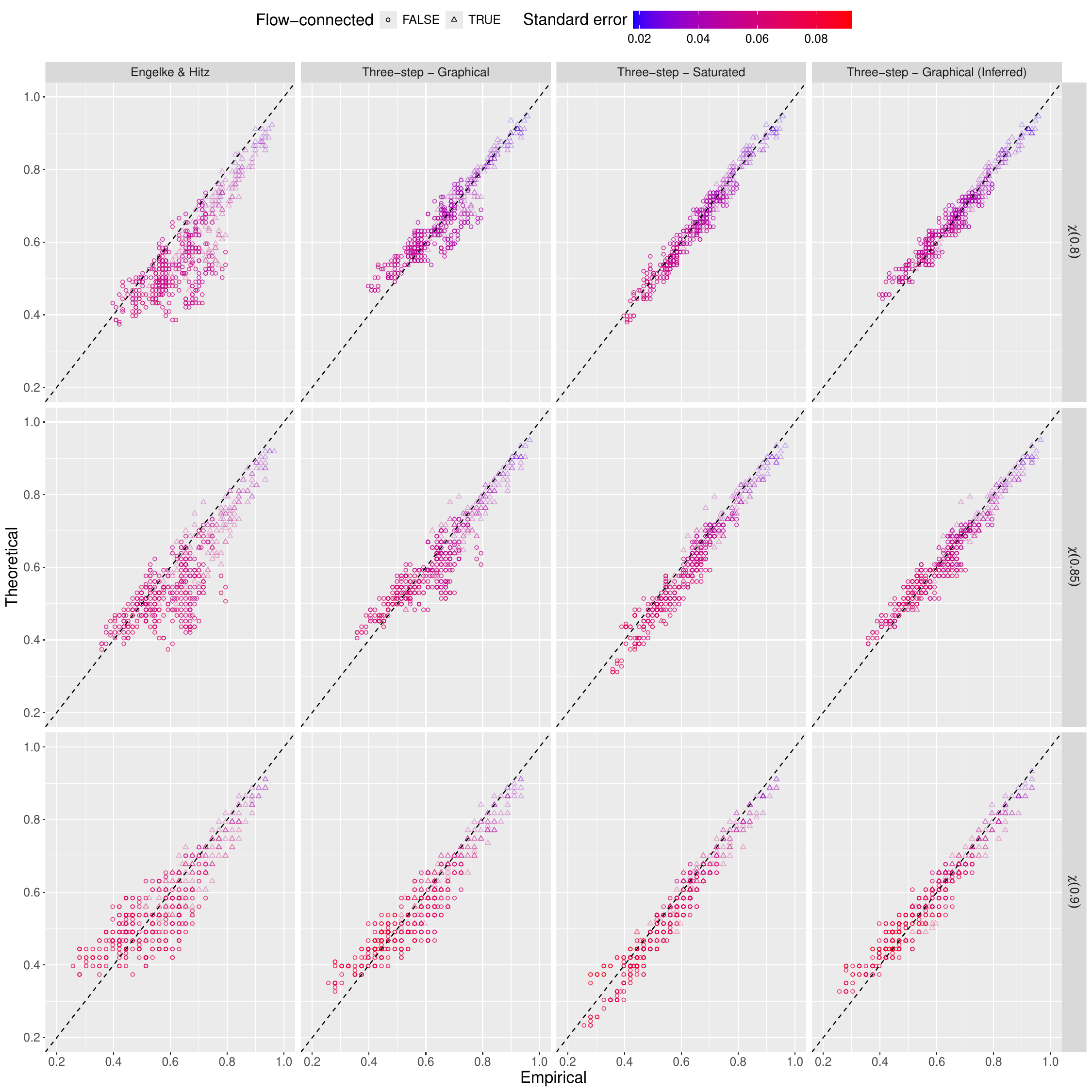}
    \caption{Empirical and model-based estimates of $\chi_{i,j}(u)$ for $u \in \{0.8, 0.85, 0.9\}$ (top to bottom), and $i, j \in V$ but $i > j$. Model-based estimates use the EHM (left) and the three-step SCMEVM with graphical covariance (centre left), with structure given in Figure 1 (left panel) of the main text, the three-step SCMEVM with saturated covariance (centre right) and graphical covariance (right) with structure given in Figure 6 (right panel) of the main text. Black dashed lines show $y = x$. Circles (triangles) show flow-connected (flow-unconnected). The colour shows the standard error of the model-based estimates.}
    \label{fig:Danube_Chi_Comp}
\end{figure}

\begin{figure}[t!]
    \centering
    \includegraphics[width = \textwidth]{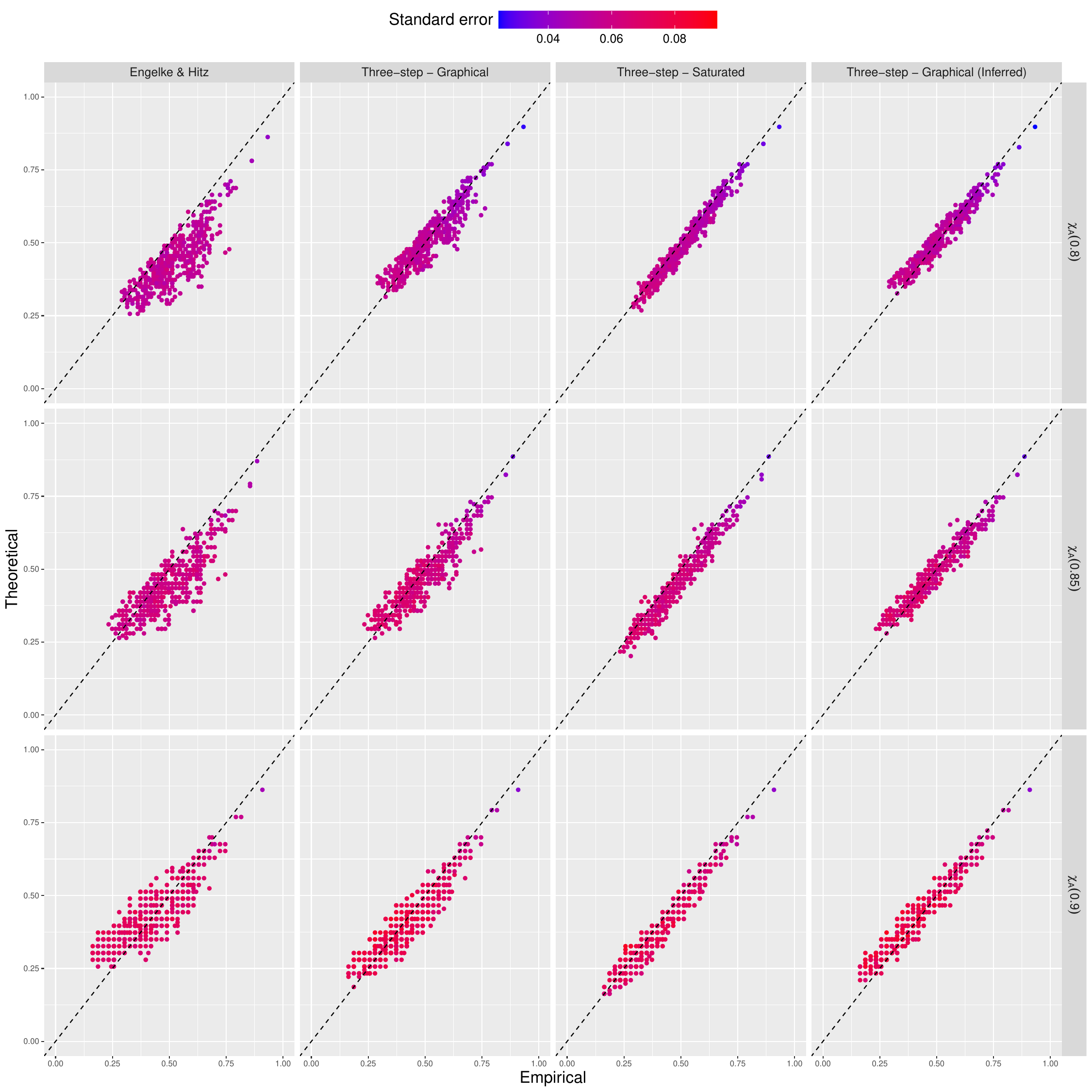}
    \caption{Empirical and model-based estimates of $\chi_{A}(u)$ for $u \in \{0.8, 0.85, 0.9\}$ (top to bottom) for 500 randomly selected triplets of $A \subset V$. Model-based estimates use the EHM (left) and the three-step SCMEVM with graphical covariance (centre left), both with structure given in Figure 1 (left panel) of the main text, the three-step SCMEVM with saturated covariance (centre right) and graphical covariance (right) with structure given in Figure 6 (right panel) of the main text. Black dashed lines show $y = x$. The colour shows the standard error of the model-based estimates.}
    \label{fig:Danube_Chi_Multi_Comp}
\end{figure}

For each fitted model and bootstrapped dataset, we obtain a single simulation, which is used for prediction. Empirical and model-based estimated for $\chi_{i,j}(u)$ 
are obtained for $i,j \in V$, $i > j$, and $u \in \{0.8, 0.85, 0.9\}$, where $V = \{1, \hdots, 31\}$. The point estimates in Figure \ref{fig:Danube_Chi_Comp} 
are the median estimates over the two sets of estimates for $\chi_{i,j}(u)$. 
As in Figure 7 of the main text, the SCMEVMs better capture the extremal dependence structure than the EHM. 
Figure \ref{fig:Danube_Chi_Multi_Comp} shows a similar comparison but for $\chi_{A}(u)$ where $A \subset V$ are 500 randomly sampled triplets, and $u \in \{0.8, 0.85, 0.9\}$. Again, the SCMEVMs better capture the extremal dependence in the upper Danube River basin. Although for higher thresholds the EHM is less biased for moderately dependent triplets, the bias for low dependent triplets increases as the threshold increases. Similar conclusions can be made for (iv), however, the magnitude of the bias is much smaller. Model (iv) appears to perform better than (ii) and rectifies the systemic underestimation in (iii). Thus, (iv) appears to have the best predictive performance across multiple components.

\subsection{Comparison with EGlearn}

\begin{figure}[t!]
    \centering
    \includegraphics[width=0.3\linewidth]{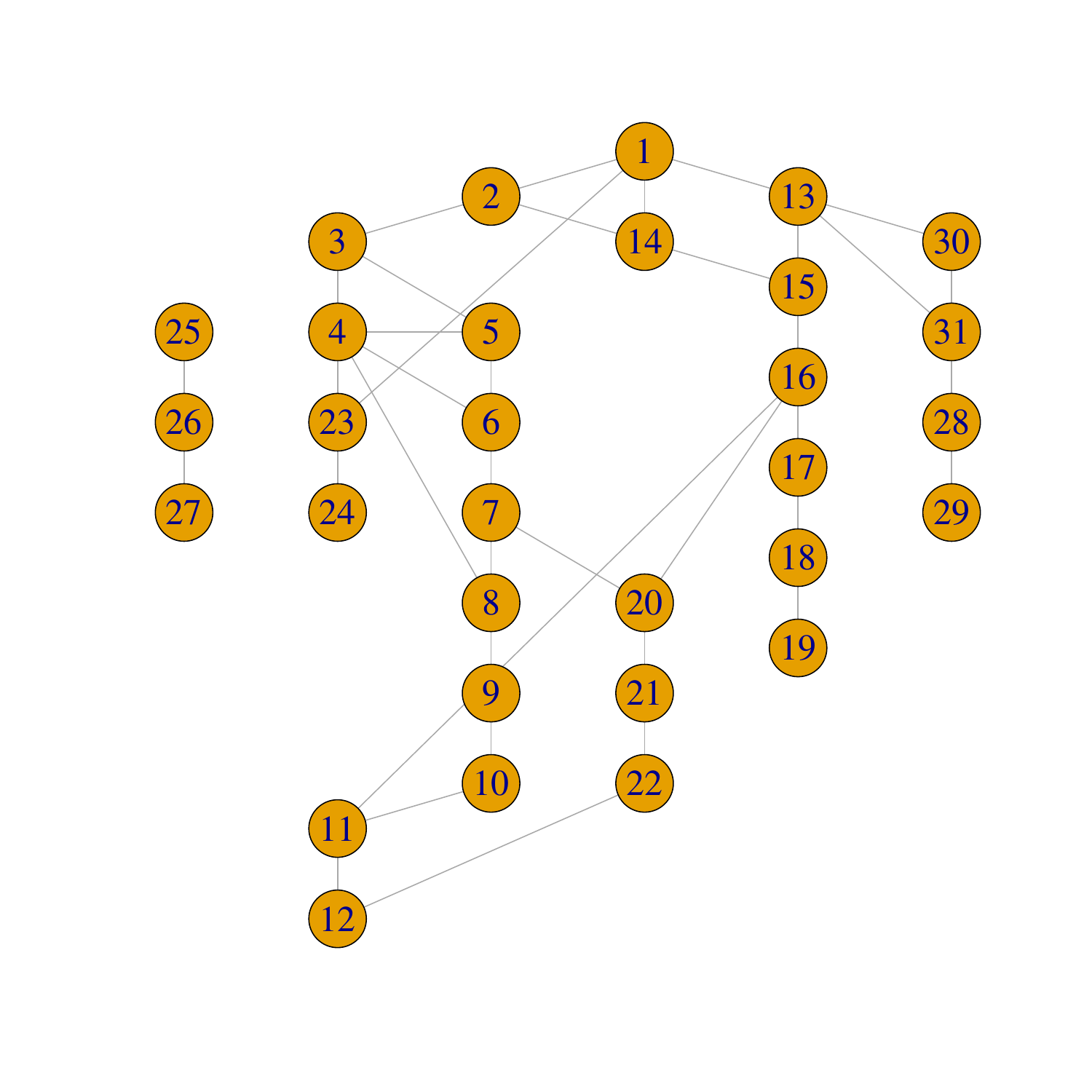} \quad
    \includegraphics[width=0.3\linewidth]{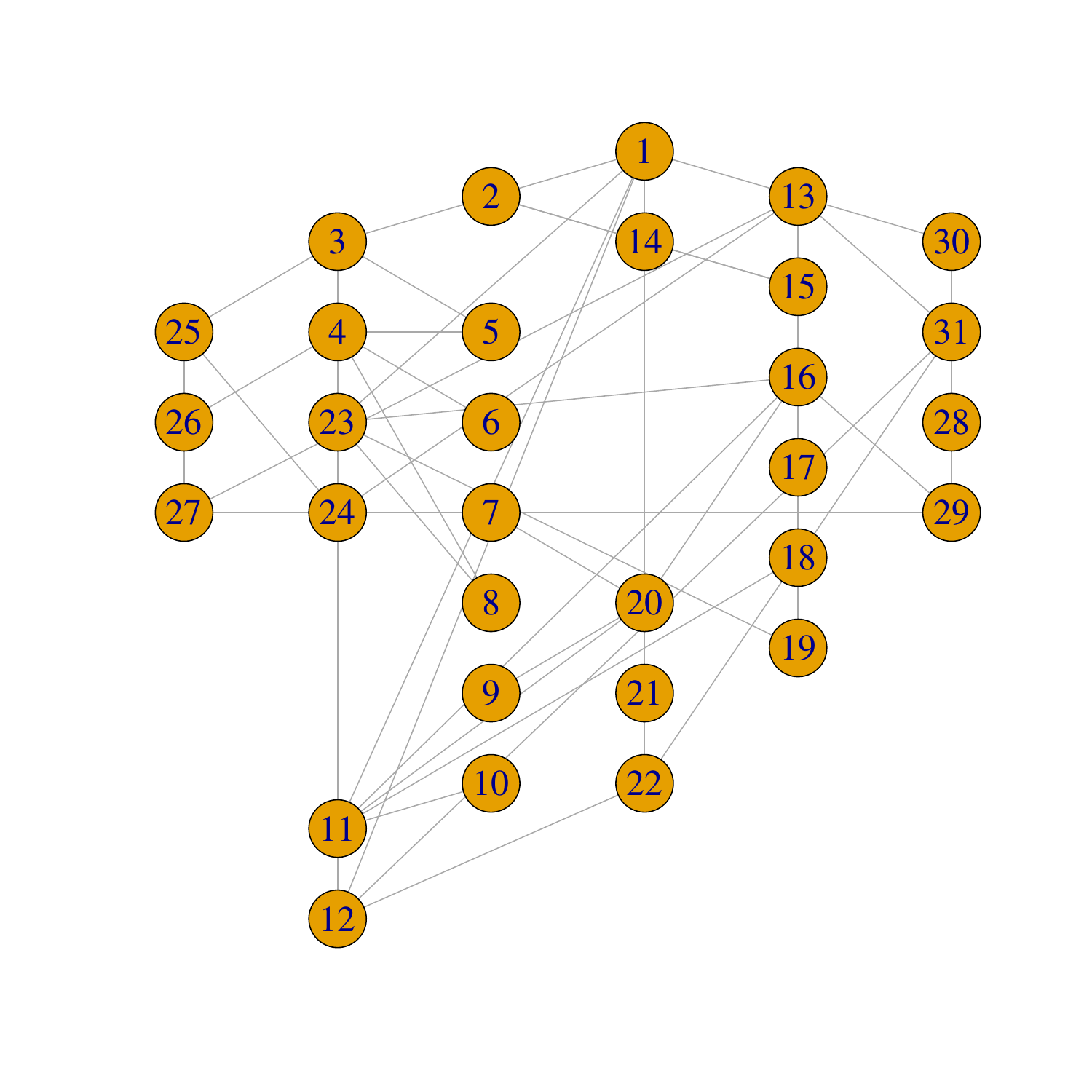} \quad
    \includegraphics[width=0.3\linewidth]{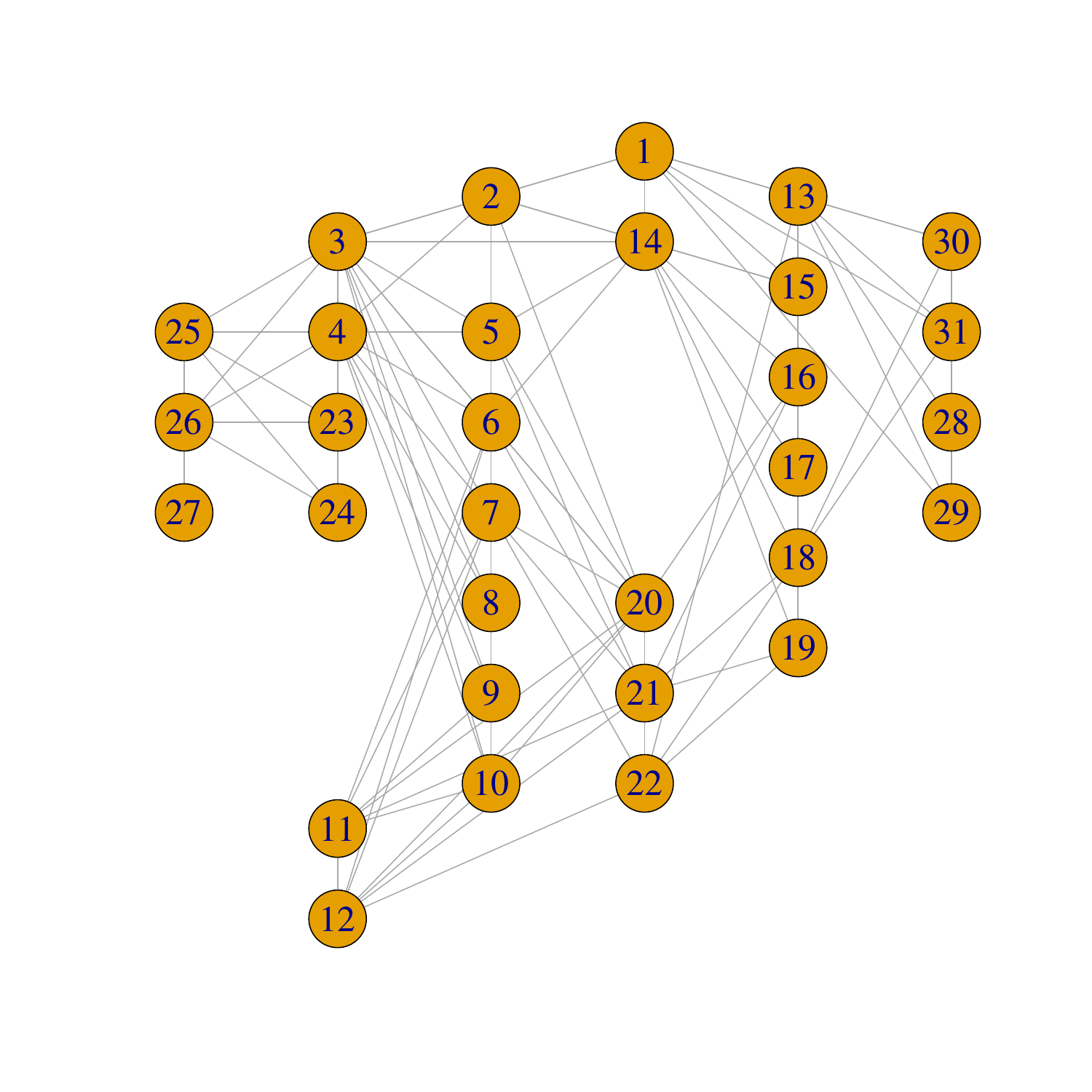}
    \caption{Inferred graphical structure of the upper Danube River basin using EGlearn, where the model selection criterion is MBIC (left) and AIC (centre), and the method proposed in the main text (right).}
    \label{fig:Danube_Graphs_Comparison}
\end{figure}

In their seminal paper, \citetsupp{Engelke_2020_supp} focus on learning block graphical structures. Since then, the literature on learning the graphical structure for data that is assumed to follow a H\"{u}sler-Reiss distribution has exploded; see \citetsupp{Engelke_2024_B_supp} for a thorough review. One such method is EGlearn \citepsupp{Engelke_2025_supp}. Figure \ref{fig:Danube_Graphs_Comparison} compares the inferred graphical structures for the upper Danube River basin using EGlearn with model selection criterion MBIC (left) and AIC (centre), and the method proposed in the main text (right).

Using EGlean with MBIC results in a very sparse graph. The inferred graph has 42 edges compared to 30 edges in the graph induced by the flow connections in the river structure. However, the inferred graph is disconnected. This is likely because stations 23-27 are asymptotically independent due to their latitude (north of the main Danube River) \citepsupp{Engelke_2025_supp}. Thus, \citetsupp{Engelke_2025_supp} remove these stations from their analysis as they do not satisfy the assumption underlying the model. Since we wish to make a like-for-like comparison between EGlearn and our method when we predict from the model, we therefore use the inferred graph using AIC. This graph is much denser (71 edges in total), with clear connections between the tributaries due to shared rainfall events that are not captured by the graph induced by the river flow. This compares to 127 edges in the graph using our method. While there are 56 more edges, the connected structures in the centre and right panel of Figure \ref{fig:Danube_Graphs_Comparison} are not dissimilar, as they both: capture the underlying structure of the river; connect geographically close but disconnected tributaries; and create additional connections along existing tributaries.

Using the same data and methods as in Section~\ref{subsec:Danube_River_Figures}, we fit the EHM to the AIC EGlearn-inferred graphical structure. For comparison, we fit the SCMEVM with the same graphical structure. Furthermore, we fit the SCMEVM to the inferred graphical structure obtained with our method. 

\begin{figure}[t!]
    \centering
    \includegraphics[width = \textwidth]{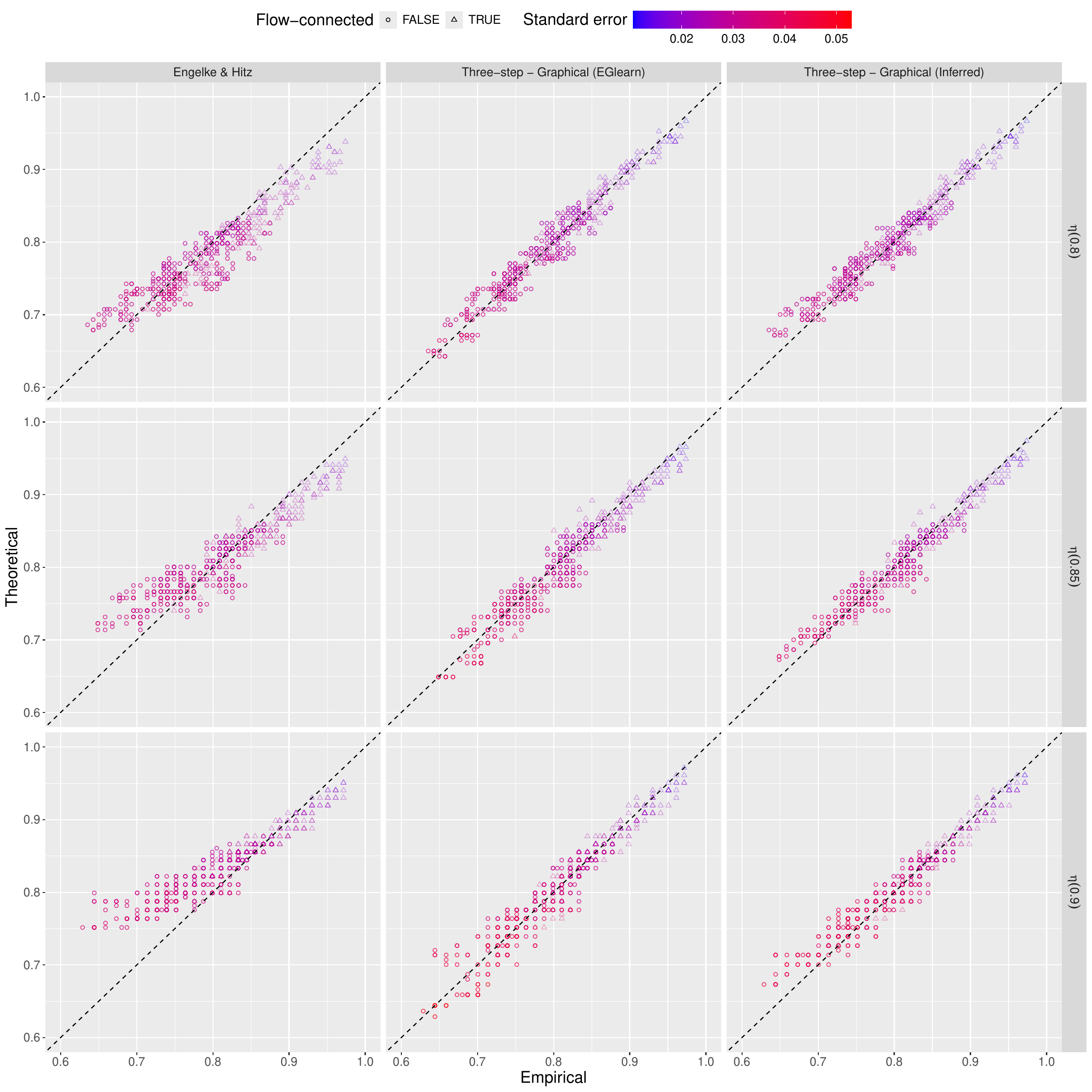}
    \caption{Empirical and model-based estimates of $\eta_{i,j}(u)$ for $u \in \{0.8, 0.85, 0.9\}$ (top to bottom), and $i, j \in V$, but $i > j$. Model-based estimates use the EHM (left) and the three-step SCMEVM with graphical covariance (centre), both with structure given in Figure \ref{fig:Danube_Graphs_Comparison} (centre panel), and the three-step SCMEVM graphical covariance (right), with structure given in Figure \ref{fig:Danube_Graphs_Comparison} (right panel). Black dashed lines show $y = x$. Circles (triangles) show flow-connected (flow-unconnected). The colour shows the standard error of the model-based estimates.}
    \label{fig:Danube_ETA_Comp}
\end{figure}

For each fitted model and for each bootstrapped dataset, we obtain a single simulation, which is used for prediction. Empirical and model-based estimated for $\eta_{i,j}(u)$ are obtained for $i,j \in V$, $i > j$, and $u \in \{0.8, 0.85, 0.9\}$, where $V = \{1, \hdots, 31\}$. The point estimates in Figure \ref{fig:Danube_ETA_Comp} are the median estimates over the two sets of estimates for $\eta_{i,j}(u)$. The left panels show predictions from the EHM are improved compared to using the graph induced by the flow connections, with closer alignment to the $y = x$ line for $u < 0.9$ and less variability across all values of $u$. However, there are deviations in the left tail where the dependence is weaker and closer to complete independence, particularly as $u$ tends towards 1. The majority of deviations can be attributed to when one or more of the sites north of the main Danube River (23-27) is included in the pair, supporting the claim that these stations tend to exhibit AI with other stations \citepsupp{Engelke_2025_supp}. Since the EHM assumes complete AD, it is unsurprising that the model overestimates in these cases. The deviation is resolved by using the three-step SCMEVM (centre panels). In addition, using the SCMEVM reduces the bias for pairs of sites with strong positive dependence. For a like-for-like comparison of the methods, we compare the left and right panels, where the EHM is fit using the EGlearn-inferred graph and the three-step SCMEVM is fit using the SCMEVM-inferred graph. The SCMEVM exhibits lower variability, has closer agreement to the $y = x$ line, and resolves the overestimation of the left-tail in the EHM, highlighting the need for a flexible model that can capture all extremal dependence classes. Interestingly, the SCMEVM performs better in the left tail with the EGlearn-inferred graph than the SCMEVM-inferred graph, suggesting the proposed method for learning the graph may require refinement.

\bibliographystylesupp{apalike}
\bibliographysupp{Supplementary/Supplementary_Library.bib}

\end{document}